\documentclass[a4paper, traditabstract, longauth]{aa} 
\usepackage{graphicx}
\usepackage{color}
\usepackage{natbib}
\usepackage{longtable,lscape}
\usepackage{txfonts}
\usepackage{amssymb}
\usepackage{mathrsfs}
\usepackage{amsfonts}
\usepackage{newlfont}
\usepackage{textcomp}
\usepackage{times}
\usepackage{units}
\usepackage{url}
\usepackage[breaklinks, colorlinks, citecolor=blue]{hyperref}
\usepackage{booktabs}
\usepackage{subfig}
\usepackage{sidecap}

\usepackage{fixltx2e}

\def\setsymbol#1#2{\expandafter\def\csname #1\endcsname{#2}}
\def\getsymbol#1{\csname #1\endcsname}

\def\Planck{\textit{Planck}}



\def\all2013resultspapers{\nocite{planck2013-p01, planck2013-p02, planck2013-p02a, planck2013-p02d, planck2013-p02b, planck2013-p03, planck2013-p03c, planck2013-p03f, planck2013-p03d, planck2013-p03e, planck2013-p01a, planck2013-p06, planck2013-p03a, planck2013-pip88, planck2013-p08, planck2013-p11, planck2013-p12, planck2013-p13, planck2013-p14, planck2013-p15, planck2013-p05b, planck2013-p17, planck2013-p09, planck2013-p09a, planck2013-p20, planck2013-p19, planck2013-pipaberration, planck2013-p05, planck2013-p05a, planck2013-pip56, planck2013-p06b}}

\newbox\tablebox    \newdimen\tablewidth
\def\leaderfil{\leaders\hbox to 5pt{\hss.\hss}\hfil}
%
%
\def\endPlancktable{\tablewidth=\columnwidth 
    $$\hss\copy\tablebox\hss$$
    \vskip-\lastskip\vskip -2pt}
\def\endPlancktablewide{\tablewidth=\textwidth 
    $$\hss\copy\tablebox\hss$$
    \vskip-\lastskip\vskip -2pt}
\def\tablenote#1 #2\par{\begingroup \parindent=0.8em
    \abovedisplayshortskip=0pt\belowdisplayshortskip=0pt
    \noindent
    $$\hss\vbox{\hsize\tablewidth \hangindent=\parindent \hangafter=1 \noindent
    \hbox to \parindent{$^#1$\hss}\strut#2\strut\par}\hss$$
    \endgroup}
\def\doubleline{\vskip 3pt\hrule \vskip 1.5pt \hrule \vskip 5pt}

%
\def\L2{\ifmmode L_2\else $L_2$\fi}

\def\DeltaT{\ifmmode \Delta T\else $\Delta T$\fi}
\def\deltat{\ifmmode \Delta t\else $\Delta t$\fi}
\def\fknee{\ifmmode f_{\rm knee}\else $f_{\rm knee}$\fi}
\def\Fmax{\ifmmode F_{\rm max}\else $F_{\rm max}$\fi}
\def\solar{\ifmmode{\rm M}_{\mathord\odot}\else${\rm M}_{\mathord\odot}$\fi}
\def\Msolar{\ifmmode{\rm M}_{\mathord\odot}\else${\rm M}_{\mathord\odot}$\fi}
\def\Lsolar{\ifmmode{\rm L}_{\mathord\odot}\else${\rm L}_{\mathord\odot}$\fi}

\def\inv{\ifmmode^{-1}\else$^{-1}$\fi}
\def\mo{\ifmmode^{-1}\else$^{-1}$\fi}
\def\sup#1{\ifmmode ^{\rm #1}\else $^{\rm #1}$\fi}
\def\expo#1{\ifmmode \times 10^{#1}\else $\times 10^{#1}$\fi}
\def\,{\thinspace}
\def\lsim{\mathrel{\raise .4ex\hbox{\rlap{$<$}\lower 1.2ex\hbox{$\sim$}}}}
\def\gsim{\mathrel{\raise .4ex\hbox{\rlap{$>$}\lower 1.2ex\hbox{$\sim$}}}}

\def\simprop{\mathrel{\raise .4ex\hbox{\rlap{$\propto$}\lower 1.2ex\hbox{$\sim$}}}}
\def\deg{\ifmmode^\circ\else$^\circ$\fi}
\def\pdeg{\ifmmode $\setbox0=\hbox{$^{\circ}$}\rlap{\hskip.11\wd0 .}$^{\circ}
          \else \setbox0=\hbox{$^{\circ}$}\rlap{\hskip.11\wd0 .}$^{\circ}$\fi}
\def\arcs{\ifmmode {^{\scriptstyle\prime\prime}}
          \else $^{\scriptstyle\prime\prime}$\fi}
\def\arcm{\ifmmode {^{\scriptstyle\prime}}
          \else $^{\scriptstyle\prime}$\fi}
\newdimen\sa  \newdimen\sb
\def\parcs{\sa=.07em \sb=.03em
     \ifmmode \hbox{\rlap{.}}^{\scriptstyle\prime\kern -\sb\prime}\hbox{\kern -\sa}
     \else \rlap{.}$^{\scriptstyle\prime\kern -\sb\prime}$\kern -\sa\fi}
\def\parcm{\sa=.08em \sb=.03em
     \ifmmode \hbox{\rlap{.}\kern\sa}^{\scriptstyle\prime}\hbox{\kern-\sb}
     \else \rlap{.}\kern\sa$^{\scriptstyle\prime}$\kern-\sb\fi}
\def\ra[#1 #2 #3.#4]{#1\sup{h}#2\sup{m}#3\sup{s}\llap.#4}
\def\dec[#1 #2 #3.#4]{#1\deg#2\arcm#3\arcs\llap.#4}
\def\deco[#1 #2 #3]{#1\deg#2\arcm#3\arcs}
\def\rra[#1 #2]{#1\sup{h}#2\sup{m}}

\def\dots{\relax\ifmmode \ldots\else $\ldots$\fi}
%
%
\def\WHzsr{\ifmmode $W\,Hz\mo\,sr\mo$\else W\,Hz\mo\,sr\mo\fi}
\def\mHz{\ifmmode $\,mHz$\else \,mHz\fi}
\def\GHz{\ifmmode $\,GHz$\else \,GHz\fi}
\def\mKs{\ifmmode $\,mK\,s$^{1/2}\else \,mK\,s$^{1/2}$\fi}
\def\muKs{\ifmmode \,\mu$K\,s$^{1/2}\else \,$\mu$K\,s$^{1/2}$\fi}
\def\muKRJs{\ifmmode \,\mu$K$_{\rm RJ}$\,s$^{1/2}\else \,$\mu$K$_{\rm RJ}$\,s$^{1/2}$\fi}
\def\muKHz{\ifmmode \,\mu$K\,Hz$^{-1/2}\else \,$\mu$K\,Hz$^{-1/2}$\fi}
\def\MJysr{\ifmmode \,$MJy\,sr\mo$\else \,MJy\,sr\mo\fi}
\def\MJysrmK{\ifmmode \,$MJy\,sr\mo$\,mK$_{\rm CMB}\mo\else \,MJy\,sr\mo\,mK$_{\rm CMB}\mo$\fi}
\def\microns{\ifmmode \,\mu$m$\else \,$\mu$m\fi}

\def\muK{\ifmmode \,\mu$K$\else \,$\mu$\hbox{K}\fi}
\def\microK{\ifmmode \,\mu$K$\else \,$\mu$\hbox{K}\fi}
\def\muW{\ifmmode \,\mu$W$\else \,$\mu$\hbox{W}\fi}
\def\kms{\ifmmode $\,km\,s$^{-1}\else \,km\,s$^{-1}$\fi}
\def\kmsMpc{\ifmmode $\,\kms\,Mpc\mo$\else \,\kms\,Mpc\mo\fi}
%
%


\setsymbol{LFI:center:frequency:70GHz:units}{70.3\,GHz}
\setsymbol{LFI:center:frequency:44GHz:units}{44.1\,GHz}
\setsymbol{LFI:center:frequency:30GHz:units}{28.5\,GHz}

\setsymbol{LFI:center:frequency:70GHz}{70.3}
\setsymbol{LFI:center:frequency:44GHz}{44.1}
\setsymbol{LFI:center:frequency:30GHz}{28.5}

\setsymbol{LFI:center:frequency:LFI18:Rad:M:units}{71.7\GHz}
\setsymbol{LFI:center:frequency:LFI19:Rad:M:units}{67.5\GHz}
\setsymbol{LFI:center:frequency:LFI20:Rad:M:units}{69.2\GHz}
\setsymbol{LFI:center:frequency:LFI21:Rad:M:units}{70.4\GHz}
\setsymbol{LFI:center:frequency:LFI22:Rad:M:units}{71.5\GHz}
\setsymbol{LFI:center:frequency:LFI23:Rad:M:units}{70.8\GHz}
\setsymbol{LFI:center:frequency:LFI24:Rad:M:units}{44.4\GHz}
\setsymbol{LFI:center:frequency:LFI25:Rad:M:units}{44.0\GHz}
\setsymbol{LFI:center:frequency:LFI26:Rad:M:units}{43.9\GHz}
\setsymbol{LFI:center:frequency:LFI27:Rad:M:units}{28.3\GHz}
\setsymbol{LFI:center:frequency:LFI28:Rad:M:units}{28.8\GHz}
\setsymbol{LFI:center:frequency:LFI18:Rad:S:units}{70.1\GHz}
\setsymbol{LFI:center:frequency:LFI19:Rad:S:units}{69.6\GHz}
\setsymbol{LFI:center:frequency:LFI20:Rad:S:units}{69.5\GHz}
\setsymbol{LFI:center:frequency:LFI21:Rad:S:units}{69.5\GHz}
\setsymbol{LFI:center:frequency:LFI22:Rad:S:units}{72.8\GHz}
\setsymbol{LFI:center:frequency:LFI23:Rad:S:units}{71.3\GHz}
\setsymbol{LFI:center:frequency:LFI24:Rad:S:units}{44.1\GHz}
\setsymbol{LFI:center:frequency:LFI25:Rad:S:units}{44.1\GHz}
\setsymbol{LFI:center:frequency:LFI26:Rad:S:units}{44.1\GHz}
\setsymbol{LFI:center:frequency:LFI27:Rad:S:units}{28.5\GHz}
\setsymbol{LFI:center:frequency:LFI28:Rad:S:units}{28.2\GHz}

\setsymbol{LFI:center:frequency:LFI18:Rad:M}{71.7}
\setsymbol{LFI:center:frequency:LFI19:Rad:M}{67.5}
\setsymbol{LFI:center:frequency:LFI20:Rad:M}{69.2}
\setsymbol{LFI:center:frequency:LFI21:Rad:M}{70.4}
\setsymbol{LFI:center:frequency:LFI22:Rad:M}{71.5}
\setsymbol{LFI:center:frequency:LFI23:Rad:M}{70.8}
\setsymbol{LFI:center:frequency:LFI24:Rad:M}{44.4}
\setsymbol{LFI:center:frequency:LFI25:Rad:M}{44.0}
\setsymbol{LFI:center:frequency:LFI26:Rad:M}{43.9}
\setsymbol{LFI:center:frequency:LFI27:Rad:M}{28.3}
\setsymbol{LFI:center:frequency:LFI28:Rad:M}{28.8}
\setsymbol{LFI:center:frequency:LFI18:Rad:S}{70.1}
\setsymbol{LFI:center:frequency:LFI19:Rad:S}{69.6}
\setsymbol{LFI:center:frequency:LFI20:Rad:S}{69.5}
\setsymbol{LFI:center:frequency:LFI21:Rad:S}{69.5}
\setsymbol{LFI:center:frequency:LFI22:Rad:S}{72.8}
\setsymbol{LFI:center:frequency:LFI23:Rad:S}{71.3}
\setsymbol{LFI:center:frequency:LFI24:Rad:S}{44.1}
\setsymbol{LFI:center:frequency:LFI25:Rad:S}{44.1}
\setsymbol{LFI:center:frequency:LFI26:Rad:S}{44.1}
\setsymbol{LFI:center:frequency:LFI27:Rad:S}{28.5}
\setsymbol{LFI:center:frequency:LFI28:Rad:S}{28.2}


\setsymbol{LFI:white:noise:sensitivity:70GHz:units}{134.7\muKs}
\setsymbol{LFI:white:noise:sensitivity:44GHz:units}{164.7\muKs}
\setsymbol{LFI:white:noise:sensitivity:30GHz:units}{143.4\muKs}

\setsymbol{LFI:white:noise:sensitivity:70GHz}{134.7}
\setsymbol{LFI:white:noise:sensitivity:44GHz}{164.7}
\setsymbol{LFI:white:noise:sensitivity:30GHz}{143.4}


\setsymbol{LFI:white:noise:sensitivity:LFI18:Rad:M:units}{512.0\muKs}
\setsymbol{LFI:white:noise:sensitivity:LFI19:Rad:M:units}{581.4\muKs}
\setsymbol{LFI:white:noise:sensitivity:LFI20:Rad:M:units}{590.8\muKs}
\setsymbol{LFI:white:noise:sensitivity:LFI21:Rad:M:units}{455.2\muKs}
\setsymbol{LFI:white:noise:sensitivity:LFI22:Rad:M:units}{492.0\muKs}
\setsymbol{LFI:white:noise:sensitivity:LFI23:Rad:M:units}{507.7\muKs}
\setsymbol{LFI:white:noise:sensitivity:LFI24:Rad:M:units}{462.2\muKs}
\setsymbol{LFI:white:noise:sensitivity:LFI25:Rad:M:units}{413.6\muKs}
\setsymbol{LFI:white:noise:sensitivity:LFI26:Rad:M:units}{478.6\muKs}
\setsymbol{LFI:white:noise:sensitivity:LFI27:Rad:M:units}{277.7\muKs}
\setsymbol{LFI:white:noise:sensitivity:LFI28:Rad:M:units}{312.3\muKs}
\setsymbol{LFI:white:noise:sensitivity:LFI18:Rad:S:units}{465.7\muKs}
\setsymbol{LFI:white:noise:sensitivity:LFI19:Rad:S:units}{555.6\muKs}
\setsymbol{LFI:white:noise:sensitivity:LFI20:Rad:S:units}{623.2\muKs}
\setsymbol{LFI:white:noise:sensitivity:LFI21:Rad:S:units}{564.1\muKs}
\setsymbol{LFI:white:noise:sensitivity:LFI22:Rad:S:units}{534.4\muKs}
\setsymbol{LFI:white:noise:sensitivity:LFI23:Rad:S:units}{542.4\muKs}
\setsymbol{LFI:white:noise:sensitivity:LFI24:Rad:S:units}{399.2\muKs}
\setsymbol{LFI:white:noise:sensitivity:LFI25:Rad:S:units}{392.6\muKs}
\setsymbol{LFI:white:noise:sensitivity:LFI26:Rad:S:units}{418.6\muKs}
\setsymbol{LFI:white:noise:sensitivity:LFI27:Rad:S:units}{302.9\muKs}
\setsymbol{LFI:white:noise:sensitivity:LFI28:Rad:S:units}{285.3\muKs}

\setsymbol{LFI:white:noise:sensitivity:LFI18:Rad:M}{512.0}
\setsymbol{LFI:white:noise:sensitivity:LFI19:Rad:M}{581.4}
\setsymbol{LFI:white:noise:sensitivity:LFI20:Rad:M}{590.8}
\setsymbol{LFI:white:noise:sensitivity:LFI21:Rad:M}{455.2}
\setsymbol{LFI:white:noise:sensitivity:LFI22:Rad:M}{492.0}
\setsymbol{LFI:white:noise:sensitivity:LFI23:Rad:M}{507.7}
\setsymbol{LFI:white:noise:sensitivity:LFI24:Rad:M}{462.2}
\setsymbol{LFI:white:noise:sensitivity:LFI25:Rad:M}{413.6}
\setsymbol{LFI:white:noise:sensitivity:LFI26:Rad:M}{478.6}
\setsymbol{LFI:white:noise:sensitivity:LFI27:Rad:M}{277.7}
\setsymbol{LFI:white:noise:sensitivity:LFI28:Rad:M}{312.3}
\setsymbol{LFI:white:noise:sensitivity:LFI18:Rad:S}{465.7}
\setsymbol{LFI:white:noise:sensitivity:LFI19:Rad:S}{555.6}
\setsymbol{LFI:white:noise:sensitivity:LFI20:Rad:S}{623.2}
\setsymbol{LFI:white:noise:sensitivity:LFI21:Rad:S}{564.1}
\setsymbol{LFI:white:noise:sensitivity:LFI22:Rad:S}{534.4}
\setsymbol{LFI:white:noise:sensitivity:LFI23:Rad:S}{542.4}
\setsymbol{LFI:white:noise:sensitivity:LFI24:Rad:S}{399.2}
\setsymbol{LFI:white:noise:sensitivity:LFI25:Rad:S}{392.6}
\setsymbol{LFI:white:noise:sensitivity:LFI26:Rad:S}{418.6}
\setsymbol{LFI:white:noise:sensitivity:LFI27:Rad:S}{302.9}
\setsymbol{LFI:white:noise:sensitivity:LFI28:Rad:S}{285.3}


\setsymbol{LFI:knee:frequency:70GHz:units}{29.5\mHz}
\setsymbol{LFI:knee:frequency:44GHz:units}{56.2\mHz}
\setsymbol{LFI:knee:frequency:30GHz:units}{113.7\mHz}

\setsymbol{LFI:knee:frequency:70GHz}{29.5}
\setsymbol{LFI:knee:frequency:44GHz}{56.2}
\setsymbol{LFI:knee:frequency:30GHz}{113.7}

\setsymbol{LFI:knee:frequency:LFI18:Rad:M:units}{16.3\mHz}
\setsymbol{LFI:knee:frequency:LFI19:Rad:M:units}{15.1\mHz}
\setsymbol{LFI:knee:frequency:LFI20:Rad:M:units}{18.7\mHz}
\setsymbol{LFI:knee:frequency:LFI21:Rad:M:units}{37.2\mHz}
\setsymbol{LFI:knee:frequency:LFI22:Rad:M:units}{12.7\mHz}
\setsymbol{LFI:knee:frequency:LFI23:Rad:M:units}{34.6\mHz}
\setsymbol{LFI:knee:frequency:LFI24:Rad:M:units}{46.2\mHz}
\setsymbol{LFI:knee:frequency:LFI25:Rad:M:units}{24.9\mHz}
\setsymbol{LFI:knee:frequency:LFI26:Rad:M:units}{67.6\mHz}
\setsymbol{LFI:knee:frequency:LFI27:Rad:M:units}{187.4\mHz}
\setsymbol{LFI:knee:frequency:LFI28:Rad:M:units}{122.2\mHz}
\setsymbol{LFI:knee:frequency:LFI18:Rad:S:units}{17.7\mHz}
\setsymbol{LFI:knee:frequency:LFI19:Rad:S:units}{22.0\mHz}
\setsymbol{LFI:knee:frequency:LFI20:Rad:S:units}{8.7\mHz}
\setsymbol{LFI:knee:frequency:LFI21:Rad:S:units}{25.9\mHz}
\setsymbol{LFI:knee:frequency:LFI22:Rad:S:units}{15.8\mHz}
\setsymbol{LFI:knee:frequency:LFI23:Rad:S:units}{129.8\mHz}
\setsymbol{LFI:knee:frequency:LFI24:Rad:S:units}{100.9\mHz}
\setsymbol{LFI:knee:frequency:LFI25:Rad:S:units}{38.9\mHz}
\setsymbol{LFI:knee:frequency:LFI26:Rad:S:units}{58.9\mHz}
\setsymbol{LFI:knee:frequency:LFI27:Rad:S:units}{104.4\mHz}
\setsymbol{LFI:knee:frequency:LFI28:Rad:S:units}{40.7\mHz}

\setsymbol{LFI:knee:frequency:LFI18:Rad:M}{16.3}
\setsymbol{LFI:knee:frequency:LFI19:Rad:M}{15.1}
\setsymbol{LFI:knee:frequency:LFI20:Rad:M}{18.7}
\setsymbol{LFI:knee:frequency:LFI21:Rad:M}{37.2}
\setsymbol{LFI:knee:frequency:LFI22:Rad:M}{12.7}
\setsymbol{LFI:knee:frequency:LFI23:Rad:M}{34.6}
\setsymbol{LFI:knee:frequency:LFI24:Rad:M}{46.2}
\setsymbol{LFI:knee:frequency:LFI25:Rad:M}{24.9}
\setsymbol{LFI:knee:frequency:LFI26:Rad:M}{67.6}
\setsymbol{LFI:knee:frequency:LFI27:Rad:M}{187.4}
\setsymbol{LFI:knee:frequency:LFI28:Rad:M}{122.2}
\setsymbol{LFI:knee:frequency:LFI18:Rad:S}{17.7}
\setsymbol{LFI:knee:frequency:LFI19:Rad:S}{22.0}
\setsymbol{LFI:knee:frequency:LFI20:Rad:S}{8.7}
\setsymbol{LFI:knee:frequency:LFI21:Rad:S}{25.9}
\setsymbol{LFI:knee:frequency:LFI22:Rad:S}{15.8}
\setsymbol{LFI:knee:frequency:LFI23:Rad:S}{129.8}
\setsymbol{LFI:knee:frequency:LFI24:Rad:S}{100.9}
\setsymbol{LFI:knee:frequency:LFI25:Rad:S}{38.9}
\setsymbol{LFI:knee:frequency:LFI26:Rad:S}{58.9}
\setsymbol{LFI:knee:frequency:LFI27:Rad:S}{104.4}
\setsymbol{LFI:knee:frequency:LFI28:Rad:S}{40.7}


\setsymbol{LFI:slope:70GHz:units}{$-1.03$\mHz}
\setsymbol{LFI:slope:44GHz:units}{$-0.89$\mHz}
\setsymbol{LFI:slope:30GHz:units}{$-0.87$\mHz}

\setsymbol{LFI:slope:70GHz}{$-1.03$}
\setsymbol{LFI:slope:44GHz}{$-0.89$}
\setsymbol{LFI:slope:30GHz}{$-0.87$}

\setsymbol{LFI:slope:LFI18:Rad:M:units}{$-1.04$\mHz}
\setsymbol{LFI:slope:LFI19:Rad:M:units}{$-1.09$\mHz}
\setsymbol{LFI:slope:LFI20:Rad:M:units}{$-0.69$\mHz}
\setsymbol{LFI:slope:LFI21:Rad:M:units}{$-1.56$\mHz}
\setsymbol{LFI:slope:LFI22:Rad:M:units}{$-1.01$\mHz}
\setsymbol{LFI:slope:LFI23:Rad:M:units}{$-0.96$\mHz}
\setsymbol{LFI:slope:LFI24:Rad:M:units}{$-0.83$\mHz}
\setsymbol{LFI:slope:LFI25:Rad:M:units}{$-0.91$\mHz}
\setsymbol{LFI:slope:LFI26:Rad:M:units}{$-0.95$\mHz}
\setsymbol{LFI:slope:LFI27:Rad:M:units}{$-0.87$\mHz}
\setsymbol{LFI:slope:LFI28:Rad:M:units}{$-0.88$\mHz}
\setsymbol{LFI:slope:LFI18:Rad:S:units}{$-1.15$\mHz}
\setsymbol{LFI:slope:LFI19:Rad:S:units}{$-1.00$\mHz}
\setsymbol{LFI:slope:LFI20:Rad:S:units}{$-0.95$\mHz}
\setsymbol{LFI:slope:LFI21:Rad:S:units}{$-0.92$\mHz}
\setsymbol{LFI:slope:LFI22:Rad:S:units}{$-1.01$\mHz}
\setsymbol{LFI:slope:LFI23:Rad:S:units}{$-0.95$\mHz}
\setsymbol{LFI:slope:LFI24:Rad:S:units}{$-0.73$\mHz}
\setsymbol{LFI:slope:LFI25:Rad:S:units}{$-1.16$\mHz}
\setsymbol{LFI:slope:LFI26:Rad:S:units}{$-0.79$\mHz}
\setsymbol{LFI:slope:LFI27:Rad:S:units}{$-0.82$\mHz}
\setsymbol{LFI:slope:LFI28:Rad:S:units}{$-0.91$\mHz}

\setsymbol{LFI:slope:LFI18:Rad:M}{$-1.04$}
\setsymbol{LFI:slope:LFI19:Rad:M}{$-1.09$}
\setsymbol{LFI:slope:LFI20:Rad:M}{$-0.69$}
\setsymbol{LFI:slope:LFI21:Rad:M}{$-1.56$}
\setsymbol{LFI:slope:LFI22:Rad:M}{$-1.01$}
\setsymbol{LFI:slope:LFI23:Rad:M}{$-0.96$}
\setsymbol{LFI:slope:LFI24:Rad:M}{$-0.83$}
\setsymbol{LFI:slope:LFI25:Rad:M}{$-0.91$}
\setsymbol{LFI:slope:LFI26:Rad:M}{$-0.95$}
\setsymbol{LFI:slope:LFI27:Rad:M}{$-0.87$}
\setsymbol{LFI:slope:LFI28:Rad:M}{$-0.88$}
\setsymbol{LFI:slope:LFI18:Rad:S}{$-1.15$}
\setsymbol{LFI:slope:LFI19:Rad:S}{$-1.00$}
\setsymbol{LFI:slope:LFI20:Rad:S}{$-0.95$}
\setsymbol{LFI:slope:LFI21:Rad:S}{$-0.92$}
\setsymbol{LFI:slope:LFI22:Rad:S}{$-1.01$}
\setsymbol{LFI:slope:LFI23:Rad:S}{$-0.95$}
\setsymbol{LFI:slope:LFI24:Rad:S}{$-0.73$}
\setsymbol{LFI:slope:LFI25:Rad:S}{$-1.16$}
\setsymbol{LFI:slope:LFI26:Rad:S}{$-0.79$}
\setsymbol{LFI:slope:LFI27:Rad:S}{$-0.82$}
\setsymbol{LFI:slope:LFI28:Rad:S}{$-0.91$}


\setsymbol{LFI:FWHM:70GHz:units}{13\parcm01}
\setsymbol{LFI:FWHM:44GHz:units}{27\parcm92}
\setsymbol{LFI:FWHM:30GHz:units}{32\parcm65}

\setsymbol{LFI:FWHM:70GHz}{13.01}
\setsymbol{LFI:FWHM:44GHz}{27.92}
\setsymbol{LFI:FWHM:30GHz}{32.65}

\setsymbol{LFI:FWHM:LFI18:units}{13\parcm39}
\setsymbol{LFI:FWHM:LFI19:units}{13\parcm01}
\setsymbol{LFI:FWHM:LFI20:units}{12\parcm75}
\setsymbol{LFI:FWHM:LFI21:units}{12\parcm74}
\setsymbol{LFI:FWHM:LFI22:units}{12\parcm87}
\setsymbol{LFI:FWHM:LFI23:units}{13\parcm27}
\setsymbol{LFI:FWHM:LFI24:units}{22\parcm98}
\setsymbol{LFI:FWHM:LFI25:units}{30\parcm46}
\setsymbol{LFI:FWHM:LFI26:units}{30\parcm31}
\setsymbol{LFI:FWHM:LFI27:units}{32\parcm65}
\setsymbol{LFI:FWHM:LFI28:units}{32\parcm66}

\setsymbol{LFI:FWHM:LFI18}{13.39}
\setsymbol{LFI:FWHM:LFI19}{13.01}
\setsymbol{LFI:FWHM:LFI20}{12.75}
\setsymbol{LFI:FWHM:LFI21}{12.74}
\setsymbol{LFI:FWHM:LFI22}{12.87}
\setsymbol{LFI:FWHM:LFI23}{13.27}
\setsymbol{LFI:FWHM:LFI24}{22.98}
\setsymbol{LFI:FWHM:LFI25}{30.46}
\setsymbol{LFI:FWHM:LFI26}{30.31}
\setsymbol{LFI:FWHM:LFI27}{32.65}
\setsymbol{LFI:FWHM:LFI28}{32.66}



\setsymbol{LFI:FWHM:uncertainty:LFI18:units}{0.170\arcm}
\setsymbol{LFI:FWHM:uncertainty:LFI19:units}{0.174\arcm}
\setsymbol{LFI:FWHM:uncertainty:LFI20:units}{0.170\arcm}
\setsymbol{LFI:FWHM:uncertainty:LFI21:units}{0.156\arcm}
\setsymbol{LFI:FWHM:uncertainty:LFI22:units}{0.164\arcm}
\setsymbol{LFI:FWHM:uncertainty:LFI23:units}{0.171\arcm}
\setsymbol{LFI:FWHM:uncertainty:LFI24:units}{0.652\arcm}
\setsymbol{LFI:FWHM:uncertainty:LFI25:units}{1.075\arcm}
\setsymbol{LFI:FWHM:uncertainty:LFI26:units}{1.131\arcm}
\setsymbol{LFI:FWHM:uncertainty:LFI27:units}{1.266\arcm}
\setsymbol{LFI:FWHM:uncertainty:LFI28:units}{1.287\arcm}

\setsymbol{LFI:FWHM:uncertainty:LFI18}{0.170}
\setsymbol{LFI:FWHM:uncertainty:LFI19}{0.174}
\setsymbol{LFI:FWHM:uncertainty:LFI20}{0.170}
\setsymbol{LFI:FWHM:uncertainty:LFI21}{0.156}
\setsymbol{LFI:FWHM:uncertainty:LFI22}{0.164}
\setsymbol{LFI:FWHM:uncertainty:LFI23}{0.171}
\setsymbol{LFI:FWHM:uncertainty:LFI24}{0.652}
\setsymbol{LFI:FWHM:uncertainty:LFI25}{1.075}
\setsymbol{LFI:FWHM:uncertainty:LFI26}{1.131}
\setsymbol{LFI:FWHM:uncertainty:LFI27}{1.266}
\setsymbol{LFI:FWHM:uncertainty:LFI28}{1.287}


\setsymbol{HFI:center:frequency:100GHz:units}{100\,GHz}
\setsymbol{HFI:center:frequency:143GHz:units}{143\,GHz}
\setsymbol{HFI:center:frequency:217GHz:units}{217\,GHz}
\setsymbol{HFI:center:frequency:353GHz:units}{353\,GHz}
\setsymbol{HFI:center:frequency:545GHz:units}{545\,GHz}
\setsymbol{HFI:center:frequency:857GHz:units}{857\,GHz}

\setsymbol{HFI:center:frequency:100GHz}{100}
\setsymbol{HFI:center:frequency:143GHz}{143}
\setsymbol{HFI:center:frequency:217GHz}{217}
\setsymbol{HFI:center:frequency:353GHz}{353}
\setsymbol{HFI:center:frequency:545GHz}{545}
\setsymbol{HFI:center:frequency:857GHz}{857}


\setsymbol{HFI:Ndetectors:100GHz}{8}
\setsymbol{HFI:Ndetectors:143GHz}{11}
\setsymbol{HFI:Ndetectors:217GHz}{12}
\setsymbol{HFI:Ndetectors:353GHz}{12}
\setsymbol{HFI:Ndetectors:545GHz}{3}
\setsymbol{HFI:Ndetectors:857GHz}{4}


\setsymbol{HFI:FWHM:Maps:100GHz:units}{9\parcm88}
\setsymbol{HFI:FWHM:Maps:143GHz:units}{7\parcm18}
\setsymbol{HFI:FWHM:Maps:217GHz:units}{4\parcm87}
\setsymbol{HFI:FWHM:Maps:353GHz:units}{4\parcm65}
\setsymbol{HFI:FWHM:Maps:545GHz:units}{4\parcm72}
\setsymbol{HFI:FWHM:Maps:857GHz:units}{4\parcm39}
\setsymbol{HFI:FWHM:Maps:100GHz}{9.88}
\setsymbol{HFI:FWHM:Maps:143GHz}{7.18}
\setsymbol{HFI:FWHM:Maps:217GHz}{4.87}
\setsymbol{HFI:FWHM:Maps:353GHz}{4.65}
\setsymbol{HFI:FWHM:Maps:545GHz}{4.72}
\setsymbol{HFI:FWHM:Maps:857GHz}{4.39}


\setsymbol{HFI:beam:ellipticity:Maps:100GHz}{1.15}
\setsymbol{HFI:beam:ellipticity:Maps:143GHz}{1.01}
\setsymbol{HFI:beam:ellipticity:Maps:217GHz}{1.06}
\setsymbol{HFI:beam:ellipticity:Maps:353GHz}{1.05}
\setsymbol{HFI:beam:ellipticity:Maps:545GHz}{1.14}
\setsymbol{HFI:beam:ellipticity:Maps:857GHz}{1.19}


\setsymbol{HFI:FWHM:Mars:100GHz:units}{9\parcm37}
\setsymbol{HFI:FWHM:Mars:143GHz:units}{7\parcm04}
\setsymbol{HFI:FWHM:Mars:217GHz:units}{4\parcm68}
\setsymbol{HFI:FWHM:Mars:353GHz:units}{4\parcm43}
\setsymbol{HFI:FWHM:Mars:545GHz:units}{3\parcm80}
\setsymbol{HFI:FWHM:Mars:857GHz:units}{3\parcm67}

\setsymbol{HFI:FWHM:Mars:100GHz}{9.37}
\setsymbol{HFI:FWHM:Mars:143GHz}{7.04}
\setsymbol{HFI:FWHM:Mars:217GHz}{4.68}
\setsymbol{HFI:FWHM:Mars:353GHz}{4.43}
\setsymbol{HFI:FWHM:Mars:545GHz}{3.80}
\setsymbol{HFI:FWHM:Mars:857GHz}{3.67}


\setsymbol{HFI:beam:ellipticity:Mars:100GHz}{1.18}
\setsymbol{HFI:beam:ellipticity:Mars:143GHz}{1.03}
\setsymbol{HFI:beam:ellipticity:Mars:217GHz}{1.14}
\setsymbol{HFI:beam:ellipticity:Mars:353GHz}{1.09}
\setsymbol{HFI:beam:ellipticity:Mars:545GHz}{1.25}
\setsymbol{HFI:beam:ellipticity:Mars:857GHz}{1.03}


\setsymbol{HFI:CMB:relative:calibration:100GHz}{$\lsim 1\%$}
\setsymbol{HFI:CMB:relative:calibration:143GHz}{$\lsim 1\%$}
\setsymbol{HFI:CMB:relative:calibration:217GHz}{$\lsim 1\%$}
\setsymbol{HFI:CMB:relative:calibration:353GHz}{$\lsim 1\%$}
\setsymbol{HFI:CMB:relative:calibration:545GHz}{}
\setsymbol{HFI:CMB:relative:calibration:857GHz}{}


\setsymbol{HFI:CMB:absolute:calibration:100GHz}{$\lsim 2\%$}
\setsymbol{HFI:CMB:absolute:calibration:143GHz}{$\lsim 2\%$}
\setsymbol{HFI:CMB:absolute:calibration:217GHz}{$\lsim 2\%$}
\setsymbol{HFI:CMB:absolute:calibration:353GHz}{$\lsim 2\%$}
\setsymbol{HFI:CMB:absolute:calibration:545GHz}{}
\setsymbol{HFI:CMB:absolute:calibration:857GHz}{}


\setsymbol{HFI:FIRAS:gain:calibration:accuracy:statistical:100GHz}{}
\setsymbol{HFI:FIRAS:gain:calibration:accuracy:statistical:143GHz}{}
\setsymbol{HFI:FIRAS:gain:calibration:accuracy:statistical:217GHz}{}
\setsymbol{HFI:FIRAS:gain:calibration:accuracy:statistical:353GHz}{2.5\%}
\setsymbol{HFI:FIRAS:gain:calibration:accuracy:statistical:545GHz}{1\%}
\setsymbol{HFI:FIRAS:gain:calibration:accuracy:statistical:857GHz}{0.5\%}


\setsymbol{HFI:FIRAS:gain:calibration:accuracy:systematic:100GHz}{}
\setsymbol{HFI:FIRAS:gain:calibration:accuracy:systematic:143GHz}{}
\setsymbol{HFI:FIRAS:gain:calibration:accuracy:systematic:217GHz}{}
\setsymbol{HFI:FIRAS:gain:calibration:accuracy:systematic:353GHz}{}
\setsymbol{HFI:FIRAS:gain:calibration:accuracy:systematic:545GHz}{7\%}
\setsymbol{HFI:FIRAS:gain:calibration:accuracy:systematic:857GHz}{7\%}


\setsymbol{HFI:FIRAS:zero:point:accuracy:100GHz:units}{0.8\MJysr}
\setsymbol{HFI:FIRAS:zero:point:accuracy:143GHz:units}{}
\setsymbol{HFI:FIRAS:zero:point:accuracy:217GHz:units}{}
\setsymbol{HFI:FIRAS:zero:point:accuracy:353GHz:units}{1.4\MJysr}
\setsymbol{HFI:FIRAS:zero:point:accuracy:545GHz:units}{2.2\MJysr}
\setsymbol{HFI:FIRAS:zero:point:accuracy:857GHz:units}{1.7\MJysr}

\setsymbol{HFI:FIRAS:zero:point:accuracy:100GHz}{0.8}
\setsymbol{HFI:FIRAS:zero:point:accuracy:143GHz}{}
\setsymbol{HFI:FIRAS:zero:point:accuracy:217GHz}{}
\setsymbol{HFI:FIRAS:zero:point:accuracy:353GHz}{1.4}
\setsymbol{HFI:FIRAS:zero:point:accuracy:545GHz}{2.2}
\setsymbol{HFI:FIRAS:zero:point:accuracy:857GHz}{1.7}


\setsymbol{HFI:unit:conversion:100GHz:units}{0.2415\MJysrmK}
\setsymbol{HFI:unit:conversion:143GHz:units}{0.3694\MJysrmK}
\setsymbol{HFI:unit:conversion:217GHz:units}{0.4811\MJysrmK}
\setsymbol{HFI:unit:conversion:353GHz:units}{0.2883\MJysrmK}
\setsymbol{HFI:unit:conversion:545GHz:units}{0.05826\MJysrmK}
\setsymbol{HFI:unit:conversion:857GHz:units}{0.002238\MJysrmK}

\setsymbol{HFI:unit:conversion:100GHz}{0.2415}
\setsymbol{HFI:unit:conversion:143GHz}{0.3694}
\setsymbol{HFI:unit:conversion:217GHz}{0.4811}
\setsymbol{HFI:unit:conversion:353GHz}{0.2883}
\setsymbol{HFI:unit:conversion:545GHz}{0.05826}
\setsymbol{HFI:unit:conversion:857GHz}{0.002238}


\setsymbol{HFI:colour:correction:alpha=-2:V1.01:100GHz}{0.9893}
\setsymbol{HFI:colour:correction:alpha=-2:V1.01:143GHz}{0.9759}
\setsymbol{HFI:colour:correction:alpha=-2:V1.01:217GHz}{1.0007}
\setsymbol{HFI:colour:correction:alpha=-2:V1.01:353GHz}{1.0028}
\setsymbol{HFI:colour:correction:alpha=-2:V1.01:545GHz}{1.0019}
\setsymbol{HFI:colour:correction:alpha=-2:V1.01:857GHz}{0.9889}


\setsymbol{HFI:colour:correction:alpha=0:V1.01:100GHz}{1.0008}
\setsymbol{HFI:colour:correction:alpha=0:V1.01:143GHz}{1.0148}
\setsymbol{HFI:colour:correction:alpha=0:V1.01:217GHz}{0.9909}
\setsymbol{HFI:colour:correction:alpha=0:V1.01:353GHz}{0.9888}
\setsymbol{HFI:colour:correction:alpha=0:V1.01:545GHz}{0.9878}
\setsymbol{HFI:colour:correction:alpha=0:V1.01:857GHz}{1.0014}

\providecommand{\sorthelp}[1]{}

\def\reff@jnl#1{{#1\/}}
\def\apj{\reff@jnl{ApJ}}       
\def\apjs{\reff@jnl{ApJS}}     
\def\aaps{\reff@jnl{A\&AS}}    
\def\mnras{\reff@jnl{MNRAS}}   
\def\prd{\reff@jnl{Phys.\ Rev.\ D}}    

\newcommand{\angstrom}{\buildrel _\circ \over{\mathrm{A}} }

\newcommand{\beq}{\begin{equation}}
\newcommand{\eeq}{\end{equation}}

\newcommand{\be}{\begin{equation}}
\newcommand{\ee}{\end{equation}}
\newcommand{\bea}{\begin{eq}}
\newcommand{\eea}{\end{equation}}

\newcommand{\yfrfh}{{Y_{\mathrm{5R_{500}}}}}

{\begin{enumerate}\setlength{\itemsep}{0mm}}%
{\end{enumerate}}
{\begin{enumerate}\setlength{\itemsep}{0mm}}%
{\end{enumerate}}


\newcommand{\bc}{\begin{center}}
\newcommand{\ec}{\end{center}}
\newcommand{\bi}{\begin{itemize}}
\newcommand{\ei}{\end{itemize}}
\newcommand{\ben}{\begin{enumerate}}
\newcommand{\een}{\end{enumerate}}

\newfont{\gwpfont}{cmssq8 scaled 1000}
\newcommand{\rexcess}{{REXCESS}}
\newcommand{\macs}{{MACS}}
\newcommand{\reflex}{{REFLEX}}

\def\xmm{{\it XMM-Newton}}
\def\msol {\mathrm{M}_{\odot}}
\def\YX {Y_\mathrm{X}} 
\def\Yv {Y_\mathrm{500}}
\def \Rv {R_{500}} 
\def\keV {\mathrm{keV}} 
\def\plck{{\it Planck}}
\def\TX {T_\mathrm{X}}
\def\Mgv{M_\mathrm{g,500}}
\def\qsdss{Q_\mathrm{SDSS}}
\def\YSZ{Y_\mathrm{500}}
\def\Yall{Y_\mathrm{5R_{500}}}
\def\YSZ {Y_{500}}
\def\Mv {M_{500}}
\def\LXv {L_{\mathrm{X}, 500}}
\def\LX {L_{\mathrm{X}}}
\def\tv {\theta_{500}}
\def\YSZYX{$\YSZ$--$\YX$}
\def\ML {$\Mv$--$\LXv$}
\def\MY {$\Mv$--$\YSZ$}
\def\MYX {$\Mv$--$\YX$}

\def\YM {$\YSZ$--$\Mv$}

\def\keV {\mathrm{keV}}
\def\ne {n_{\mathrm{e}}}
\def\da {D_{\mathrm A}^{2}}
\def\MYSZ {$\Mv$--$\da\YSZ$}

\def\lesssim{\mathrel{\hbox{\rlap{\hbox{\lower4pt\hbox{$\sim$}}}\hbox{$<$}}}}
\def\gtrsim{\mathrel{\hbox{\rlap{\hbox{\lower4pt\hbox{$\sim$}}}\hbox{$>$}}}}

\newcommand{\propsim}{\lower 3pt \hbox{$\, \buildrel {\textstyle
     \propto}\over {\textstyle \sim}\,$}}

\citestyle{aa}


\def\xmm{{\it XMM-Newton}}
\def\chandra{{\it Chandra}}
\def\planck{\Planck}
\def\rosat{{ ROSAT}}
\def\rass{{\mathrm{RASS}}}

\def\lesssim{\mathrel{\hbox{\rlap{\hbox{\lower4pt\hbox{$\sim$}}}\hbox{$<$}}}}
\def\gtrsim{\mathrel{\hbox{\rlap{\hbox{\lower4pt\hbox{$\sim$}}}\hbox{$>$}}}}


\begin{document}
%

\title{\Planck\ 2013 results. XXIX. The \Planck\ catalogue of
Sunyaev--Zeldovich sources\thanks{Available at \url{http://www.sciops.esa.int}}}
\author{\small
Planck Collaboration:
P.~A.~R.~Ade\inst{99}
\and
N.~Aghanim\inst{68}\thanks{{\footnotesize Corresponding author: N.~Aghanim \url{nabila.aghanim@ias.u-psud.fr}}}
\and
C.~Armitage-Caplan\inst{104}
\and
M.~Arnaud\inst{81}
\and
M.~Ashdown\inst{78, 7}
\and
F.~Atrio-Barandela\inst{21}
\and
J.~Aumont\inst{68}
\and
H.~Aussel\inst{81}
\and
C.~Baccigalupi\inst{97}
% ...
}
\institute{\small
APC, AstroParticule et Cosmologie, Universit\'{e} Paris Diderot, CNRS/IN2P3, CEA/lrfu, Observatoire de Paris, Sorbonne Paris Cit\'{e}, 10, rue Alice Domon et L\'{e}onie Duquet, 75205 Paris Cedex 13, France\\
\and
Aalto University Mets\"{a}hovi Radio Observatory, Mets\"{a}hovintie 114, FIN-02540 Kylm\"{a}l\"{a}, Finland\\
\and
Academy of Sciences of Tatarstan, Bauman Str., 20, Kazan, 420111, Republic of Tatarstan, Russia\\
\and
African Institute for Mathematical Sciences, 6-8 Melrose Road, Muizenberg, Cape Town, South Africa\\
\and
Agenzia Spaziale Italiana Science Data Center, Via del Politecnico snc, 00133, Roma, Italy\\
\and
Agenzia Spaziale Italiana, Viale Liegi 26, Roma, Italy\\
\and
Astrophysics Group, Cavendish Laboratory, University of Cambridge, J J Thomson Avenue, Cambridge CB3 0HE, U.K.\\
\and
Astrophysics \& Cosmology Research Unit, School of Mathematics, Statistics \& Computer Science, University of KwaZulu-Natal, Westville Campus, Private Bag X54001, Durban 4000, South Africa\\
\and
Atacama Large Millimeter/submillimeter Array, ALMA Santiago Central Offices, Alonso de Cordova 3107, Vitacura, Casilla 763 0355, Santiago, Chile\\
\and
CITA, University of Toronto, 60 St. George St., Toronto, ON M5S 3H8, Canada\\
\and
CNRS, IRAP, 9 Av. colonel Roche, BP 44346, F-31028 Toulouse cedex 4, France\\
\and
California Institute of Technology, Pasadena, California, U.S.A.\\
\and
Centre for Theoretical Cosmology, DAMTP, University of Cambridge, Wilberforce Road, Cambridge CB3 0WA, U.K.\\
\and
Centro de Astrof\'{\i}sica, Universidade do Porto, Rua das Estrelas, 4150-762 Porto, Portugal\\
\and
Centro de Estudios de F\'{i}sica del Cosmos de Arag\'{o}n (CEFCA), Plaza San Juan, 1, planta 2, E-44001, Teruel, Spain\\
\and
Computational Cosmology Center, Lawrence Berkeley National Laboratory, Berkeley, California, U.S.A.\\
\and
Consejo Superior de Investigaciones Cient\'{\i}ficas (CSIC), Madrid, Spain\\
\and
DSM/Irfu/SPP, CEA-Saclay, F-91191 Gif-sur-Yvette Cedex, France\\
\and
DTU Space, National Space Institute, Technical University of Denmark, Elektrovej 327, DK-2800 Kgs. Lyngby, Denmark\\
\and
D\'{e}partement de Physique Th\'{e}orique, Universit\'{e} de Gen\`{e}ve, 24, Quai E. Ansermet,1211 Gen\`{e}ve 4, Switzerland\\
\and
Departamento de F\'{\i}sica Fundamental, Facultad de Ciencias, Universidad de Salamanca, 37008 Salamanca, Spain\\
\and
Departamento de F\'{\i}sica, Universidad de Oviedo, Avda. Calvo Sotelo s/n, Oviedo, Spain\\
\and
Department of Astronomy and Astrophysics, University of Toronto, 50 Saint George Street, Toronto, Ontario, Canada\\
\and
Department of Astronomy and Geodesy, Kazan Federal University,  Kremlevskaya Str., 18, Kazan, 420008, Russia\\
\and
Department of Astrophysics/IMAPP, Radboud University Nijmegen, P.O. Box 9010, 6500 GL Nijmegen, The Netherlands\\
\and
Department of Electrical Engineering and Computer Sciences, University of California, Berkeley, California, U.S.A.\\
\and
Department of Physics \& Astronomy, University of British Columbia, 6224 Agricultural Road, Vancouver, British Columbia, Canada\\
\and
Department of Physics and Astronomy, Dana and David Dornsife College of Letter, Arts and Sciences, University of Southern California, Los Angeles, CA 90089, U.S.A.\\
\and
Department of Physics and Astronomy, University College London, London WC1E 6BT, U.K.\\
\and
Department of Physics and Astronomy, University of Sussex, Brighton BN1 9QH, U.K.\\
\and
Department of Physics, Florida State University, Keen Physics Building, 77 Chieftan Way, Tallahassee, Florida, U.S.A.\\
\and
Department of Physics, Gustaf H\"{a}llstr\"{o}min katu 2a, University of Helsinki, Helsinki, Finland\\
\and
Department of Physics, Princeton University, Princeton, New Jersey, U.S.A.\\
\and
Department of Physics, University of California, Berkeley, California, U.S.A.\\
\and
Department of Physics, University of California, One Shields Avenue, Davis, California, U.S.A.\\
\and
Department of Physics, University of California, Santa Barbara, California, U.S.A.\\
\and
Department of Physics, University of Illinois at Urbana-Champaign, 1110 West Green Street, Urbana, Illinois, U.S.A.\\
\and
Dipartimento di Fisica e Astronomia G. Galilei, Universit\`{a} degli Studi di Padova, via Marzolo 8, 35131 Padova, Italy\\
\and
Dipartimento di Fisica e Scienze della Terra, Universit\`{a} di Ferrara, Via Saragat 1, 44122 Ferrara, Italy\\
\and
Dipartimento di Fisica, Universit\`{a} La Sapienza, P. le A. Moro 2, Roma, Italy\\
\and
Dipartimento di Fisica, Universit\`{a} degli Studi di Milano, Via Celoria, 16, Milano, Italy\\
\and
Dipartimento di Fisica, Universit\`{a} degli Studi di Trieste, via A. Valerio 2, Trieste, Italy\\
\and
Dipartimento di Fisica, Universit\`{a} di Roma Tor Vergata, Via della Ricerca Scientifica, 1, Roma, Italy\\
\and
Discovery Center, Niels Bohr Institute, Blegdamsvej 17, Copenhagen, Denmark\\
\and
Dpto. Astrof\'{i}sica, Universidad de La Laguna (ULL), E-38206 La Laguna, Tenerife, Spain\\
\and
European Southern Observatory, ESO Vitacura, Alonso de Cordova 3107, Vitacura, Casilla 19001, Santiago, Chile\\
\and
European Space Agency, ESAC, Planck Science Office, Camino bajo del Castillo, s/n, Urbanizaci\'{o}n Villafranca del Castillo, Villanueva de la Ca\~{n}ada, Madrid, Spain\\
\and
European Space Agency, ESTEC, Keplerlaan 1, 2201 AZ Noordwijk, The Netherlands\\
\and
Finnish Centre for Astronomy with ESO (FINCA), University of Turku, V\"{a}is\"{a}l\"{a}ntie 20, FIN-21500, Piikki\"{o}, Finland\\
\and
GEPI, Observatoire de Paris, Section de Meudon, 5 Place J. Janssen, 92195 Meudon Cedex, France\\
\and
Helsinki Institute of Physics, Gustaf H\"{a}llstr\"{o}min katu 2, University of Helsinki, Helsinki, Finland\\
\and
INAF - Osservatorio Astrofisico di Catania, Via S. Sofia 78, Catania, Italy\\
\and
INAF - Osservatorio Astronomico di Padova, Vicolo dell'Osservatorio 5, Padova, Italy\\
\and
INAF - Osservatorio Astronomico di Roma, via di Frascati 33, Monte Porzio Catone, Italy\\
\and
INAF - Osservatorio Astronomico di Trieste, Via G.B. Tiepolo 11, Trieste, Italy\\
\and
INAF Istituto di Radioastronomia, Via P. Gobetti 101, 40129 Bologna, Italy\\
\and
INAF/IASF Bologna, Via Gobetti 101, Bologna, Italy\\
\and
INAF/IASF Milano, Via E. Bassini 15, Milano, Italy\\
\and
INFN, Sezione di Bologna, Via Irnerio 46, I-40126, Bologna, Italy\\
\and
INFN, Sezione di Roma 1, Universit\`{a} di Roma Sapienza, Piazzale Aldo Moro 2, 00185, Roma, Italy\\
\and
IPAG: Institut de Plan\'{e}tologie et d'Astrophysique de Grenoble, Universit\'{e} Joseph Fourier, Grenoble 1 / CNRS-INSU, UMR 5274, Grenoble, F-38041, France\\
\and
ISDC Data Centre for Astrophysics, University of Geneva, ch. d'Ecogia 16, Versoix, Switzerland\\
\and
IUCAA, Post Bag 4, Ganeshkhind, Pune University Campus, Pune 411 007, India\\
\and
Imperial College London, Astrophysics group, Blackett Laboratory, Prince Consort Road, London, SW7 2AZ, U.K.\\
\and
Infrared Processing and Analysis Center, California Institute of Technology, Pasadena, CA 91125, U.S.A.\\
\and
Institut N\'{e}el, CNRS, Universit\'{e} Joseph Fourier Grenoble I, 25 rue des Martyrs, Grenoble, France\\
\and
Institut Universitaire de France, 103, bd Saint-Michel, 75005, Paris, France\\
\and
Institut d'Astrophysique Spatiale, CNRS (UMR8617) Universit\'{e} Paris-Sud 11, B\^{a}timent 121, Orsay, France\\
\and
Institut d'Astrophysique de Paris, CNRS (UMR7095), 98 bis Boulevard Arago, F-75014, Paris, France\\
\and
Institute for Space Sciences, Bucharest-Magurale, Romania\\
\and
Institute of Astronomy and Astrophysics, Academia Sinica, Taipei, Taiwan\\
\and
Institute of Astronomy, University of Cambridge, Madingley Road, Cambridge CB3 0HA, U.K.\\
\and
Institute of Theoretical Astrophysics, University of Oslo, Blindern, Oslo, Norway\\
\and
Instituto de Astrof\'{\i}sica de Canarias, C/V\'{\i}a L\'{a}ctea s/n, La Laguna, Tenerife, Spain\\
\and
Instituto de F\'{\i}sica de Cantabria (CSIC-Universidad de Cantabria), Avda. de los Castros s/n, Santander, Spain\\
\and
Jet Propulsion Laboratory, California Institute of Technology, 4800 Oak Grove Drive, Pasadena, California, U.S.A.\\
\and
Jodrell Bank Centre for Astrophysics, Alan Turing Building, School of Physics and Astronomy, The University of Manchester, Oxford Road, Manchester, M13 9PL, U.K.\\
\and
Kavli Institute for Cosmology Cambridge, Madingley Road, Cambridge, CB3 0HA, U.K.\\
\and
LAL, Universit\'{e} Paris-Sud, CNRS/IN2P3, Orsay, France\\
\and
LERMA, CNRS, Observatoire de Paris, 61 Avenue de l'Observatoire, Paris, France\\
\and
Laboratoire AIM, IRFU/Service d'Astrophysique - CEA/DSM - CNRS - Universit\'{e} Paris Diderot, B\^{a}t. 709, CEA-Saclay, F-91191 Gif-sur-Yvette Cedex, France\\
\and
Laboratoire Traitement et Communication de l'Information, CNRS (UMR 5141) and T\'{e}l\'{e}com ParisTech, 46 rue Barrault F-75634 Paris Cedex 13, France\\
\and
Laboratoire de Physique Subatomique et de Cosmologie, Universit\'{e} Joseph Fourier Grenoble I, CNRS/IN2P3, Institut National Polytechnique de Grenoble, 53 rue des Martyrs, 38026 Grenoble cedex, France\\
\and
Laboratoire de Physique Th\'{e}orique, Universit\'{e} Paris-Sud 11 \& CNRS, B\^{a}timent 210, 91405 Orsay, France\\
\and
Lawrence Berkeley National Laboratory, Berkeley, California, U.S.A.\\
\and
MPA Partner Group, Key Laboratory for Research in Galaxies and Cosmology, Shanghai Astronomical Observatory, Chinese Academy of Sciences, Nandan Road 80, Shanghai 200030, China\\
\and
Max-Planck-Institut f\"{u}r Astrophysik, Karl-Schwarzschild-Str. 1, 85741 Garching, Germany\\
\and
Max-Planck-Institut f\"{u}r Extraterrestrische Physik, Giessenbachstra{\ss}e, 85748 Garching, Germany\\
\and
McGill Physics, Ernest Rutherford Physics Building, McGill University, 3600 rue University, Montr\'{e}al, QC, H3A 2T8, Canada\\
\and
MilliLab, VTT Technical Research Centre of Finland, Tietotie 3, Espoo, Finland\\
\and
Moscow Institute of Physics and Technology, Dolgoprudny, Institutsky per., 9, 141700, Russia\\
\and
National University of Ireland, Department of Experimental Physics, Maynooth, Co. Kildare, Ireland\\
\and
Niels Bohr Institute, Blegdamsvej 17, Copenhagen, Denmark\\
\and
Observational Cosmology, Mail Stop 367-17, California Institute of Technology, Pasadena, CA, 91125, U.S.A.\\
\and
Optical Science Laboratory, University College London, Gower Street, London, U.K.\\
\and
SB-ITP-LPPC, EPFL, CH-1015, Lausanne, Switzerland\\
\and
SISSA, Astrophysics Sector, via Bonomea 265, 34136, Trieste, Italy\\
\and
SUPA, Institute for Astronomy, University of Edinburgh, Royal Observatory, Blackford Hill, Edinburgh EH9 3HJ, U.K.\\
\and
School of Physics and Astronomy, Cardiff University, Queens Buildings, The Parade, Cardiff, CF24 3AA, U.K.\\
\and
Space Research Institute (IKI), Russian Academy of Sciences, Profsoyuznaya Str, 84/32, Moscow, 117997, Russia\\
\and
Space Sciences Laboratory, University of California, Berkeley, California, U.S.A.\\
\and
Special Astrophysical Observatory, Russian Academy of Sciences, Nizhnij Arkhyz, Zelenchukskiy region, Karachai-Cherkessian Republic, 369167, Russia\\
\and
Stanford University, Dept of Physics, Varian Physics Bldg, 382 Via Pueblo Mall, Stanford, California, U.S.A.\\
\and
Sub-Department of Astrophysics, University of Oxford, Keble Road, Oxford OX1 3RH, U.K.\\
% ...
}

\abstract {\tiny We describe the all-sky \Planck\ catalogue of clusters and
  cluster candidates derived from Sunyaev--Zeldovich (SZ) effect
  detections using the first 15.5 months of \Planck\ satellite
  observations. The catalogue contains 1227 entries, making it over
  six times the size of the \Planck\ Early SZ (ESZ) sample and the
  largest SZ-selected catalogue to date. It contains 861 confirmed
  clusters, of which 178 have been confirmed as clusters, mostly
  through follow-up observations, and a further 683 are
  previously-known clusters. The remaining 366 have the status of
  cluster candidates, and we divide them into three classes according
  to the quality of evidence that they are likely to be true
  clusters. The \planck\ SZ catalogue is the deepest all-sky cluster
  catalogue, with redshifts up to about one, and spans the broadest
  cluster mass range from $(0.1$ to $1.6)\times
  10^{15}\,\msol$. Confirmation of cluster candidates through
  comparison with existing surveys or cluster catalogues is
  extensively described, as is the statistical characterization of the
  catalogue in terms of completeness and statistical reliability. The
  outputs of the validation process are provided as additional
  information. This gives, in particular, an ensemble of 813 cluster
  redshifts, and for all these \planck\ clusters we also include a
  mass estimated from a newly-proposed SZ-mass proxy. A refined
  measure of the SZ Compton parameter for the clusters with X-ray
  counter-parts is provided, as is an X-ray flux for all the
  \Planck\ clusters not previously detected in X-ray surveys.  }


\authorrunning{Planck Collaboration}
\titlerunning{\Planck\ catalogue of Sunyaev--Zeldovich sources}

\maketitle


\clearpage 

\section{Introduction}

This paper, one of a set associated with the 2013 release of data from
the \Planck\footnote{\Planck\ (\url{http://www.esa.int/Planck}) is a
  project of the European Space Agency (ESA) with instruments provided
  by two scientific consortia funded by ESA member states (in
  particular the lead countries France and Italy), with contributions
  from NASA (USA) and telescope reflectors provided by a collaboration
  between ESA and a scientific consortium led and funded by Denmark.}
mission \citep{planck2013-p01}, describes the construction and
properties of the \Planck\ catalogue of SZ sources (PSZ).

Clusters of galaxies play several important roles in astrophysics and
cosmology. As rare objects, their number density is especially
sensitive to properties of the cosmological model such as the
amplitude of primordial density perturbations \citep{pee80}, and their
development with redshift probes the growth of cosmic structure,
hence perhaps helping to distinguish between dark energy and modified gravity
explanations for cosmic acceleration \citep[e.g., see reviews by
][]{bor09,all11}. The galaxies, hot gas and dark matter held in their
gravitational potential wells provide a sample of the universal
abundance of these components \citep[e.g.,][]{voi05}, while the
thermal state of the gas probes both the cluster formation mechanism
and physical processes within the cluster such as cooling and
energy-injection feedback \citep[e.g., reviews by][]{fab12,mcn12}. The
study of the constituent galaxies, including the brightest cluster
galaxies normally found at their centres, allows sensitive tests of
galaxy formation models.

Because of these uses, there is considerable interest in developing
large galaxy cluster catalogues that can be used for population and
cosmological studies \citep[e.g.,][]{sch03,boe04}. Clusters are
genuinely multi-wavelength objects that can be selected in several
ways: optical/infrared (IR) imaging of the galaxy populations; X-ray
imaging of bremsstrahlung radiation from the hot cluster gas; and
through the Sunyaev--Zeldovich (SZ) effect \citep{sun72,sun80}
whereby scattering of cosmic microwave background (CMB) photons from
that hot gas distorts the spectral shape of the CMB along lines of
sight through clusters and groups.

Construction of cluster catalogues in the optical/IR and in the X-ray
are relatively mature activities. The first large optical cluster
survey is now over 50 years old \citep{abe58,abe89}, and current
catalogues constructed from the Sloan Digital Sky Survey data contain over
a hundred thousand clusters \citep[e.g.,][]{koe07, wen12}. In X-rays,
large samples first became available via {ROSAT} satellite
observations \citep[e.g.,][]{vik98,boe00,gio03,boe04,bur07,ebe07}, but
also more recently for instance from dedicated or serendipitous survey
with \xmm\ \citep[][]{pac07,fas11,tak11,meh12}. Currently several
thousand X-ray selected clusters are known \citep[see for instance the
 meta-catalogue MCXC by][]{pif11}. By contrast, although proposed
about fifteen years ago \citep[e.g.,][]{bar96,agh97}, it is only very
recently that SZ-selected samples have reached a significant size,
with publication of samples containing several hundred clusters from
the Early SZ (ESZ) catalogue from the \Planck\ Satellite
\citep{planck2011-5.1a}, the South Pole Telescope \citep[SPT,
][]{rei13} and the Atacama Cosmology Telescope
\citep[ACT,][]{has13}.

\begin{figure}[t]
\begin{center}
\includegraphics[width=8.8cm]{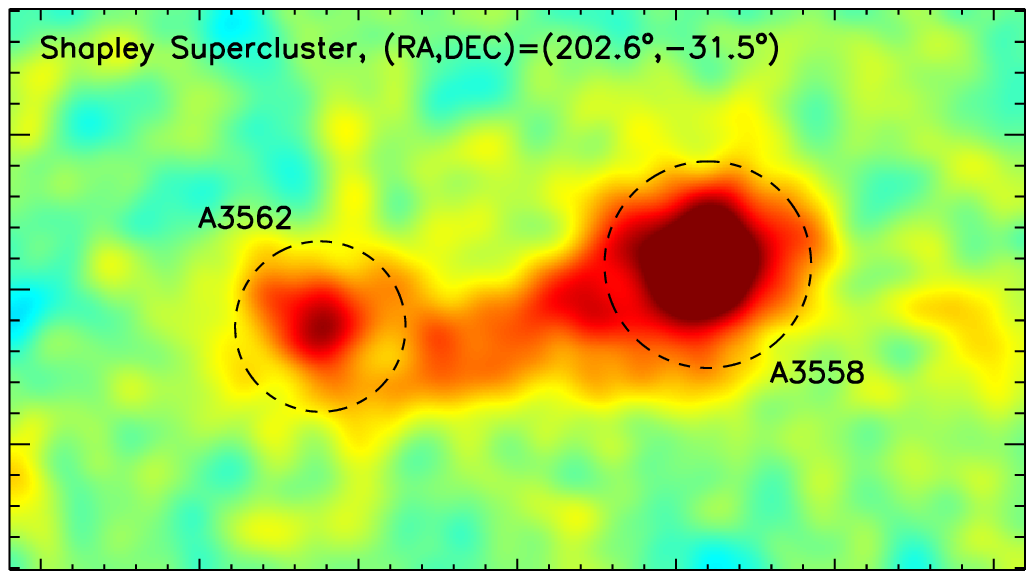}\\%
\vspace*{-1ex}\includegraphics[width=8.8cm]{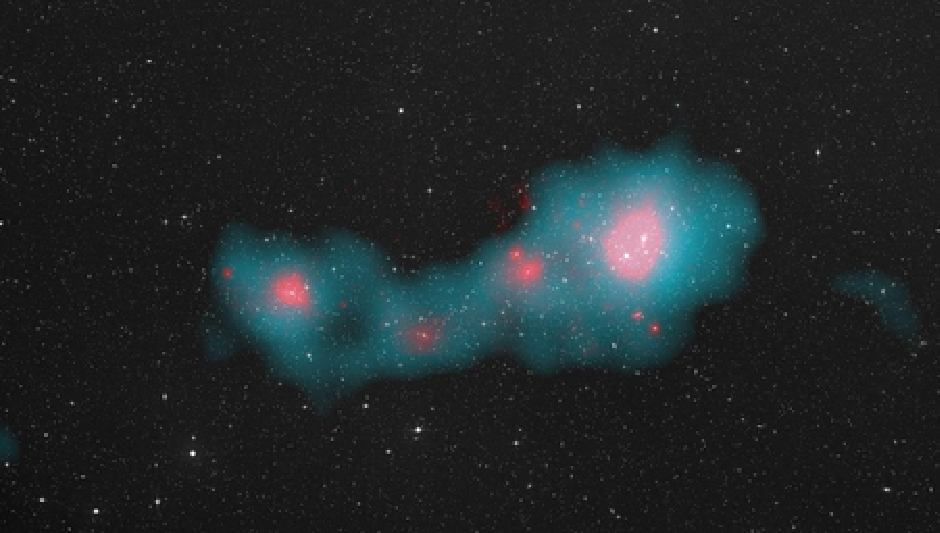}
\caption{The Shapley super-cluster as seen in the
 \Planck\ survey. \emph{Upper panel:} reconstructed thermal SZ map
 $3.2^{\circ} \times 1.8^{\circ}$ from \texttt{MILCA}
 \citep{hur13}. The dotted circles represent apertures of
 $\theta_{500}$ from the MCXC meta-catalogue around the resolved
 clusters. \emph{Lower panel:} composite view of the optical from
 DSS images (white), X-rays from {ROSAT} (pink) survey and of
 the thermal SZ effect as seen in \planck\ (blue). }
\label{fig:shapley}
\end{center}
\end{figure}

The usefulness of the different selection methods, particularly for
cosmology, depends not just on the total number of clusters identified
but also on how readily the selection function of the survey can be
modelled, and on how well the observed cluster properties can be
related to quantities such as the total cluster mass that are most
readily predicted from theory \citep[e.g., see][]{voi05}. It has
proven difficult to capitalize on the large size of optical/IR cluster
samples because the observable, the number of galaxies in each
cluster, exhibits large scatter with respect to the total cluster mass
\citep[e.g.,][]{joh07}. In this regard the X-ray selected samples are
considerably more powerful, due to the tighter correlations of X-ray
properties with mass
\citep{arn05,vik06,pra09,rei11,mau12}. Simulations predict that
SZ-selected surveys may do even better, with a very tight relation
between SZ signal and mass \citep[e.g.,][]{das04,mot05,nag06,wik08,agh09,
ang13}. Moreover, this relation, except at low redshifts,
corresponds to a nearly redshift-independent mass limit, thus allowing
such surveys to reach to high redshift and provide a strong lever arm
on growth of structure.

We report on the construction and properties of the PSZ catalogue,
which is to date the largest SZ-selected cluster catalogue and has
value added through compilation of ancillary information. It contains
1227 entries including many multiple systems, e.g., the Shapley
super-cluster displayed in Fig.~\ref{fig:shapley} together with a
composite image. Of these 861 are confirmed, amongst which 178 are new
discoveries, whilst amongst the 366 candidate clusters 54 are of high
reliability ({\sc class1} in our terminology), 170 are reliable, and
the remaining 142 are in the lowest reliability class.  In
Sect.~\ref{s:construc} we start with a description of the
\Planck\ data used to provide cluster candidates, and the two
different methodologies (one of which has two independent
implementations) used to carry out the extraction of the SZ sources.
In Sect.~\ref{sec:qa} we provide a characterization of the PSZ
catalogue in terms of completeness, statistical reliability, and
accuracy of cluster parameters including size and
photometry. Section~\ref{s:exval} extensively describes validation of
cluster candidates through pre-existing surveys and cluster catalogues
in many wavebands, while Sect.~\ref{sec:fu} describes the follow-up
campaigns conducted by the \Planck\ collaboration to confirm new
cluster discoveries. This leads to a description of the catalogue
properties in Sect.~\ref{s:results}.  The physical properties of the
clusters are exploited in Sect.~\ref{s:physprop}. These include an
update of the SZ--X-ray scaling relations from the \Planck\ data, the
measure of the X-ray flux for all SZ detections, and the production of
homogenized SZ-mass estimates for 813 clusters with measured redshifts
that are provided to the community as a value-added element to the
\Planck\ SZ catalogue.

Throughout the article, the quantities $M_{500}$ and $R_{500}$ stand
for the total mass and radius where the mean enclosed density is 500
times the critical density at the cluster redshift. The SZ flux is
denoted $Y_{500}$, where $Y_{500}$ $D^2_{\mathrm{A}}$ is the
spherically-integrated Compton parameter within $R_{500}$, and
$D_{\mathrm{A}}$ is the angular-diameter distance to the cluster. The
physical cluster quantities are computed with a fiducial $\Lambda$CDM
cosmology with $H_0 = 70 \,
\mathrm{km}\,\mathrm{s}^{-1}\,\mathrm{Mpc}^{-1}$,
$\Omega_{\mathrm{m}}=0.3$ and $\Omega_{\Lambda}=0.7$. Furthermore, all
the fits are undertaken in the log-log plane using the
BCES orthogonal regression method of \citet{akr96}, with bootstrap
resampling, which allows for intrinsic scatter as well as
uncertainties in both variables. All uncertainties are given at 68 per
cent confidence level and all dispersions are given in log$_{10}$.

\section{Construction of the \planck\ SZ Catalogue}
\label{s:construc}

\subsection{Input \Planck\ data}\label{ss_preprocessing}

The \planck\ catalogue of SZ sources is constructed from the total
intensity data taken during the first 15.5 months of \Planck\ survey
observations. Raw data were first processed to produce cleaned
time-lines (time-ordered information) and associated flags
correcting for different systematic effects; channel maps were then
produced for all the observing frequencies \citep[see details
 in][]{planck2013-p03,planck2013-p02}. These maps, together with the
associated beam characteristics, are the main inputs for the SZ-finder
algorithms presented in Sect.~\ref{sec:det_meth}. Following
\citet{planck2011-5.1a}, we used the six highest-frequency
\Planck\ channel maps, from 100 to 857~GHz, to produce the catalogue
of SZ detections. This optimizes the signal-to-noise (S/N) of the
extracted SZ detections and the usable sky fraction; see
Appendix~\ref{Sec:freq_cho} for the choice of channel maps.

\begin{figure*}[t]
\begin{center}
\includegraphics[width=16cm]{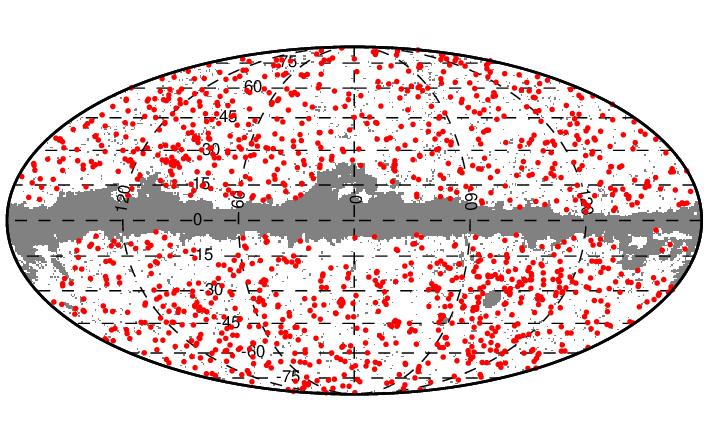}
\caption{Sky distribution of the 1227 \Planck\ clusters and candidates
  (red dots), in a Mollweide projection with the Galactic plane
  horizontal and centred at longitude zero. Small {grey} dots show the
  positions of masked point sources, and {grey} shading shows the mask
  used to exclude the Magellanic clouds and the Galactic plane
  mask. The mask covers 16.3\% of the sky.
}
\label{fig:sz_dist}
\end{center}
\end{figure*}

In order to optimize the SZ detection, together with avoiding
contamination of the PSZ catalogue by bright point sources (PS), the
latter are masked from the channel maps prior to the SZ detection as
detailed in the following.  To construct the PS mask, we use the
\Planck\ Catalogue of Compact Sources (PCCS).  The PCCS
\citep{planck2013-p05} is a collection of single-frequency source
catalogues, one for each of the nine \Planck\ frequency channels. The
six single \Planck-HFI frequency PS catalogues are used to first
produce individual-frequency masks constructed by masking a radius
equivalent to 1.28 FWHM (3$\,\sigma_{\mathrm{beam}}$) around every point
source detected with (S/N)$_{\mathrm{PS}}\ge 10$. Then a single common
PS mask (see Fig.~\ref{fig:sz_dist}), which is the union of the six
individual HFI-frequency channel masks, is constructed. It is applied
to all six highest-frequency \Planck\ channel-maps to mask the point
sources prior to running the algorithms to detect SZ signal. The
masked regions are filled using a harmonic in-painting method based on
that of \citet{baj05}, which has the advantage of eliminating the
discontinuities caused by the masking.  In order to avoid any possible
artificial spurious detections at the edges of the in-painted area, we
further reject detections within an expanded common mask, constructed
using the same procedure as described above, but using a masking
radius equivalent to 2.13 FWHM ($5\,\sigma_{\mathrm{beam}}$) and
covering less than $2.9$\% of the sky.

Bright radio sources are known to exist at the centre of galaxy
clusters, but they generally have steep spectra and hence their flux
is significantly reduced at the six highest \planck\ frequencies where
the PS mask is constructed and where the clusters are detected. The
Perseus cluster (see Fig.~\ref{milcaext} later and the associated
discussion) is one exception, with a point source that is so bright
that the cluster is masked and thus not included in the \planck\ SZ
catalogue.

\subsection{Detection methods}\label{sec:det_meth}

The catalogue of SZ sources is the result of a blind multi-frequency
search, i.e., no prior positional information on known clusters is
used as input to the detection, by three detection algorithms briefly
described below. These algorithms were described and tested using
simulations \citep{mel12}. They were used to construct the Early SZ
(ESZ) \Planck\ sample by \citet{planck2011-5.1a}. All three assume
priors on the cluster spectral and spatial characteristics, which
optimize the SZ detection by enhancing the SZ contrast over a set of
observations containing contaminating signals. In the following we
present the cluster model used as a template by the SZ-finder
algorithms and we briefly describe the three detection methods
\citep[for details we refer the reader
  to][]{her02,mel06,car09,car11b,mel12}.

\subsubsection{Cluster model}\label{sec:clus_mod}

The baseline pressure profile model used in the detection methods is
the generalized NFW \citep{nav97} profile of \citet{arn10}. This
profile model was constructed by combining the observed, scaled, X-ray
pressure profile of 31 clusters from the \rexcess\ sample
\citep{boe07} for $R < R_{500}$,\footnote{$R_{500}$ relates to the
  characteristic cluster scale $R_{\mathrm{s}}$ through the NFW
  concentration parameter $c_{500}=1.177$ for the baseline profile
  ($R_{\mathrm{s}}=R_{500}/c_{500}$).} with the mean pressure profile
from three sets of numerical simulations \citep{bor04,nag07,pif08} for
$R_{500} < R < 5\,R_{500}$. New observational constraints on the
pressure distribution at $R > R_{500}$ have become available.
\citet{planck2012-V} constrained the detection of the thermal pressure
distribution out to about $3\, R_{500}$ through stacking of the
observed SZ profiles of 62 nearby massive clusters detected with high
significance in the \Planck\ ESZ sample. The resulting profile is in
agreement with that derived for the Coma cluster
\citep{planck2012-X}. Both show a slightly flatter distribution in the
outer parts (i.e., beyond $\Rv$) with respect to the predictions from
the numerical simulations.  These results are further confirmed by
independent measurements from Bolocam in a smaller radial range
\citep[$r< 2\, \Rv$,][]{say13}. {Pressure profiles different from
  the generalized NFW and consistent with the observations can be
  devised, e.g., the SuperModel used by \citet{lap12} for SPT stacked
  clusters or by \citet{fus13} for the Coma cluster.} Using the profile
of \citet{planck2012-V} does not affect the detection yield (see
Sect.~\ref{sec:qa}) and only slightly modifies the measure of the SZ
flux density (see Sect.~\ref{sec:scal}) as compared to the generalized
NFW (GNFW) profile adopted in the three cluster.  The fiducial model
parameters for the GNFW profile are given by the parameterization of
the pressure profile in Eq.~12 of \citet{arn10}. It states
\begin{equation}
\mathbf{p}(x) = \frac{P_{0}}
       {(c_{500}x)^{\gamma}\left[1+(c_{500}x)^\alpha\right]^{(\beta-\gamma)/\alpha}}
       \,,   
\label{eq:pgnfw}
\end{equation}
with the parameters
\begin{equation}
[P_{0} ,c_{500},\gamma,\alpha,\beta] =
[8.40\,h_{70}^{-3/2},1.18,0.308,1.05,5.49]\,. 
\label{eq:pargnfw}
\end{equation}
The (weak) mass
dependence of the profiles is neglected. Within the SZ-finder
algorithms, the size and amplitude of the profile are allowed to vary
but all other parameters are fixed. The cluster model is thus
equivalent to a shape function characterized by two free parameters,
its amplitude and a characteristic scale $\theta_{\mathrm s} =
\theta_{500}/c_{500}$.

\subsubsection{Matched multi-filter (MMF)}
\label{ss_mmf}

Two different implementations of the matched multi-frequency filter
algorithm (\texttt{MMF1} and \texttt{MMF3}) are used to detect SZ
clusters.  Both are extensions, over the whole sky, of the \texttt{MMF}
algorithm \citep{her02,mel06}. The matched filter optimizes the
cluster detection using a linear combination of maps (which requires
an estimate of the statistics of the contamination) and uses spatial
filtering to suppress both foregrounds and noise (making use of the
prior knowledge of the cluster pressure profile and thermal SZ
spectrum).

\begin{figure}[t]
\begin{center}
\includegraphics[width=8.8cm]{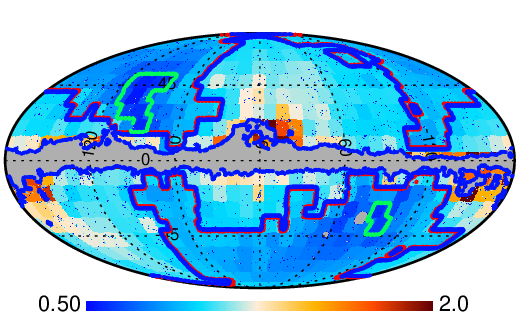}
\caption{Noise maps per detection patch of \texttt{MMF3} method
  measured for a 6$^{\prime}$ filter. The noise ranges from 0.5 to 2
  times the average noise of the map, which is $\sigma_Y=2.4\times
  10^{-4}$ arcmin$^2$. The Ecliptic polar regions, delimited by green
  contours, with increased redundancy in the observations define a
  deep survey zone covering in total 2.7\% of the sky. It is less
  noisy than the areas near the Galactic plane, where the dust emission
  is higher. Two other zones are defined: a medium-deep survey zone of
  41.3\% coverage delimited by the red contours and with higher noise
  level; and a shallow-survey zone covering 56\% of the sky and with
  the highest noise levels including regions near the Galactic plane.}
\label{fig:noise_map}
\end{center}
\end{figure}

The \texttt{MMF1} algorithm divides the full-sky \Planck\ frequency
maps into 640 patches, each $14.66^{\circ}\times14.66^{\circ}$,
covering 3.33 times the sky. The \texttt{MMF3} algorithm divides the
maps into a smaller set of 504 overlapping square patches of area
$10^{\circ}\times10^{\circ}$ with the sky covered 1.22 times. The
smaller redundancy of \texttt{MMF3} with respect to \texttt{MMF1}
implies a potentially lower reliability of the SZ detections. In order
to increase the reliability of the detections, the \texttt{MMF3}
algorithm is thus run in two iterations. After a first detection of
the SZ candidates, a subsequent run centred on the positions of the
candidates refines the estimated S/N and candidate properties. If the
S/N of a detection falls below the threshold at the second iteration,
it is removed from the catalogue.  For both implementations, the
matched multi-frequency filter optimally combines the six frequencies
of each patch. Auto- and cross-power spectra are directly estimated
from the data and are thus adapted to the local instrumental noise and
astrophysical contamination, which constitutes the dominant noise
contribution. Figure~\ref{fig:noise_map} illustrates, for a 6$^{\prime}$
filter size, the ensemble noise maps as measured by \texttt{MMF3} in
each of the patches.  For both \texttt{MMF1} and \texttt{MMF3}, the
detection of the SZ-candidates is performed on all the patches, and
the resultant sub-catalogues are merged together to produce a single
SZ-candidate catalogue per method.

The candidate size in both algorithms is estimated by filtering the
patches over the range of potential scales, and finding the scale
that maximizes the S/N of the detected candidate. When merging the
sub-catalogues produced from the analysis of individual patches, it is
also the S/N of the detection (the refined S/N estimate for
\texttt{MMF3}) which is used when deciding which detection of the
candidate is kept. Furthermore, both \texttt{MMF1} and \texttt{MMF3}
can also be run with fixed cluster size and position to estimate the
SZ signal. This version of the algorithms is used to assess the
reliability of the association with known clusters and/or to refine
the measurement of the integrated Compton parameters of known X-ray
clusters, as presented in Sect.~\ref{sec:sizeF}.

\begin{figure*}[t]
\begin{center}
\includegraphics[width=8.44cm]{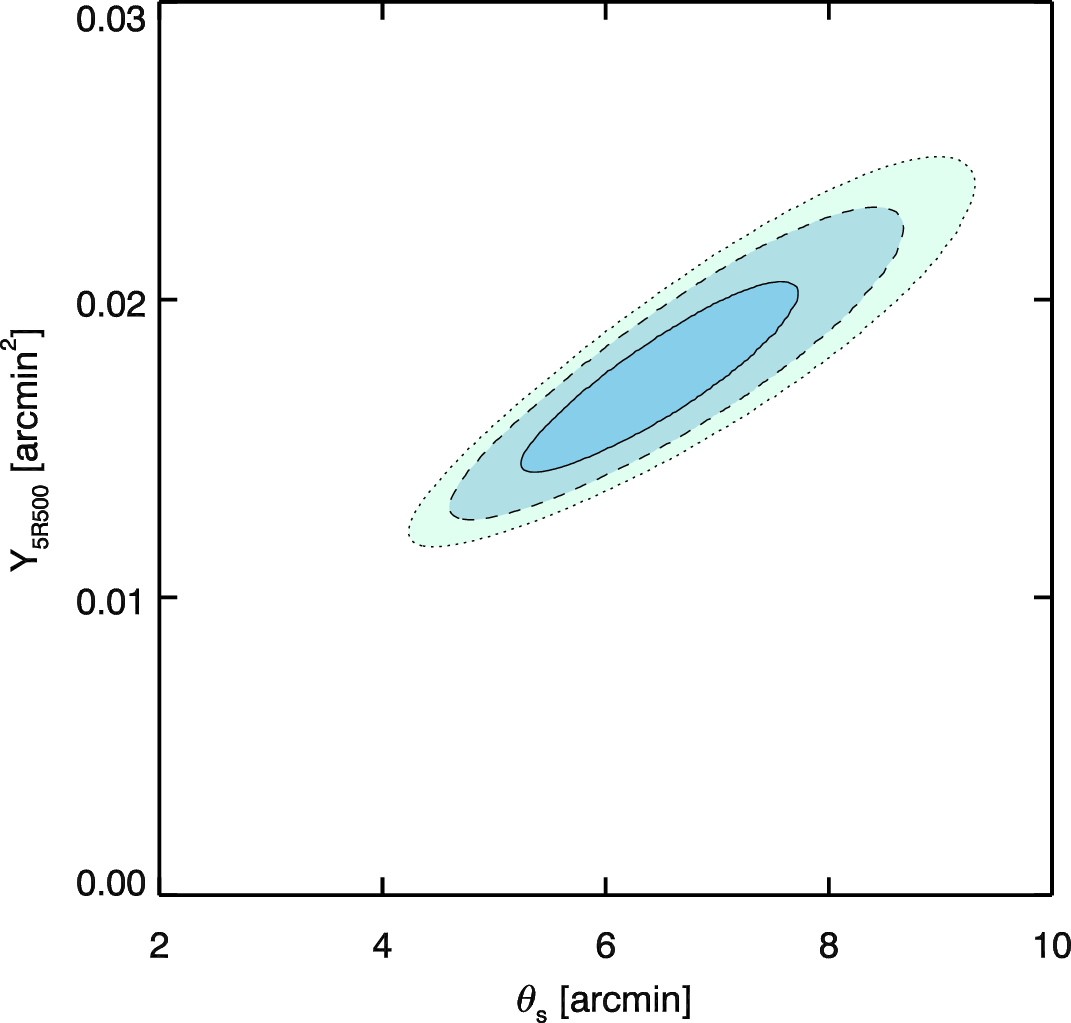}\hspace*{2em}
\includegraphics[width=8.8cm]{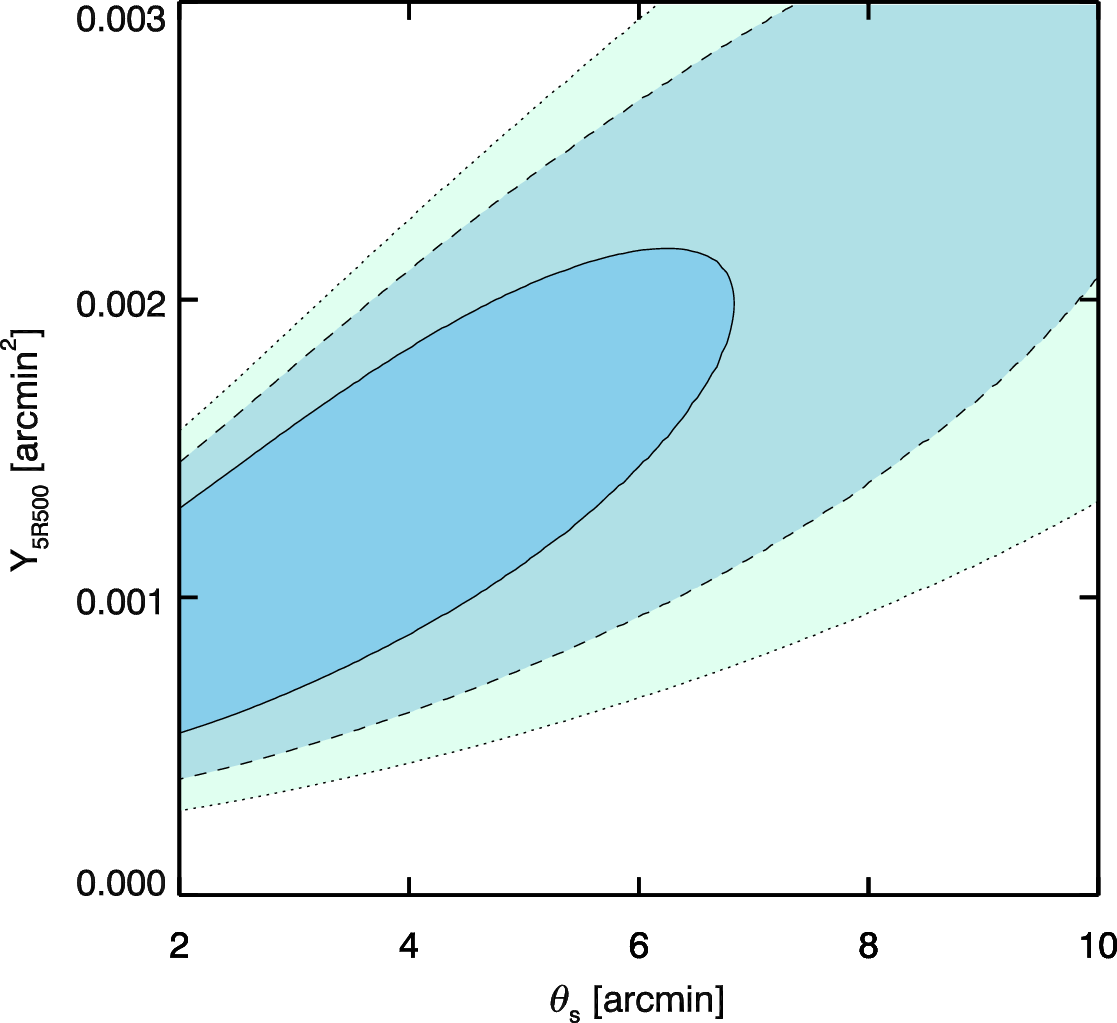}
\caption{Illustration of the SZ size--flux degeneracy for two
 clusters detected by \Planck. \emph{Right:} Abell 2163 (S/N $=27$)
 and \emph{left:} PSZ1 G266.6-27.3 (S/N $=6$ at $z\simeq 1$). The
 contours show the 68, 95, and 99 percent confidence levels.}
\label{fig:dege}
\end{center}
\end{figure*}

\subsubsection{PowellSnakes}

PowellSnakes (\texttt{PwS}) is different from the \texttt{MMF}
methods. It is a fast Bayesian multi-frequency detection algorithm
designed to identify and characterize compact objects buried in a
diffuse background. The detection process is grounded in a statistical
model comparison test.  The statistical foundations of \texttt{PwS}
are described in \citet{car09}, and more recently in \citet{car11b}
with a greater focus on the \Planck\ setup.  \texttt{PwS} may be run
either based on a Generalized Likelihood Ratio Test or in full
Bayesian mode. This duality allows \texttt{PwS} measured quantities to
be consistently compared with those of the \texttt{MMF} algorithms.

\texttt{PwS} also operates in square flat patches of $14.66^{\circ}\times
14.66^{\circ}$.  The total number of patches employed, of order
2800, varies with sky area but always guarantees a very large overlap;
on average each cluster is detected about $4.7$ times.  \texttt{PwS}
detects putative clusters and at the same time it computes the
evidence ratio and samples from the posterior distributions of the
cluster parameters.  Then, it merges all intermediate sub-catalogues
and applies the criterion of acceptance/rejection \citep{car11b}.
\texttt{PwS} computes the cross-channel covariance matrix directly
from the pixel data. To reduce the contamination of the background by
the SZ signal itself, the estimation of the covariance matrix is
performed iteratively.  After an initial estimate, all detections in
the patch with S/N higher than the current target detection are
subtracted from the data using their best-fit values and the
cross-channel covariance matrix is re-estimated.  This is \texttt{PwS}
`\emph{native}' mode of background estimation that produces, on
average, an S/N estimate about $20\%$ higher than
\texttt{MMF}. However, in order to produce a homogeneous \Planck\ SZ
catalogue from the three algorithms, it is possible to run
\texttt{PwS} in `\emph{compatibility}' mode, skipping the
re-estimation step to mimic more closely the evaluation of the
background noise cross-power spectrum of the \texttt{MMF} algorithms
and thus their evaluation of the S/N.  In this mode, \texttt{PwS} is a
maximum likelihood estimator like the \texttt{MMF}.

In the following, unless stated otherwise, all quoted or plotted S/N
values from \texttt{PwS} are obtained in `\emph{compatibility}' mode in order
to ensure homogeneity across the catalogue entries and in order to
ease the comparison with the \texttt{MMF} outputs.

\subsection{Outputs of the detection methods}\label{sec:dege}
Each of the three detection algorithms outputs a catalogue of SZ
detections above S/N $=4.5$ outside the highest-emitting Galactic
regions (this corresponds to a mask of about 15\% of the sky, see
masked area in Fig.~\ref{fig:sz_dist}) and the Small and Large
Magellanic Clouds and outside the PS mask described in
Sect.~\ref{ss_preprocessing}. The union PS-Galactic mask covers 16.3\%
of the sky.  The survey area used for the SZ detections in \Planck\ is
thus 83.7\% of the sky coverage.  The three individual lists of SZ
candidates are cleaned by removal of obvious false detections. These
are spurious sources that pass the \texttt{MMF} and \texttt{PwS}
filters despite the pre-processing step applied to the
\Planck\ channel maps, see Sect.~\ref{ss_preprocessing}. In order to
identify them, we cross-match the SZ detections with an intermediate,
low S/N cut of $4$, catalogue of point sources detected at
the highest frequencies of \Planck. Galactic sources in dense and cold
regions at high latitudes also contaminate the SZ detections outside
the Galactic mask. These cold Galactic sources \citep[CGS,
  see ][]{planck2011-7.7b,planck2011-7.7a} are detected in the
\Planck\ channel maps following an optimized method proposed by
\citet{mon10}. The SZ detections matching with PS at both 545 and
857~GHz, or with CGS sources, all show a rising spectrum at high
frequencies, indicating that they are false detections. The SZ
detections corresponding to such PCCS or CG sources are removed from
the individual lists and from the published \Planck\ catalogue of SZ
sources.

\begin{figure}
\centering
\includegraphics[width=8.8cm]{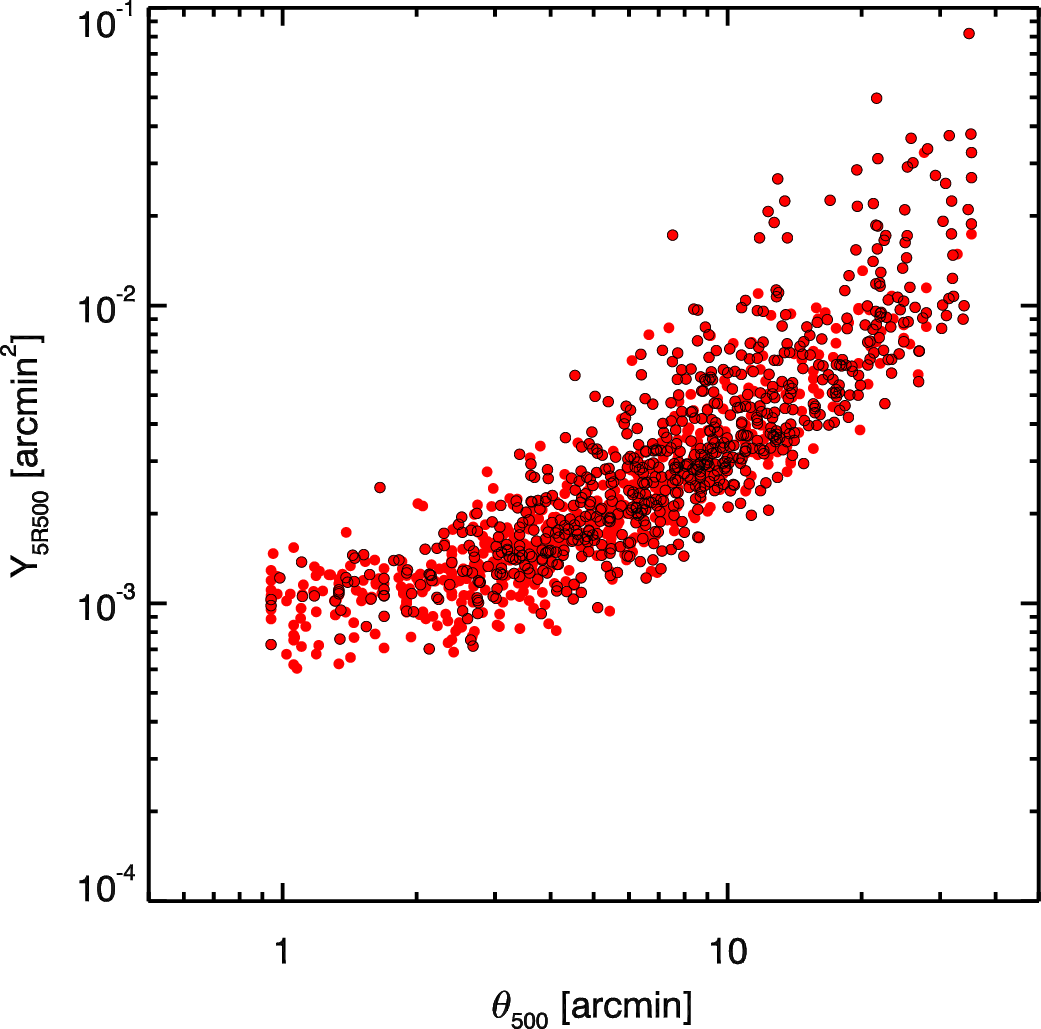}
\caption{Distribution of the maximum likelihood SZ flux $\Yall$ and
  size $\theta_{500}$ for \Planck\ SZ detections in the union
  catalogue down to S/N
  $=4.5$. Detections associated with known or new confirmed clusters
  are shown as open black circles. SZ cluster candidates are shown as
  filled red circles.}
\label{fig:y_th_distrib}
\end{figure}

The three detection algorithms used in the present study deploy the
GNFW cluster profile to detect SZ signal with the two parameters of
the shape function, the central value and the characteristic scale
$\theta_{\mathrm{s}}$ let free, with $\theta_{\mathrm s} =
\theta_{500}/c_{500}$. Each of the three algorithms therefore assigns,
to each detected SZ candidate, a position with estimated uncertainty,
a S/N value, and an estimated size, $\theta_{\mathrm{s}}$
or equivalently $\theta_{500}$, with its uncertainty.  The detection
likelihood or the posterior probability of the integrated Compton
parameter within 5$\theta_{500}$, denoted $\Yall$, exhibits a large
correlation with the size. Figure~\ref{fig:dege} illustrates the
likelihood plots for two cases: a spatially-resolved cluster detected
with a high S/N, Abell~2163; and a non-resolved cluster at
high redshift ($z\simeq 1$), PSZ1~G266.6-27.3 \citep[also known as
 PLCK~G266.6-27.3 in][]{Planck2011-5.1c}.  We also show in
Fig.~\ref{fig:y_th_distrib} the distribution of maximum likelihood SZ
fluxes ($\Yall$) and sizes ($\theta_{500}$) for the \texttt{MMF3}
detections.

This ``degeneracy'' between cluster size and SZ flux propagates the
size uncertainty to the SZ flux estimate, increasing and biasing its
value dramatically. This effect being so detrimental, both the SZ
blind flux and size best-fit estimates, and respective error bars, are
not quoted in the catalogue outputs to avoid their misuse. Only the
full joint $\Yall$--$\,\theta_{\mathrm{s}}$, or equivalently
$\Yall$--$\,\theta_{500}$, posterior probability contours provide a
complete description of the information output by each detection
method. They are thus provided for each detection.  In order to use
the flux measure, one ought to break the size--flux degeneracy. This
can be achieved by a joint analysis with a high-resolution observation
of the same objects, or by assuming a prior on, or fixing, the cluster
size e.g., to the X-ray size.  The SZ signal can then be re-extracted
with an uncertainty much smaller than the variation of the joint
$Y$--$\theta$ probability distribution.

\begin{figure*}[t]
\begin{center}
\includegraphics[width=18cm]{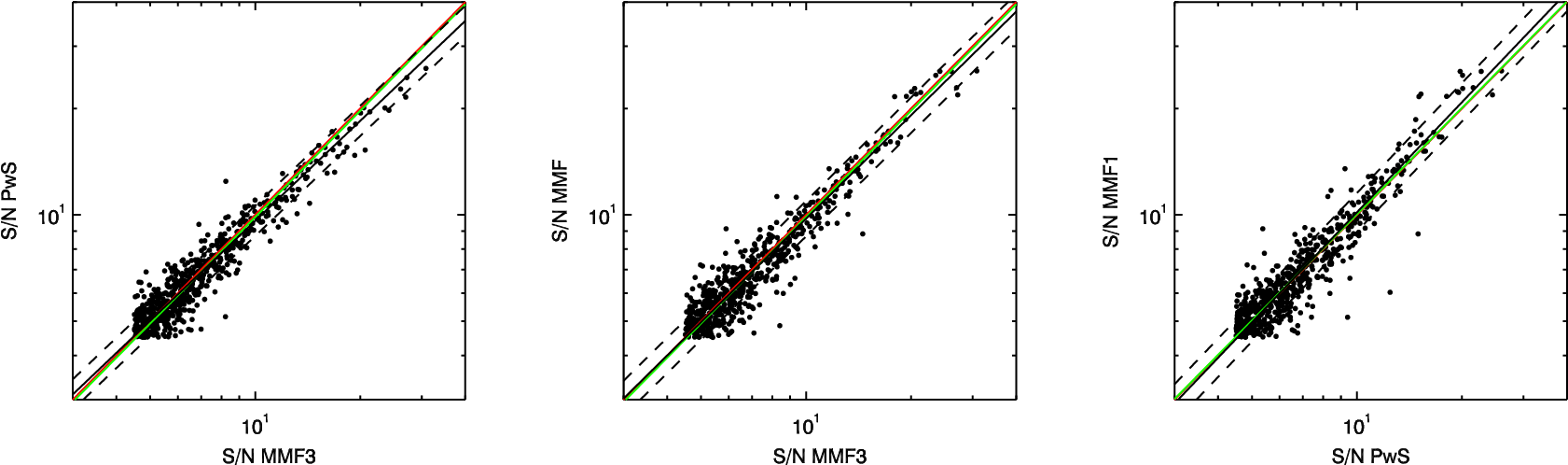}\\[1em]
\includegraphics[width=18cm]{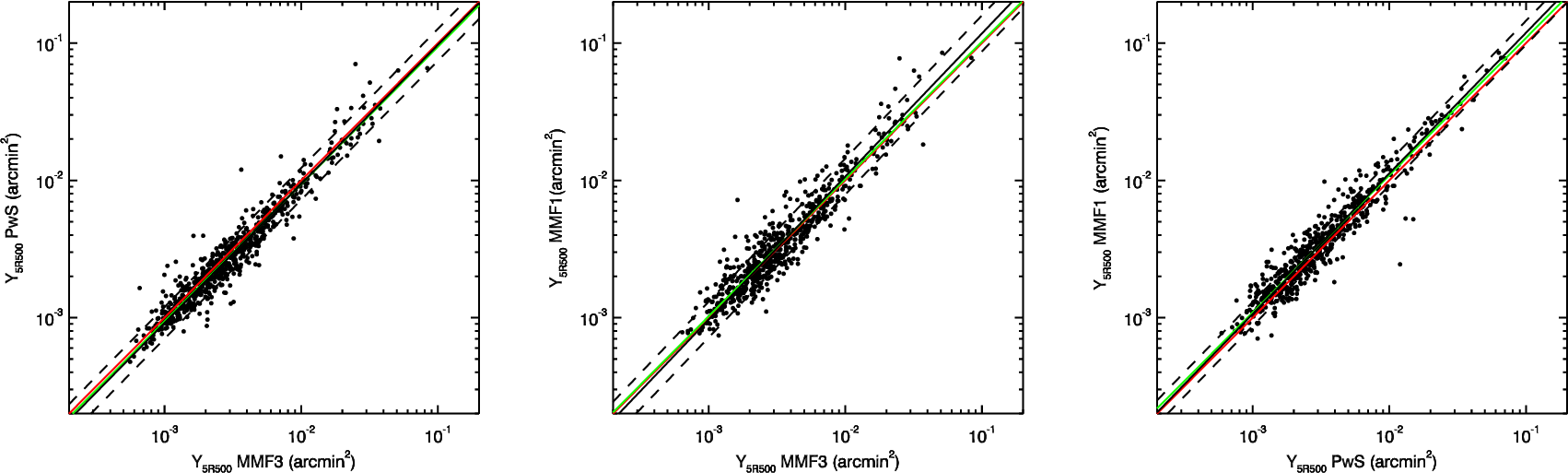}
\caption{Comparison of S/N (top panels) and maximum likelihood
  Compton-parameter values (bottom panels) from the three detection
  algorithms, \texttt{MMF}s and \texttt{PwS}, down to S/N $=4.5$ after
  removing obvious false detections (see Sect.~\ref{sec:dege}). In
  each panel, the red line denotes the equality line. The black line
  is the best fit to the data, and the dashed lines correspond to the
  $\pm 1\sigma$ dispersion about the fit relation. For clarity, error
  bars are omitted on $\Yall$ values in the plot, but are taken into
  account in the fit. The green line of slope fixed to unity
  corresponds to the mean offset between the two quantities. Numerical
  results for the fits are given in Table~\ref{tab:compysnr}}
\label{fig:compysnr}
\end{center}
\end{figure*}

We now perform a systematic comparison of the outputs of the three
algorithms and we compare the S/N. In addition and for purposes of
illustration, we compare the best-fit blind $Y$ value from
maximum-likelihood or posterior probability outputs, namely
$\Yall$.\footnote{$\Yall$ can be rescaled to $\Yv$ for the fiducial
 GNFW model as $\Yall=1.79\times \Yv$ \citep{arn10}.}  We show the
comparison in Fig.~\ref{fig:compysnr}, considering detections down to
S/N $= 4.5$. We quantify the difference between a given quantity
estimated by two different algorithms, $Q_2$ and $Q_1$, by fitting a
power law to the data in the form $Q_2/Q_{\mathrm{p}} =
10^A\,(Q_1/Q_{\mathrm{p}})^\alpha $ with a pivot $Q_{\mathrm{p}}=6$
for S/N and $Q_{\mathrm{p}}=4\times 10^{-3}$arcmin$^2$ for $\Yall$.
The results are given in Table~\ref{tab:compysnr}, including the
scatter estimates. The raw scatter was estimated using the
error-weighted vertical distances to the regression line. The
intrinsic scatter on $\YSZ$ was computed from the quadratic difference
between the raw scatter and that expected from the statistical
uncertainties. Table~\ref{tab:compysnr} also lists the mean difference
in logarithm, $\Delta(\log Q)$=$\log(Q_2/Q_1)$, computed taking into
account both statistical errors and intrinsic scatter, estimated
iteratively.

\begin{table*}[t]
\begingroup
\caption{{Parameters of the fitted lines in Fig.~\ref{fig:compysnr}.
  The function $Q_2/Q_{\mathrm{p}} =
  10^A\,(Q_1/Q_{\mathrm{p}})^\alpha$ is fitted using BCES orthogonal
  regression, with pivot $Q_{\mathrm{p}}=6$ for S/N and
  $Q_{\mathrm{p}}=4\times 10^{-3}$ arcmin$^2$ for $\Yall$.  The
  intrinsic and raw scatter (see text) around the fit are given by
  $\sigma^{\mathrm{log}}_{\mathrm{int}}$ and $\sigma^{\mathrm{log}}_{\mathrm{raw}}$.  The
  mean offset is given by $\Delta\log{Q}$=$\log(Q_2/Q_1)$.}}
\label{tab:compysnr}
\nointerlineskip
\vskip -3mm
\footnotesize
\setbox\tablebox=\vbox{
 \newdimen\digitwidth 
 \setbox0=\hbox{\rm 0} 
 \digitwidth=\wd0 
 \catcode`*=\active 
 \def*{\kern\digitwidth}
 \newdimen\signwidth 
 \setbox0=\hbox{+} 
 \signwidth=\wd0 
 \catcode`!=\active 
 \def!{\kern\signwidth}
\halign{\hbox to 1.2in{#\leaderfil}\tabskip 2em&
    \hfil$#$\hfil\tabskip=0.8em&
    \hfil$#$\hfil&
    \hfil$#$\hfil&
    \hfil$#$\hfil\tabskip=2em&
    \hfil$#$\hfil\tabskip=0.8em&
    \hfil$#$\hfil&
    \hfil$#$\hfil\tabskip=0pt\cr 
\noalign{\doubleline}
\omit&\multispan4\hfil Power-law\hfil&\multispan3\hfil Offset\hfil\cr
\noalign{\vskip -3pt}
\omit\hfil Quantity and\hfil&\multispan4\hrulefill&\multispan3\hrulefill\cr
\omit\hfil Algorithms\hfil&A&\alpha&\sigma^{\mathrm{log}}_{\mathrm{int}}&\sigma^{\mathrm{log}}_{\mathrm{raw}}&\Delta\log{Q}&\sigma^{\mathrm{log}}_{\mathrm{int}}&\sigma^{\mathrm{log}}_{\mathrm{raw}}\cr
\noalign{\vskip 3pt\hrule\vskip 5pt}
\omit S/N\hfil\cr
\hglue 2em{\tt MMF3}-{\tt PwS}& -0.003\pm0.002& 0.94\pm0.01& 0.043\pm0.002&        \dots& -0.006\pm0.002& 0.045\pm0.002&\dots\cr 
\hglue 2em{\tt MMF3}-{\tt MMF1}&-0.005\pm0.002& 0.97\pm0.01& 0.050\pm0.002&        \dots& -0.006\pm0.002& 0.051\pm0.002&\dots\cr 
\hglue 2em{\tt PwS}-{\tt MMF1}& -0.000\pm0.002& 1.04\pm0.02& 0.054\pm0.003&        \dots& +0.002\pm0.002& 0.054\pm0.002&\dots\cr 
\noalign{\vskip 7pt}
\omit$\Yall$\hfil\cr 
\hglue 2em{\tt MMF3}-{\tt PwS}&  -0.030\pm0.004& 1.01\pm0.01& 0.08\pm0.03& 0.116\pm0.018& -0.027\pm0.004& 0.065\pm0.006& 0.102\cr 
\hglue 2em{\tt MMF3}-{\tt MMF1}& +0.011\pm0.005& 1.04\pm0.02& 0.11\pm0.02& 0.131\pm0.014& +0.010\pm0.005& 0.085\pm0.006& 0.118\cr 
\hglue 2em{\tt PwS}-{\tt MMF1}&  +0.041\pm0.004& 1.02\pm0.01& 0.04\pm0.01& 0.088\pm0.005& +0.038\pm0.004& 0.040\pm0.007& 0.079\cr 
\noalign{\vskip 3pt\hrule\vskip 3pt}}}
\endPlancktablewide
\endgroup
\end{table*}

\begin{figure*}[t]
\begin{center}
\includegraphics[width=16cm]{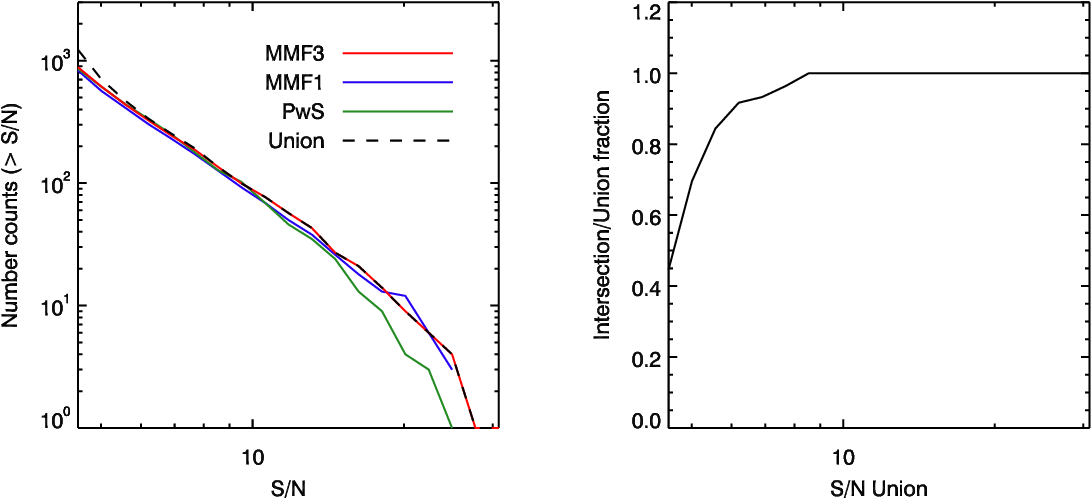}
\caption{{\it Left:} detection number counts as a function of S/N of
  the individual algorithms. The S/N value in the union catalogue is
  that of the \texttt{MMF3} detections when available, followed by
  that of \texttt{PwS} followed by \texttt{MMF1} (see
  Sect.~\ref{sec:psz_def}). See text for discussion on the lower S/N
  values of \texttt{PwS} compared to the \texttt{MMF}-based
  algorithms. {\it Right:} fraction of common detection over counts
  from the union catalogue. Sources with S/N $>8.5$ are detected by
  all methods.}
\label{fig:count}
\end{center}
\end{figure*}

\subsubsection{Signal-to-noise}

A crucial ingredient of the SZ detection algorithms, either the
\texttt{MMF}s or \texttt{PwS}, is the background cross-power spectrum
used to estimate the noise level. It is evaluated from the data
locally on a per-patch basis (see Fig.~\ref{fig:noise_map} for an
example of the noise per patch across the sky). The algorithms, and
implementations, slightly differ with respect to the stabilization
assumptions (e.g., smoothing) of the background noise cross-power
spectrum and to the treatment of the background SZ signal, now acting
as a contaminant.  These differences translate into variations in the
S/N values per method. In particular, when operated in
``\emph{compatibility}'' mode (without background cluster
subtraction), \texttt{PwS} estimation of the background cross-power
spectrum is more affected than the \texttt{MMF} by SZ signal
contamination. The SZ signal adds an extra component to the background
noise producing lower S/N estimates. This is particularly noticeable
when the SZ signal is very strong compared with background (typically
S/N $\ge 15$).

Despite the differences in background estimates, the yields from the
three algorithms agree.  In the left panel of Fig.~\ref{fig:count}, we
show that the detection counts as a function of S/N for each detection
method are in good overall agreement. The right panel of
Fig.~\ref{fig:count} shows the fraction of common detections over the
union of detections from all three algorithms as a function of
S/N. Sources with S/N $>8.5$ are detected by all three methods.
However, we note that \texttt{PwS} number counts decrease more rapidly
than \texttt{MMF} counts above S/N $=15$. This reflects the behaviour
of \texttt{PwS} in ``\emph{compatibility}'' mode described above, which
estimates a higher background than the \texttt{MMF} methods at high
S/N. Figure~\ref{fig:compysnr} shows the comparison of the S/N
estimates from all three methods. The agreement is good on
average. The mean ratio (or the normalization at the pivot of the
power-law relation) deviates from unity by less than $2\%$ and at less
than $3\,\sigma$ significance. Here again at high S/N values, we note
the tendency for lower S/N in \texttt{PwS} as compared to \texttt{MMF}
(Fig.~\ref{fig:compysnr}), and indeed the slope of the power-law
relation is smaller than unity ($\alpha=0.94\pm0.01$ for
\texttt{MMF3}).

\subsubsection{Photometry}

\begin{figure}[t]
\begin{center}
\includegraphics[width=8.8cm]{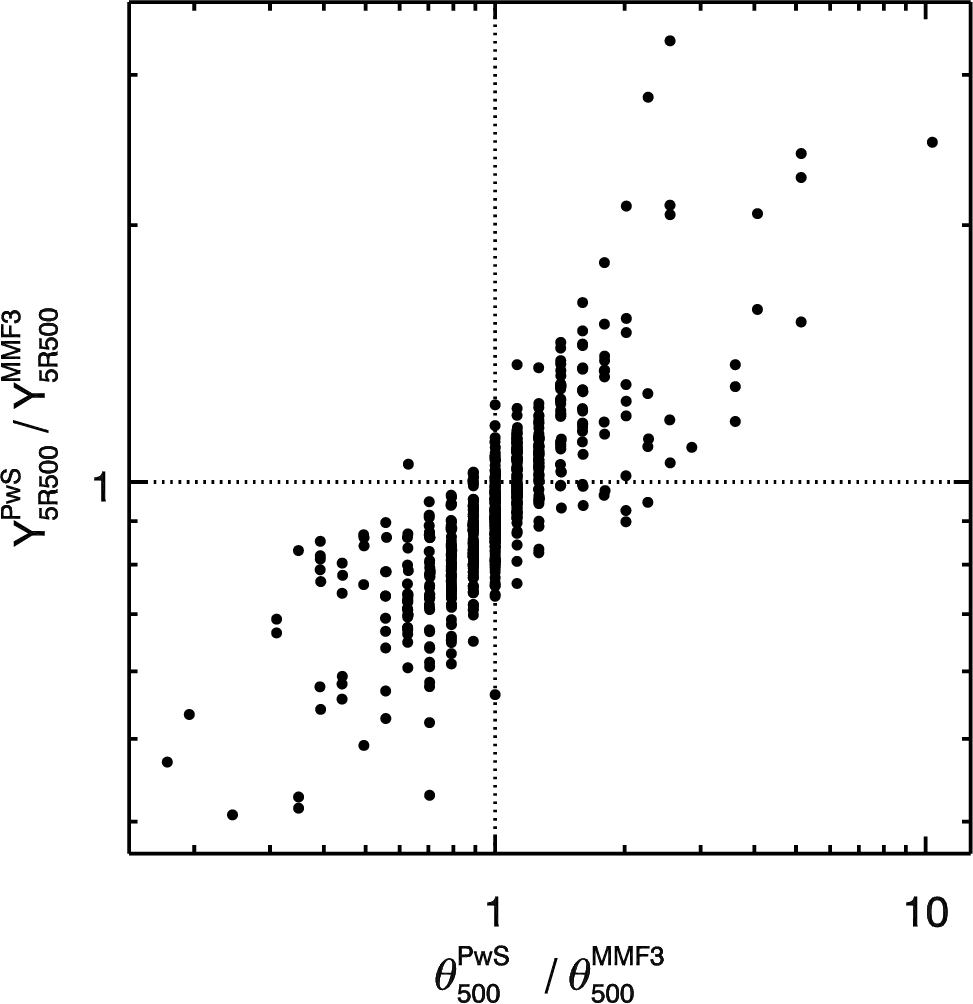}
\caption{Correlation between the ratio of $\Yall$ estimates with
 \texttt{PwS} and \texttt{MMF3} and the ratio of size estimates,
 shown on a grid of sizes.}
\label{fig:compysize}
\end{center}
\end{figure}

We now compare the best-fit $Y$ values (from maximum likelihood and
posterior probability) for the three detection algorithms. The
comparison (Fig.~\ref{fig:compysnr}, lower panels) shows a systematic
bias with \texttt{PwS}, yielding slightly smaller values than
\texttt{MMF}, typically by $10\%$.  However, the slope is consistent
with unity, showing that this bias is not flux dependent.  The
\texttt{MMF} values differ from each other by less than $3\%$ on
average. The scatter between $Y$ estimates is dominated by the
intrinsic scatter (Table~\ref{tab:compysnr}).  It is clearly related
to the size--flux degeneracy, the ratio between $Y$ estimates for a
given candidate being correlated with the size estimate ratio, as
illustrated by Fig.~\ref{fig:compysize}.  The scatter becomes
compatible with the statistical scatter when a prior on the size is
used, e.g., size fixed to the X-ray size.

\subsection{Definition of the \planck\ SZ catalogue}\label{sec:psz_def}

As discussed above, the processing details of each
algorithm/implementation differ in the computation of the background
noise. The significance of the detections in terms of S/N, although in
overall agreement, differs from one algorithm to the other and
translates into different yields for the candidate lists from the
three algorithms. We choose to construct a catalogue of SZ candidates
that ensures, through redundant detections, an increased reliability
of the low S/N sources, when they are detected by two methods at
least, together with maximizing the yield of the catalogue.

\begin{table*}[tmb]
\begingroup
\newdimen\tblskip \tblskip=5pt
\caption{{Characteristics of the catalogues.  The union catalogue
    contains SZ detections found by at least one of the three
    extraction algorithms; the intersection catalogue contains
    detections found by all three.  $Y_{500}$ at a given completeness
    {C} is estimated by marginalizing over $\theta_{500}$,
    weighting each ($Y_{500},\theta_{500}$) bin by the
    theoretically-expected cluster counts.  Position error is the
    median angular separation between real and estimated positions.}}
\label{qa:summary_table}
\nointerlineskip
\vskip -3mm
\footnotesize
\setbox\tablebox=\vbox{
 \newdimen\digitwidth 
 \setbox0=\hbox{\rm 0} 
 \digitwidth=\wd0 
 \catcode`*=\active 
 \def*{\kern\digitwidth}
 \newdimen\signwidth 
 \setbox0=\hbox{+} 
 \signwidth=\wd0 
 \catcode`!=\active 
 \def!{\kern\signwidth}
\halign{\hbox to 1.2in{#\leaderfil}\tabskip 1em&
    \hfil#\hfil\tabskip=2em&
    \hfil#\hfil\tabskip=0.5em&
    \hfil#\hfil&
    \hfil#\hfil\tabskip=2.2em&
    \hfil#\hfil\tabskip=0pt\cr 
\noalign{\doubleline}
\omit&&\multispan3\hfil $Y_{500}$\,[$10^{-3}{\mathrm{arcmin}}^{2}$]\hfil\cr
\noalign{\vskip -3pt}
\omit&&\multispan3\hrulefill\cr
\omit\hfil Catalogue\hfil&Reliability[\%]&\textbf{C}=50\%&\textbf{C}=80\%& \textbf{C}=95\%&Position error\cr 
\noalign{\vskip 3pt\hrule\vskip 5pt}
Union& 84& 0.61& 1.2& 3.2& 1\parcm2\cr
Intersection& 98& 0.85& 1.8& 6.6& 1\parcm1\cr 
\texttt{MMF1}& 87& 0.75& 1.6& 4.7& 1\parcm2\cr
\texttt{MMF3}& 91& 0.71& 1.5& 3.8& 1\parcm2\cr 
\texttt{PwS}& 92& 0.65& 1.4& 3.2& 0\parcm9\cr
\noalign{\vskip 3pt\hrule\vskip 3pt}}}
\endPlancktablewide          
\endgroup
\end{table*}

The \Planck\ SZ cluster catalogue described in the following is thus
constructed from the union of the cleaned SZ-candidate lists produced
at S/N $\ge 4.5$ by all three algorithms.  It contains in total 1227
SZ detections above S/N $= 4.5$. Note that in order to ensure
homogeneity, in terms of detection significance, the S/N values of
\texttt{PwS} quoted in the union catalogue are obtained in
\emph{compatibility} mode, whereas the S/N obtained from \texttt{PwS}
\emph{native} mode are quoted in the \texttt{PwS} individual list.
The union catalogue is constructed by merging detections from the
three methods within an angular separation of at most 5$^{\prime}$, in
agreement with \planck\ position accuracy shown later in
Fig.~\ref{qa:position_recovery}.  As mentioned, no reference
photometry is provided. However a reference position for the SZ
detection is needed. For compatibility with the ESZ \Planck\ sample,
in the case of matching detection between methods we arbitrarily
choose to take the coordinates from the \texttt{MMF3} detection as the
fiducial position (\texttt{MMF3} was the reference method used to
construct the ESZ \Planck\ sample). When no detection by \texttt{MMF3}
above S/N $=4.5$ is reported, we took the \texttt{PwS} coordinates as
fiducial, and the \texttt{MMF1} coordinates elsewhere. The S/N values
in the union catalogue are taken following the same order, which
explains why the \texttt{MMF3} curve in Fig.~\ref{fig:compysnr}
coincides with the union curve.  The cluster candidates in the union
catalogue are cross-referenced with the detections in the individual
lists.  The reference positions and the S/N values are reported in the
union catalogue. Given the size--flux degeneracy, the full information
on the degeneracy between size and flux is provided with each
individual list in the form of the two-dimensional marginal
probability distribution for each cluster candidate as discussed
above. It is specified on a grid of $256 \times 256$ values in
$\theta_{\mathrm{s}}$ and $\Yall$ centred at the best-fit values found by each
algorithm for each SZ detection.

An extract of the \planck\ SZ catalogue is given in
Appendix~\ref{ap:cat_cont}. The full online table for union
\planck\ catalogue, the individual lists of SZ detections, and the
union mask used by the SZ-finder algorithms together with comments
assembled in an external file are available at ESA's \planck\ Legacy
Archive (PLA).\footnote{\url{http://www.sciops.esa.int/index.php?page=}
  \\ \url{Planck\_Legacy\_Archive\&project=planck}.} 

\section{Statistical Characterization}\label{sec:qa}

The statistical characterization of the PSZ catalogue is achieved
through a process of Monte Carlo quality assessment (MCQA) that can
be applied to each individual catalogue and to the merged union
catalogues.  The statistical quantities produced include completeness,
fraction of detections associated with true clusters called,
statistical reliability or purity, positional accuracy, and accuracy
of parameter estimation. Together, these statistics describe the
quality of detections in the catalogue. The quality of the parameter
estimation, including astrometry (cluster position and extent), is
determined through comparison with the parameters of the input
clusters.  The statistical characteristics of the different lists are
summarized in Table~\ref{qa:summary_table}.

\subsection{MCQA Pipeline and simulations} \label{qa: sims section}

The MCQA pipeline contains a common segment producing simulated input
catalogues and processed, source-injected maps, which are then fed
into the detection pipeline. In summary, the pipeline steps per
MonteCarlo loop are:
\begin{enumerate}
\item creation of an input cluster catalogue;
\item injection of clusters into common simulated diffuse frequency
maps, including beam convolution;
\item injection of multi-frequency point sources;
\item pre-processing of maps, including masking and filling point sources;
\item detection and construction of individual cluster-candidate catalogues;
\item construction of a union catalogue given merging criteria;
\item collation of input and output
catalogues, producing detection truth-tables and catalogues of
unmatched spurious detections.\footnote{A cluster is considered to be matched
if there is a detection within 5$^{\prime}$ of its position.}
\end{enumerate}

To estimate the completeness, clusters are injected into
the real data. In this case, steps 3 and 4 are skipped and each
detection algorithm estimates noise statistics on the real data prior
to injection in order to avoid artificially raising the S/N and
biasing the completeness estimates.  The pressure profiles of the
injected clusters follow that described in Sect.~\ref{sec:clus_mod}.
To account for the profile variation across the cluster population,
the profile parameters are drawn from the covariance matrix of the 62
measured pressure profiles from \citet{planck2012-V}, ensuring that
the injected profiles are consistent with measured dispersion and
consistent, on average, with the extraction filter.  The injected
clusters are convolved with effective beams in each pixel including
asymmetry computed following \citet{mitra2010}.

The simulated input cluster catalogues differ for statistical
reliability and completeness determination. For completeness, clusters
injected in real data are drawn from a uniform distribution in
$(Y_{500},\theta_{500})$ so as to provide equal statistics in each
completeness bin. To avoid an over-contamination of the signal,
injected clusters are constrained to lie outside an exclusion radius
of $5R_{500}$ around a cluster, either detected in the data or
injected.

For the statistical reliability estimation of the input cluster
distribution injected in simulations is such that cluster masses and
redshifts are drawn from a \citet{tin08} mass function and converted
into the observable parameters $(Y_{500},\theta_{500})$ using the
\Planck\ ESZ $Y_{500}$--$M_{500}$ scaling relation
\citep{planck2011-5.2a}. The simulated maps consist of CMB
realizations, diffuse Galactic components and instrumental noise
realizations, including realistic power spectra and inter-detector
correlations, from the FFP6 simulations
\citep{planck2013-p06,planck2013-p28}. Residual extragalactic point
sources are included by injecting, mock-detecting, masking and filling
realistic multi-frequency point sources using the same process as for
the real data (see Sect.~\ref{ss_preprocessing}).

\subsection{Completeness}

The completeness is the probability that a cluster with given
intrinsic parameters ($Y_{500},\theta_{500}$) is detected given a
selection threshold (here in S/N).

If the Compton-$Y$ estimates are subject to Gaussian errors, the
probability of detection per cluster follows the error function and is
parameterized by $\sigma_{Yi}(\theta_{500})$, the standard deviation
of pixels in the multi-frequency matched-filtered maps for a given
patch $i$ at the scale $\theta_{500}$, the \emph{intrinsic}
Compton $Y_{500}$, and the detection threshold $q$:
\begin{equation}
P\left(d | Y_{500}, \sigma_{Yi}(\theta_{500}), q\right) = \frac{1}{2} \left[1 +
\mathrm{erf}\left(\frac{Y_{500} - q
  \sigma_{Yi}(\theta_{500})}{\sqrt{2}
  \sigma_{Yi}(\theta_{500})}\right)\right]\,,
\label{qa:erf_compl}
\end{equation}
where $\mathrm{erf}(x)= (2/\pi)\int_0^x \exp\left(-t^2\right) dt$ and $d$ is
the Boolean detection state.

The completeness of the catalogue, thresholded at S/N $q$, is expected
to follow the integrated per-patch error function completeness
\begin{equation}
C(Y_{500},\theta_{500}) = \sum_{i} f_{{\mathrm{sky}},i} P\big( d | Y_{500},
\sigma_{Yi}(\theta_{500}),q\big)\,, 
\label{qa:int_erf_compl}
\end{equation}
where $f_{{\mathrm{sky}},i}$ is the fraction of the unmasked sky in
the patch $i$.  The true completeness departs from this theoretical
limit.  This is due to the non-Gaussian nature of the noise dominated
by the astrophysical, namely Galactic, contamination. This is
also the case when the actual cluster pressure profile deviates from
the GNFW used in the SZ-finder algorithms, or when the effective beams
deviate from constant symmetric Gaussians, and also when the detection
algorithm includes extra steps of rejection of spurious sources not
formulated in Eq.~\ref{qa:erf_compl}. This is why an MCQA-based
assessment of the completeness is essential to characterize the
\planck\ detections.

\begin{figure}
\begin{center}
\includegraphics[angle=0, width= 8.8cm]{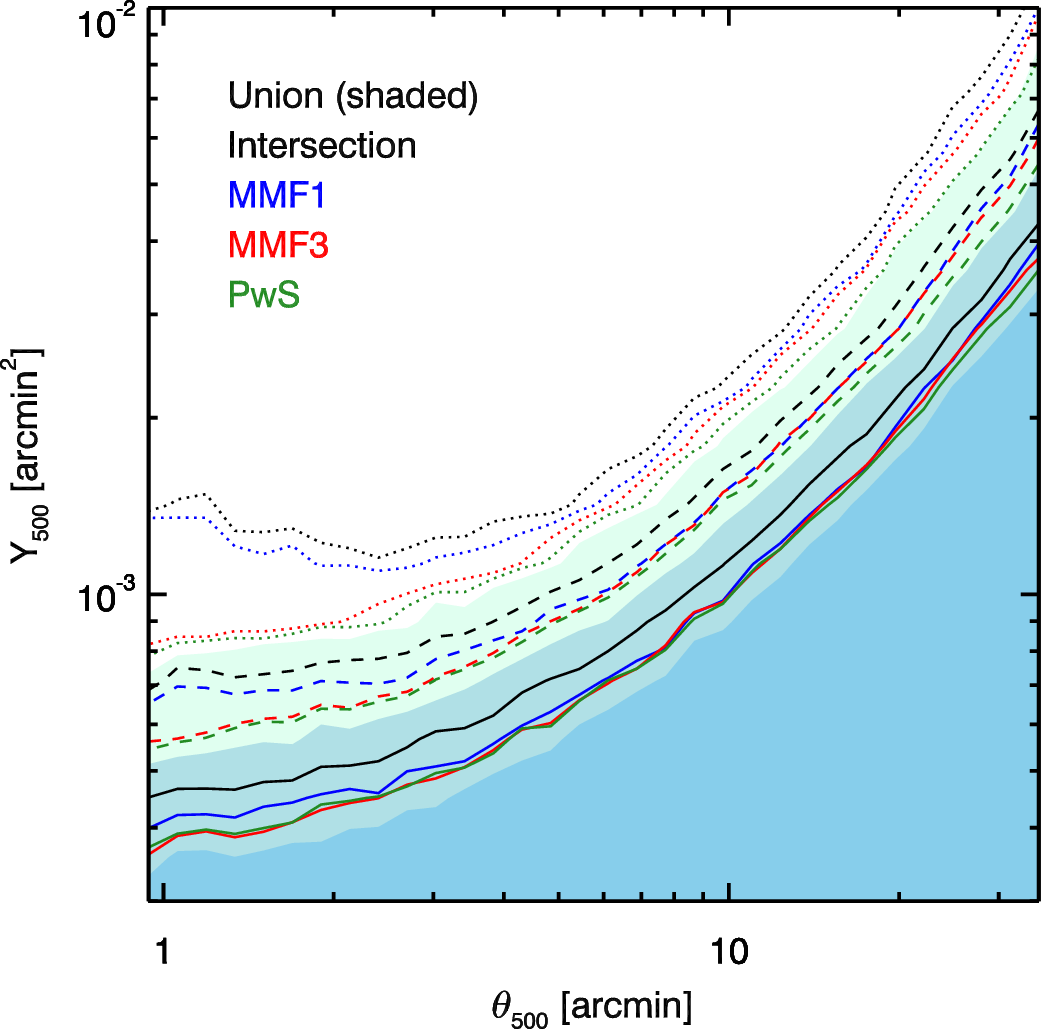}\\
\vspace*{1.5em}
\includegraphics[angle=0, width= 8.8cm]{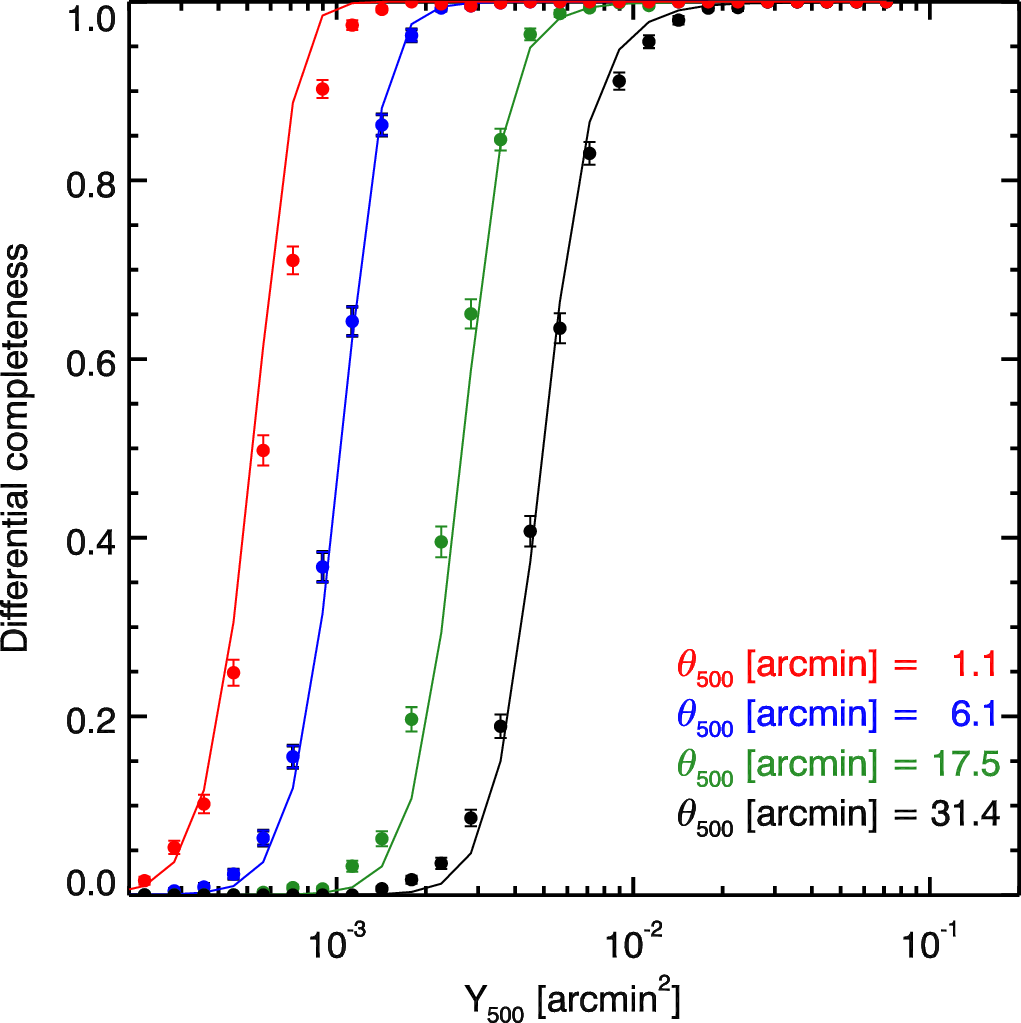}
\caption{\emph{Top panel:} differential completeness as a function of
  ($Y_{500},\theta_{500}$) for each detection algorithm (\texttt{MMF1}
  in blue, \texttt{MMF3} in red, and \texttt{PwS} in green) and for
  the union (shaded area) and intersection (black) catalogues.  From
  bottom to top, the solid, dashed, and dotted lines show 15\%, 50\%
  and 85\% completeness, respectively. \emph{Bottom panel:} slices
  through the MCQA-based completeness function at various
  $\theta_{500}$ for \texttt{MMF3} compared to the error function
  approximation (solid curves). }
\label{qa:compl_plot}
\end{center}
\end{figure}

The MonteCarlo completeness of each of the individual lists and the
union catalogue are shown in Fig.~\ref{qa:compl_plot}.  The
\texttt{MMF} lists are consistent with one another at $\theta_{500} >
4^{\prime}$, but \texttt{MMF3} is more complete at lower radii.  This is
due to an extra step implemented in \texttt{MMF1} that rejects as
spurious the detections estimated to be point-like. The union improves
upon the completeness of each of the individual catalogues, because it
includes the faint real detections by one method alone.  In contrast,
the intersection of the lists from the three algorithms, while more
robust, is markedly less complete than the union and each of the
individual catalogues.  The intersection and union catalogues
represent the extremes of the trade-off between statistical
reliability and completeness.  The quantities for each of the
catalogues, plus the union and intersection, are summarized in
Table~\ref{qa:summary_table}. Figure~\ref{qa:compl_plot} shows four
constant $\theta_{500}$ slices through the completeness contours for
\texttt{MMF3}, comparing the MCQA-based completeness with the
integrated error function completeness.  At radii smaller than 
6$^{\prime}$, the MCQA-based completeness is systematically less complete,
and the drop-off of the completeness function shallower, than the
theoretical expectation.  This effect is a consequence of the
variation of intrinsic cluster profiles from the GNFW profile assumed
for extraction.

\subsection{Statistical reliability}

The fraction of detections above a given S/N that are associated with
a real cluster is characterized by injecting clusters into
high-fidelity simulations of the \Planck\ channels.  Unassociated
detections from these simulations define the fraction of spurious
detections.  We have verified that the simulations produced detection
noise $\sigma_{Y_{500}}$ consistent with the real data and that the
simulated detection counts match the real data.

\begin{figure}
\begin{center}
\includegraphics[angle=0, width= 8.8cm]{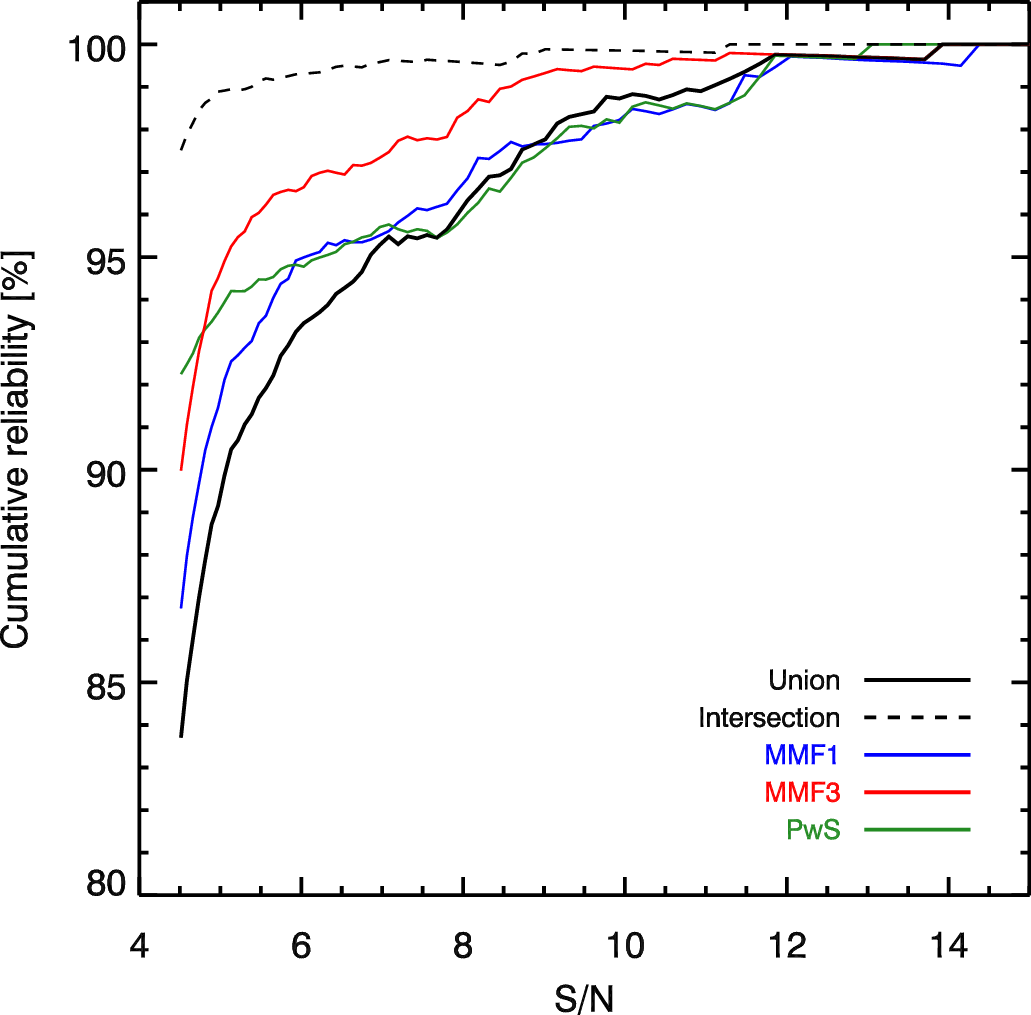}
\caption{Cumulative statistical reliability, defined as the fraction
  of sources above a given S/N associated with a ``real'' cluster from
  the simulated input catalogue. }
\label{qa:purity}
\end{center}
\end{figure}

The cumulative fraction of true clusters, as characterized by the
simulations, is shown for the output of each detection algorithm and
for the union catalogue in Fig.~\ref{qa:purity}.  The union
catalogue is less pure than any of the individual lists because it
includes all the lower-reliability, individual-list detections, in
addition to the more robust detections made by all three SZ-finder
algorithms.  The union catalogue constructed over 83.7\% of the sky at
S/N of 4.5 is 84\% pure. 

\begin{figure}
\begin{center}
\includegraphics[angle=0, width= 8.8cm]{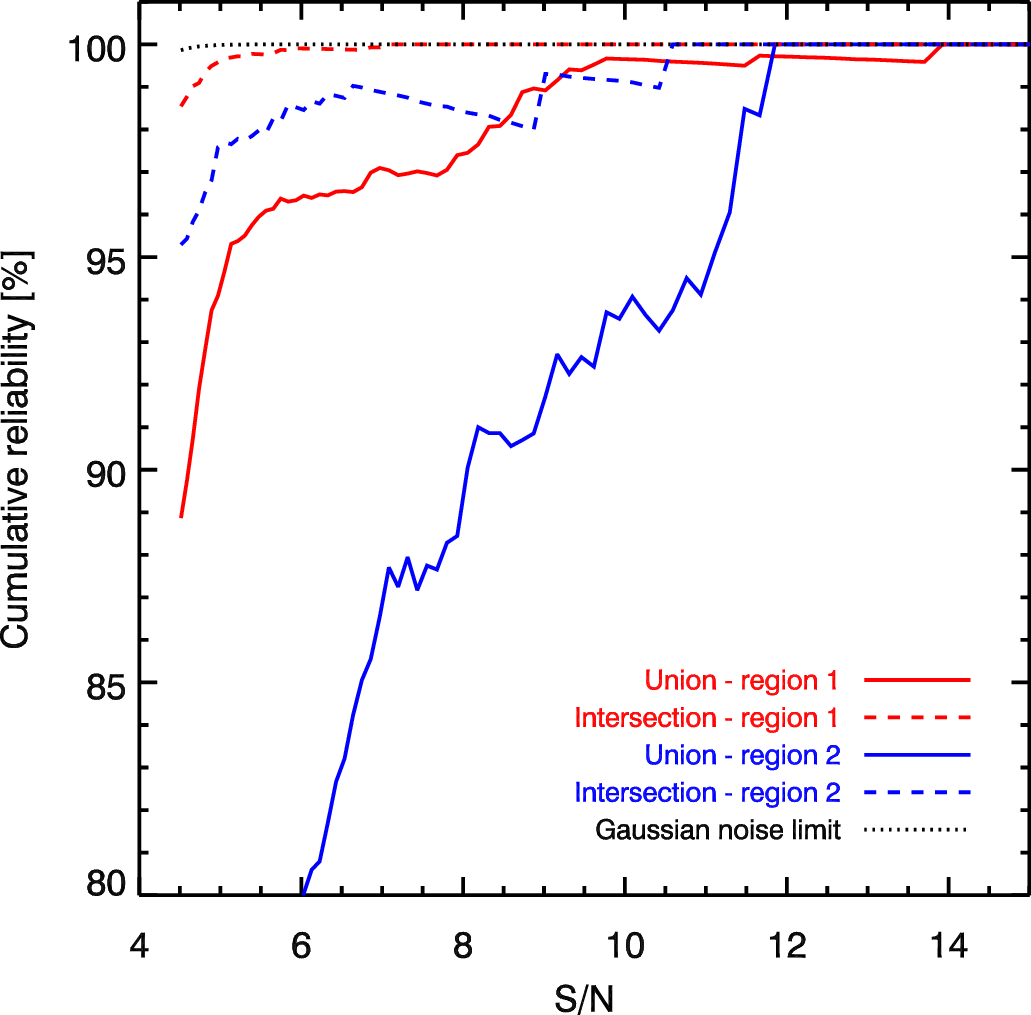}\\[1em]
\includegraphics[angle=0, width= 8.8cm]{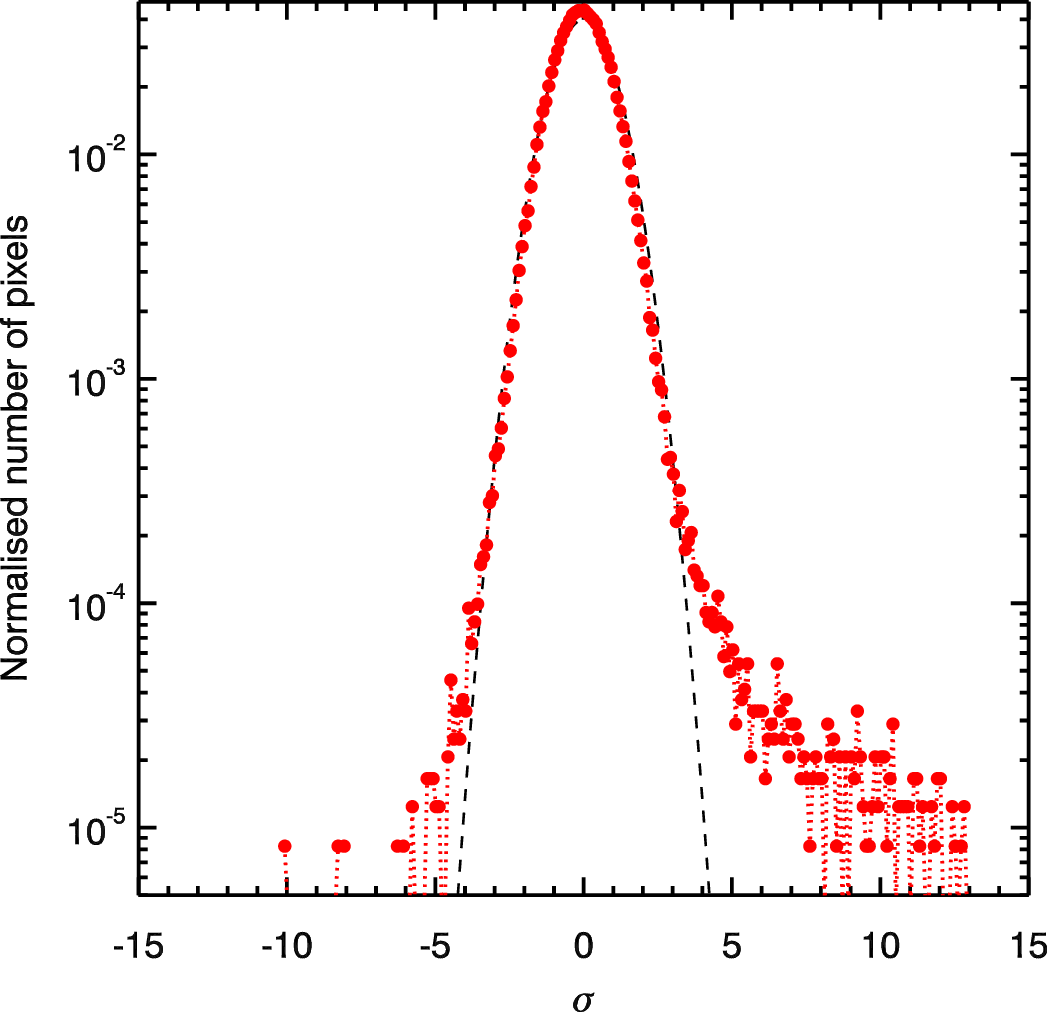}
\caption{\emph{Top panel:} cumulative reliability for the union and
intersection catalogues, as a function of dust contamination.
Region 1 is the low-dust contamination region, being the 65\% of the
sky outside the Galactic dust mask, and region 2 is the
complementary dustier region added to this when the smaller 15\%
dust mask is applied.  The Gaussian noise limit is the expected
reliability from purely Gaussian fluctuations. \emph{Bottom panel:}
histogram of the $y$-signal in a typical filtered patch from a
null-test simulation, compared to the best-fit Gaussian (black
dashed line). The distribution of $y$-noise is non-Gaussian.
\label{qa:non-gaussianity}}
\end{center}
\end{figure}

The fraction of false detections is dominated by systematic foreground
signals, in particular Galactic dust emission. This is illustrated in
Fig.~\ref{qa:non-gaussianity} by the effect of dust contamination on
the cumulative reliability. We define two sky regions by the level of
dust contamination: ``region~1'' is the low dust-contamination region
outside of the \Planck\ Galactic dust, and PS, mask that excludes 35\%
of the sky. This mask is used in \citet{planck2013-p15} for
cosmological analysis of SZ counts.  ``Region~2'' is the complementary
region included by the smaller 15\% dust mask but excluded by the 35\%
mask.  When the larger Galactic dust mask is applied leaving 65\% of
the \planck\ sky survey in which to detect SZ signal, the statistical
reliability increases from 84\% in 83.7\% of the sky to 88\% in 65\%
of the sky. As seen in Fig.~\ref{qa:non-gaussianity} upper panel, the
reliability of the detections deteriorates markedly in ``region~2''
relative to ``region~1''. The noisy behaviour of the curves in
Fig.~\ref{qa:non-gaussianity} upper panel is due to the reduced size
of sky area used in the analysis.

In both regions, the spurious count much higher than is predicted by
Gaussian fluctuations. This reflects the non-Gaussian nature of the
filtered patches.  The bottom panel of Fig.~\ref{qa:non-gaussianity}
illustrates this for a typical mid-latitude patch from a null-test
simulation with no injected clusters.  The patches are well
approximated as Gaussian at deviations smaller than $3\,\sigma$
(consistent with the assumptions of Eq.~\ref{qa:erf_compl}), but show
enhanced numbers of high significance deviations, which can translate
into spurious detections.

\subsection{Positional Accuracy}

\begin{figure}
\begin{center}
\includegraphics[angle=0, width=8.8cm]{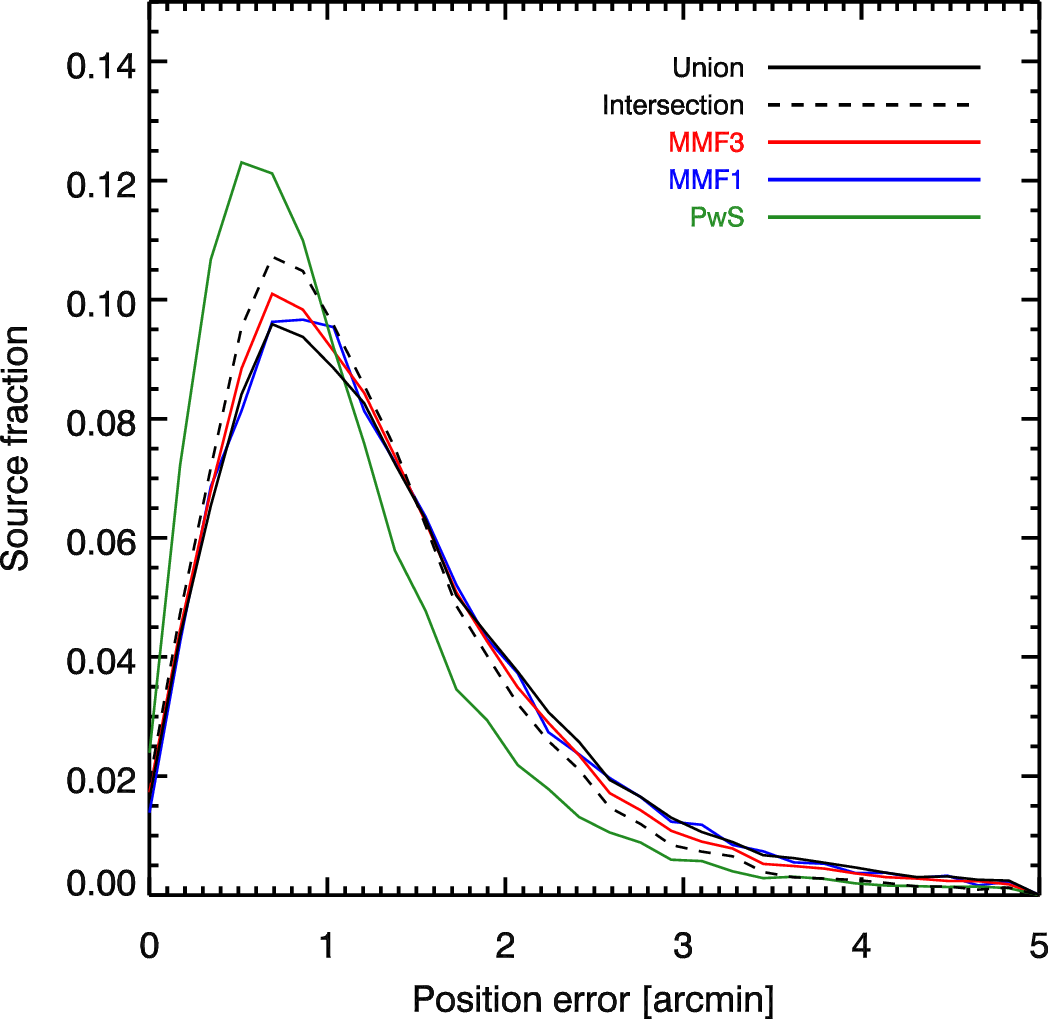}
\caption{Distributions of positional error for each catalogue, normalized
by the total number of detections in the catalogue.  By
construction, the positional error is defined to be less than 5$^{\prime}$.}
\label{qa:position_recovery}
\end{center}
\end{figure}

Positional accuracy is characterized by the radial offset between
estimated and injected positions.  The distribution of position error
is shown in Fig.~\ref{qa:position_recovery}, for each individual list
and the union catalogue.  In contrast to the \texttt{MMF}s, which
estimate the maximum-likelihood position, the \texttt{PwS} position estimator
is the mean of the position posterior, which produces more accurate
positional constraints.  The union catalogue positions are taken from
\texttt{MMF3} if available, followed by \texttt{PwS} and then \texttt{MMF1}.
Its positional estimates are hence consistent with the \texttt{MMF}s.
The mode of the union distribution is consistent with a characteristic
position error scale of half an HFI map pixel ($0.86^{\prime}$).

\subsection{Parameter Recovery}

\begin{figure}
\begin{center}
\includegraphics[angle=0, width=8.8cm]{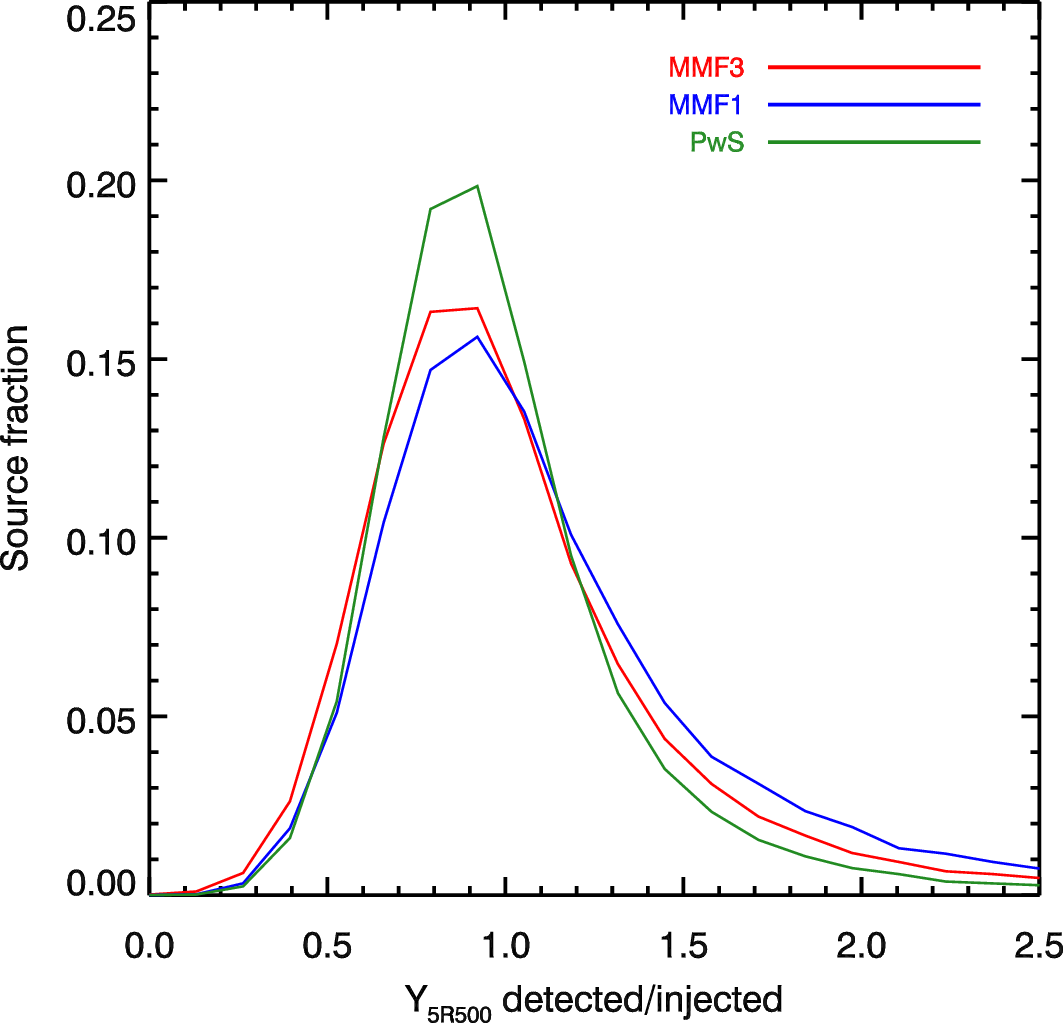}\\
\vspace*{1em}
\includegraphics[angle=0, width=8.8cm]{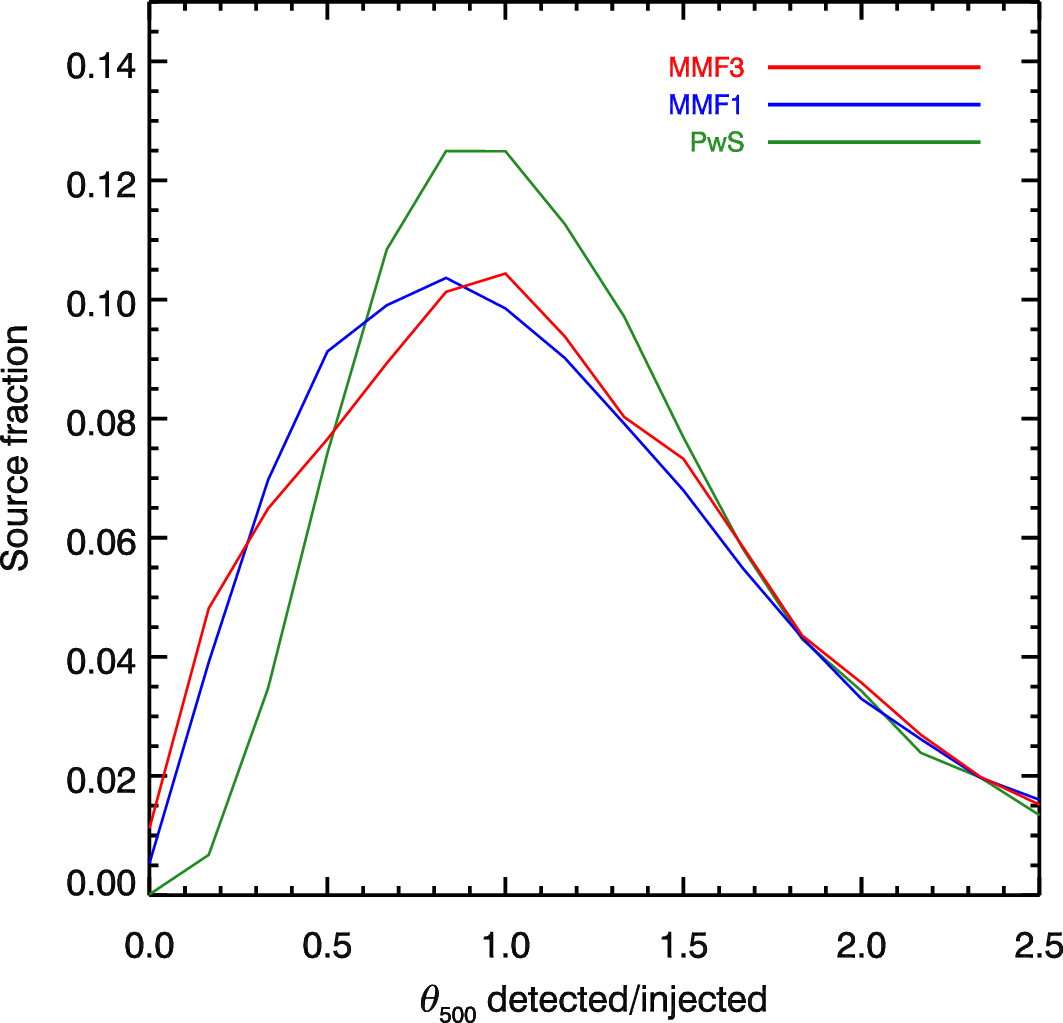}
\caption{
Distributions of the ratio of detected over injected parameters for
$\Yall$ and $\theta_{500}$. }
\label{qa:param_distr}
\end{center}
\end{figure}

The Compton $\Yall$ is characterized by comparing detected and input
values for matched detections from the injection of clusters into the
real data (see Fig.~\ref{qa:param_distr}).  The injection follows the
scheme outlined above with one exception: input cluster parameters are
drawn using the Tinker mass function and the scaling relations
discussed above for reliability simulations. This ensures a realistic
distribution of parameters and S/N values.

What we characterize is slightly different for each catalogue.  For
the \texttt{MMF}s, we characterize the maximum-likelihood point of the 2-D
degeneracy contours provided in the individual lists.  For \texttt{PwS}, we
characterize the mean of the marginal distribution for each parameter.
In each case, the 2D $(\Yall,\theta_\mathrm{s})$ are marginalized over
position.  The contours are scaled for each cluster and are time
consuming to compute, so we characterize the parameters from a
lower-resolution grid that is better suited to Monte-Carlo
analysis.\footnote{\texttt{PwS} does not resort to a low-resolution scale grid
and always works at the full resolution.}

\begin{figure*}
\begin{center}
\includegraphics[angle=0, width=8.8cm]{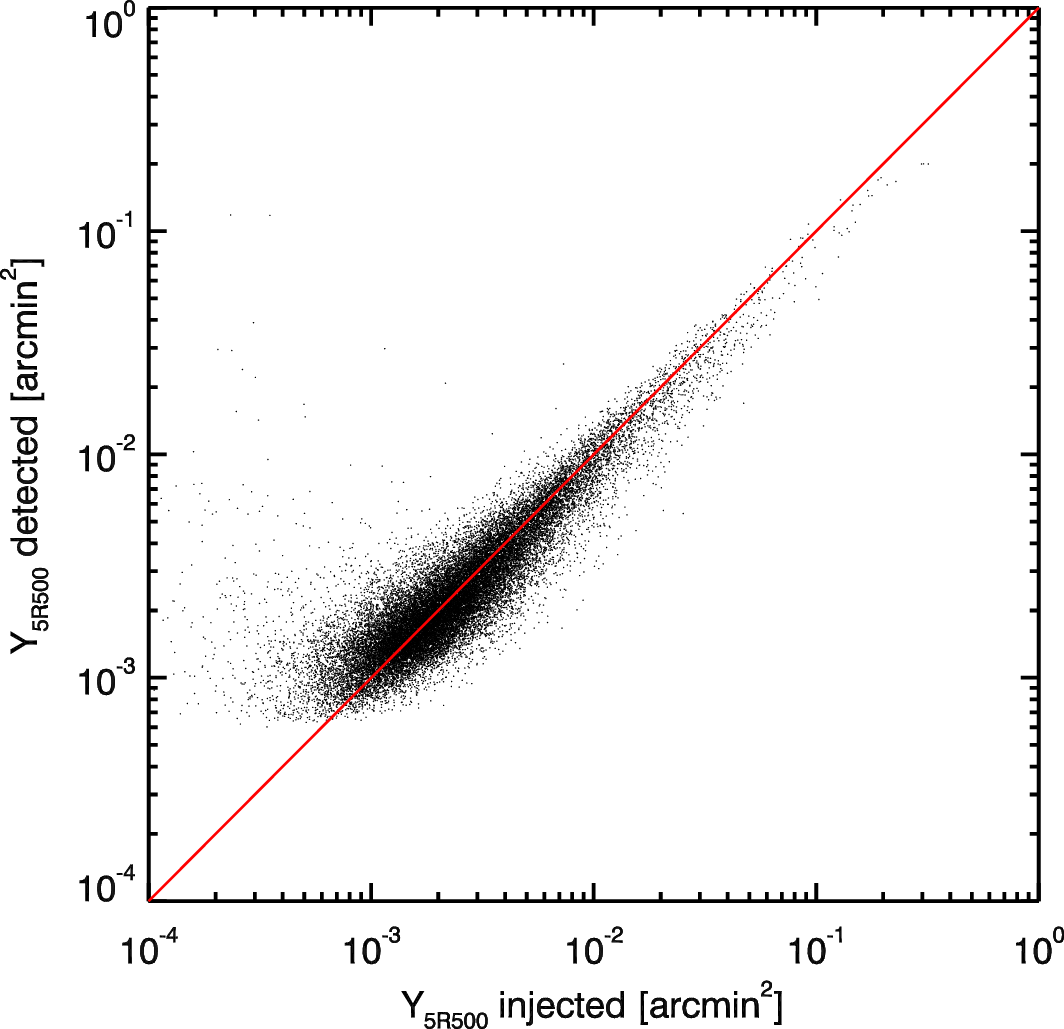}\hspace*{1em}
\includegraphics[angle=0, width=8.8cm]{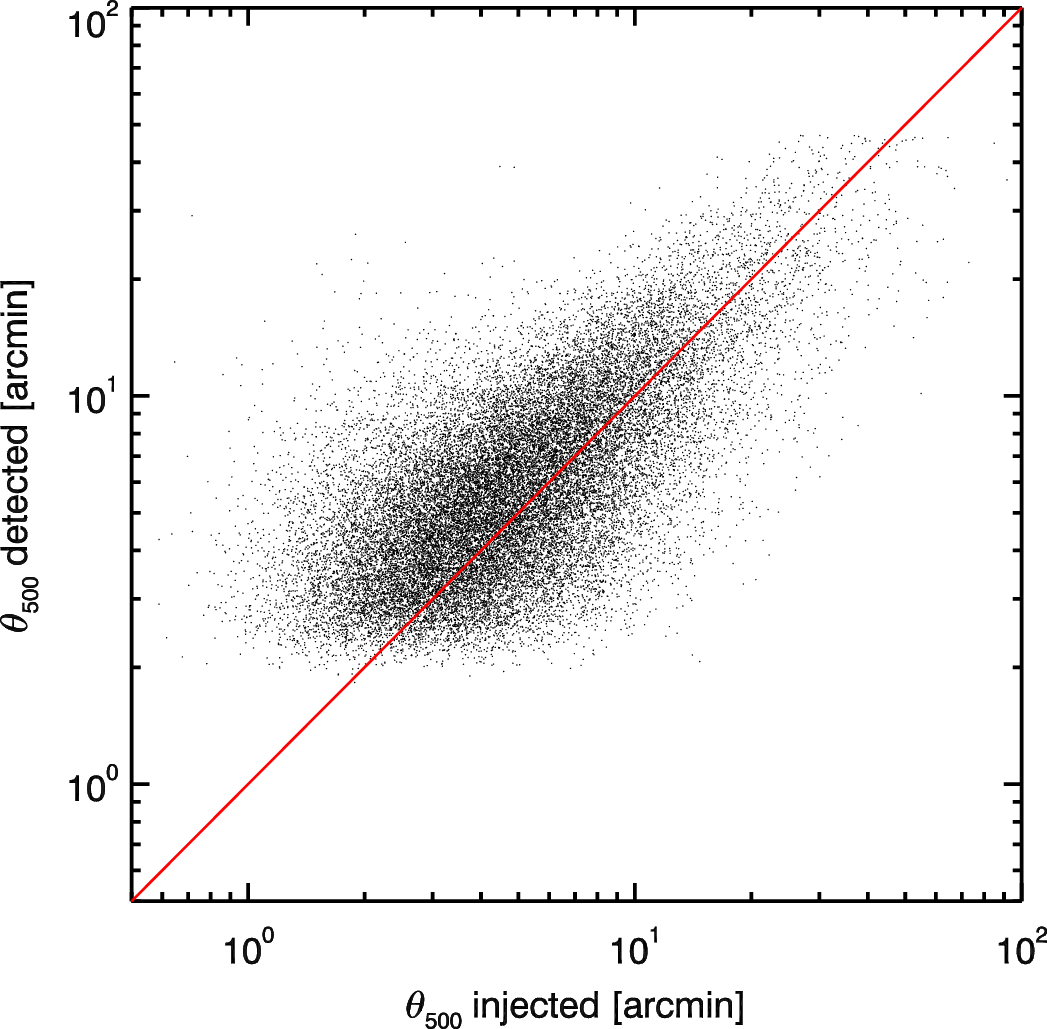}
\caption{Injected versus detected values of $\Yall$ (left panel) and
$\theta_{500}$ (right-panel),
illustrated for \texttt{PwS}.}
\label{qa:param_scatter}
\end{center}
\end{figure*}

The scatter between input and detected parameters is shown in
Fig.~\ref{qa:param_scatter} as an example for \texttt{PwS}.  Biases
are evident at both the low and high end for $\Yall$.  The low-flux
bias is the Malmquist bias related to the S/N $\ge 4.5$ threshold.
The high-flux bias is due to a hard prior on the upper limit for
cluster radius.  Figure~\ref{qa:param_scatter} also shows the
distribution of the ratio of estimated over injected parameters.  The
median and median absolute deviation of these ratios are shown in
Table~\ref{qa:param_table}.

The distributions for flux are positively skewed due to Malmquist
bias.  The median ratios of the flux recoveries are consistent with
unity for \texttt{MMF3} and \texttt{PwS} and are slightly higher for
\texttt{MMF1}.  The recovery of $\theta_\mathrm{s}$ is biased high in
the median by about 20\% for each of the codes.  This bias is a
consequence of the intrinsic cluster profile variation and disappears
when the injected profiles match the detection filter.  The
$Y_{5R_{500}}$ estimate by contrast is relatively unaffected by
profile variation.  The parameter constraints from \texttt{PwS} are
tighter than the \texttt{MMF}s due to the \texttt{PwS} priors and the
definition of the estimator as the expected value of the parameters
rather than the maximum likelihood.

\begin{table}[tmb]   
\begingroup
\newdimen\tblskip \tblskip=5pt
\caption{{Median and median absolute deviation (MAD) of the ratio
    of detected to injected parameters.}}
\label{qa:param_table}                   
\nointerlineskip
\vskip -3mm
\footnotesize
\setbox\tablebox=\vbox{
 \newdimen\digitwidth 
 \setbox0=\hbox{\rm 0} 
 \digitwidth=\wd0 
 \catcode`*=\active 
 \def*{\kern\digitwidth}
 \newdimen\signwidth 
 \setbox0=\hbox{+} 
 \signwidth=\wd0 
 \catcode`!=\active 
 \def!{\kern\signwidth}
\halign{\hbox to 0.85in{#\leaderfil}\tabskip 2em&
    \hfil#\hfil\tabskip=1em&
    \hfil#\hfil\tabskip=2em&
    \hfil#\hfil\tabskip=1em&
    \hfil#\hfil\tabskip=0pt\cr
\noalign{\doubleline}
\omit& \multispan2\hfil $\Yall$\hfil& \multispan2\hfil $\theta_\mathrm{s}$\hfil\cr
\noalign{\vskip -3pt}
\omit&\multispan2\hrulefill&\multispan2\hrulefill\cr
\omit\hfil Catalogue\hfil& median& MAD& median& MAD\cr
\noalign{\vskip 3pt\hrule\vskip 5pt}
\texttt{MMF1}& 1.09& 0.39& 1.17& 0.70\cr
\texttt{MMF3}& 1.02& 0.34&1.19& 0.69\cr
\texttt{PwS}& 0.99& 0.27&1.21& 0.56\cr
\noalign{\vskip 3pt\hrule\vskip 3pt}}}
\endPlancktable            
\endgroup
\end{table}

\section{External Validation}
\label{s:exval}

The cluster-candidate catalogue constructed from the union of all
three SZ-finder algorithms undergoes a thorough validation process
that permits us to identify previously-known clusters and to assess
the reliability of the \Planck\ SZ candidates not associated with
known clusters. In order to achieve this, we make use of the existing
cluster catalogues and we also search in optical, IR, and X-ray
surveys for counter-parts at the position of the \Planck\ SZ
sources. { In practice, we search within 5$^{\prime}$ of the SZ
  position, in agreement with \Planck\ position errors shown in
  Fig.~\ref{qa:position_recovery}.} In Sect.~\ref{sec:fu}, we present
the follow-up programmes that were undertaken by the
\Planck\ collaboration in order to confirm and measure the redshifts
of the \Planck\ candidate new-clusters.

The first step of the validation of the PSZ catalogue is to identify
among the \Planck\ SZ candidates those associated with known
clusters. For this purpose, we use existing X-ray, optical or SZ
cluster catalogues. A positional matching is not sufficient to decide
on the association of a \Planck\ SZ source with a previously-known
cluster, and a consolidation of the association is needed. For the
X-ray associations, a mass proxy can be built and used to estimate the
SZ flux, S/N, etc, that are compared with measured quantities for the
\Planck\ cluster candidates.  In contrast to the X-ray clusters,
optical clusters either have no reliable mass estimates or suffer from
large uncertainties in the mass--richness relations. In this case, the
consolidation cannot be performed uniquely through the coherence of
measured versus predicted properties.  It rather relies on extra
information from surveys in the X-ray, optical, or IR at the
\Planck\ cluster-candidate positions.

In the following, we detail the search for counter-parts in optical,
IR, and X-ray surveys; list the cluster catalogues used for the
identification; and finally present the identification procedure
followed to associate \Planck\ SZ detections with bona fide
clusters. In this process, we define quality flags for the association
of \planck\ SZ detections with external data. We set $Q=1$ for
high-reliability associations, i.e., very clear cluster signatures,
$Q=2$ for reliable associations, and $Q=3$ for low-reliability
associations, i.e., unclear cluster signature.

\subsection{Search for counter-parts of \Planck\ detections in surveys}

We made use of the {ROSAT} All Sky Survey \citep[RASS,
][]{vog99}, the all-sky survey with the Wide-field Infrared Survey
Explorer \citep[WISE,][]{wri10}, and the Sloan Digital Sky Survey
\citep[SDSS,][]{yor00} to search for counter-parts of the \Planck\ SZ
detections. This information was used in two ways. When
\planck\ detections were associated with known clusters from
catalogues, in particular in the optical, the counter-parts in RASS,
WISE, or SDSS helped in consolidating the association, increasing the
confidence in the identification of \planck\ candidates with known
clusters.  When no association between \Planck\ detections and
previously-known clusters was found, the information on the
counter-parts, in the surveys, of \planck\ SZ detections was used to
assess the reliability of the \Planck\ cluster candidates, i.e., clear
or unclear cluster signatures.

\subsubsection{Search in RASS data} \label{sec:rass}

As detailed in \citet{planck2012-IV}, the validation follow-up with
\xmm\ has shown the importance of the RASS data to assess the
reliability of the \Planck\ sources. In particular,
\citet{planck2012-IV} showed that a large fraction of
\Planck\ clusters are detectable in RASS maps, but this depends on the
region of the sky and on the ratio $Y_{500}/S_{\mathrm{X}}$ which
exhibits a large scatter (see later in Fig.~\ref{fig:fxsz} the case of
the PSZ sources).  We therefore exploit the RASS data to consolidate
the identification with clusters from optical catalogues (see
below Sect.~\ref{sec:ideno}) and to assess the reliability of the
\Planck\ SZ candidates.

We first perform a cross-match with the RASS bright source catalogue
\citep[BSC,][]{vog99} and the faint source catalogue
\citep[FSC,][]{vog00} within a 5$^{\prime}$ radius of the position of
each of the \Planck\ SZ detections. We then perform a reanalysis of
the RASS data following the methodology and prescriptions given by
\citet{boe00,boe04} and \citet{rei02}. We compute count-rate growth
curves in order to check for the extension of the signal. We estimate
the source flux from both the growth curve (when adequate) and from a
fixed 5$^{\prime}$ aperture radius with respect to the surrounding
background (after PS subtraction). We then derive the associated
S/N in RASS, (S/N)$_{\mathrm{RASS}}$. For this, we make
use of the RASS hard-band, [0.5--2] keV, data that maximize the S/N of
the detections. We furthermore computed
the source density map of the BSC and FSC catalogues and the
associated probability that a \Planck\ cluster candidate will be
associated with a B/FSC source within a radius of 5$^{\prime}$. For the
BSC, the probability of chance association is relatively low, with a
median $<$1\,\%. As detailed in \citet{planck2012-IV}, the
correspondence of a \Planck\ SZ-candidate with a RASS-BSC source is a
semi-certain association with a real cluster, whereas for the FSC
catalogue the probability of chance association is larger, 5.2\%.

We define a quality flag, $Q_{\mathrm{RASS}}$, for the association of
\Planck\ candidates with RASS counter-parts using both the
S/N in RASS and the association with B/FSC sources. This
is of particular importance for the \Planck\ candidate new
clusters. Based on the results from \citet{planck2012-IV}, the quality
of the association with RASS counter-parts is high,
$Q_{\mathrm{RASS}}=1$, for \Planck\ cluster candidates matching a
RASS-BSC source or with (S/N)$_{\mathrm{RASS}}\ge 2$. We find a total
of 887 out of 1227 \Planck\ SZ detections in this
category, with mean and median S/N of 7.4 and 5.8,
respectively.  The quality is poor, $Q_{\mathrm{RASS}}=3$, for RASS
counter-parts with (S/N)$_{\mathrm{RASS}}< 0.5$ in regions of
reasonable depth (quantified by the probability of chance association
with FSC sources being larger than 2.5\% \citep{planck2012-IV}).

\subsubsection{Search in SDSS data}\label{sec:sdss}

We performed a systematic search for counter-parts in the SDSS Data
Release DR9 \citep{sds12} at the position of all the \Planck\ SZ
detections. This was performed based on a cluster-finder algorithm
developed by (Fromenteau et al. 2014, in prep) to search for red
galaxy over-densities in the SDSS galaxy catalogues.

For each associated counter-part within a 5$^{\prime}$ circle
centred at the position of the \Planck\ SZ detection, a quality
criterion is defined on the basis of a fit to the luminosity function
and the associated mass limit, and on the number of galaxies within
5$^{\prime}$, $N_{\mathrm{gal}}$, such that we have
$Q_{\mathrm{SDSS, dat}}= 1$, i.e., high quality, for cases where
$N_{\mathrm{gal}}\ge 40$ and for masses $M_{200}\ge
5.7\times10^{14}\,\msol$, $Q_{\mathrm{SDSS, dat}}= 2$, i.e., good
quality, for $N_{\mathrm{gal}}$ between 40 and 20 for masses between
$1.5\times10^{14}\,\msol$ and $5.7\times10^{14}\,\msol$, and
$Q_{\mathrm{SDSS, dat}}= 3$ otherwise. 

The cluster-finder algorithm outputs the position of the counter-part
(Brightest Cluster Galaxy (BCG) and barycentre) and the estimated
photometric redshift. When spectroscopic data are available for the
brightest selected galaxy a spectroscopic redshift is also
reported. The outputs of the cluster-finder algorithm are compared to
those obtained by (Li \& White 2014, in prep) from different method
based on the analysis of the full photometric-redshift probability
distribution function \citep{cun08}. In this approach, the position
and redshift in the SDSS data that maximizes the S/N are considered as
the best estimates for the counter-parts of the \Planck\ SZ
detections.

\subsubsection{Search in WISE data} \label{sec:wise}

WISE provides an all-sky survey at 3.4, 4.6, 12, and 22 $\mu$m (W1,
W2, W3, W4) with an angular resolution of 6.1 to 12.0 arcsec in the
four bands.

We search for counter-parts of the \Planck\ SZ detections in the WISE
source catalogue in two ways. On the one hand, we run an adaptive
matched filter cluster finder developed by (Aussel et al. 2014, in
prep), similar to the one described by \citet{kep99}, using the
cluster members' luminosity function of \citet{lin12}. The background
counts were determined from the neighbouring square degree in the
vicinity of the \Planck\ cluster candidate, excluding regions of
fifteen arcmin centred on candidate positions. On the other hand, we
use a method developed by (Aghanim \& Fromenteau 2014, in prep) based
on a search for overdensities of bright (W1 $\le$ 17) and red (W1$-$W2
$>0$) sources within a 5$^{\prime}$ radius circle centred on the
position of \Planck\ detections with respect to a background computed
in a 15$^{\prime}$ radius area.

Aghanim \& Fromenteau (2014, in prep) find that a good-quality
association between a \Planck\ SZ-detection and a counter-part
overdensity in WISE data is reached when there are at least ten
galaxies above $2\,\sigma$ in the 5$^{\prime}$ search region, and when
the corresponding fraction of galaxies is at least 30\% of the total
number of galaxies retained in the 15$^{\prime}$ circle.  Performing
the search for counter-parts of an ensemble of random positions on the
sky, we compute the purity of the detections, i.e., the probability of
a \Planck\ candidate having a real counter-part in the WISE data as
opposed to a chance association.  The quality criterion for the
association between \Planck\ detection and WISE overdensity is high,
$Q_{\mathrm{WISE}}=1$, for a purity larger than 90\%. When it lies
between 90\% and 80\% the association of \Planck\ SZ-detections and
WISE overdensities is assigned a lower quality criterion
$Q_{\mathrm{WISE}}=2$. We set the quality of the association to
$Q_{\mathrm{WISE}}=3$, bad, when the purity is below 80\%. We find 856
\Planck\ SZ detections with high or good quality counter-parts in WISE
data, including 658 $Q_{\mathrm{WISE}}=1$ detections.

\subsubsection{DSS images}

Finally for each \planck\ cluster candidate, the second Digitized Sky
Survey\footnote{\url{http://stdatu.stsci.edu/dss/}.} (DSS) database
was queried for a field of $5^{\prime}\times 5^{prime}$ centred at the
position of the \planck\ SZ detections in the $r$ and $ir$ bands.  The
DSS images were used for visual inspection.\footnote{Images from the
  RASS, SDSS and WISE surveys at the position of the \planck\ SZ
  detections were also inspected.} Clusters and rich groups out to
$z\simeq 0.3$ to $0.4$ can easily be identified in these plates as an
obvious concentration of galaxies. This qualitative information was
thus used: (i) to consolidate some identifications of \Planck\ SZ
detections with previously-known clusters; (ii) to optimize our
strategy for the follow-up observations of \Planck\ candidates (see
Sect.~\ref{sec:fu}); and (iii) to qualitatively assess the reliability
or significance of the \Planck\ SZ detections.

\subsection{Cluster catalogues}\label{sec:clus_cat}
We now present the ensemble of catalogues that were used to identify
the \Planck\ SZ detections with previously-known clusters. In the case
of the {ROSAT}- and SDSS-based catalogues, we have used homogenized
quantities, see below, that allowed us to perform the identification
with comparable association criteria, which ensures homogeneity in the
output results.

\paragraph{MCXC meta-catalogue} -- For the association of \Planck\ SZ
candidates with previously-known X-ray clusters, we use the Meta-Catalogue of
X-ray detected Clusters of galaxies \citep[MCXC,][and reference
therein]{pif11} constructed from the publicly available \rosat\ All
Sky Survey-based and serendipitous cluster catalogues, as well as the
{\it Einstein} Medium Sensitivity Survey.  For each cluster in the MCXC
several properties are available, including the X-ray coordinates,
redshift, identifiers, and standardized luminosity, $\LXv$, measured
within $\Rv$. The MCXC compilation includes only clusters with
available redshift information (thus X-ray luminosity) in the original
catalogues.  We updated the MCXC, considering the first release of the
\reflex-II survey \citep{cho12}, the third public release of clusters
from the \macs\ sample \citep{man12}, individual \macs\ cluster
publications and a systematic search in NED and SIMBAD for
spectroscopic redshift for clusters without this information in the
\rosat\ catalogues.  This yields an ensemble of 1789 clusters with $z$
and $\LXv$ values, adding 20 \macs\ clusters, 21 \reflex-II clusters
and 5 SGP clusters to the MCXC.  For these clusters, the expected
Compton-parameter, $\YSZ^{\LX}$, and size, $\tv^{\LX}$, are estimated
combining the \ML\ relation of \citet{pra09} and the \MY\ relation
given by \citet{arn10}. The expected S/N,
(S/N)$^{\LX}$, is computed taking into account the noise within
$\tv^{\LX}$ at the cluster location.  We furthermore supplement the
updated MCXC with 74 clusters from \rosat\ catalogues without redshift
information and 43 unpublished \macs\ clusters observed by \xmm\ or
\chandra. For these 117 objects, only centroid positions are
available.  Finally, we considered the published catalogues from
\xmm\ serendipitous cluster surveys with available redshifts, the XCS
catalogue \citep{meh12}, the 2XMMi/SDSS catalogue \citep{tak11} and
the XDCP catalogue \citep{fas11}. However, these catalogues mostly
extend the MCXC to lower masses and only two \planck\ candidates were
found to be associated with these new clusters.

\begin{figure}
\includegraphics[width=8.8cm]{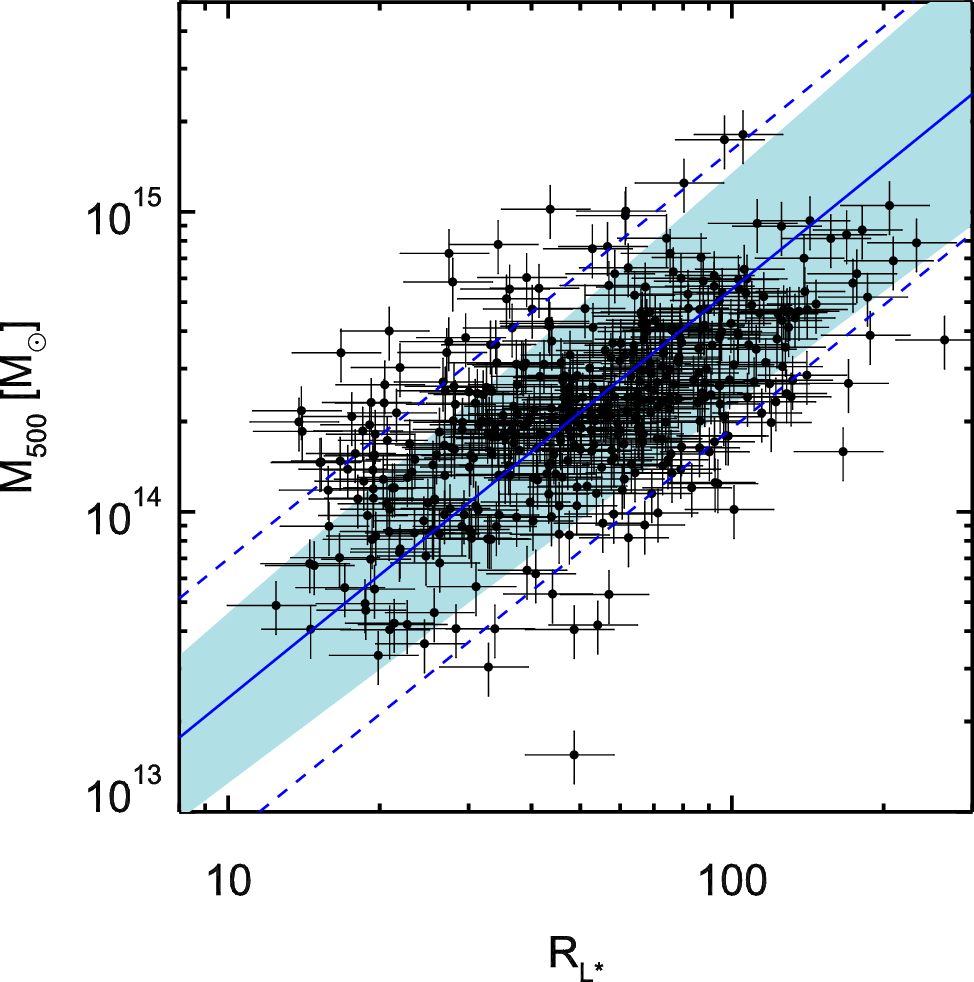}%
\caption{Mass-to-richness scaling relation,
$\Mv$--$R_{\mathrm{L}^\star}$, for the 444 MCXC clusters included in
the WHL12 catalogue \citep{wen12}. The best-fit relation, from BCES
fit, is given by the solid blue line.  We adopted 15\%
uncertainties on the MCXC masses as prescribed in \citet{pif11}. As
no uncertainty is provided for the WHL12's richness, we arbitrarily
assumed a 20\% uncertainty for all richness values. The blue shaded
area shows the associated errors on the best-fit, while the dashed
line marks the intrinsic scatter.}
\label{fig:SDSS_M-R}
\end{figure}

\begin{figure*}[t]
\centering
\begin{minipage}[t]{\textwidth}
\resizebox{\hsize}{!} {
\includegraphics[]{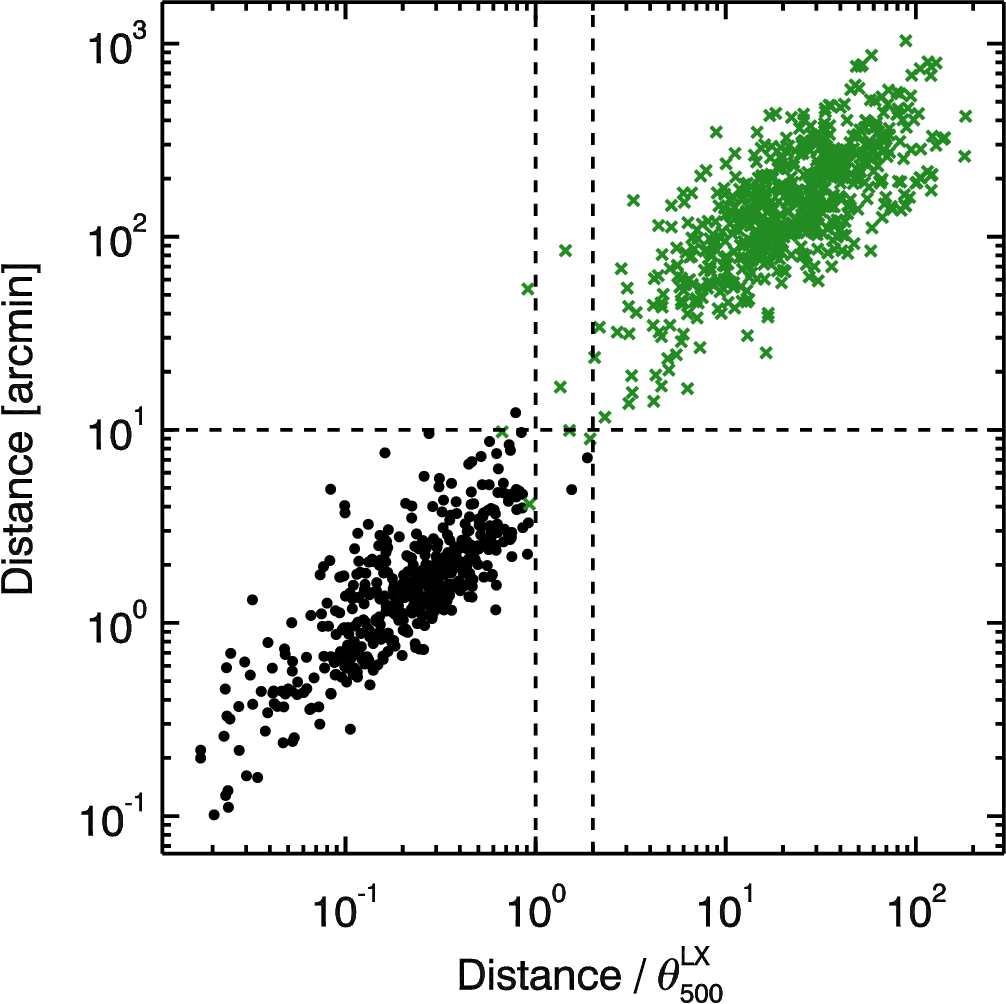} 
\hspace{5mm}
\includegraphics[]{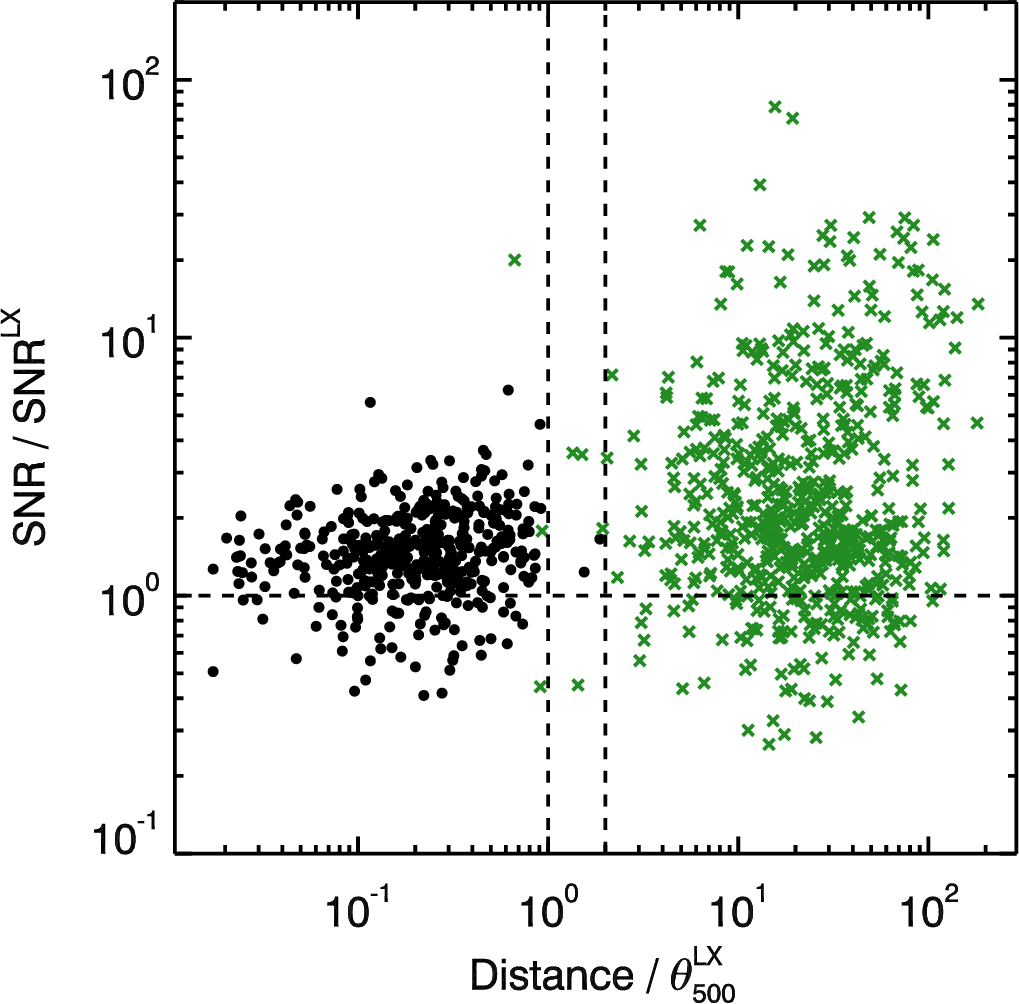}  
\hspace{5mm}
\includegraphics[]{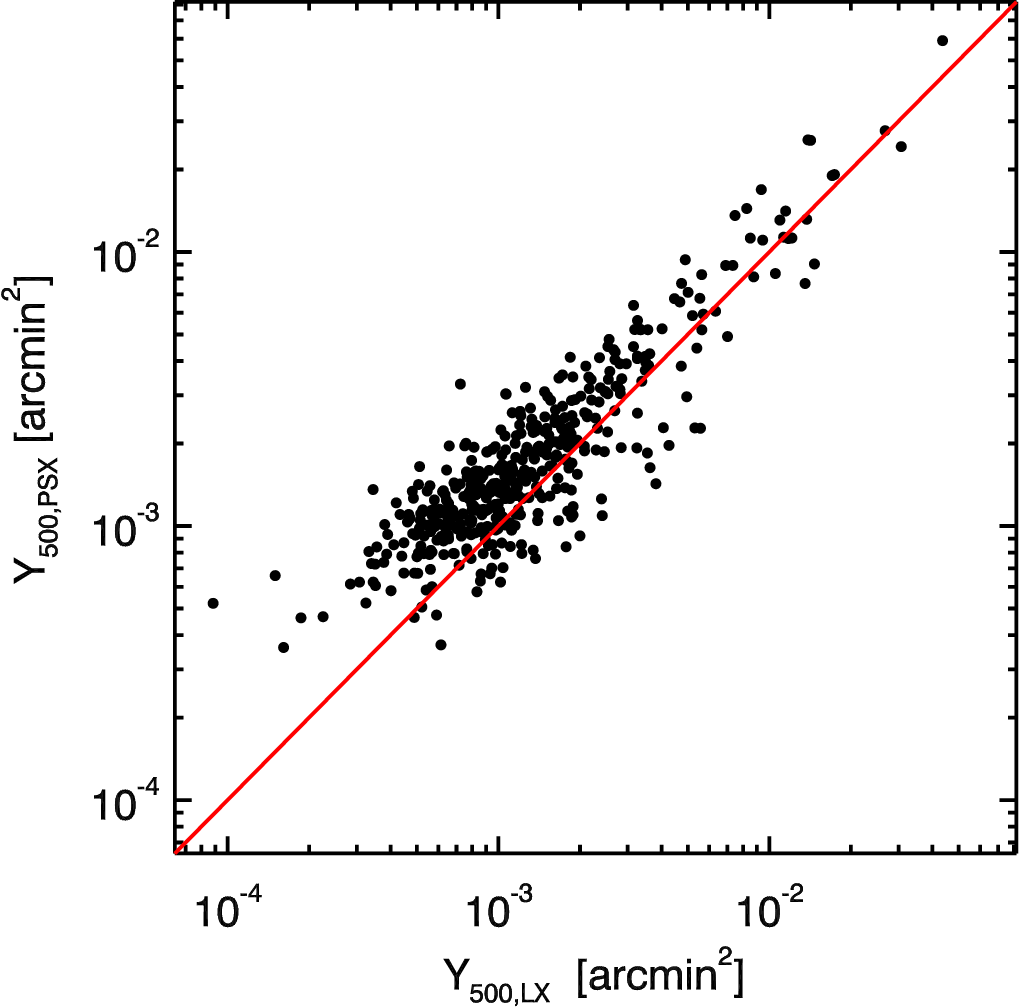}  
}
\end{minipage}
\caption{Identification of the \planck\ cluster candidates with X-ray
  clusters from the MCXC catalogue.  Black points are candidates
  firmly identified with MCXC clusters, while green points are
  candidates with no association.  {\it Left panel:} distance of the
  \Planck\ position to the position of the closest MCXC cluster as a
  function of the distance normalized to the cluster size
  $\tv^{L_{\mathrm{X}}}$.  {\it Middle panel:} S/N normalized to the
  expected value as a function of normalized distance.  {\it Right
    panel:} SZ flux, $Y_{500,{\mathrm{PSX}}}$, re-extracted fixing the
  position and size to the X-ray value, as a function of expected
  values. The red line is the equality line.  In all panels,
  $\YSZ^{L_{\mathrm{X}}}$, and $\tv^{L_{\mathrm{X}}}$ are estimated
  from the cluster X-ray luminosity used as mass proxy (see text).}
\label{fig:pxccid}
\end{figure*}

\paragraph{Optical-cluster catalogues} -- The identification of the
\Planck\ SZ candidates with clusters known in the optical is based on
the Abell \citep{abe58} and the Zwicky \citep{zwi61} cluster
catalogues. Furthermore, we have used four different catalogues of
clusters based on the Sloan Digital Sky Survey (SDSS, \citealt{yor00})
data: (1) the MaxBCG catalogue \citep[13,\,823 objects,][]{koe07}; (2)
the GMBCG catalogue \citep[55,\,424 objects,][]{hao10}; (3) the AMF
catalogue \citep[69,\,173 objects,][]{sza11}; and (4) the WHL12 catalogue
\citep[132,\,684 objects,][]{wen12}. We refer the reader to
\citet{wen12} for a comparison of the existing SDSS-based catalogues
of clusters and groups. Each of the SDSS-based catalogues provides an
estimated richness; we first start by homogenizing the richness
estimates to that of WHL12. For each catalogue, we compute the median
ratio of WHL12's richness to that of the considered catalogue over its
intersection with WHL12's. We then renormalize the individual
richness by the corresponding ratio. The correcting factors applied to
the richness estimators\footnote{Field NGALS\_R200 for MaxBCG,
GM\_SCALED\_NGALS for GMBCG and LAM200 for AMF.} are respectively
$1.52$, $1.75$, and $0.74$ for MaxBCG, GMBCG, and AMF, obtained from
$7627$, $17245$, and $1358$ common clusters.\footnote{We considered
the associations of clusters with positions matching within 6~arcsec
radius and with $\Delta z\leq 0.05$ (typical uncertainty for
photometric redshifts in SDSS).} The richness is then related to the
halo mass, $\Mv$, by extending the \citet{wen12} richness--mass
relation provided on about $40$ clusters\footnote{Their $M_{200}$ are
taken from the literature either from weak lensing or X-ray
measurements \citep{wen10}.} to 444 MCXC clusters, with masses
estimated from the X-ray luminosities. The data points and the
best-fit scaling relation are presented in
Fig.~\ref{fig:SDSS_M-R}. The derived $\Mv$--$R_{L^\star}$ and
$L_{\mathrm X, 200}$--$R_{L^\star}$ relations are compatible with the
findings of \citet{wen12}. We find $\log{(\Mv/10^{14}\,\msol}) =
(-2.00\pm 0.17) + (1.37\pm0.10) \times
\log{R_{\mathrm{L}^\star}}$. The relation presents a large intrinsic
log-scatter, $\sigma_\mathrm{int}=0.27\pm0.02$, hampering any accurate
estimation of the cluster mass. This is further illustrated by the
richest clusters with $R_{\mathrm{L}^\star}>110$ having MCXC masses
systematically below the best-fit $M_{200}$--$R_{{\mathrm{L}}^\star}$
relation (although within the $1\sigma$ intrinsic scatter).

\paragraph{SZ catalogues} -- At millimetre wavelengths, we
cross-check the \Planck\ SZ catalogue with the recent ACT and SPT
samples \citep{men10,van10,wil11}, including the most recent data that
increased the number of SZ detections and updated the redshift
estimates for the clusters \citep{rei13,has13}. We have furthermore
identified the \Planck\ SZ detections associated with previous SZ
observations of galaxy clusters from the literature. We used a
compilation of SZ observations conducted with the numerous
experiments developed during the last 30 years (Ryle, OVRO, BIMA,
MITO, Nobeyama, SZA, APEX-SZ, AMI, Diabolo, Suzie, Ryle, AMIBA, ACBAR,
etc.).

\subsection{Identification with previously-known clusters}\label{sec:idi}

\subsubsection{Identification with X-ray clusters}\label{sec:idenx}
The \Planck\ SZ candidates are cross-checked against previously-known
X-ray clusters from the updated version of the MCXC. For a given
\Planck\ candidate-cluster we identify the closest MCXC
cluster.\footnote{The information of the second closest is also kept
  to identify potential confusion or duplicate associations.}  The
reliability of the association is assessed based on distance, $D$,
compared to the cluster size and on the measured $\YSZ$ and S/N values
compared with the expected values (see Fig.~\ref{fig:pxccid}).  Two
clouds of points stand out in the scatter plot of absolute versus
relative distance, $D/\tv^{\LX}$ (Fig.~\ref{fig:pxccid}, left panel).
They correspond to two clouds in the scatter plot of the measured over
expected S/N versus $D/\tv^{\LX}$ (Fig.~\ref{fig:pxccid}, middle
panel).

The association process follows three main steps. First, we
provisionally assign an X-ray identification flag based on distance:
\begin{itemize}
\item $Q_{\mathrm{X}}=3$ if $D\,>\,2\tv^{\LX}$ and $D>\,10^{\prime}$.
Those are considered as definitively not associated with an MCXC
cluster in view of \planck\ positional accuracy and cluster extent.
\item $Q_{\mathrm{X}}=1$ if $D\,<\,\tv^{\LX}$ and $D\,<\,10^{\prime}$.  
Those are associated with an MCXC cluster.
\item $Q_{\mathrm{X}}=2$ otherwise, corresponding to uncertain
associations.
\end{itemize}

We then refine the classification. In the $Q_{\mathrm{X}}=1$ category,
we identify outliers in terms of the ratio of measured to expected
S/N and $\YSZ$, taking into account the scatter and the size--flux
degeneracy. Their flags are changed to $Q_{\mathrm{X}}=2$.  In some
cases, two distinct $Q_{\mathrm{X}}>1$ candidates are associated with
the same MCXC cluster. The lowest S/N detection is flagged as
$Q_{\mathrm{X}}=2$.

In the final step, we consolidate the status of $Q_{\mathrm{X}}<3$
candidates.  We first re-extract the SZ signal at the X-ray position,
both leaving the size free and fixing it at the X-ray value. The
$\YSZ$ obtained with the cluster and size fixed to the X-ray values
are compared to the expected values, $\YSZ^{L_{\mathrm{X}}}$, in the
right panel of Fig.~\ref{fig:pxccid}.  For bona fide association, we
expect no major change of $\YSZ$ and S/N, with, on average, a better
agreement with the expected $\YSZ$ value and some decrease of S/N.
\begin{itemize}
\item For $Q_{\mathrm{X}}=1$ candidates, the re-extracted $\YSZ$ and
S/N values are compared to both blind and expected values (as a
function of distance, S/N, etc.) to identify potential problematic
cases, e.g., important decrease of S/N or outliers in terms of
measured-over-expected $\YSZ$ ratio. We found only one such case,
whose flag is changed to $Q_{\mathrm{X}}=2$. The identification of
other candidates is considered as consolidated, with definitive
flag $Q_{\mathrm{X}}=1$.
\item We then examine the $Q_{\mathrm{X}}=2$ candidates. We consider
the re-extracted $\YSZ$ and S/N, but also perform a visual
inspection of the SZ maps and spectra and ancillary data, including
$\rass$\ and DSS images. The $Q_{\mathrm{X}}=2$ candidates were
identified as clearly identified as multiple
detections of extended clusters or duplicate detections of the same
clusters by different methods that were not merged (the former are
flagged as false detections, the latter are merged with the
corresponding candidate in the union catalogue) or not associated
(e.g., SZ sources clearly distinct from the MCXC clusters with no
significant re-extracted signal at the cluster position and size).
\end{itemize}

Finally, for MCXC clusters without redshift and luminosity
information, the association was only based on distance, setting
$D_{\mathrm{X}}\,<5^{\prime}$, and the consolidated based on visual
inspection of SZ, $\rass$\ and DSS images and other ancillary
information. Two cases were found to be a mis-identification.  The SZ
candidate was closer by chance to a faint XCS cluster, in the vicinity
of the real counter-part (another MCXC cluster and an Abell cluster,
respectively).

\subsubsection{Identification with optical clusters}\label{sec:ideno}
The \Planck\ SZ candidates are associated with known clusters from
optical catalogues (Abell, Zwicky, SDSS-based catalogues) on the basis
of distance with a positional matching within a search radius set to
5$^{\prime}$. The consolidation of the association was performed
using the RASS information as described below, which allows us to
mitigate the chance associations with poor optical galaxy groups and
clusters.

\paragraph{SDSS-based catalogues} -- We have considered the 
four catalogues listed in Sect.~\ref{sec:clus_cat}.  We define a
quality criterion for the association, $\qsdss$, in terms of cluster
richness as a proxy of the cluster mass \citep[see for
  instance][]{joh07,roz09}.  We set the quality criterion, $\qsdss$,
to 3 for low reliability {(richness below 70)}, to 2 for good
reliability {(richness ranging from 70 to 110)} and to 1 for high
reliability {(richness above 110)}.

\begin{figure}
\includegraphics[width=8.8cm]{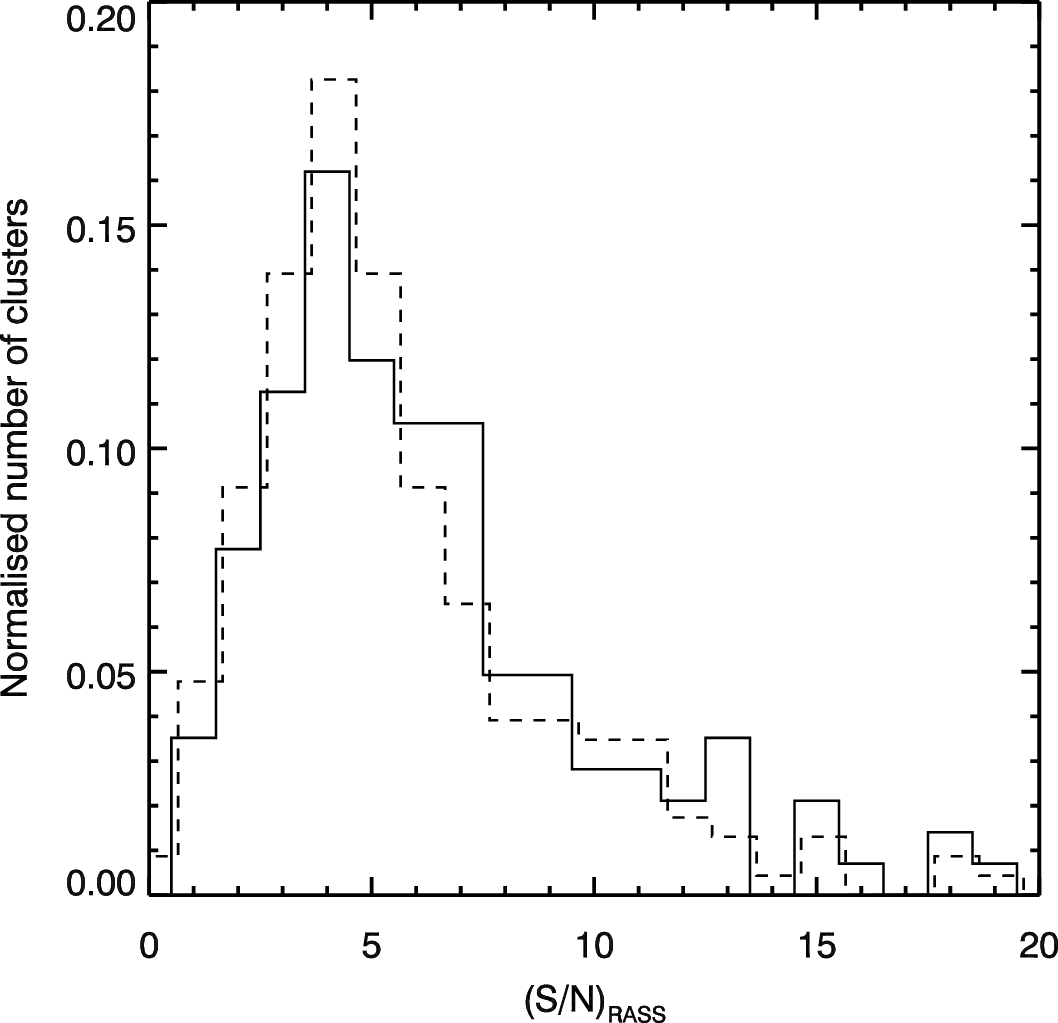}
\caption{Normalized distribution of the S/N in RASS
at the position of \Planck\ SZ detections with SDSS richness-based
quality $\qsdss=1$ (solid line) and $\qsdss=2$ (dashed line).}
\label{fig:SDSSqual}
\end{figure}

The corresponding estimated masses
(given the $\Mv$--$R_{L^\star}$ relation) are $\Mv > 6.5\times
10^{14}\,\msol$ and $\Mv > 3.5 \times 10^{14}\,\msol$. However
due to the large scatter and associated uncertainty in the mass
estimate from the mass--richness relation, we consolidate the
association of the \Planck\ candidates with SDSS clusters by combining
the $\qsdss$ with the RASS signal at the \Planck-candidate position
(see Sect.~\ref{sec:rass}).  In practice, only associations with
$\qsdss=1$ or 2 and a S/N, measured at the
\Planck\ position in an aperture of 5$^{\prime}$ in the RASS survey,
(S/N)$_{\mathrm{RASS}}\ge 1$ are retained as firm identifications. We
stress that our choice of richness thresholds is relatively
conservative on average. Indeed, our $\qsdss=1$ and 2 matched
candidates are found with high (S/N)$_{\mathrm{RASS}}$ values as shown
in Fig.~\ref{fig:SDSSqual}, with mean (S/N)$_{\mathrm{RASS}}=7.1$ and
6.6 and median (S/N)$_{\mathrm{RASS}}=5.9$ and 5.4 for $\qsdss=1$ and
2 matches, respectively.

\paragraph{Abell and Zwicky catalogues} --  The \Planck\ candidates
are associated with Abell and Zwicky clusters on the basis of a
positional matching within five arcmin. In the present case, we do
not make use of any richness information in order to consolidate the
association. We rather use here solely the RASS signal,
(S/N)$_{\mathrm{RASS}}$, at the SZ-candidate
position. \Planck-candidates associated with Abell or Zwicky clusters
and with (S/N)$_{\mathrm{RASS}}\ge 1$ are retained as firmly
identified. For associations with \mbox{(S/N)$_{\mathrm{RASS}}< 1$},
we decided on a firm identification only after checking the status of
the counter-part in the WISE data and performing a visual inspection
of the SZ signal and of the images from ancillary data, including DSS
images.

\subsubsection{Identification with SZ clusters}\label{sec:idensz}
The association with known SZ
clusters was performed within a 5$^{\prime}$ radius. A visual
inspection of the ancillary data and an \textit{a posteriori} check of
the RASS signal at the position of the \Planck\ candidates associated
with clusters from SZ catalogues is performed. It confirms that the
values of (S/N)$_{\mathrm{RASS}}$, when the coverage is significant,
are high with an average value of 5.4.

\subsubsection{Identifications from NED and SIMBAD}\label{sec:idendb}

The information provided from querying NED and SIMBAD databases is
mainly redundant with cross-checks with cluster catalogues. However,
it lets us avoid missing a few associations.  We therefore performed a
systematic query in SIMBAD and NED with an adopted search radius set
to 5$^{\prime}$. Similarly to the association with clusters in optical
catalogues, the positional association is consolidated using the
results of the search in RASS data. Furthermore, the
\Planck-candidates solely matching NED or SIMBAD entries were
inspected and the identification was confirmed or discarded using the
information from WISE counter-parts and the DSS images.

\begin{table*}
\begingroup
\newdimen\tblskip \tblskip=5pt
\caption{{Observing facilities used for the confirmation of
    clusters discovered by \Planck, and for the measurement of their
    redshifts.} }
\label{tab:fu}
\nointerlineskip \vskip 1mm \footnotesize
\setbox\tablebox=\vbox{
\newdimen\digitwidth
\setbox0=\hbox{\rm 0}
\digitwidth=\wd0
\catcode`*=\active
\def*{\kern\digitwidth}
\newdimen\signwidth
\setbox0=\hbox{+}
\signwidth=\wd0
\catcode`!=\active
\def!{\kern\signwidth}
\halign{\hbox to 1.0in{#\leaderfil}\tabskip=2em&
    \hfil#\hfil& 
    \hfil#\hfil& 
    \hfil#\hfil& 
    \hfil#\hfil& 
    #\hfil\tabskip=0pt\cr
\noalign{\doubleline}
\omit&&Aperture&&&\omit\hfil Type of\hfil\cr
\omit\hfil Site\hfil& Telescope&[m]& Instrument& Filters&\omit\hfil redshift\hfil\cr
\noalign{\vskip 3pt\hrule\vskip 5pt}
Earth orbit& {\it XMM-Newton}&& EPIC/MOS, PN& $\dots$& Fe K\cr
La Palma& NOT& *2.56& ALFOSC& $\dots$&Spectroscopic\cr
La Palma& INT& *2.5*& WFC& {\it griz}& Photometric\cr
La Palma& GTC& 10.4*& OSIRIS& $\dots$&Spectroscopic\cr
La Palma& TNG& *3.5*& DOLORES& $\dots$&Spectroscopic\cr
La Palma& WHT& *4.2*& ACAM& {\it griz}& Photometric\cr
La Silla& NTT& *3.7*& EFOSC2& $\dots$&Spectroscopic\cr
La Silla& MPG/ESO-2.2m& *2.2*& WFI& VRI& Photometric\cr
MRAO& AMI& 3.7,13& SA, LA& 13.5--18\,GHz&\omit\hfil$\dots$\hfil\cr
Tenerife& IAC80& *0.82& CAMELOT& {\it griz}& Photometric\cr
Tubitak Nat. Obs. & RTT150& 1.5& TFOSC& {\it gri}&Spectroscopic\cr
\noalign{\vskip 5pt\hrule\vskip 3pt}}}
\endPlancktable 
\endgroup
\end{table*}

\section{Follow-up programme for confirmation of
\Planck\ candidates}\label{sec:fu} 

We have undertaken, since Spring 2010, an extensive follow-up
programme in order to perform a cluster-by-cluster confirmation of the
\Planck\ cluster candidates and obtain a measurement of their
redshifts. A total of 276 \Planck\ candidates, selected
down to S/N $=4$ from intermediate versions of the \Planck\ SZ catalogue,
were observed in pursuit of their redshift measurement. We have
constructed our strategy for the selection of the
\Planck\ targets primarily on the successful results of the series of
follow-up observations in X-rays based on Director's discretionary
time on the \xmm\ observatory
\citep{planck2011-5.1b,planck2012-I,planck2012-IV}.  Snapshot
observations, sufficient to detect extended X-ray emission associated
with \planck\ clusters and to estimate redshifts from the Fe line for
the brightest clusters, were conducted sampling the SZ detections down
to S/N $=4$.  These observations allowed us to better understand the SZ
signal measured by \Planck\ and hence to refine the criteria to select
targets, especially for further optical follow-up.

We have engaged numerous campaigns on optical facilities, which now
constitute our main means of confirmation of \Planck\ SZ detections.
\plck\ candidates with low-quality DSS images or without SDSS
information, or low (S/N)$_{\mathrm{RASS}}$, were primarily sent for
deeper multi-band imaging observations. They were followed-up to the
depth needed for the confirmation, i.e., finding an optical
counter-part, and for the determination of a photometric
redshift. Candidates with galaxy concentrations in DSS or with
counter-parts in SDSS, and/or with high (S/N)$_{\mathrm{RASS}}$, were
preferentially sent for spectroscopic confirmation. The priority being
to confirm the clusters and to secure the largest number of robust
redshifts, no systematic spectroscopic confirmation of photometric
redshifts was performed for low-redshift clusters
($z_{\mathrm{phot}}<0.4$).  For higher-redshift clusters,
spectroscopic confirmation of the photometric redshifts is more
crucial. As a result, we have made use of telescopes of different
sizes, from 1-m to 10-m class telescopes, optimizing the selection of
targets sent to the different observatories (Table~\ref{tab:fu} gives
the list of the main telescopes). Telescopes of 8- and 10-m classes,
e.g., GTC, GEMINI and VLT, were used to spectroscopically confirm
redshifts above 0.5 for already confirmed clusters.

Our efforts to confirm the \Planck\ cluster candidates, measure
redshifts, and characterize cluster physical properties relies on
ongoing follow-up of a large number of cluster candidates in the
optical (ENO, RTT150, WFI), in the infrared ({\it Spitzer}\footnote{Under
  {\it Spitzer} programs 80162 and 90233.}) and at SZ wavelengths
(Arcminute Microkelvin Imager, AMI). The output of the confirmation
and redshift measurements from the observing campaigns is summarized
in Sect.~\ref{sec:confi}.  Companion publications, in preparation,
will detail the observing campaigns and their results.

\subsection{{\it XMM-Newton} observatory}

The X-ray validation follow-up programme of 500\,ks observations
undertaken in \xmm\ DDT is detailed in
\citet{planck2011-5.1b}, \citet{planck2012-I}, and
\citet{planck2012-IV}. It consisted of observing 51 \planck\ targets
and led to the confirmation of 43 \Planck\ cluster candidates, two
triple systems and four double systems. There were eight false
candidates. This follow-up programme has constituted the backbone of
the \Planck\ cluster confirmation and most importantly has allowed us
to better understand the SZ signal measured by \Planck\ and thus to
better master the criteria for confirmation (or pre-confirmation) of
the \Planck\ cluster candidates.  By providing us with the physical
properties and redshift estimates of the confirmed clusters, it has
furthermore given us a first view on the physical characteristics of the
newly discovered \Planck\ clusters. Snapshot observations
(around 10\,ks) of the \Planck\ candidates took place between May 2010
and October 2011. All the results from the four observing campaigns
were published in \citet{planck2011-5.1b}, \citet{planck2012-I}, and
\citet{planck2012-IV}. Calibrated event lists were produced with v11.0
of \xmm-SAS, and used to derive redshifts and global physical
parameters for the confirmed clusters \citep{planck2011-5.1b}. The
redshifts were estimated by fitting an absorbed redshifted thermal
plasma model to the spectrum extracted within a circular region
corresponding to the maximum X-ray detection significance. Most of the
redshifts were confirmed using optical observations.  Additional
observations at VLT were conducted to confirm spectroscopically the
highest redshifts.\footnote{Observations are conducted under programme
090A-0925.}

\subsection{Optical observation in the northern hemisphere}
\subsubsection{ENO telescopes}
In total 64 cluster candidates from \Planck\ were observed at
European Northern Observatory (ENO\footnote{ENO:
\url{http://www.iac.es/eno.php?lang=en}. }) telescopes, both for
imaging (at IAC80, INT and WHT) and spectroscopy (at NOT, GTC, INT
and TNG), between June 2010 and January 2013.\footnote{The observations
were obtained as part of proposals for the Spanish CAT time
(semesters 2010A, 2010B, 2011A, 2011B, 2012A and 2012B), and an {\it
  International Time Programme (ITP)}, accepted by the International
Scientific Committee of the Roque de los Muchachos (ORM, La
Palma) and Teide (OT, Tenerife) observatories (reference ITP12\_2).}
The aims of these observations were the confirmation, photometric
redshift measurement, and spectroscopic confirmation of redshifts above
$z=0.3$.

\paragraph{INT, WHT and IAC80} -- The optical imaging 
observations were taken either with the Wide-Field Camera (WFC) on the
2.5-m {\it Isaac Newton} Telescope (INT), the auxiliary-port camera (ACAM) at
the 4.2-m {\it William-Herschel} Telescope (WHT), or with CAMELOT, the optical
camera at the 0.82-m telescope (IAC80). The targets
were observed in the Sloan $gri$ filters.
For the majority of fields, either Sloan $z$ or Gunn $Z$ images are
also available.  Images were reduced using the publicly-available
software {\tt Iraf} and {\tt SExtractor} \citep{ber96}.  The data
reduction included all standard steps, i.e., bias and flat field
corrections, astrometric and photometric calibrations.  The
photometric calibration is based either on standard star observations
or, if available, on data from the SDSS.  Finally, all magnitudes were
corrected for interstellar extinction, based on the dust maps by
\citet{sch98}.  We obtained photometric redshifts using the BPZ code
\citep{ben00}, using a prior based on SDSS data, and fitting a set of
galaxy templates. The BPZ code provides the Bayesian posterior
probability distribution function for the redshift of each object,
which is later used in the process of cluster identification. The
identification of the galaxy overdensity located near the
\Planck\ positions and the estimate of the photometric redshifts of
the associated clusters were performed using a modified version of the
cluster-algorithm described in Sect.~\ref{sec:sdss}.

\paragraph{GTC and TNG} -- Spectroscopic observations were performed
using the 10-m Gran Telescopio Canarias (GTC) telescope and the 3.6-m
Telescopio Nazionale {\it Galileo} (TNG) telescope. 
The OSIRIS spectrograph at GTC was used in long-slit mode to observe a
total of eight targets with two slit positions per candidate. We
used the R500R grism and a binning $2\times 1$, which provides a
resolution $R=300$ with a slit width 1 arcsec, and a wavelength
coverage $4800$--$10000\,\angstrom$. We retrieved three exposures of 1200\,s
each. The final spectra present a S/N of about $20$ in galaxies with
$r'=20$\,mag. We used the DOLORES multi-object spectrograph (MOS) at
TNG to observe 9 candidates. The masks were designed to contain
more than 30 slitlets, 1.5 arcsec width, placed within an area about
$6^{\prime}\times 8^{\prime}$ in order to cover the target field. We
used the LR-B grism, which provides a dispersion of $2.7\,\angstrom$/pixel,
and a wavelength coverage between 4000 and $8000\,\angstrom$. We carried out
three acquisitions of 1800\,s each and obtained spectra with S/N $\simeq 15$
in galaxies with $r'=20$\,mag using a total integration time of
5400\,s.

\paragraph{Nordic Optical Telescope (NOT)} -- Spectroscopic 
redshift measurements were obtained using the Andalucia Faint Object
Spectrograph and Camera (ALFOSC) at the NOT.\footnote{The observing
runs took place on June 28 - July 3, 2011, January 20-25, 2012, July
16-21, 2012 and January 9-14, 2013.}
Most targets were observed in MOS mode, targeting typically ten to
fifteen galaxies per ALFOSC field (covering 
$6.4^{\prime} \times 6.4^{\prime}$, with an image scale of $0.188$ arcsec/pixel). 
One or two unfiltered 300s
pre-imaging exposures were obtained per candidate cluster, in addition
to a single 300s exposure in each of the SDSS $g$-and $i$ bands.  The
de-biased and flat field calibrated pre-imaging data were used to
select spectroscopy targets.  The final mask design\footnote{The MOS
masks were cut at the Niels Bohr Institute, Copenhagen University.}
was carved out using custom software, generating slits of fixed width
$1.5$ arcsec and of length typically $15$ arcsec.  Grism No. 5 of
ALFOSC was used, covering a wavelength range 5000 --
$10250\,\angstrom$ with a resolution of about $R=400$ and dispersion
$3.1\,\angstrom$/pixel. Redwards of $7200\,\angstrom$ strong fringing is
present in the ALFOSC CCD. It was effectively suppressed using dither
pattern alternating the placement of the spectroscopy targets between
these sets of slits. 

In addition to the MOS observations, spectroscopic observations in
single-slit mode were conducted for some \planck\ candidates.  For
these observations, a long slit covering the entire $6.4^{\prime}$
length of the ALFOSC field and a width of $1.3$ arcsec was employed,
with the same grism and wavelength coverage as for the MOS
observations. The field angle was rotated to place the long slit over
multiple targets, to include the apparent BCG as well as two to three
other bright cluster galaxies within the ALFOSC field.

\subsubsection{RTT150}
A total of 88 \Planck\ cluster candidates were followed up with
the Russian Turkish Telescope
(RTT150\footnote{\url{http://hea.iki.rssi.ru/rtt150/en/index.php}.})
from July 2011 to December 2012 within the Russian quota of
observational time.  In total, about 50 dark nights, provided by Kazan
Federal University and Space Research Institute (IKI, Moscow), were
used for these observations. Direct images and spectroscopic redshift
measurements were obtained using T\"UB\.ITAK Faint Object Spectrograph
and Camera
(TFOSC\footnote{\url{http://hea.iki.rssi.ru/rtt150/en/}\\
\url{index.php?page=tfosc}.}),
similar in layout to ALFOSC at NOT (see above) and to
other instruments of this series.

The TFOSC CCD detector cover a $13.3^{\prime} \times 13.3^{\prime}$
area with $0.39$ arcsec per pixel image scale. Direct images of
cluster candidates were obtained in Sloan \emph{gri} filters, in
series of $600$s exposures with small ($\approx 10$--$30$ arcsec)
shifts of the telescope pointing direction between the exposures. All
standard CCD calibrations were applied using {\tt Iraf} software,
individual images in each filter were then aligned and combined. The
total of 1800~s exposure time in each filter was typically obtained
for each field, longer exposures were used for more distant cluster
candidates.  Deep multi-filter observations were obtained for all
candidates, except those unambiguously detected in SDSS. With these
data, galaxy clusters can be efficiently identified at redshifts up to
$z\approx1$.

Galaxy clusters were identified as enhancements of surface number
density of galaxies with similar colours. Cluster red sequences were
then identified in the colour--magnitude diagram of galaxies near the
optical centre of the identified cluster. The detected red sequence
was used to identify the BCG and cluster member galaxies. Using the
measured red-sequence colour photometric redshift estimates were
obtained, which were initially calibrated using the data on optical
photometry for galaxy clusters from the 400SD X-ray galaxy cluster survey
\citep{bur07}.

For spectroscopy we used the long-slit mode of the instrument with
grism No. 15, which covers the 3900--$9100\,\angstrom$ wavelength range with
$\approx12\,\angstrom$ resolution when a slit of $1.8$ arcsec width is
used. Galaxy redshifts were measured through the cross-correlation of
obtained spectra with a template spectrum of an elliptical galaxy.
Spectroscopic redshifts were typically obtained for the spectra of a
few member galaxies, including the BCG, selected from their red
sequence in the imaging observations. These data allow us to
efficiently measure spectroscopic redshifts for clusters up to
$z\approx0.4$. For the highest-redshift clusters, complementary
spectroscopic observations were performed with the BTA 6-m telescope
of SAO RAS using SCORPIO focal reducer and spectrometer \citep{afa05}.

\subsection{Optical observation in the southern hemisphere}
\subsubsection{MPG/ESO 2.2-m Telescope} 
Optical imaging of 94 \Planck\ cluster candidates in the southern
hemisphere was performed under MPG programmes at the MPG/ESO 2.2-m
telescope using the Wide-Field Imager (WFI).\footnote{Based on
  observations under MPG programmes 086.A-9001, 087.A- 9003,
  088.A-9003, 089.A-9010, and 090.A-9010. The observations were
  conducted during the periods of November 27 - December 3, 2010,
  March 8-19, May 21 - June 3, and November 30 - December 4, 2011,
  December 30, 2011 - January 7, 2012, June 10-18, 2012, and January
  6-13 2013.} The WFI detector is a mosaic of 8
$2{\mathrm{k}}\times4{\mathrm{k}}$ CCDs, covering a total area of $33^{\prime}
\times 34^{\prime}$ on the sky, with an image scale of $0.238$
arcsec/pixel. Each field was observed in the $V$-, $R$-, and $I$-bands
with a default exposure time of $1800$~s (with five dithered sub-exposures)
per passband.  The basic data calibration, including de-biasing and
flat-field frame calibration, followed standard techniques. The
individual exposures were re-registered and WCS calibrated using the
USNO-B1 catalogue as an astrometric reference before being stacked
into a combined frame for each filter, covering the entire WFI
field. Photometric redshifts of the observed clusters were then
determined from an algorithm that searches for a spatial galaxy
overdensity located near the position of the SZ cluster candidate that
also corresponds to an overdensity in $V-R$ versus $R-I$
colour--colour space. The median colour of galaxies located in this
overdensity was then compared to predicted colours of early-type
galaxies at different redshifts by convolving a redshifted elliptical
galaxy spectral energy distribution template with the combined
filter+telescope+detector response function.

\subsubsection{New Technology Telescope (NTT)} 
Observations\footnote{The observations were performed during three
spectroscopic observing campaigns, 087.A-0740, 088.A-0268 and
089.A-0452.} were conducted at the 3.5-m NTT at the ESO observatory
at La Silla to measure spectroscopic redshifts of 33
\planck\ clusters with the EFOSC2 instrument in the MOS mode.
A clear BCG was identified in the clusters in pre-imaging data,
and besides the BCG a redshift was measured for at least one other
member of the cluster.  In the following a brief outline of the
observations and the data reduction are given \citep[see ][for
details]{cho12}.

Each field of the \Planck\ target candidates was optically imaged in
Gunn $r$ band for target selection and mask making. The imaging
resolution is $0.12^{\prime\prime} \times 0.12^{\prime\prime}$, and
the field of view is $4.1^{\prime} \times 4.1^{\prime}$ for both
imaging and spectroscopic observations. When necessary, the field was
rotated to optimize target selection. We used the grism that covers
the wavelength range between $4085\,\angstrom$ and $7520\,\angstrom$,
with $1.68\,\angstrom$ per pixel at resolution $13.65\,\angstrom$ per
arcsec. We typically applied 10 to 15 slitlets per field with a fixed
width of 1.5 arcsec for the MOS and of 2.0 arcsec for the long-slit
observations. Including at least three bright objects, preferably
stars, to orient the field, the slitlets were allocated to the
candidate member galaxies.  The exposure times for the clusters range
from 3600~s to 10800~s.

The data were reduced with the standard reduction pipeline of
{\tt Iraf}. The redshifts from the emission lines were determined
separately after correlation with the passive galaxy templates. We use
the {\tt rvsao} package, which applies the cross-correlation
technique to the input templates of galaxy spectra to measure the
object redshift. The REFLEX templates were used for this analysis,
which include 17 galaxy and stellar templates. We confirmed a
spectroscopic cluster detection if at least three galaxies have their
R-value greater than 5, and lie within $\pm\,3000$km/s of the mean
velocity of the cluster members. We then took the median of those
galaxy redshifts as the cluster redshift. For the long-slit
observations, the cluster was confirmed with the redshift of the BCG
and another galaxy at similar redshift within the aforementioned
criteria.

\subsection{Observations in the SZ domain with AMI}\label{sec:ami}
An ensemble of 60 \Planck\ blind SZ candidates, spanning a
range of S/N between $4$ and $9$ and meeting  the Arcminute
Microkelvin Imager (AMI) observability
criteria, was observed with AMI. The
goal of this programme was to confirm \Planck\ cluster candidates
through higher-resolution SZ measurements with AMI and to refine the
position of confirmed clusters in order to optimize the subsequent
optical follow-up observations aiming at redshift measurement.  AMI
comprises of two arrays: the Small Array (SA); and the Large Array
(LA). Further details of the instrument are given in
\citet{ami08}. Observations carried out with the SA provide
information that is well coupled to the angular scales of the SZ
effect in clusters, whereas snapshot observations obtained with the LA
provide information on the discrete radio-source environment. The
latter allowed us to detect the presence of nearby, bright radio
sources, helping in further selecting the targets for observation with
the SA. Details of the AMI data reduction pipeline and mapping are
described in \citet{planck2012-II}.

\section{Results of the validation and follow-up}
\label{s:results}

\begin{table*}
\begingroup
\newdimen\tblskip \tblskip=5pt
\caption{{Numbers of previously-known clusters, new confirmed clusters, and new candidate
    SZ clusters.  Previously-known clusters can be found in the
    catalogues indicated.  Confirmations from follow-up do not cover
    the observations performed by the \Planck\ collaboration to
    measure the missing redshifts of known clusters.  Confirmation
    from archival data covers X-ray data from {\it Chandra}, {\it XMM-Newton}, and
    {ROSAT} PSPC pointed observations only.}}
\label{tab:valid_sum}
\nointerlineskip \vskip -3mm \footnotesize
\setbox\tablebox=\vbox{
\newdimen\digitwidth
\setbox0=\hbox{\rm 0}
\digitwidth=\wd0
\catcode`*=\active
\def*{\kern\digitwidth}
\newdimen\signwidth
\setbox0=\hbox{+}
\signwidth=\wd0
\catcode`!=\active
\def!{\kern\signwidth}
\halign{\hbox to 4.5cm{#\leaderfil}\tabskip=2em&
   \hfil#\hfil& 
   \hfil#\hfil\tabskip=0pt& 
   #\hfil\tabskip=0pt\cr
\noalign{\doubleline}
\omit\hfil Category\hfil&$N$&&\omit\hfil\hglue 2em Catalogue, telescope, or reliability\hfil\cr
\noalign{\vskip 3pt\hrule\vskip 3pt}
Previously known&*683&$\Biggl\{ \;$&$\vcenter{\hbox{472\quad \hbox to 0.5in{X-ray:\strut\hfil}   MCXC meta-catalogue\hfill}
                                           \hbox{182\quad \hbox to 0.5in{Optical:\strut\hfil} Abell, Zwicky, SDSS catalogues\hfill}
                                           \hbox{*16\quad \hbox to 0.5in{SZ:\strut\hfil}      SPT, ACT\hfill}
                                           \hbox{*13\quad \hbox to 0.5in{Misc:\strut\hfil}    NED or SIMBAD\hfill}}$\cr
\noalign{\vskip 8pt}
New confirmed&*178&&\hbox{***\quad Follow-up, archival data, SDSS survey}\cr
\noalign{\vskip 8pt}
New candidate&*366&   $\Biggl\{ \;$&$\vcenter{\hbox{*54\quad High reliability}
                                           \hbox{170\quad Medium reliability}
                                           \hbox{142\quad Low reliability}}$\cr
\noalign{\vskip 8pt}
\omit&\hrulefill\cr
\bf Total \Planck\ SZ catalogue&\bf 1227\cr
\noalign{\vskip 5pt\hrule\vskip 3pt}}}
\endPlancktable
\endgroup
\end{table*}

The external validation allows us to identify \Planck\ SZ detections
with previously-known clusters and to assemble crucial information on
the identified clusters such as their redshifts. The validation steps
corresponding to the association with known clusters were performed
following a chosen hierarchy: X-ray clusters from the updated MCXC
meta-catalogue; then optical clusters from Abell and Zwicky
catalogues; then optical clusters from the SDSS-based catalogues;
followed by SZ clusters from SPT and ACT samples; and finally clusters
from NED and SIMBAD queries. The first identifiers of the \Planck\ SZ
detections given in Table \ref{tab:PSZ} reflect the validation
hierarchy.

\begin{figure}[t]
\begin{center}
\includegraphics[width=8.8cm]{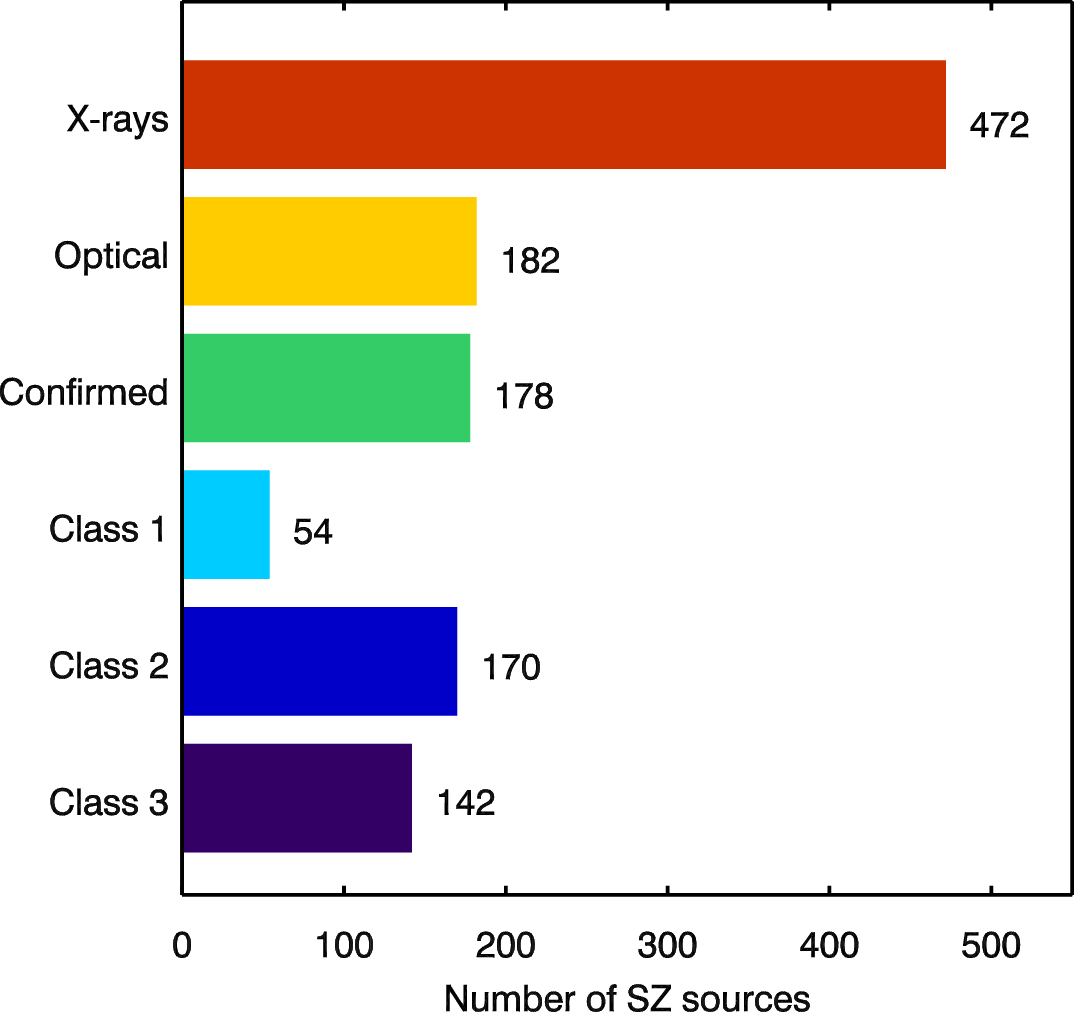}
\caption{Distribution of the \planck\ clusters and candidates in the
  different categories defined in the external validation process.
  The validation follows the order of association with MCXC clusters,
  then Abell and Zwicky clusters, then SDSS clusters, then SZ
  clusters, and finally clusters from NED/SIMBAD.}
\label{fig:pie}
\end{center}
\end{figure}

In the following, we present the results of the external validation
process and of the follow-up
campaigns for confirmation of \Planck\ candidates and measurement of
their redshifts (see Table~\ref{tab:valid_sum} and
Fig.~\ref{fig:pie}). We also present the confirmation from SDSS galaxy
catalogues and from X-ray archival data. We further discuss the
unconfirmed candidate new clusters detected by \planck, which we
classify into three categories of different reliability. 

\subsection{\Planck\ clusters associated with known clusters}
A total of 683 out of 1227 SZ detections in the \Planck\ catalogue,
i.e., 55.7\%, are associated with previously-known clusters from
X-ray, optical, or SZ catalogues, or with clusters found in the NED or
SIMBAD databases. We give the number of clusters identified in each
category and we discuss notable cases of known clusters that are not
included in the \Planck\ SZ catalogue.

\subsubsection{Identification with known X-ray clusters}
A total of 472 \Planck\ SZ-candidates are identified with known X-ray
clusters from the MCXC meta-catalogue, which represents 38.5\% of the
\Planck\ SZ detections and 69.1\% of the identifications with
previously-known clusters. These identifications of course account for
many Abell clusters in the RASS-based catalogues of X-ray clusters.

Using the cluster properties reported in the MCXC and the
\planck\ noise maps at the cluster positions, we
computed the expected SZ signal and the expected S/N for a
measurement with \Planck.  We have compared the number of detected
clusters in the \Planck\ catalogue with S/N $\ge 4.5$ to the number MCXC
clusters at an expected significance of 4.5. Only 68 clusters expected
to be detected at S/N $>4.5$ are not included in the \Planck\ catalogue,
including 16 with predicted S/N between 4 and 4.5. Of the 52 clusters
with expected S/N $\ge 4.5$, only 41 are outside the masked regions
and could thus be in the PSZ catalogue. Our computation of the
expected SZ signal and S/N were based on scaling relations for
X-ray-selected clusters, not accounting for the dispersion in the
relations. We therefore focus on the non-detected MCXC clusters that
significantly depart from the expected S/N value, namely by more than
5$\,\sigma$. A total of 13 clusters are in this category. The two
objects RXCJ2251.7-3206 and RXCJ0117.8-5455 show emission in
high-resolution \textit{Chandra} imaging that is point-like rather
than extended and are likely not clusters of galaxies
\citep{man10,mag07}. Of the other eleven missing MCXC clusters, some
present AGN contamination. This is the case for RXC J1326.2+1230
\citep{mag07}, RXJ1532.9+3021 \citep{hla12}, RXCJ1958.2-3011,
RXCJ2251.7-3206, and RXCJ0117.8-5455 \citep{mag07}, Abell 689
\citep{gil12}, ZwCl2089 \citep{raw12}, PKS 0943-76 \citep{abd10}, and
Abell 2318 \citep{cra99}. In these cases, the presence of the AGN
affects the X-ray luminosity measure leading to an overprediction of
the SZ signal. Some exhibit significant radio contamination,
e.g., RXCJ1253.6-3931 \citep{pla10} and RXCJ1958.2-3011 \citep{mag07},
which hampers the SZ detection. Cool-core clusters for which the X-ray
luminosity is boosted due to the central density peak have an
over-estimated expected SZ signal. This is the case for
RXCJ0425.8-0833 \citep{hud10}, ZwCl2701 \citep{raw12}, Abell 1361
\citep{raf08}, and RBS 0540 \citep{eck11,bel05}. Other ``missing''
clusters are CIZA clusters: RXC J0643.4+4214, RXC J1925.3+3705, RXC
J2042.1+2426 and RXC J0640.1-1253, REFLEX cluster RXCJ2149.9-1859,
APMCC 699, Abell 3995, Abell 2064 and RBS 171.

\begin{figure*}[htbp]
\begin{center}
\includegraphics[angle=0,width=0.3\textwidth]{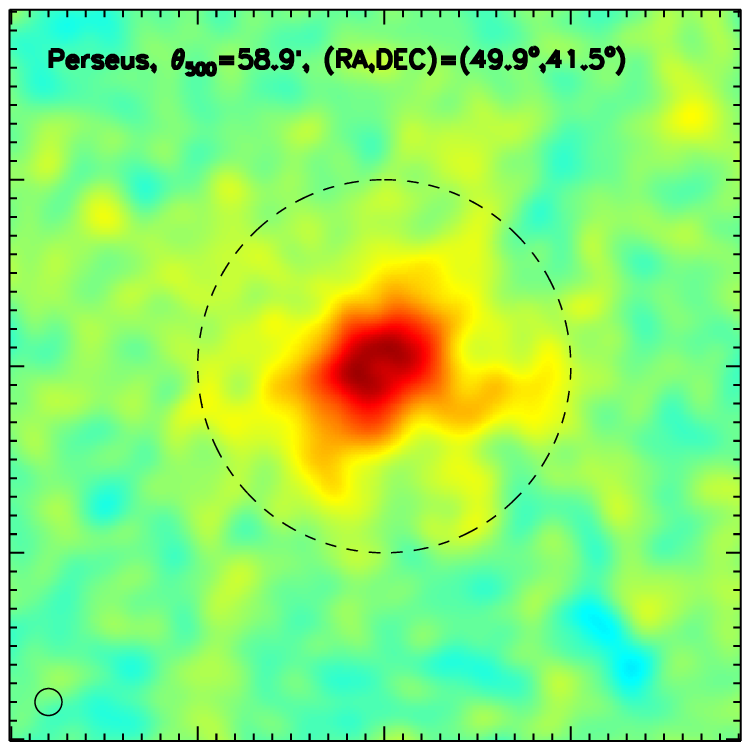}
\includegraphics[angle=0,width=0.3\textwidth]{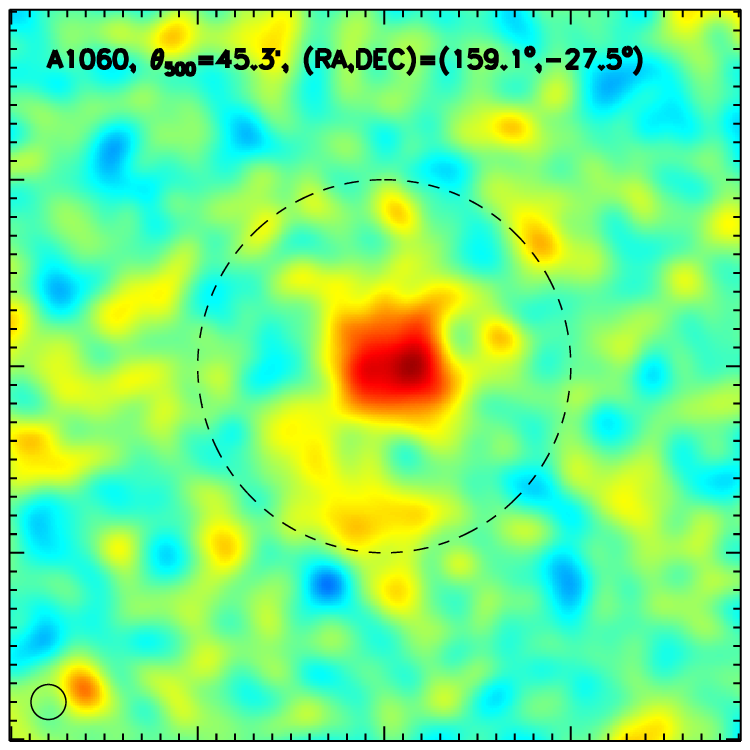}
\includegraphics[angle=0,width=0.3\textwidth]{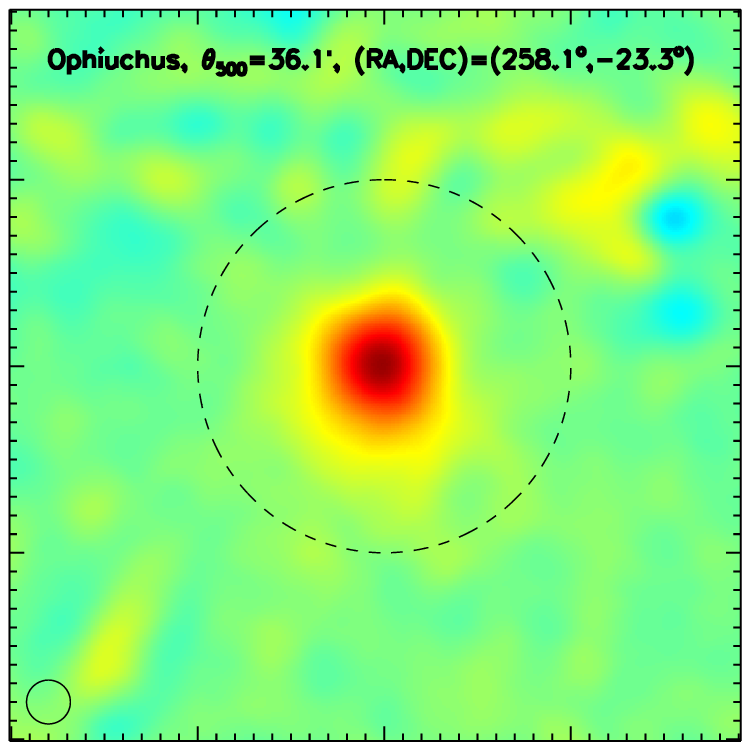}\\[-1ex]
\includegraphics[angle=0,width=0.3\textwidth]{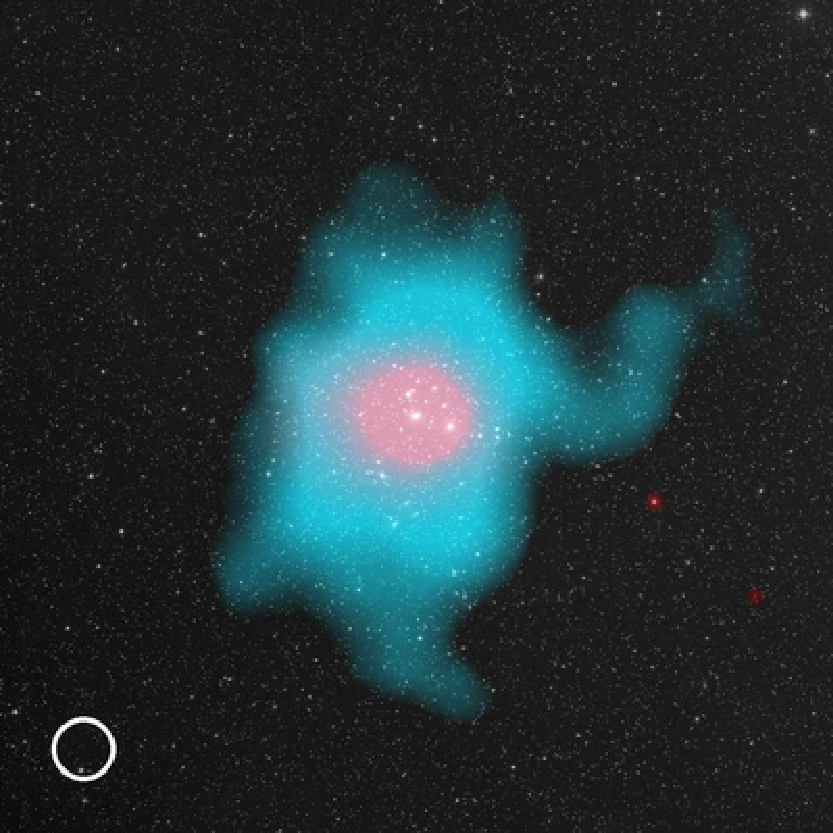}
\includegraphics[angle=0,width=0.3\textwidth]{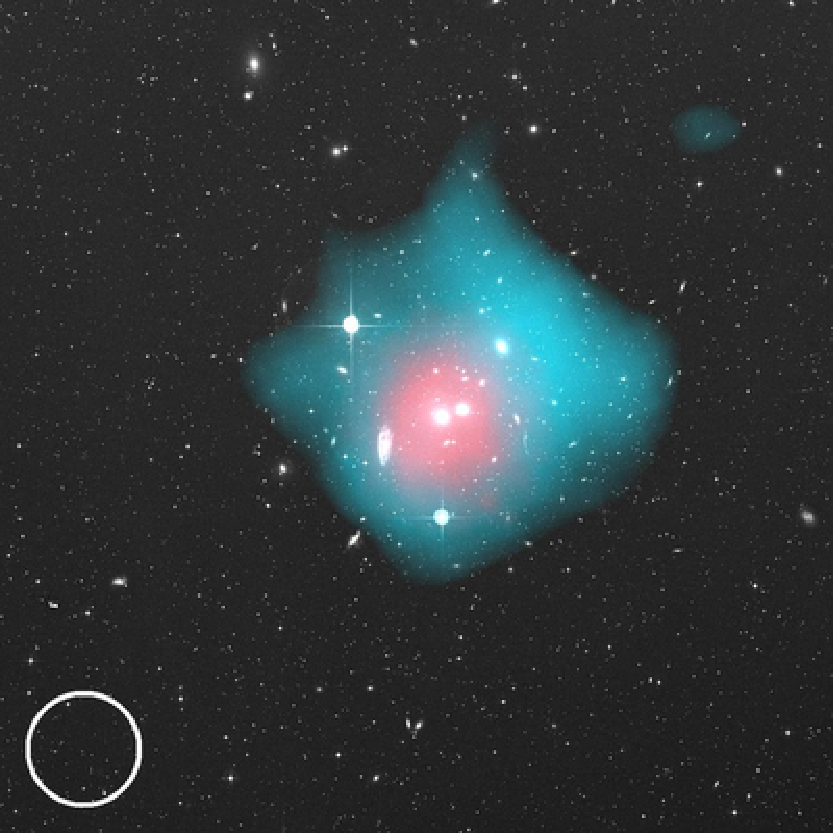}
\includegraphics[angle=0,width=0.3\textwidth]{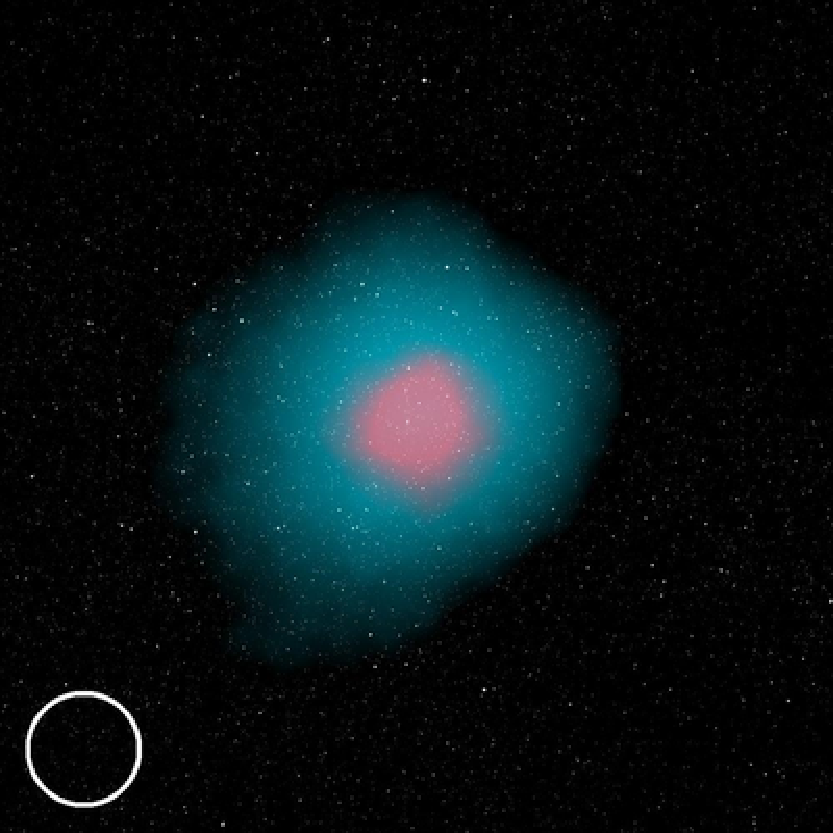}\\[1em]
\includegraphics[angle=0,width=0.3\textwidth]{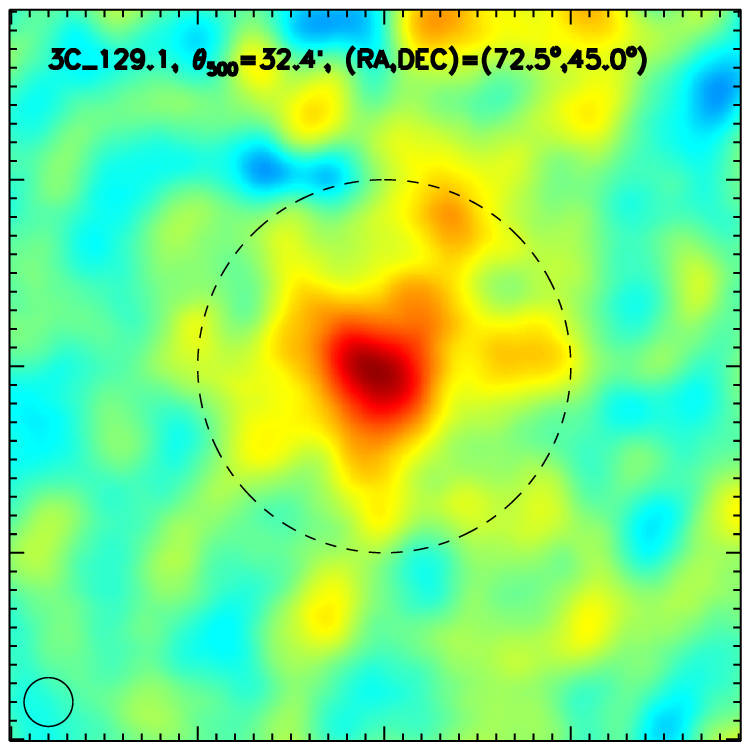}
\includegraphics[angle=0,width=0.3\textwidth]{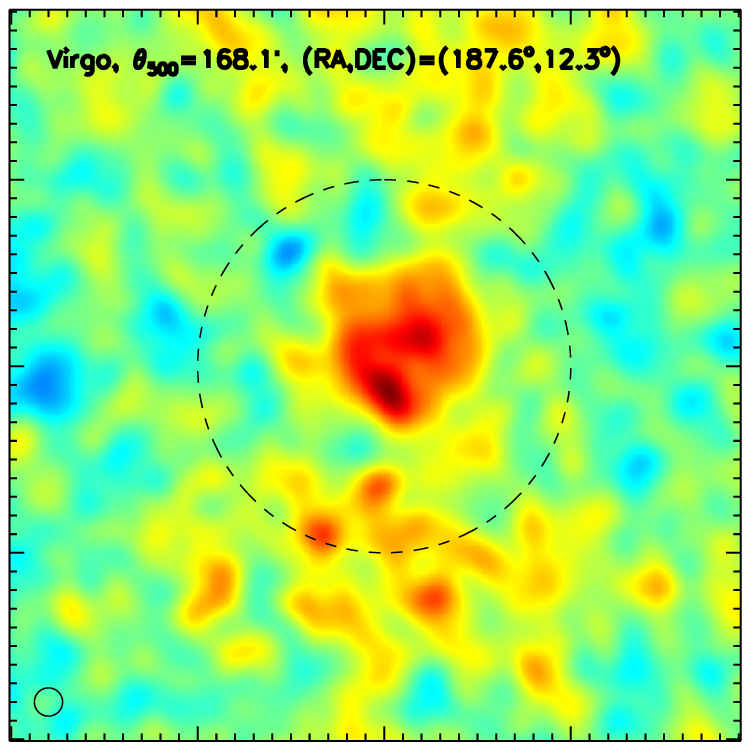}\\[-1ex]
\includegraphics[angle=0,width=0.3\textwidth]{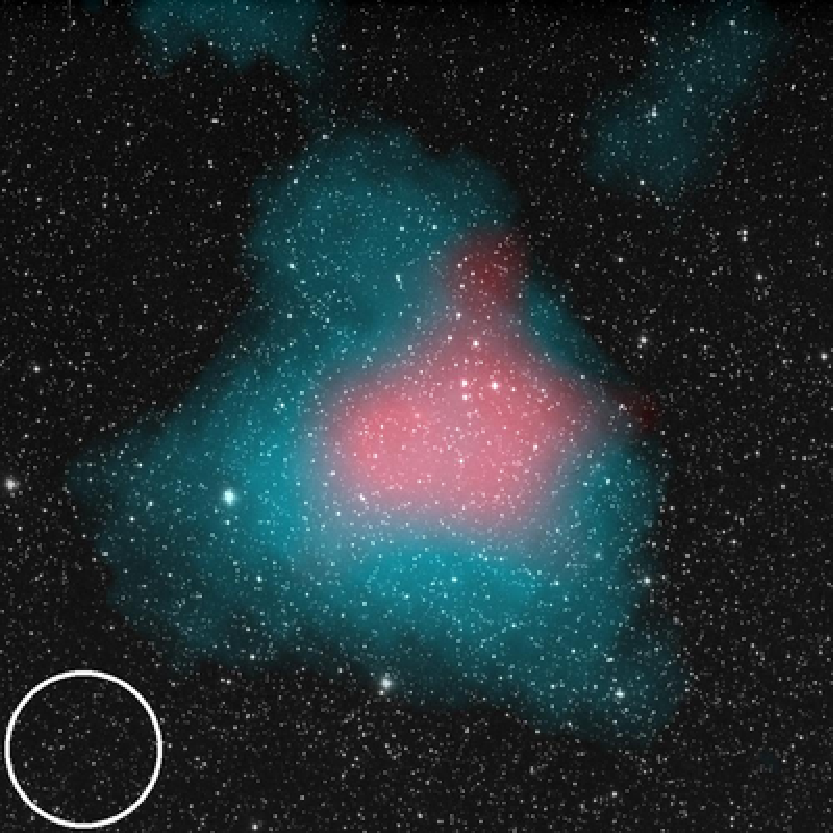}
\includegraphics[angle=0,width=0.3\textwidth]{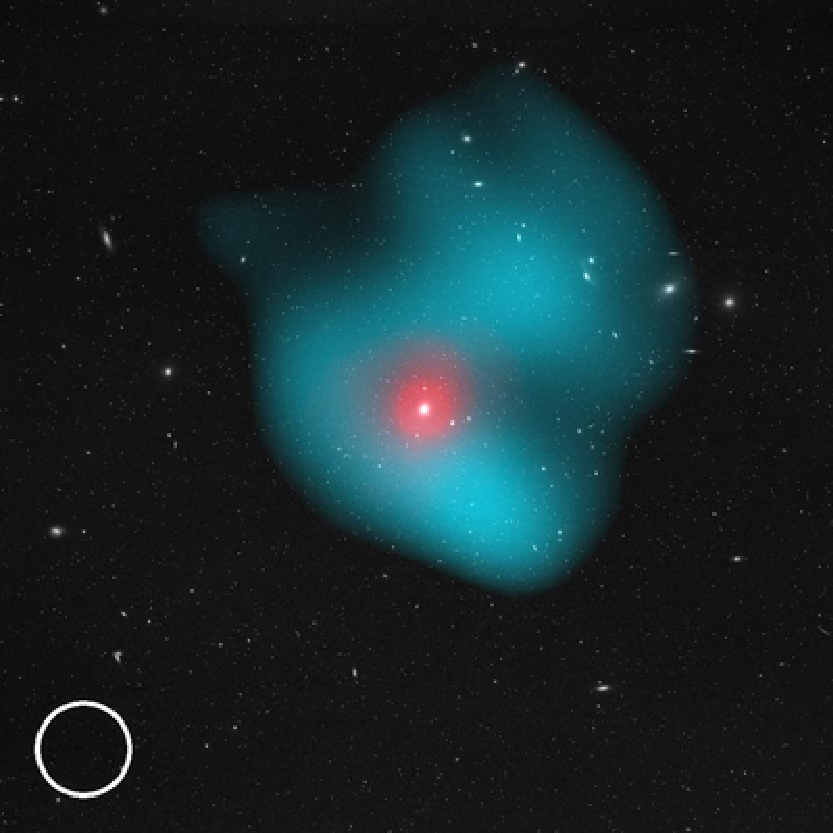}

\caption{Five nearby and extended clusters not included in
  the PSZ catalogue: 
  the Perseus cluster and Abell 1060 (in the point-source mask);
  Ophiuchus cluster and 3C~129.1 (in the Galactic mask); and Virgo
  cluster (below the S/N threshold of the catalogue).  \emph{Top
    panels:} reconstructed thermal SZ maps from the \texttt{MILCA}
  algorithm \citep{hur13}. The dashed circles represent the apertures
  of $\theta_{500}$ from the MCXC catalogue. Each SZ-map covers an
  area of $4\theta_{500} \times 4\theta_{500}$. \emph{Bottom panels:}
  composite images of the optical (DSS, white), X-ray (ROSAT, pink)
  and SZ signal (\planck\, blue). The sizes of the composite images are
  $2^{\circ} \times 2^{\circ}$ for Perseus; $1^{\circ}\times
  1^{\circ}$ for A1060; $1^{\circ} \times 1^{\circ}$ for Ophiuchus;
  $0.77^{\circ}\times 0.77^{\circ}$ for 3C~129.1 and $3.84^{\circ} \times
  3.84^{\circ}$ for Virgo. The black and white circles picture a
  10~arcmin aperture, but for Virgo for which the aperture is
  30~arcmin.}
\label{milcaext}
\end{center}
\end{figure*}

In addition to the clusters discussed above which are not included in
the catalogue due to contamination by AGN or presence of cool-cores
etc., we note that some notable nearby extended clusters are also not
included in the \planck\ SZ catalogue. Indeed, the detection methods
used to detect the SZ effect are not optimized for the detection of
sources with scale radius $\theta_{500}$ in excess of 30$^{\prime}$. Of
the 25 clusters in this category (with $z<0.03$) in the MCXC
meta-catalogue, six are included in the \Planck\ catalogue. The
remaining 19 fall into the masked areas (seven out of 19,
among which Perseus and Abell 1060 lie in the PS mask
(Fig.~\ref{milcaext}, first two panels), and Ophiuchus and 3C~129.1 lie
in the Galactic mask (Fig.~\ref{milcaext}, second two panels) and/or
have a S/N below the PSZ catalogue threshold S/N $= 4.5$.  This
is the case of Virgo cluster (Fig.~\ref{milcaext}, lowest panel), which
is detected in the \Planck\ survey but with a S/N at
its position of  about 3.9. Virgo's extension on the sky
($\theta_{500}=168$ arcmin) further hampers its blind detection.

We show in Fig.~\ref{milcaext} the reconstructed SZ signal from the
{\tt MILCA} algorithm \citep{hur13} for five of the ``missing''
extended clusters. These clusters, despite not being part of the
\Planck\ catalogue of SZ sources, are well detected in the
\Planck\ survey. They all are included in the thermal SZ map
constructed from the \Planck\ channel maps and presented in
\citet{planck2013-p05b}.

\subsubsection{Identification with known optical clusters}
A total of 182 \Planck\ SZ detections are identified exclusively with
optical clusters from Abell and Zwicky catalogues, and from the
SDSS-based published catalogues, i.e., 26.6\% of the known clusters in
the \planck\ catalogue.

The \Planck\ SZ candidates at S/N $\ge 4.5$ have 111 exclusive
associations with Abell or Zwicky clusters, i.e., with clusters not in
any of the catalogues compiled in the MCXC meta-catalogue. In addition
to these associations, 72 \Planck\ detections are solely identified
with clusters from the SDSS-based catalogues. These are either rich
and massive systems ($R_{L^\star}$ greater than 110, $\qsdss=1$
clusters) or moderately low-richness systems ($\qsdss =2$ clusters,
exhibiting hot gas as indicated by their S/N value in the RASS
survey). However, not all the rich $\qsdss =1$ clusters in SDSS-based
catalogues are found in the \Planck\ catalogue. A total of 213 $\qsdss
=1$ clusters from all four SDSS-based catalogues (201 outside the
\planck\ union PS and Galactic mask) are not included in the
\Planck\ catalogue.

We explore why these rich clusters are not detected blindly by the
SZ-finder algorithms. We first compare the richness-based masses
against the X-ray luminosity-based masses of 26 of these ``missing''
clusters found in the MCXC meta-catalogue.  We find a median ratio of
$2.6\pm 1.2$ for the richness-to-X-ray based masses, indicating that
the richness-based masses seem to be systematically overestimated.
Unlike the X-ray clusters, we thus cannot compute a reliable estimate
of the expected S/N value for SZ detection of these optical
clusters. We therefore directly search for the SZ signal at the
positions of the 201 ``missing'' SDSS-clusters and found that all of
them have S/N values below the \planck\ threshold, with a mean
S/N of 1.6, except for three clusters.  Two of these three
``missing'' SDSS-clusters have their S/N value from the extraction at
the cluster position slightly higher than 4.5.  The increase in S/N
value is due to the difference in estimated background noise when
centring the extraction at the cluster position as opposed to the
blind detection. The third missing rich cluster is affected by
contamination from CMB anisotropy, which results in a bad estimate of
its size and consequently of its SZ signal.

\subsubsection{Identification with known SZ clusters}

The majority of the SZ clusters, from SPT or ACT, used in the
validation process are low-mass systems
($M_{500}^{\mathrm{median}}$ around $2.3\times 10^{14}\,\msol$).
\Planck\ is particularly sensitive to massive rich clusters and thus
only a total of 56 of these clusters match \Planck\ SZ detections, out
of which 16 candidates are exclusively associated with SZ
clusters\footnote{Six \planck\ clusters were confirmed from \xmm\ or
 NTT observations and are also published in \citet{rei13}.} from ACT or
SPT. Nine more ACT and SPT clusters are associated with \Planck\ SZ
detections between S/N $= 4$ and 4.5. We have searched for the SZ
signal in the \Planck\ data at the position of the remaining
non-observed ACT/SPT clusters by extracting the SZ signal at their
positions. We found that all had S/N values lower than 4.

\begin{figure}
\includegraphics[width=8.8cm]{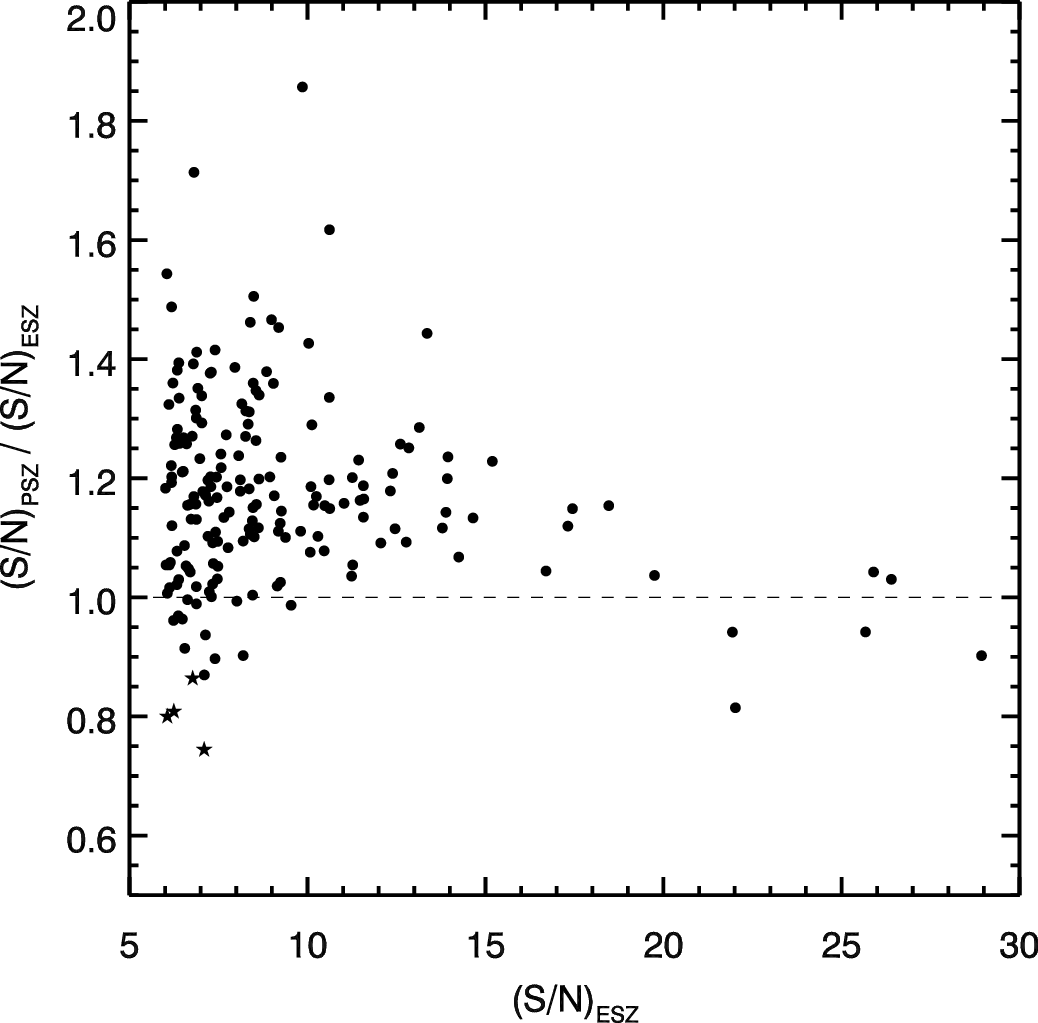}
\caption{Ratio of S/N in the 
  \Planck\ catalogue, (S/N)$_{\mathrm{PSZ}}$, to that in the ESZ
  sample \citep{planck2011-5.1a}, (S/N)$_{\mathrm{ESZ}}$, for the 184
  confirmed ESZ clusters included in the \Planck\ catalogue. Four
  clusters whose S/N in the PSZ catalogue is significantly
  smaller than the ESZ threshold ((S/N)$_{\mathrm{ESZ}}=6$) are shown
  as stars. }
\label{fig:ESZ_in_DX9}
\end{figure}

We have also checked the redundancy of SZ detections within
\Planck\ by comparing the ESZ sample, constructed from 10 months of
survey with a cut at Galactic latitudes of $\pm$ 14$^{\circ}$, with the
present \planck\ catalogue. Of the 189 high significance
((S/N)$_{\mathrm{ESZ}}\ge 6$ ESZ detections, 184 ESZ confirmed
clusters are included the present \Planck\ catalogue within a distance
of 5$^{\prime}$ from their ESZ position. The mean separation between
the ESZ and present positions is of order 1.35$^{\prime}$, within
\Planck's positional accuracy.  Their S/N values were increased by a
factor 1.17 on average with respect to their (S/N)$_{\mathrm{ESZ}}$,
(Fig.~\ref{fig:ESZ_in_DX9}) and only four out of six of the ESZ
clusters have new S/N values significantly lower than ESZ
S/N threshold (S/N)$_{\mathrm{ESZ}}=6$. They are displayed
as stars in Fig.~\ref{fig:ESZ_in_DX9}.  Four ESZ clusters are not
included the present \Planck\ catalogue, they fall in, or nearby, the
PS mask used for the pre-processing of the channel maps prior to
running the detection algorithms. Such a mask was not utilized for the
construction of ESZ sample. We choose not to a posteriori include
these four ``missing'' ESZ clusters in the present \planck\ SZ
catalogue.

\subsubsection{Identification with clusters from NED or SIMBAD}

As expected only a small number of clusters are identified from
querying the databases, supplying identifiers for thirteen SZ
\Planck\ detections. This is because the information in NED and SIMBAD
is redundant with that in the X-ray, optical, or SZ catalogues used
for the external validation. The thirteen clusters found solely from
querying the databases are found in the RASS survey but not in
dedicated cluster catalogues, and thus not included in the MCXC; they
are found in serendipitous {\it Chandra} surveys, or they are part of
miscellaneous cluster catalogues.

\subsection{Newly-discovered \Planck\ clusters and candidates}\label{sec:confi}

Among the 544 \Planck\ SZ sources, we distinguish two categories: (1)
confirmed clusters, i.e., those that have been confirmed by the
follow-up programmes of the \Planck\ collaboration\footnote{A handful
  of new \Planck\ clusters from the ESZ sample were confirmed
  independently from the \planck\ collaboration by SPT \citep{sto11},
  AMI \citep{ami11}, Bolocam \citep{say12} and CARMA \citep{muc12}.}
or using the SDSS galaxy catalogues, plus also add eight confirmations
from X-ray archival data {(one of those, PSZ1 G292.00-43.64,
  coincides with the XCLASS cluster candidate, J023303.4-711630
  \citep{cle12})}; (2) Candidate clusters with different levels of
reliability, namely, {\sc class1} cluster candidates, that fulfil
high-quality criteria for the SZ detection and for the associations
and/or counterparts in ancillary data, {\sc class2} candidate
clusters, i.e., those that fulfil, on average, good-quality criteria,
and {\sc class3}, low-reliability cluster candidates.

\paragraph{{\bf Confirmation from \Planck\ collaboration follow-up
  programmes}} 

At S/N $\ge 4.5$, a total of 233 \planck\ SZ detections were followed
up in X-rays, optical, and SZ at the different facilities listed
previously, with some observations targeted to the measurement of
spectroscopic redshifts for already known clusters. In total 157
\planck\ SZ detections with S/N $\ge 4.5$ were confirmed as new
clusters.  Some of the
\planck-confirmed  clusters were also reported in recent
cluster catalogues in the optical, e.g., \citet{wen12} or in the SZ
e.g., \citet{rei13}. \\
The analysis of the
observations of \planck\ sources by AMI yielded ten sources with strong
Bayesian evidences that have clearly visible decrements and were
considered as confirmed, including the confirmation of three
associations with optical clusters.\\
For the candidates confirmed by \xmm\ and by optical
telescopes, redshifts from Fe lines and from photometric or
spectroscopic data are available.  The validation of \Planck\ cluster
candidates with \xmm\ has shown its particular efficiency in
confirming SZ candidates due both to the high sensitivity of \xmm,
allowing \Planck\ clusters to be detected up to the highest redshifts
\citep{Planck2011-5.1c}, and the tight relation between X-ray and SZ
properties.  The detection of extended \xmm\ emission and a comparison
between the X-ray and SZ flux permits an unambiguous confirmation of
the candidates.  By contrast, confirmation in the optical may be
hampered by the \Planck\ positional accuracy and by the scatter
between the optical observables and the SZ signal, which increase the
chance of false associations.  The \xmm\ follow-up programme yielded
51 bona fide newly-discovered clusters, including four double
systems and two triple systems. There were eight false candidates.
Thirty-two of the 51 individual clusters have high-quality redshift
measurements from the Fe line. The relation between the X-ray and SZ
properties was used to further constrain the redshift of the other
clusters; most of these redshifts were confirmed clusters using
optical observations. Out of a total of 37 single clusters confirmed
by \xmm, 34 are reported in the \planck\ catalogue of SZ sources at
S/N $\ge 4.5$. Additionally four double systems are included in the
present PSZ catalogue and were also confirmed by \xmm. \\
The follow-up
observations conducted with optical telescopes lead to the confirmation
and to the measurement of spectroscopic or photometric redshifts
(companion publications, in preparation, will present the detailed
analysis and results from these follow-up). In the northern
hemisphere, 26 spectroscopic 
redshifts for \planck\ clusters detected at S/N $\ge 4.5$ and observed
at the RTT150 are reported, to date, in the PSZ catalogue. A dozen
additional spectroscopic redshifts were measured for known
clusters. Confirmation of 21 \planck\ SZ clusters detected above 4.5
were obtained with the ENO facilities (at INT, GTC and NOT), and
robust redshift measurements were obtained for 19 of them, including 13
spectroscopic redshifts. In the southern hemisphere, WFI observations
provided photometric redshifts for 54 clusters included in the
\planck\ catalogue at S/N $\ge 4.5$, while 19 spectroscopic redshifts
obtained with the NTT-EFOCS2 instrument are reported in the
\planck\ catalogue. 

\paragraph{{\bf Confirmation from SDSS galaxy catalogues}}

The firm confirmation of the candidates was done through the follow-up
observations for confirmation and measurement of their redshift as
detailed above. However in the case of the \planck\ candidates falling
in the SDSS footprint we also used the SDSS galaxy
catalogues to search, as presented in Sect.~\ref{sec:sdss}, for galaxy
overdensities associated with \planck\ SZ detections. This provides us
with an estimate of the photometric redshifts, and in some cases
we could retrieve spectroscopic redshifts for the BCG as well. 

In this process, the major uncertainty in the associations of
\planck\ SZ detections with galaxy overdensities is due to chance
associations with low-richness systems or associations with diffuse
concentrations of galaxies in the SDSS data. The \xmm\ confirmation
programmes (see \citet{planck2012-IV} for discussion) showed that
\Planck\ candidates with SDSS counterparts were confirmed including
PLCK~G193.3$-$46.1 at $z\simeq 0.6$. However, the X-ray analysis of
the \planck\ detections with SDSS counterparts illustrated the
difficulty in distinguishing between associations of \planck\ SZ
signals with massive clusters and with pre-virialized structures. In
particular, in the case of extended filamentary structures or
dynamically perturbed sources, an offset between the BCG position and
the concentration barycentre is noted.

We considered the \planck\ SZ candidates with counterparts in the
SDSS data taking into account diagnostics such as the richness/mass
estimates as well as the offsets between the SZ, the BCG and the
barycentre positions. We further used the outputs of the search in
WISE and in RASS data, and the associated images, in order to assess
the significance of the galaxy overdensity in SDSS at the position of
the \Planck\ candidates. For the \planck\ SZ detections where both
ancillary data and SDSS barycentre/BCG positions agreed, we set that
they are confirmed. We found a total of 13 such associations
for which we report the photometric or the spectroscopic redshifts. It
is worth noting that firm confirmation of these associations is needed
and needs to be performed using either optical spectroscopic
observations or X-ray observations of the \planck\ SZ detections.  In
the cases where the offsets between barycentre and BCG position
output by the search in SDSS data were too large, and/or when other
ancillary information was unable to discriminate between reliable or
chance associations, we have chosen to keep the status of candidate
for the \Planck\ SZ detection. These cases sometimes also coincide
with association of \Planck\ detections with clusters from the SDSS
cluster catalogues, with a quality flag $\qsdss=0$, or with confusion
in the association, i.e., with positions not in agreement between
counterpart and published SDSS clusters.  We provide a note for all
these cases in order to indicate that an overdensity in SDSS data was
found.

\paragraph{{\bf Candidate new clusters}}\label{sec:candidates}

\begin{figure*}[t]
\begin{center}
\includegraphics[width=16cm]{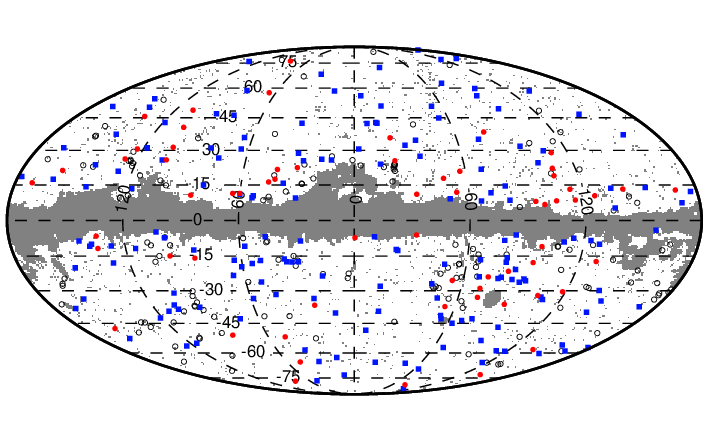}
\caption{Distribution of the \Planck\ SZ candidates across the
  sky. Blue symbols represent the {\sc class1} candidate
  clusters and red the {\sc class2} candidates. The open symbols stand
  for the {\sc class3} low-reliability SZ sources.}
\label{fig:dist_c123}
\end{center}
\end{figure*}

The remaining 366 \planck\ SZ sources, not identified with previously
known cluster nor confirmed by follow-up observation or ancillary
data, are distributed over the whole sky (Fig.~\ref{fig:dist_c123})
and are yet to be firmly confirmed by multi-wavelength follow-up
observations. They are characterized by an ensemble of quality flags
defined in Sects.~\ref{sec:rass}, \ref{sec:sdss}, and \ref{sec:wise}
based on the systematic searches for counterparts in the public
surveys during the external validation process. We
further define an empirical \Planck-internal quality flag
$Q^{\mathrm{SZ}}$.  It assesses the reliability of the SZ detection
itself from three independent visual inspections of the nine
\planck\ frequency maps, of frequency maps cleaned from Galactic
emission and CMB, and of reconstructed $y$-maps or $y$-maps produced
from component separation methods \citep[e.g.,
][]{hur13,rem11}. Moreover, we visualize the SZ spectra from the 
SZ-finder algorithms and from aperture photometry measurements at the
candidate positions. Finally we correlate, at the position of the
\Planck\ SZ candidates and within an area of 10$^{\prime}$ radius, the
$y$-map to the 857~GHz channel map, as a tracer of the dust emission,
and to the \Planck\ mono-frequency CO map at 217~GHz
\citep{planck2013-p03a}. The qualitative flag $Q^{\mathrm{SZ}}$
combines all this information into three values 1 to 3 from highest to
lowest reliability with the following criteria:
\begin{itemize}
\item $Q^{\mathrm{SZ}} = 1$, i.e., high reliability: (i) Clear compact
SZ source in the SZ maps; (ii) significant measurements of the SZ
decrement below 217\,GHz and good or reasonable detection at 353\,GHz;
(iii) no correlation with dust nor CO emission and no rise of the 545 and
857\,GHz fluxes on the thermal SZ spectrum.
\item $Q^{\mathrm{SZ}} = 2$, i.e., good reliability: (i) visible SZ
detection in the SZ map or significant detection of the
SZ signal below 217\,GHz; (ii) contamination causing rise of the
 545\,GHz and possibly 857~GHz flux on the SZ spectrum without a
 strong correlation with dust and CO signals.
\item $Q^{\mathrm{SZ}} = 3$, i.e., low reliability: (i) weak SZ signal
in the $y$-maps and/or noisy SZ maps; (ii) weak or no SZ signal in
the cleaned frequency maps (iii) strong correlation ($\ge$80\%) with
dust and CO emission contamination with rising fluxes on the SZ
spectrum at high frequencies, 353\,GHz and above.
\end{itemize}

\begin{figure*}[p]
\begin{center}
\includegraphics[height=0.9\textheight]{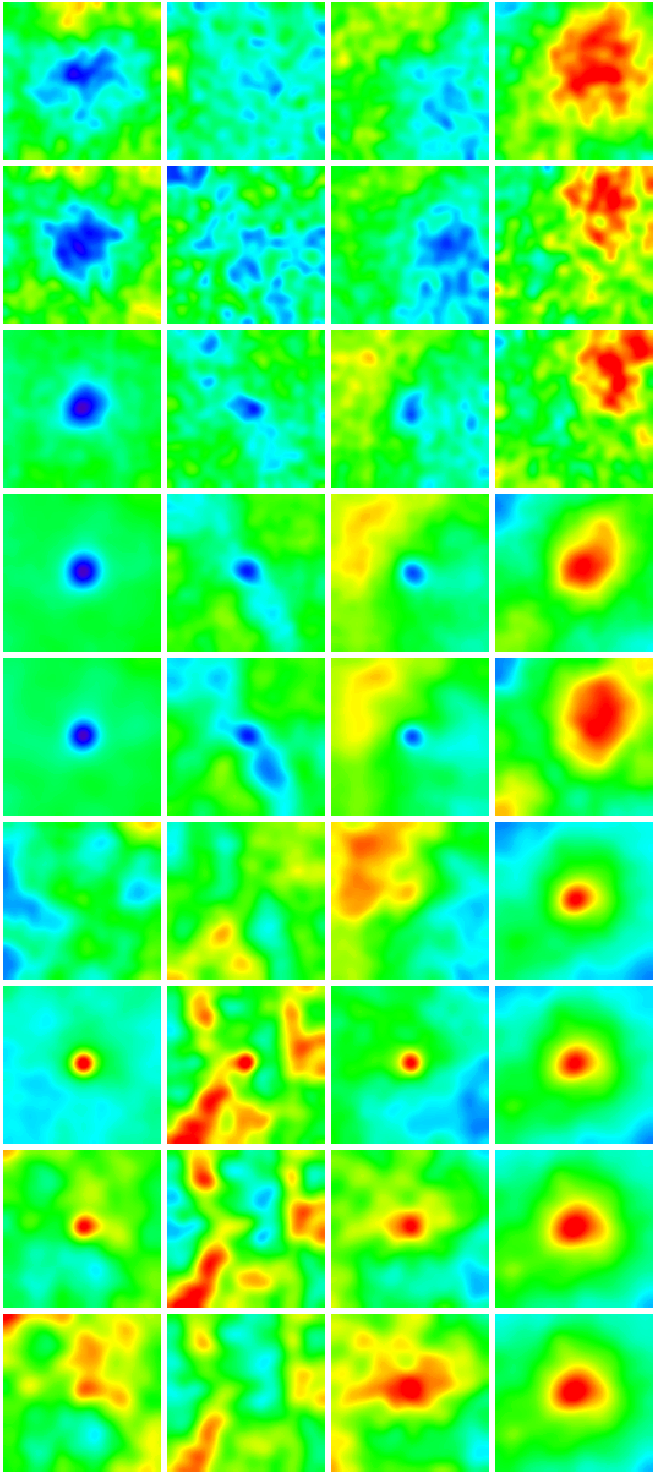}
\caption{Stacked signal in the nine \planck\ frequencies (30 to
  857\,GHz from upper to lower row). From left to right are displayed
  the \Planck\ SZ detections identified with known clusters, the {\sc
    class1} high-reliability \Planck\ SZ candidates, the {\sc class2}
  good-reliability \Planck\ SZ candidates, and finally the {\sc
    class3} low-reliability SZ sources. The three
  lowest-frequency-channel images were convolved with a 10$^{\prime}$ FWHM
  Gaussian kernel, whereas the remaining six highest-frequency-channel
  images were smoothed with a 7$^{\prime}$ FWHM Gaussian kernel. }
\label{fig:stack_999_33}
\end{center}
\end{figure*}

We combine the qualitative SZ quality flag with the information from
the search in the all-sky surveys, RASS and WISE, for counterparts of
\Planck\ candidates in order to assess the overall reliability of the
cluster candidates. We thus distinguish three classes of candidates:

\begin{itemize}
\item[$\bullet$] {\bf {\bf\sc class1}} candidates. {\it
 Highly-reliable candidates or pre-confirmed clusters:} these are the
 \Planck\ SZ detections that have a high probability of being
 associated with bona fide clusters and need to fulfil
 high-quality criteria for SZ, RASS, and WISE detections. We retain
 in this category \Planck\ SZ detections with high or good SZ quality
 flags ($Q^{\mathrm{SZ}}=1$ or 2) and with a RASS-BSC source (not
 coinciding with stars) or with (S/N)$_{\mathrm{RASS}}\ge 2$,
 i.e., SZ detections with quality flag $Q_{\mathrm{RASS}}=1$. The
 {\sc class1} candidates furthermore have to fulfil a condition of
 high or good probability ($\ge$80\%) of being associated with an
 overdensity of galaxies in the WISE survey.

We find 54 {\sc class1} \Planck\ candidates ranging from S/N of 4.5 to
6.3, with a median S/N of 4.8. The majority of them
are detected by two methods and 25.9\% of them are detected only by
one method.  They are distributed as 26 and 28 $Q^{\mathrm{SZ}}=1$ and
2 candidates, respectively. These candidates show significant X-ray
emissions with a median (S/N)$_{\mathrm{RASS}}\simeq 3.7$ and a mean
of $4.2$.
\item[$\bullet$] {\bf {\bf\sc class2}} candidates. {\it Reliable
 cluster candidates:} they represent 170 \Planck\ SZ detections that
 show good or high quality criteria either in SZ or in RASS or in
 WISE without fulfilling all of them at once.  Amongst them 61 have
 $Q^{\mathrm{SZ}}=1$ and 109 have $Q^{\mathrm{SZ}}=2$.
\item[$\bullet$] {\bf {\sc class3}} candidates. {\it Low-reliability cluster
candidates:} these \Planck\ SZ detections are the poor-quality,
$Q^{\mathrm{SZ}}=3$, detections. They can also be associated with
good quality, $Q^{\mathrm{SZ}}=2$, detections for which there are no
good indications of the presence of an X-ray counterpart
((S/N)$_{\mathrm{RASS}}<0.5$ and high probability of false
association with FSC sources $>$2.5\%) or a counterpart in the WISE
survey (probability of association $<$70\%).

This class of candidates contains 142 \Planck\ SZ detections with 27
and 115 SZ detection of quality $Q^{\mathrm{SZ}}=2$ and 3,
respectively.
\end{itemize}
It is worth noting that this definition of the {\sc class3}
\Planck\ candidates is dominated by the assessment of the SZ quality
complemented by information from ancillary data. In doing so we
assemble in this category of candidates the SZ detections that are
either false or very low quality due to contamination. Moreover,
according to the statistical characterization from simulations, about
200 false detections are expected. The number of false detections
could be smaller since the simulations do not reproduce the entire
validation procedure, in particular omitting the cleaning from obvious
false detections. Figure~\ref{fig:stack_999_33} suggests that the {\sc
  class3} candidates are likely to be dominated by false detections.
Therefore, we would like to warn against dismissing entire {\sc
  class3} of the catalogue as populated with false detections as some
{\sc class3} candidates may be real clusters. For this reason, we
choose not to remove these detections from the PSZ catalogue but
rather flag them as low-reliability candidates. Careful follow-up
programmes are needed in order to separate real clusters of galaxies
from false detections among the {\sc class2} and {\sc class3} objects.

In order to illustrate our classification defined in terms of
reliability, we stack the signal in patches of 2.51$^{\circ}$ across,
centred at the position of the \Planck\ clusters and candidates in the
nine channel maps of \Planck, removing a mean signal estimated in the
outer regions where no SZ signal is expected (see
Fig.~\ref{fig:stack_999_33} with the rows arranged from 30\,GHz, upper
row, to 857\,GHz, lower row). The stacked and smoothed images are
displayed for the \planck\ SZ detections identified with known
clusters, {\sc class1}, {\sc class2} and {\sc class3} candidates,
Fig.~\ref{fig:stack_999_33} from left to right column. We clearly see
the significant detection of both the decrement and increment of the
683 \planck\ clusters and of the \planck\ candidates of {\sc class1}
and {\sc class2}. For the \planck\ SZ detections associated with
bona fide clusters the increment is clearly seen at 353 and
545\,GHz and is detected at 857\,GHz. The smaller sample of the {\sc
  class1} highly reliable candidates shows, in addition to the
decrement at low frequency, a good detection of the increment at
353~GHz. The significance of the increment at 545\,GHz is marginal and
no signal is seen at 857\,GHz. The case of the {\sc class2} candidates
(good reliability) shows that we now have lower-quality SZ detections
(62\% of the {\sc class2} candidates have a good but not high SZ
quality flag). This is illustrated by the fact that an excess emission
is detected at 217\,GHz, most likely due to contamination by IR
sources, and both at 545 and 857\,GHz where emission from dust is
dominating. As for the stacked signal of the {\sc class3} sample of
low-reliability candidates, it does not show any significant SZ
detection across frequencies, as compared to the sample of
\planck\ detections identified with known clusters
(Fig.~\ref{fig:stack_999_33}, right column). This confirms on
statistical grounds the definition of the sample dominated by
definition by the low-quality SZ, $Q^{\mathrm{SZ}}=3$, detections
representing 84\% of the detections in this class. Not surprisingly,
the stacked signal of the {\sc class3} candidates shows a large amount
of contamination across all \Planck\ frequencies. The low-frequency
signal is dominated by radio contamination, and/or CO emission at
100~GHz, while the high-frequency signal is contaminated by emission
from dust or extragalactic point sources. A more quantitative analysis
is presented in Sect.~\ref{sec:rad}.

\subsection{Summary of the external validation and redshift assembly}

\begin{figure}
\centering
\includegraphics[width=8.8cm]{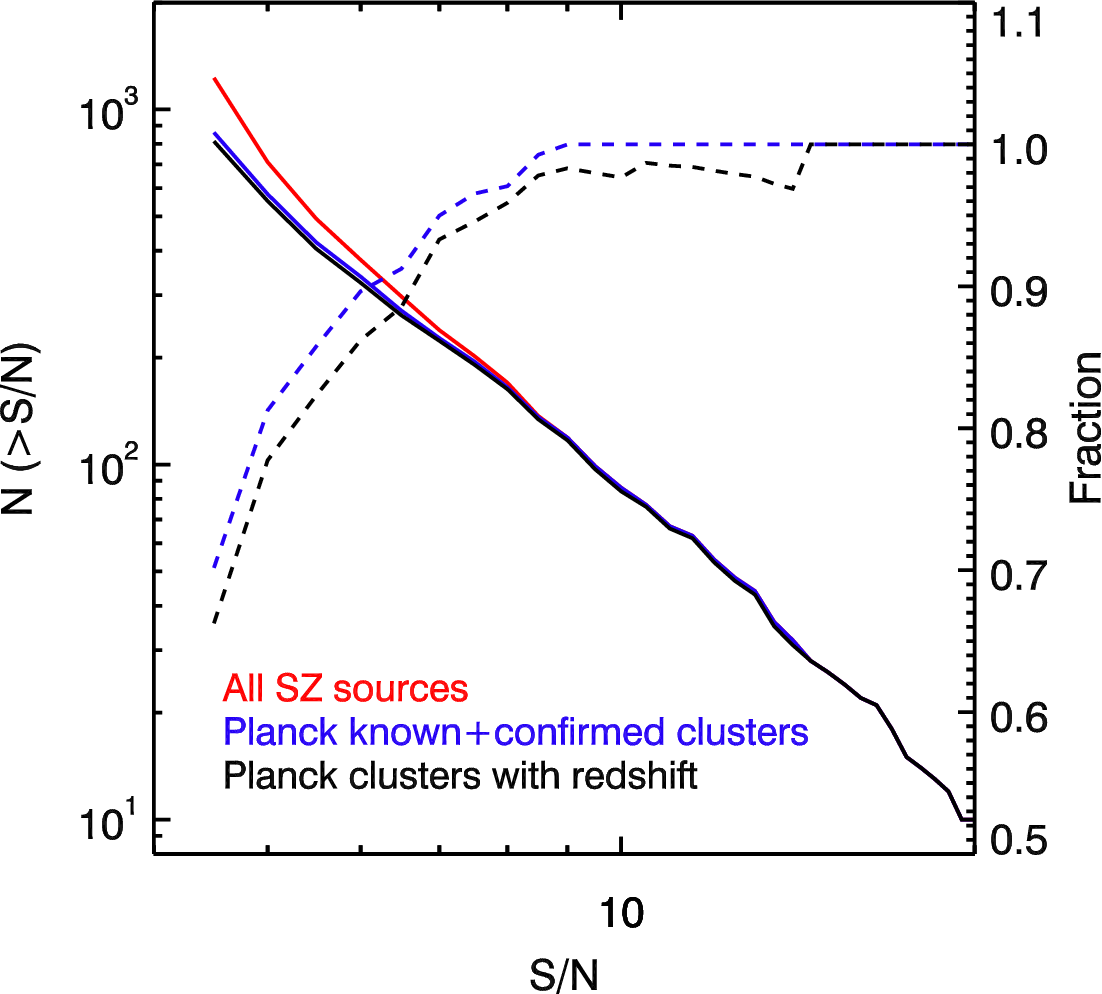}
\caption{Status of the \planck\ SZ sources. Left-hand-axis plots show
the distribution of all \planck\ sources (in red). The blue line
represents the known or new confirmed clusters and, among these, the
clusters with a reported redshift measurement in
black. Right-hand-axis cumulative distributions show, as a function
of S/N, the fraction of known or new confirmed clusters
in blue and those with a redshift in black.}
\label{fig:hist_SN}
\end{figure}

The \Planck\ catalogue of SZ sources comprises a total of 861
identified or confirmed clusters with only nine percent of them being
detected by one SZ-finder algorithm. We summarize in
Table~\ref{tab:valid_sum} and Fig.~\ref{fig:pie} the results of the
cluster identification. Figure~\ref {fig:hist_SN} illustrates the
status of the \planck\ SZ detections. In particular, 70.2\% of the
\planck\ SZ detections with S/N$\ge$4.5 have so far been associated
with clusters. The fraction increases to about 73\% at S/N $= 6$.

\begin{figure}
\centering
\includegraphics[width=8.8cm]{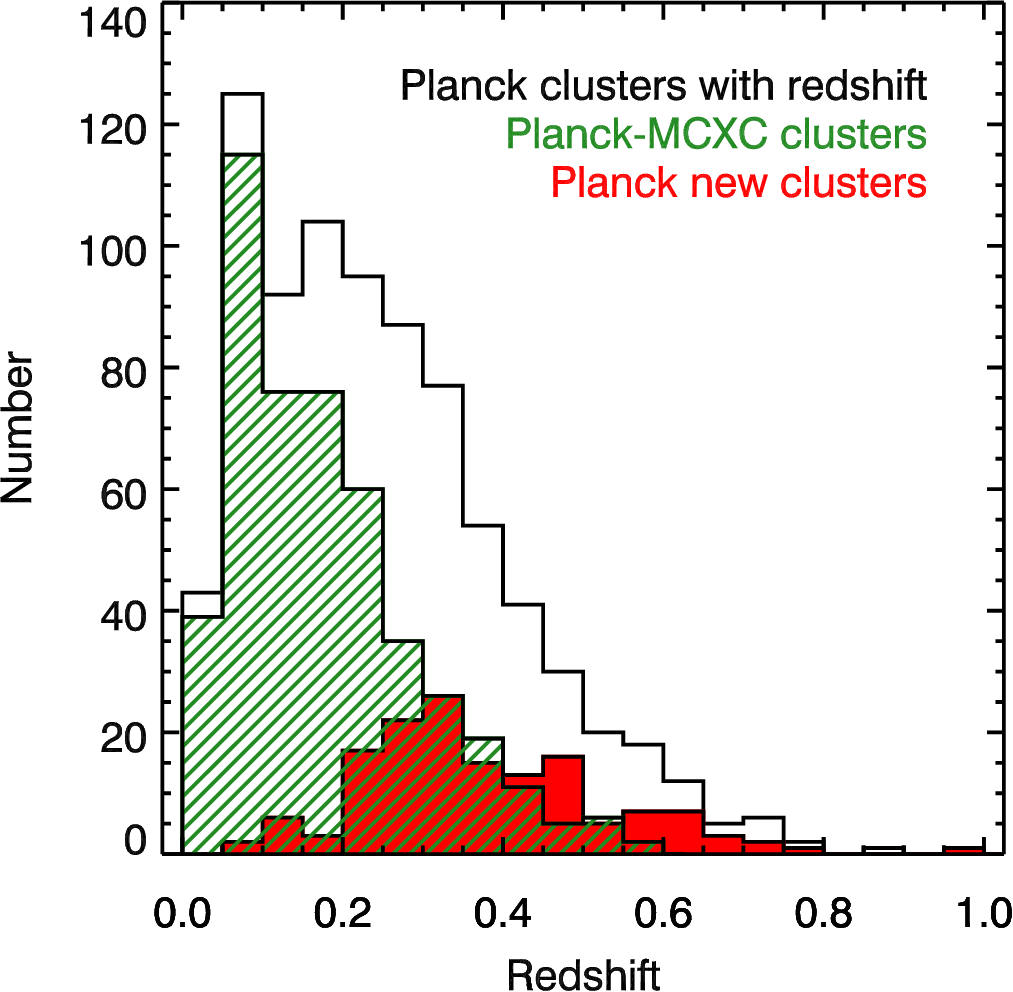}
\caption{Distribution of redshifts for the \planck\ SZ clusters (black
  line). The \planck\ clusters associated with MCXC clusters are shown
  in dashed green and the new \planck\ clusters are in
  the filled red histogram.}
\label{fig:zhist}
\end{figure}

We have assembled, at the date of submission, a total of 813 redshifts
for the 861 identified or confirmed \planck\ clusters, which we provide
together with the published \Planck\ catalogue. Their distribution is
shown in Fig.~\ref{fig:zhist}. In the process of the redshift
assembly that is summarized below, especially for the already known
clusters, we have favoured homogeneity for the sources of redshift
rather than a cluster-by-cluster assembly of the most accurate $z$
measure. A large fraction of the redshifts, 456 of them, shown as the
dashed green histogram in Fig.~\ref{fig:zhist} correspond to the
spectroscopic redshifts quoted in the updated MCXC meta-catalogue
\citep{pif11}. They are associated with the \Planck\ clusters
identified with known X-ray clusters and they are denoted
\Planck-MCXC. For the \Planck-MCXC clusters without reported redshifts
from the MCXC, we have complemented the information with the
available redshifts from NED and SIMBAD. We have further quoted when
available, mainly for the MACS clusters, the estimated photometric
redshifts from SDSS cluster catalogue of \citet{wen12}. At the end
only two \planck\ detections identified with MCXC clusters remain
without redshifts. The redshift distribution of the \planck\ clusters
identified with MCXC clusters mostly reflects that of the REFLEX/NORAS
catalogues at low and moderate redshifts and the MACS clusters at
higher redshifts.\\
For the \planck\ detections exclusively identified with Abell or
Zwicky clusters, we choose to report the redshifts published in the
NED and SIMBAD data bases rather than those quoted in the native
catalogues.  As for the \planck\ detections identified with clusters
from the SDSS-based catalogues, we choose to favour homogeneity by
reporting whenever possible the \citet{wen12} redshifts. Furthermore,
we favour when available spectroscopic redshifts over photometric
ones. The \Planck\ detections exclusively associated with ACT or SPT
clusters have published redshifts \citep{sif12,has13,rei13}. We select
in priority the spectroscopic ones when available. If not, we quote
the photometric redshifts.  \\
Finally, the follow-up observations for
confirmation of \planck\ detections started in 2010 and are still
ongoing. As mentioned earlier our priority was to assemble the largest
possible number of confirmations and redshifts. Therefore, we did not
systematically confirm the photometric redshift estimates
spectroscopically. We report the obtained redshifts when available. In
some cases, the new \planck\ clusters were confirmed from imaging or
pre-imaging observations and the analysis is still ongoing. The
spectroscopic redshifts will be updated when available. Spectroscopic
redshifts for some known clusters will also be updated. A
dozen \planck\ clusters were confirmed by a search in the SDSS galaxy
catalogues. For these clusters, only a photometric redshift estimated
by the cluster-finder algorithm of Fromenteau et al. (2014, in prep) is
available and is reported.

We show in Fig.~\ref{fig:zhist} the distribution of redshifts of the
\planck\ clusters. The mean redshift of the sample is 0.25 and its
median is 0.22. One third of the \Planck\ clusters with measured
redshifts lie above $z=0.3$. The new \planck\ clusters probe higher
redshifts and represent 40\% of the $z\ge 0.3$ clusters. Their mean
redshift is 0.38 and the median is $z=0.35$. At even higher redshifts,
$z\ge 0.5$, the \planck\ catalogue contains 65 clusters including
\Planck\ SZ clusters identified with WHL12's clusters \citep{wen12},
or with clusters from ACT and SPT, or with X-ray clusters. The
\Planck\ detections in this range of redshifts, 29 \planck\ new
clusters, almost double the number of high redshift clusters.

The \planck\ SZ catalogue has been followed up by the
\Planck\ collaboration using different facilities and only a small
fraction of the \planck\ candidates were observed to date. A
systematic follow-up effort for the confirmation of the remaining
cluster candidates will likely reveal clusters at redshifts above
0.3. As a matter of fact, very few new clusters were found below
$z=0.2$ (see Fig.~\ref{fig:zhist}). Such an observational programme
is challenging and will most likely be undertaken by the
\Planck\ collaboration and by the community. It will increase further
the value of the \planck\ SZ catalogue as the first all-sky
SZ-selected catalogue.


\section{Physical properties of \Planck\ SZ clusters}
\label{s:physprop}

The first goal of the external validation process based on the
ancillary multi-wavelength data is to assess the status of the
\planck\ SZ detections in terms of known clusters, brand new clusters
or cluster candidates. The wealth of information assembled and used
during this process also allows us to explore the properties of the
\Planck\ SZ clusters and candidates.  We present in the following some
of these properties, namely the contamination levels of the \planck\ SZ
detections, a refined measurement of the Compton $Y$ parameter for the
\planck\ clusters identified with X-ray clusters from the MCXC, an
SZ-mass estimate based on a new proxy for all the \planck\ clusters
with measured redshifts, and an estimate of the X-ray flux from the
RASS data for the \planck\ SZ detections not included in the X-ray
catalogues.  This additional information associated with the
\planck\ clusters and candidates derived from the validation process
is summarized in the form of an ensemble of outputs given in
Table~\ref{tab:valid_info}.

We further present an updated and extended study of the SZ versus
X-ray scaling relation, confirming at higher precision the strong
agreement between the SZ and X-ray measurements (within $\Rv$) of the
intra-cluster gas properties found by \citet{planck2011-5.2b}.

\subsection{Point-source contamination}\label{sec:rad}

\begin{figure}[!th]
\centering
\includegraphics[width=8.8cm]{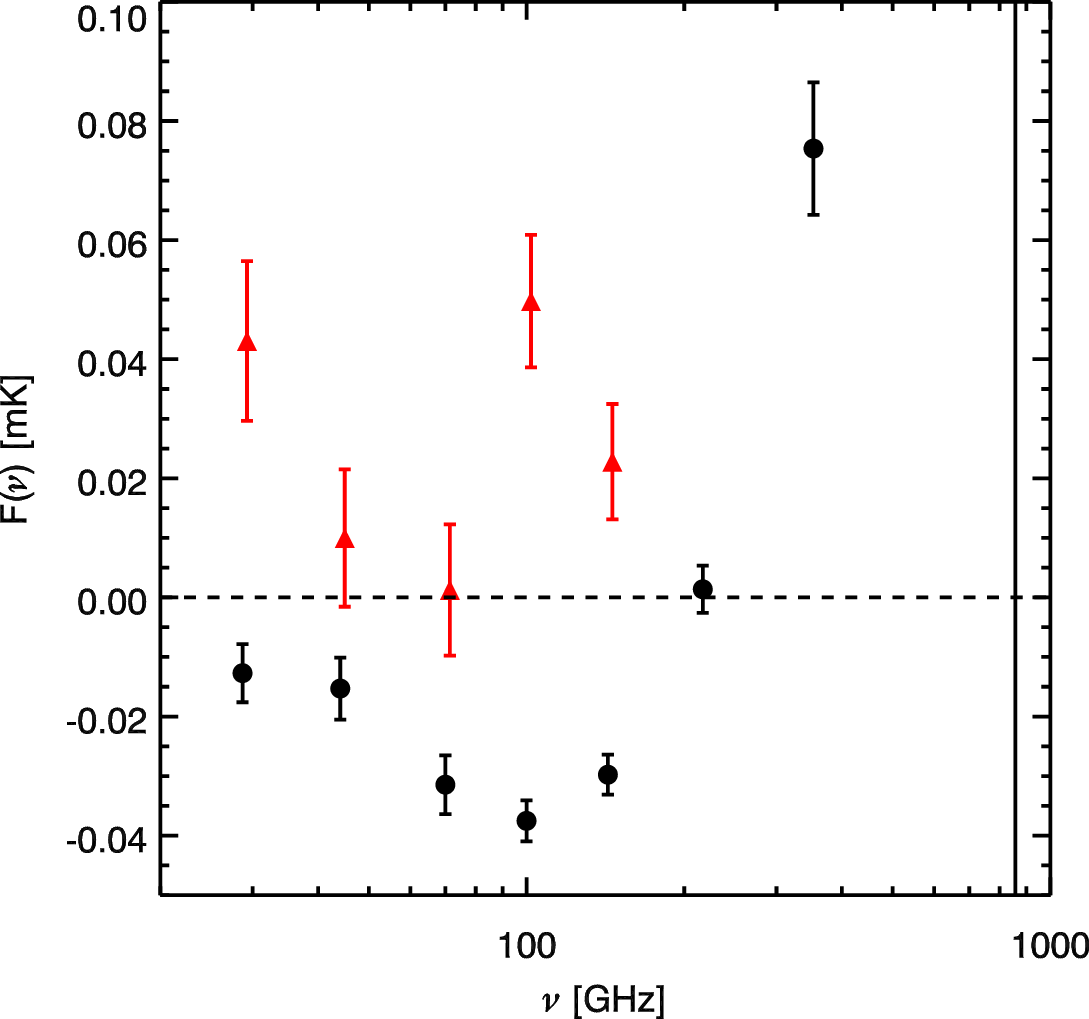}\\[1em]
\includegraphics[width=8.8cm]{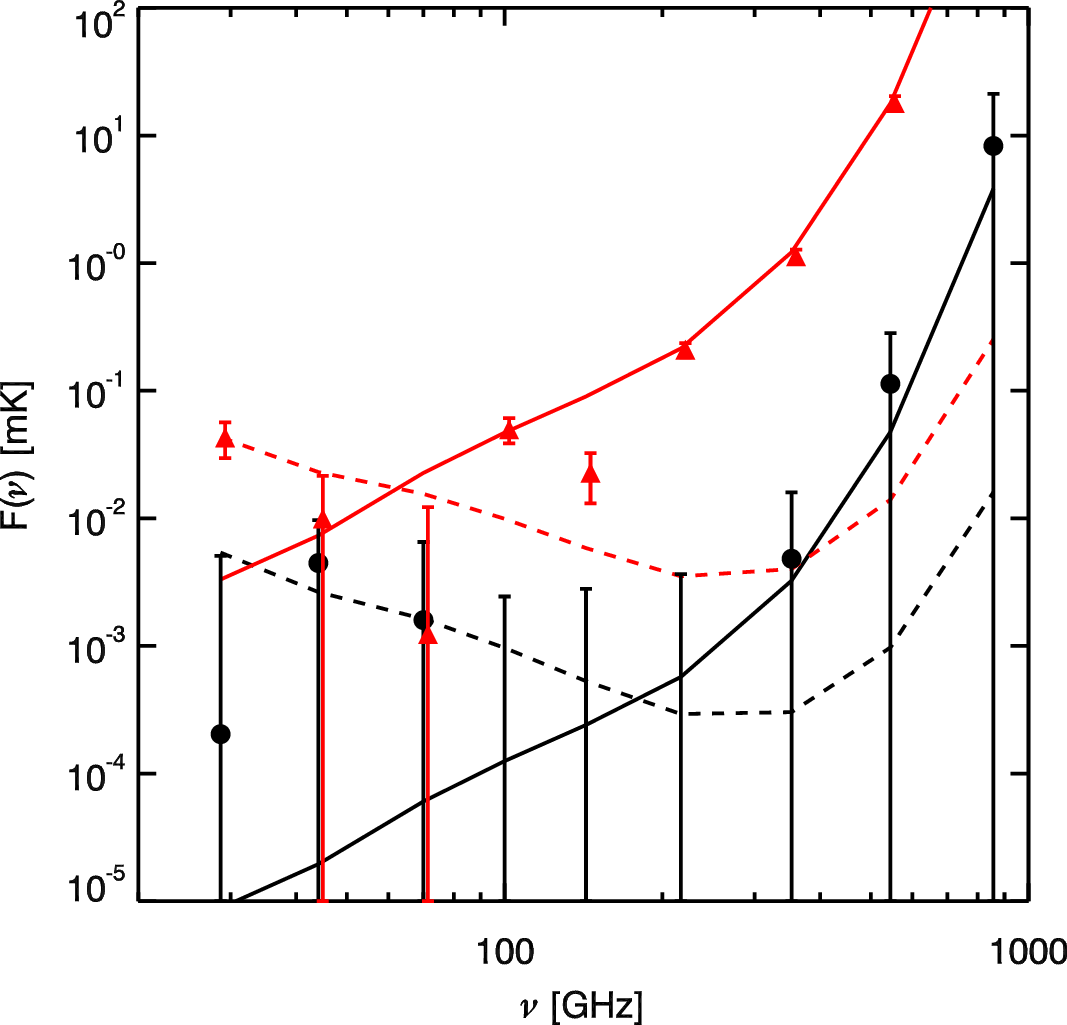}
\caption{Stacked spectrum for known clusters SZ fluxes across
  \Planck\ frequency bands.  Stacked fluxes are measured in an
  aperture equal to the FWHM of the 143~GHz channel (i.e., about 
  7$^{\prime}$) for the known clusters (black filled circles) and the
  low-reliability {\sc class3} candidates (red filled triangles). The
  associated uncertainties correspond to the fluctuation of the
  background outside the cluster region. The average signal is
  estimated in each channel before (upper panel) and after (lower
  panel) the removal of the SZ signal.  The average signals expected
  from IR and radio sources are shown as solid and dashed lines,
  respectively. Red and black lines are for {\sc class3} and bona fide
  clusters, respectively. No subtraction of an SZ signal is performed
  for the {\sc class3} candidates.}
\label{fig:PS_cont}
\end{figure}

Galactic and extragalactic sources, emitting in the radio or infrared
domain, are known to lie in galaxy clusters and hence are a possible
source of contamination for the SZ measurement
\citep[e.g.,][]{rub03,Agh05,Lin09}.  We address the possible
contamination of the SZ flux by bright radio sources that may affect
the measured signal in the direction of some of the \planck\ SZ
detections. In order to do so, we searched for known radio sources in
the vicinity of the \Planck\ cluster candidates. In particular, we use
the NVSS~1.4\,GHz survey \citep{con98} and SUMMS~0.85\,GHz survey
\citep{boc99} to identify bright radio sources within 7$^{\prime}$ of
the \planck\ cluster or candidate position. {This search radius
  corresponds to the \Planck\ resolution at 143~GHz.}  We assumed a
spectral index $\alpha=-0.5$ for these sources to extrapolate their
flux to the \Planck\ frequencies. Most bright sources in NVSS and
SUMSS have steeper spectral indexes ($-0.6$ or $-0.7$), so the value
$\alpha=-0.5$ provides us with an upper limit in most cases.  After
convolving the radio sources by \Planck's beam, we estimate the
maximum amplitude in units of $\mu$K within 5$^{\prime}$ of the
\Planck\ position. We report only those cases where this amplitude is
above $5\, \mu$K in the 143\,GHz channel and could thus contaminate
the SZ signal. Below this value, the emission from radio sources can
be considered negligible.

We find that a total of 274 \Planck\ clusters and candidates, i.e.,
22\% of the SZ detections, are affected by such emission from bright
radio sources. These clusters or candidates are identified in the PSZ
catalogue and a specific note is provided. We find that the fraction
of contaminated \planck\ SZ clusters identified with known X-ray,
optical, or SZ clusters is also 22\%. The \Planck\ candidate-clusters
of {\sc class1} and {\sc class2} are less contaminated by bright radio
sources; only a fraction of 15\% and 17\% for {\sc class1} and 2,
respectively. This is due to the definition of our quality criteria
for SZ detection, which results in less contamination for the high and
good reliability candidates.

Another approach used to assess the contamination is based on the
stacking analysis of the \Planck\ clusters and candidates described in
Sect.~\ref{sec:candidates}. This analysis is performed on the sample
of \planck\ clusters identified with known clusters and on the sample
of low-reliability {\sc class3} \planck\ candidates. To do so we fit a
GNFW pressure profile to the signal at 100\,GHz and 143\,GHz and we
subtract the associated SZ signal from the stacked maps. The residual
signal is then compared with a toy model for point sources
($F_\nu=S_{30}^{\mathrm{rad}}(\nu/30\,\mathrm{GHz})^{\alpha_{\mathrm{rad}}}$
for radio sources) and
($F_\nu=S_{857}^{\mathrm{IR}}(\nu/857\,\mathrm{GHz})^{\alpha_{\mathrm{IR}}}$
for IR point sources). Note that the residual signal at high
frequencies is a combination of possible IR sources and IR emission
from Galactic dust; the latter is not explicitly modelled in the
present analysis. The PS toy models are convolved by the beam at each
frequency and the signal is measured at a fixed aperture set to the
FWHM of the 143\,GHz channel. The average signal within this aperture
is estimated for each channel before (Fig.~\ref{fig:PS_cont}, upper
panel) and after (Fig.~\ref{fig:PS_cont}, lower panel) removal of the
SZ signal. The black filled circles are for \planck\ SZ sources
associated with known clusters and the red filled triangles stand for
{\sc class3} candidates. The average signal from the PS models is
shown in Fig.~\ref{fig:PS_cont} as solid (IR sources) and dashed
(radio sources) lines.  Red and black are for {\sc class3} and
  bona fide clusters, respectively. The error bars correspond to the
fluctuation of the background outside the cluster region. For the
sample of {\sc class3} candidates no SZ-signal removal was applied,
since no significant detection is seen at 100\,GHz or 143\,GHz.

We find that the residual signal (after SZ subtraction) in the sample
of known \Planck\ clusters is compatible with the emission from radio
sources at low frequencies with
$(S_{30}^{\mathrm{rad}},\alpha_{\mathrm{rad}})=(14.6\,{\mathrm{mJy}},-1)$
for the known clusters. It is also compatible with IR emission at high
frequencies with a spectral index 
$\alpha_{\mathrm{IR}}=2.5$, in agreement with the results of
\citet{planck2012-VII} and with $S_{857}^{\mathrm{IR}}=0.117\,$Jy. For
{\sc class3}, where no SZ signal is subtracted, it is the full
signal that is compatible with the IR emission at high frequencies,
with
$(S_{857}^{\mathrm{IR}},\alpha_{\mathrm{IR}})=(43.9\,{\mathrm{Jy}},2.5)$,
and with radio emission from point sources with
$(S_{30}^{\mathrm{rad}},\alpha_{\mathrm{rad}})=(117.1\,{\mathrm{mJy}},-0.8)$.

\subsection{Refined measurement of $Y$}

\begin{figure}
\centering
\includegraphics[width=8.8cm]{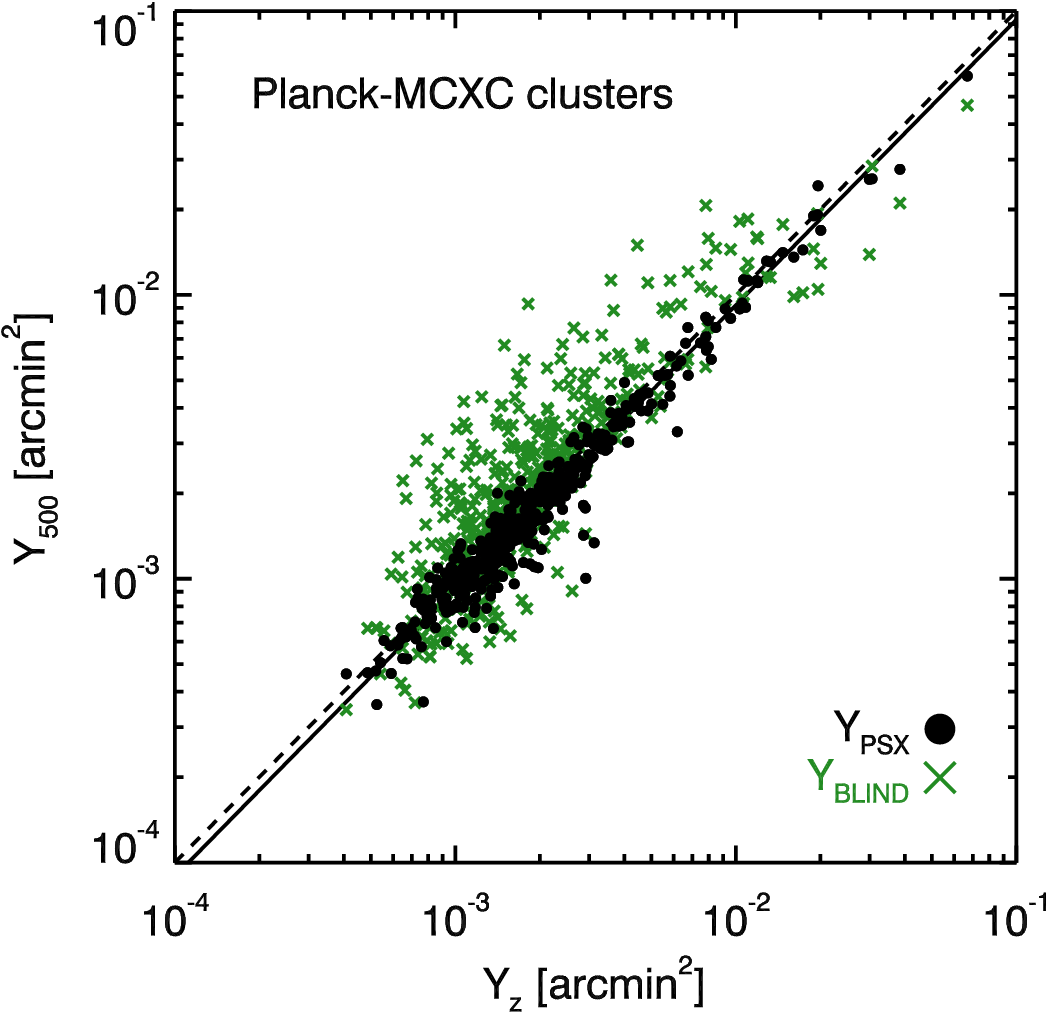}
\caption{Comparison of the different $Y$ estimates for the
\planck\ clusters identified with MCXC clusters. In green are the
blind measured $Y$ values and in black are the refined
$Y_{500,{\mathrm{PSX}}}$ measured fixing the size and positions to
the X-ray values. Both are plotted as a function of the new proxy
$Y_z$.}
\label{fig:Yb_Ypsx_Yprox}
\end{figure}

While the true $\YSZ$ is expected to be a low-scatter mass proxy, this
is not the case for the blind $\YSZ$. Without a cluster-size estimate,
$\YSZ$ cannot be accurately measured. Moreover, the blind SZ flux is
biased high on average, because the size is over-estimated on
average. This effect is amplified by the non-linear nature of the
size--flux degeneracy, with a larger effect of size over-estimation
than size under-estimation. This behaviour, first identified and
discussed in \citet{planck2011-5.1a} and \citet{planck2012-II},
hampers the direct use of the blind SZ fluxes as a mass proxy. As
shown in \citet{planck2011-5.1a}, this degeneracy calls for a refined
measurement of the SZ signal. In this section, we present two ways of
refining the $Y$ measurement. Both are based on fixing the cluster
size in two cases, by setting it equal to the X-ray estimated size or
by using the redshift information when available. The outputs of the
refined measurement are provided as additional information
complementary to the catalogue of \Planck\ SZ detections (see
Appendix~\ref{ap:valid} and Table~\ref{tab:valid_info}).

\subsubsection{$Y$ at fixed X-ray size and position}\label{sec:sizeF}

As shown by \citet{planck2011-5.1a}, the size--flux degeneracy can be
broken by introducing a higher-quality estimate of the cluster size
$\theta_{500}$. This prior is directly provided by X-ray observations
using an X-ray mass proxy such as $Y_{\mathrm{X}}$ or the luminosity
$L_{\mathrm{X}}$. Resorting to estimates of the cluster size from
optical richness is also possible, but suffers from the large scatter
in richness--mass relation, as discussed previously.

A detailed investigation of the effects of fixing the cluster size was
presented in \citet[Appendix~A]{planck2011-5.2b}.  Following this
approach, and for the \planck\ detections identified with clusters
from the MCXC meta-catalogue, we have adopted the $\Rv$ and $z$ values
reported in \citet{pif11} as priors to re-extract at the X-ray
position the SZ signal denoted $Y_{500,{\mathrm{PSX}}}$ assuming the
\citet{arn10} pressure profile (see Table~\ref{tab:valid_info}). The
comparison between the blind $\Yv$ and refined
$Y_{500,{\mathrm{PSX}}}$ (Fig.~\ref{fig:Yb_Ypsx_Yprox}) shows that
both the scatter and the offset are significantly reduced by the
refined SZ measure.  The SZ re-extraction at X-ray position and fixing
the size to the X-ray derived size provides an unbiased estimate of
the SZ signal. However, as stressed in
\citet[Appendix~A]{planck2011-5.2b}, the MCXC cluster size derivation
involves the $M_{500}$--$L_{\mathrm{X,500}}$ relation, which exhibits
a non-negligible scatter. This leads to a remaining systematic
discrepancy between the expected $Y$ value from X-ray measurements and
the actual SZ flux derived from the \Planck\ data. The use of the
$Y_{\mathrm{X}}$ proxy does not suffer from such an effect, but
high-quality X-ray data permitting the use of such a quantity are not
available for a large number of clusters (see Sect.~\ref{sec:scal} for
the presentation of a sample of \planck\ SZ clusters with high-quality
X-ray data).

\subsubsection{$Y$ from the $Y(\theta)\,$--$M$ relation}\label{sec:mprox}

The size--flux degeneracy can further be broken, as proposed by
(Arnaud et al. 2014, in prep), using the \MYSZ\ relation itself that
relates $\tv$ and $\YSZ$, when $z$ is known. Then $\YSZ$ is derived
from the intersection of the \MYSZ\ relation and the size--flux
degeneracy curve.  A detailed description of the method and the
comparison of results in terms of bias and scatter can be found in
(Arnaud et al. 2014, in prep).

The derived $\YSZ$ parameter is denoted $Y_z$ (since it involves a
measurement of the Compton $Y$ signal for clusters with measured
redshift $z$). It is the SZ mass proxy $Y_z$ that is equivalent to the
X-ray mass proxy $\YX$. $Y_z$ is computed for all the 813
\planck\ clusters with measured redshifts. We use
Malmquist-bias-corrected scaling relation between mass and $Y$ given
in \citet{planck2013-p15}
\begin{equation}
E^{-2/3}(z)\left[\frac{D_{\mathrm{ A}}^2(z) \, {Y}_{500}}
{\mathrm{ 10^{-4}\,{\mathrm{Mpc}^2}}}\right] = 10^{-0.19} \left[\frac{
  \Mv}{6\times10^{14}\,\msol}\right]^{1.79}\,,\label{scaling}
\end{equation}
with $E^2(z)=\Omega_{\mathrm m}(1+z)^3+\Omega_{\Lambda}$ computed
in the fiducial $\Lambda$CDM cosmology. 

In Fig.~\ref{fig:Yb_Ypsx_Yprox}, the refined $\Yv$ value, measured
fixing the size and position to the X-ray values
$Y_{500,{\mathrm{PSX}}}$, is compared to the blind $Y$ as a function
of the derived $Y_z$ proxy. We see that the scatter and the offset are
significantly reduced.

\begin{figure}
\centering
\includegraphics[width=8.8cm]{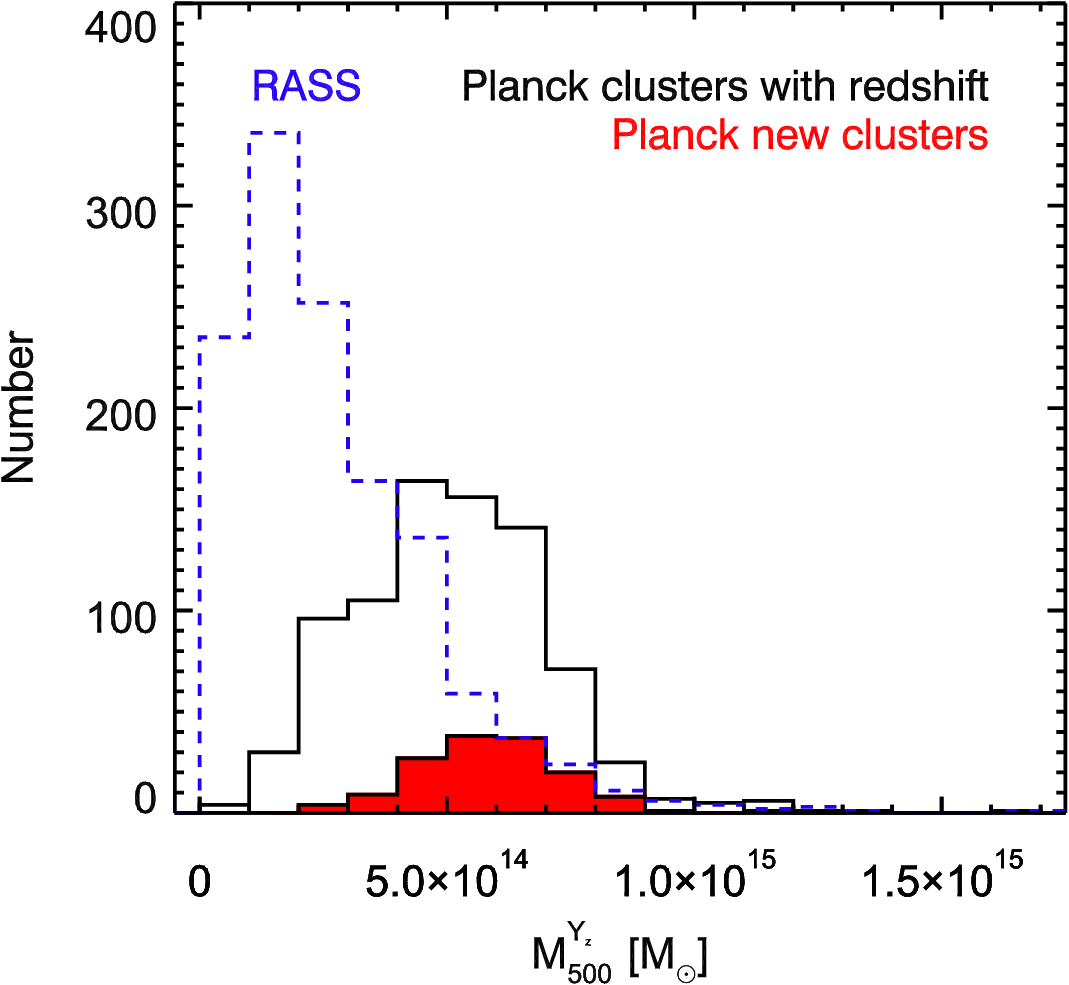}
\caption{Distribution of masses for the \planck\ SZ clusters, known or
  new confirmed clusters (solid black line), compared to the
  distribution of masses from the RASS-based cluster catalogues
  (dashed blue line). The masses for the MCXC clusters are estimated
  from the luminosity--mass relation. The masses for the
  \planck\ clusters are computed using the SZ-proxy. The filled red
  histogram shows the distribution of the newly-discovered
  \planck\ clusters.}
\label{fig:Mhist}
\end{figure}

Under the two hypotheses of cosmology and scaling relation, $Y_z$
provides the best estimate of $Y_{500}$ for the \Planck\ SZ
clusters and conversely a homogeneously-defined estimate of an SZ-mass,
X-ray calibrated, denoted $M_{500}^{Y_z}$.  For the ensemble of
\planck\ clusters with measured redshifts, the largest such sample of
SZ-selected clusters, we show in Fig.~\ref{fig:Mhist} the distribution
(black solid line) of the masses obtained from the SZ-based mass
proxy. The distribution of the SZ masses is compared with those of
the RASS clusters (dashed blue line) computed from the X-ray
luminosity $L_{\mathrm{X,500}}$.  The mean and median masses of the
\planck\ clusters are 3.3 and $3.5\times10^{14}\,\msol$,
respectively. The \planck\ SZ catalogues contains
all the massive clusters of the RASS catalogues. Interestingly, the
distribution of newly-discovered 
\Planck\ clusters extends to higher masses with a median mass of
$5.7\times10^{14}\,\msol$. Besides providing a
homogeneous estimate of the masses from an SZ proxy for the largest SZ
selected sample of clusters, we show that \planck\ detections
significantly extend the mass range in the high-mass region up to
$1.6\times10^{15}\,\msol$.

\subsection{$M$--$z$ distribution and comparison with other surveys}

\begin{figure*}[!ht]
\begin{center}
\includegraphics[width=8.6cm]{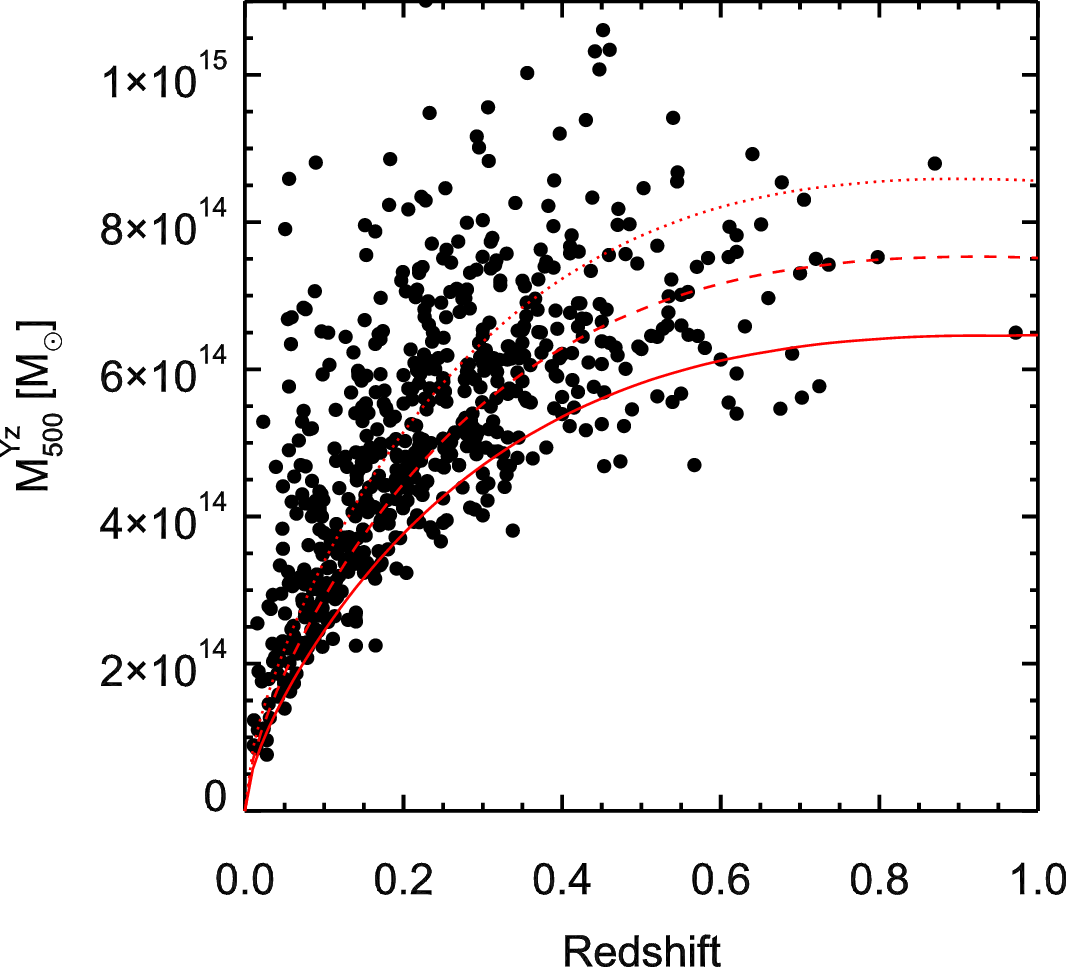}\hspace*{1em}
\includegraphics[width=8.6cm]{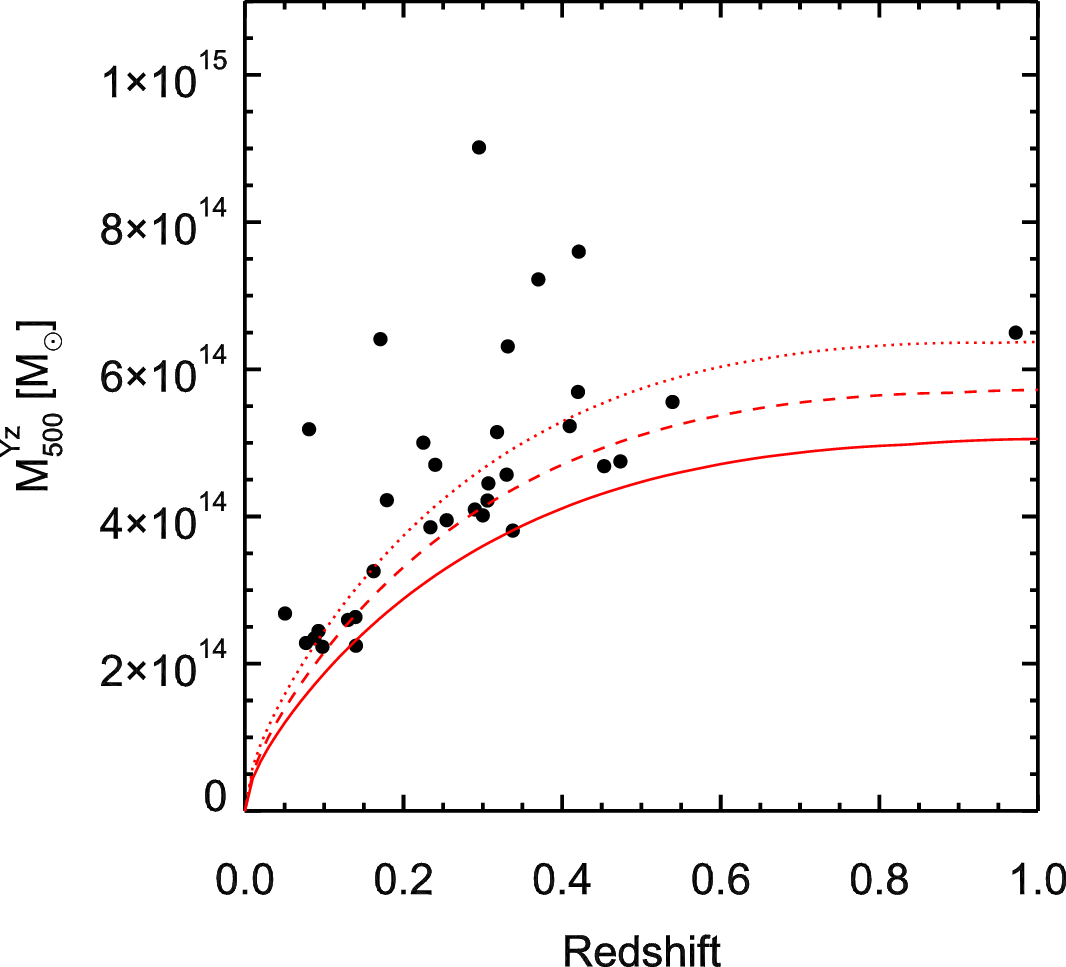}\\[1em]
\includegraphics[width=8.6cm]{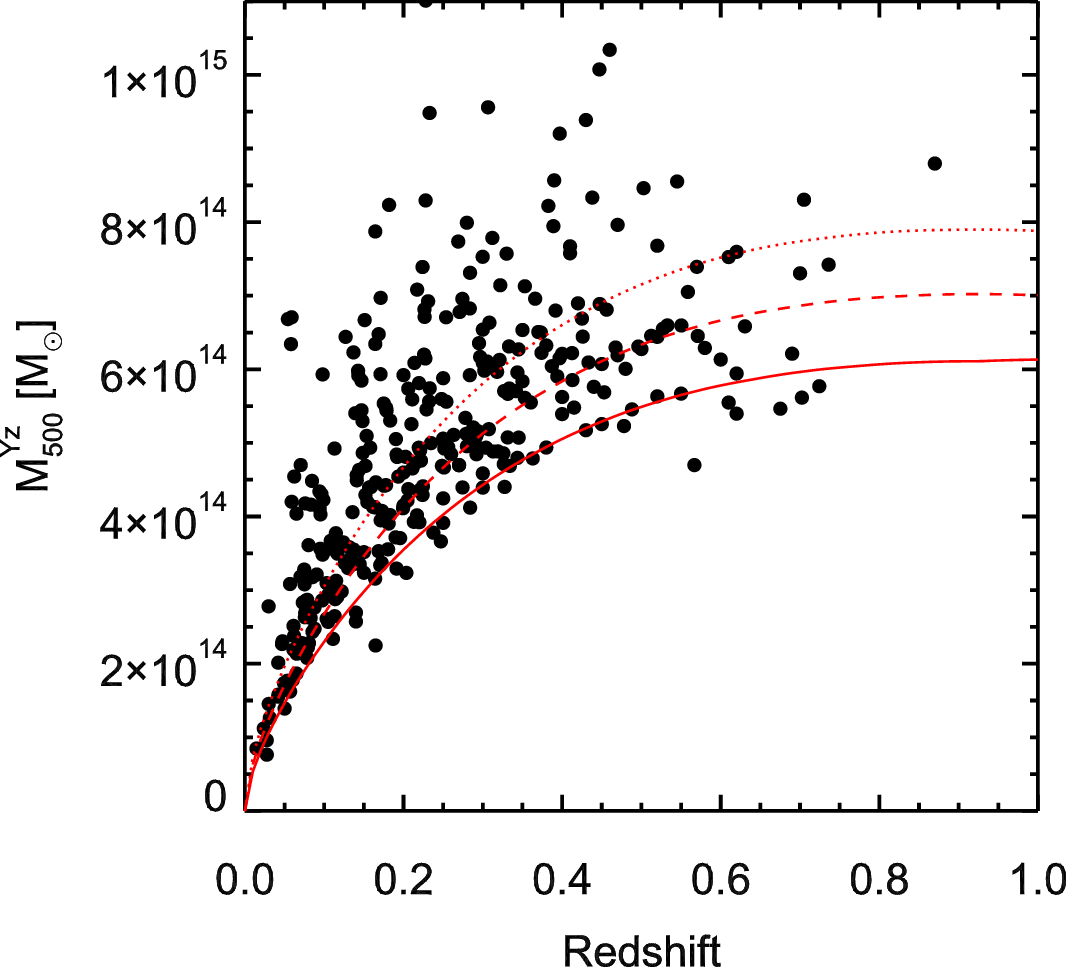}\hspace{1em}
\includegraphics[width=8.6cm]{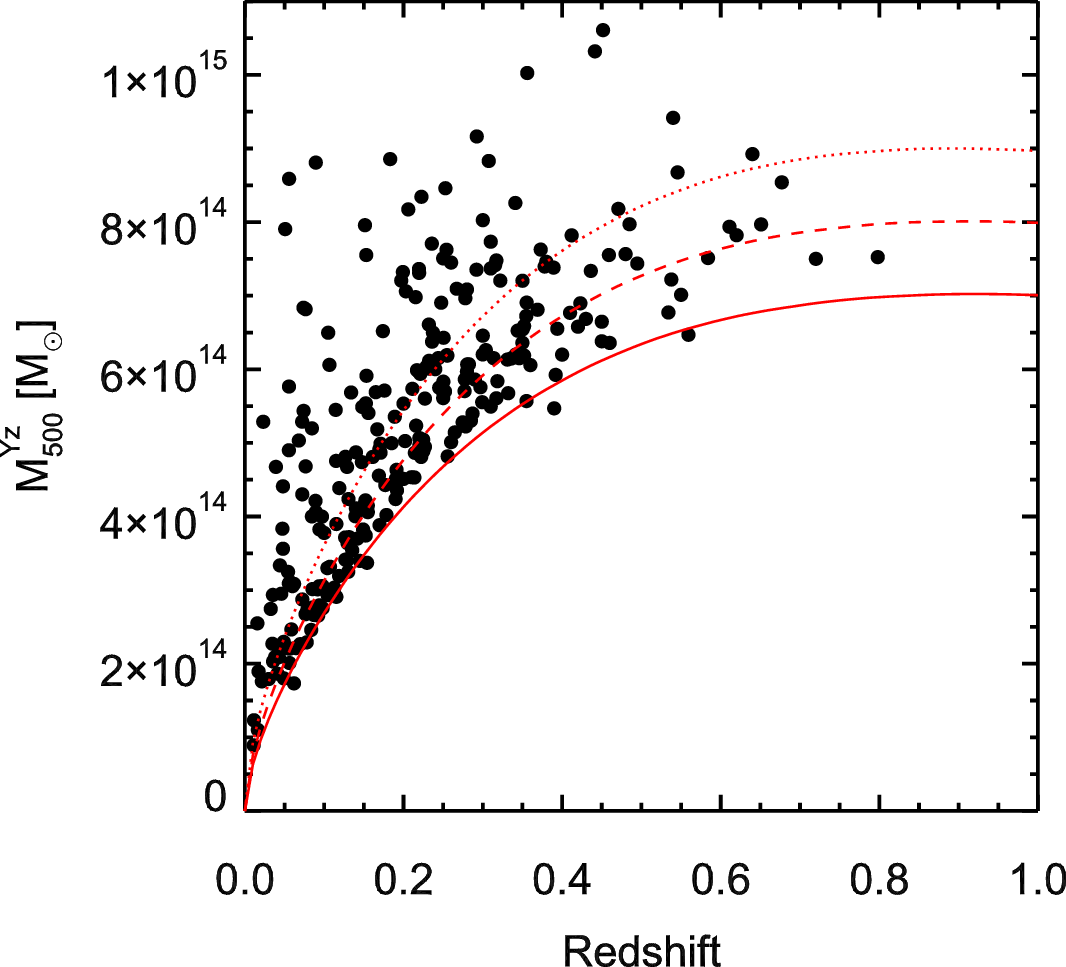}\\
\caption{Mass limit illustrated for SZ detections by \texttt{MMF3}
  algorithm. \emph{Upper left:} average mass limit computed from the
  average noise over the sky. \emph{Upper right:} same for the deep
  survey zone corresponding to 2.7\% sky coverage centred at the
  Ecliptic polar regions. \emph{Lower left:} same for the medium-deep survey
  area covering 41.3\% of the sky. \emph{Lower right:} same for the
  shallow-survey area covering 56\% of the sky. In each panel, only
  detections in the corresponding areas are plotted. The lines dotted,
  dashed and solid lines show the \planck\ mass limit at 80, 50 and
  20\% completeness, respectively.}
\label{fig:mlim}
\end{center}
\end{figure*}

Based on the masses derived from the SZ-proxy, we illustrate for
\texttt{MMF3} the $M$--$z$ distribution of \Planck\ SZ clusters
detected over 83.7\% of the sky. We show in all panels of
Fig.~\ref{fig:mlim} the limiting mass $M_{\mathrm{lim}}$ computed
following \citet{planck2013-p15} for three values of the completeness:
20\% (solid line); 50\% (dashed line); and 80\% (dotted line). The
upper left panel exhibits the \planck\ clusters, with redshifts,
detected by \texttt{MMF3} at S/N$\ge$4.5. The mass limit corresponds
to the average limit computed from the noise over the 83.7\% sky
fraction used by the SZ-finder algorithm. The resulting
$M_{\mathrm{lim}}$ is not representative of the inhomogeneity of the
noise across the sky (see Fig.~\ref{fig:noise_map}). We therefore show
the limiting mass in three areas of the sky
(Fig.~\ref{fig:noise_map}): the deep-survey area (upper right panel);
the medium-deep survey area (lower left panel); and the shallow-survey
area (lower right panel). The lines indicate the limit at which
clusters have C\% chances to be detected (C being the completeness
value). We clearly see that whereas the average $M_{\mathrm{lim}}$ at
20\% completeness does not fully represent the SZ detections by
\texttt{MMF3}, the limiting masses in different survey depths are more
representative of the detection process. We further note that except
at low redshifts, $z<0.3-0.4$, the \planck\ cluster distribution
exhibits a nearly redshift-independent mass limit with a cut that
varies according to the survey depth.

\begin{figure*}[t]
\begin{center}
\resizebox{0.9\textwidth}{!} {
\includegraphics[]{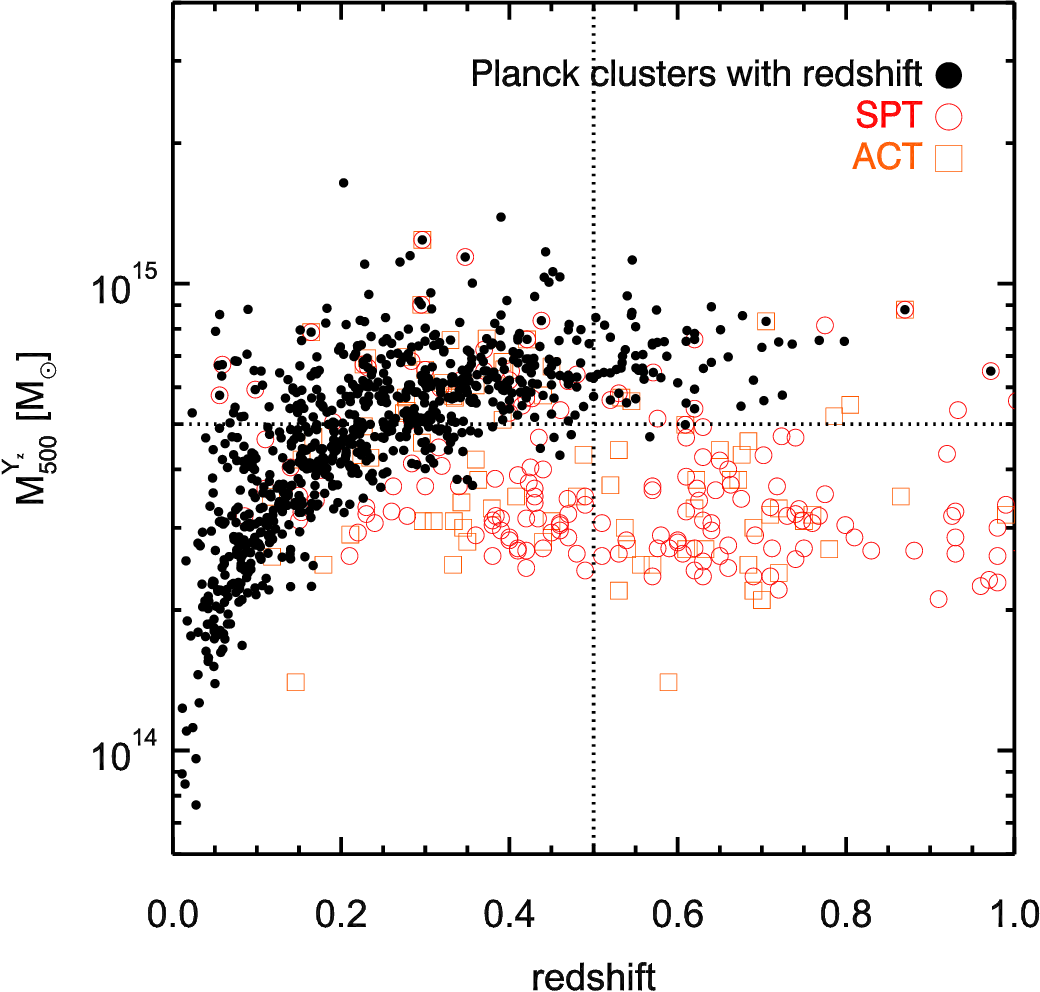}
\hspace{3cm}
\includegraphics[]{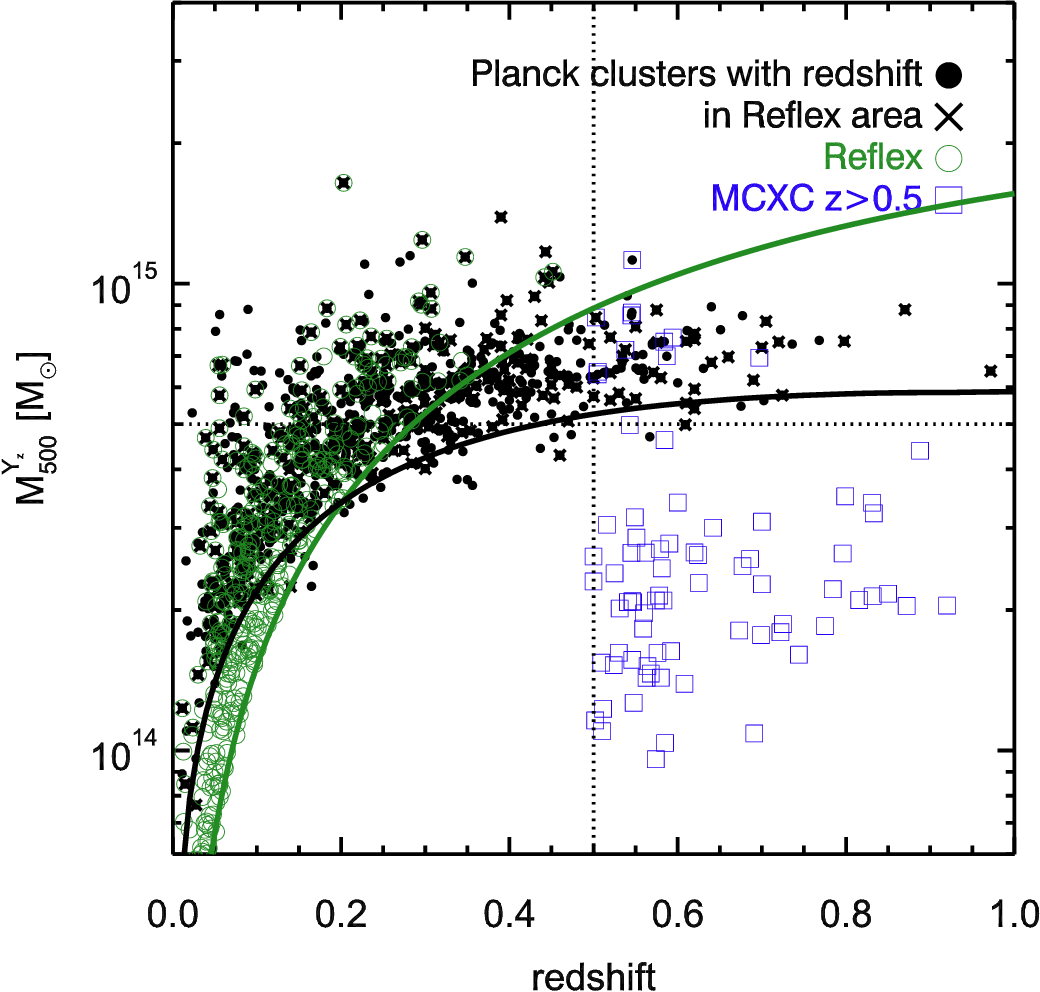}}
\caption{{\it Left panel:} distribution in the $M$--$z$ plane of the
  \planck\ clusters (filled circles) compared with the SPT clusters
  (open light red circles) from \citet{rei13} and ACT
  catalogue (open red squares) from \citet{has13}. {\it Right panel:}
  distribution in the $M$--$z$ plane of the \planck\ clusters (black
  symbols) as compared to the clusters from the REFLEX catalogue
  (green open circles) \citet{boe04}. The black crosses indicate the
  \Planck\ clusters in the REFLEX area.  The open blue squares
  represent clusters from the MCXC catalogue with redshifts above
  $z=0.5$. The green solid line shows the REFLEX detection limit
  whereas the black solid line shows the \planck\ mass limit for the
  medium-deep survey zone at 20\% completeness.}
\label{fig:mz_scat}
\end{center}
\end{figure*}

It is worth examining the distribution of the \planck\ SZ clusters in
the $M$--$z$ plane and comparing it to that of other catalogues. For
illustration, we compare to an X-ray selected sample, namely REFLEX-I,
on the one hand (Fig.~\ref{fig:mz_scat}, right panel green open
circles) and to the large-area SZ-selected cluster catalogues by ACT
\citep{has13} and SPT \citep{rei13}, on the other hand
(Fig.~\ref{fig:mz_scat}, red open symbols). In this comparison we
report, for the ACT clusters (open squares), the so-called UPP
(Universal Pressure Profile) masses given in \citet{has13}.

The range of redshifts covered by the \Planck\ SZ sample, from
$z=0.01$ to about 1 with 67\% of the clusters lying below $z=0.3$, is
quite complementary to the high redshift range explored by ACT and in
by SPT.  For the comparison of the mass distribution we take advantage
of our newly-proposed SZ-mass estimate, derived from $Y_z$, which
provides us with a homogeneous definition of the masses over the whole
range of \planck\ SZ clusters with measured redshifts.  The
\planck\ clusters populate the full redshift range and they quite
nicely fill a unique space of massive, $M\ge5\times 10^{14}\,\msol$,
and high redshift $z\ge 0.5$ clusters, as shown in
Fig.~\ref{fig:mz_scat}. This contrasts with the SZ clusters detected
in 720 square degrees of SPT observations and those of ACT
observations, which are dominated, as shown in Fig.~\ref{fig:mz_scat}
left panel, by lower-mass higher-redshift clusters (up to
$z\sim1.3$). The combination of \planck\ and SPT/ACT catalogues
samples the $M$--$z$ space in a complementary manner. Clearly the
all-sky nature of the \planck\ makes the most massive clusters
preferentially accessible to \planck\, whereas the highest redshift
clusters, $z\ge 1$, are accessible to SPT.

Very few massive high-redshift clusters exist in the X-ray catalogues,
as seen in Fig.~\ref{fig:mz_scat} (right panel open blue squares). The
all-sky NORAS/REFLEX catalogues \citep{boe00,boe04} are limited to
$z=0.45$, a result of the $(1+z)^4$ surface brightness dependence of
the X-ray detection limit (Fig.~\ref{fig:mz_scat}, right panel solid
green line). The smaller-area MACS sample, based on systematic
follow-up of {ROSAT} bright sources \citep{ebe07}, contains a dozen
clusters at $z\ge 0.5$. The 400SD sample \citep{bur07}, based on
serendipitous detections in 400$\,$deg$^2$ of ROSAT pointed
observations, contains only two clusters with $M\ge 5\times
10^{14}\,\msol$ and $z\ge 0.5$. Finally, only a couple of clusters in
the range $M\ge 5\times 10^{14}\,\msol$ are found in the {\it XMM-Newton}
based serendipitous cluster samples (XCS, \citet{meh12}; XMM-LSS,
\citet{pac07}; XDCP, \citet{fas11}). By contrast to an X-ray selected
cluster catalogue, the \planck\ detection-limit, illustrated for the
medium-deep survey zone and shown in Fig.~\ref{fig:mz_scat} (right
panel solid black line), has a much shallower dependence on redshift
and is quasi-redshift independent above $z=0.4$. The difference in
cluster selection starts at redshifts $z\ge0.2$. As a result of the
quasi-redshift independent mass-selection of SZ surveys,
\Planck\ probes deeper than the X-ray selection. This is also seen in
the overall distribution of redshifts of the \planck\ clusters,
Fig.~\ref{fig:zhist}.

This leaves the \Planck\ SZ catalogue as the deepest all-sky catalogue
spanning the broadest cluster mass range from $0.1$ to $1.6\times
10^{15}\,\msol$, and particularly adapted to the detection of rare
very massive clusters in the tail of the distribution in the range
$M\ge 5\times 10^{14}\,\msol$ and $z\ge 0.5$.

\subsection{X-ray flux of the \planck\ clusters and candidates}

\begin{figure*}[t]
\begin{center}
\includegraphics[width=\textwidth]{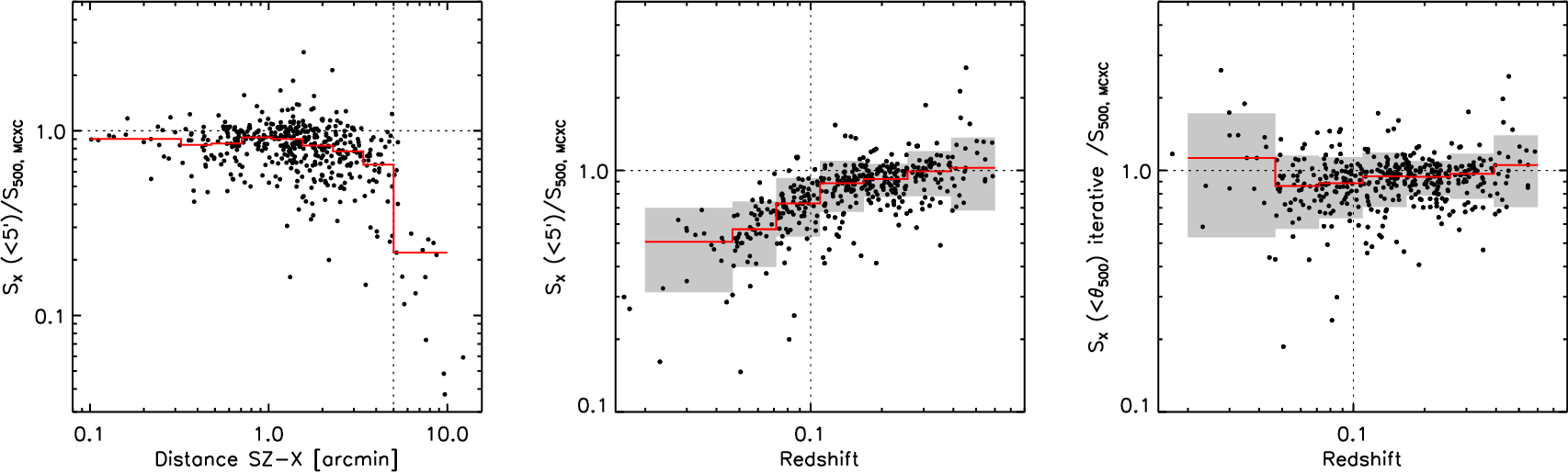}
\caption{Ratio between $\rass$\ flux, computed in an aperture of five
  arcmin in radius centred on the \planck\ position, and MCXC value
  for \planck\ candidates identified with MCXC clusters. The fluxes
  are computed in the $[0.1$--$2.4]$\,$\keV$\ band at Earth and
  corrected for absorption.  $S_{500}$ is the flux corresponding to
  the luminosity within $\Rv$ published in the MCXC catalogue. {\it
    Left panel:} the ratio is plotted as a function of distance
  between the \Planck\ and X-ray positions; {\it Middle panel:} same,
  as a function of cluster redshift, for distances smaller than five
  arcmin; {\it Right panel:} same as middle panel, for $\rass$\ flux
  within $\Rv$ derived from the aperture flux, using the MCXC
  iterative procedure based on the $L_{500}$--$\Mv$ relation and the
  \rexcess\ gas density profile \citep{pif11}.  The red line is the
  median ratio in distance or redshift bins with the grey area
  corresponding to $\pm1\,\sigma$ standard deviation in each bin. }
\label{fig:RASSfx}
\end{center}
\end{figure*}

\begin{figure}[t]
\begin{center}
\includegraphics[width=8.8cm]{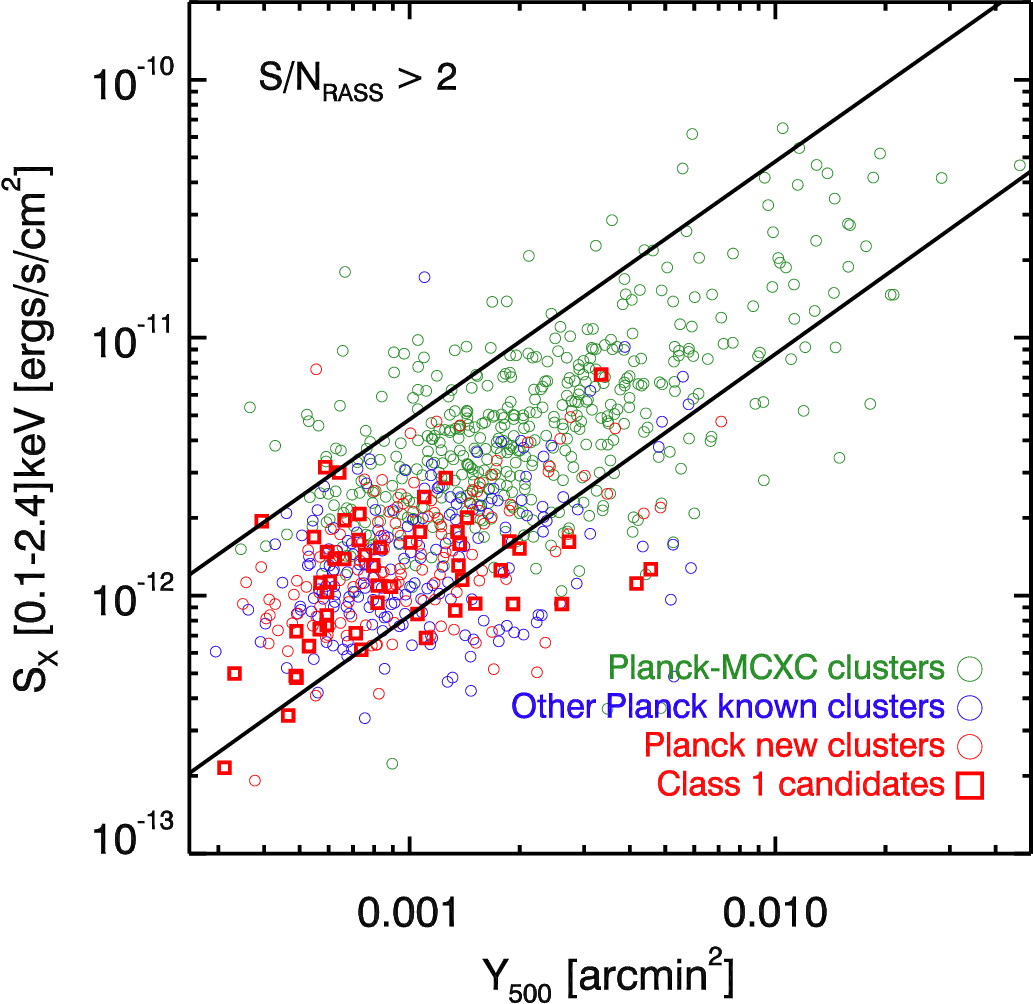}
\caption{X-ray unabsorbed flux versus SZ flux. For \Planck\ SZ
  detections identified with MCXC clusters (open green circles), the
  X-ray flux is estimated from $L_{500}$. For other \Planck\ SZ
  detections, the flux is derived from $\rass$\ count-rate in a
  five-arcmin aperture (see Sect.~\ref{sec:rass}). \planck\ new
  clusters and {\sc class1} candidates are shown as open red circles
  and squares, respectively. The two lines corresponds to the expected
  $L_{500}$--$\YSZ$ relation \citep{arn10} at $z=0.01$ and $z=1$,
  respectively. }
\label{fig:fxsz}
\end{center}
\end{figure}

For all \planck\ SZ detections, we estimated the unabsorbed fluxes at
Earth in the [0.1--2.4]\,$\keV$\ band (as in the MCXC) measured in an
aperture of five arcmin.  The aperture is centred on the
\planck\ candidate position, except for candidates associated with a
BSC source, for which we adopt the X-ray position, since the BSC
source is very likely the counterpart \citep{planck2012-IV}.  The
conversion between the $\rass$\ count rate in the hard band and flux
is performed using an absorbed thermal emission model with the
$N_{\mathrm{H}}$ value fixed to the 21\,cm value. The conversion
depends weakly on temperature and redshift and we assumed typical
values of $kT=6\,\keV$ and $z=0.5$.  \citet{planck2012-IV} compared
such flux estimates with precise \xmm\ fluxes measured within $\Rv$,
$S_{500}$, for candidates confirmed with the \xmm\ follow-up
programme. These clusters lie in the range $0.1<z<0.9$ and the
$0.3\times 10^{-12}<S_{500}<6\times 10^{-12} \, \mathrm{erg} \,
\mathrm{s}^{-1}\,\mathrm{cm}^{-2}$ flux range.  The $\rass$\ aperture
fluxes were found to underestimate the ``true'' flux by about $30\,\%$.

Figure~\ref{fig:RASSfx} extends this comparison further to all the
\Planck\ SZ detections identified with MCXC clusters. \citet{pif11}
published homogenized $L_{500}$ and $\Rv$ values derived from the flux
given in the original catalogues in various apertures, using an
iterative procedure based on the \rexcess\ \mbox{$L_{500}$--$\Mv$} relation
and gas density profile shape.  We simply computed $S_{500}$ from
$L_{500}$, taking into account the K-correction at the cluster
redshift, but neglecting its variation with temperature.

Although derived from \rosat\ survey data as our present flux
estimate, $S_{500}$ values from the MCXC are expected to be more
accurate due to: (i) optimum choice of the X-ray centre; (ii) higher
S/N detection; (iii) more sophisticated flux extraction adapted to
data quality and source extent (e.g., growth curve analysis); and (iv)
use of $\Rv$ rather than a fixed aperture. Not surprisingly, the ratio
between the present flux estimate and the MCXC value decreases with
increasing offset between the \planck\ position and X-ray position
(Fig.~\ref{fig:RASSfx}, left panel). The ratio drops dramatically when
the distance is larger than five arcmin, i.e., when the X-ray peak
lies outside the integration aperture. Those are rare cases, 18 nearby
clusters ($z<0.1$ with a median value of $z=0.05$), for which a
physical offset likely contributes to the overall offset.  When these
cases are excluded, the median ratio is $0.85$ and depends on redshift
(Fig.~\ref{fig:RASSfx}, middle panel); it significantly decreases with
decreasing redshift below $z$ of $0.1$. The median ratio is $0.65$ and
$0.92$, with a standard deviation of $0.10$ and $0.15$ dex, below and
above $z=0.1$, respectively. This is mostly due to the choice of a
fixed aperture that becomes too small as compared to $\Rv$ at low
$z$. If we apply the same iterative procedure used by \citet{pif11} to
estimate $S_{500}$ from the aperture flux, the resulting value is
consistent on average with the MCXC value at all redshifts
(Fig.~\ref{fig:RASSfx}, right panel). The dispersion is slightly
increased. The aperture unabsorbed fluxes are thus reliable estimates
of the X-ray fluxes above $z > 0.1$ on average.

Figure~\ref{fig:fxsz} shows the X-ray flux as function of $\YSZ$ for
\Planck\ candidates identified with known clusters, for the confirmed
new \planck\ clusters and for the {\sc class1} candidates.  For
\planck\ detections identified with MCXC clusters we plot the more
precise published $S_{500}$ value. All three categories of sources
behave in a similar manner in good agreement with the range of
redshifts probed by the sample. In this respect {\sc class1}
candidates do not exhibit any departure with respect to the known or
confirmed clusters. We provide the X-ray fluxes for the
\planck\ clusters and candidates that are not identified with MCXC
clusters (see Appendix \ref{ap:valid} and
Table~\ref{tab:valid_info}). For the \Planck\ cluster with MCXC
identifier, we refer the reader to the RASS catalogue outputs or to
the homogenized MCXC meta-catalogue.  The main limitation of the
aperture unabsorbed fluxes is the statistical precision on the
$\rass$\ estimate (most of the \planck\ SZ detections not identified
with MCXC clusters have low (S/N)$_{\mathrm{RASS}}$ values) and the
relatively large scatter ($\pm30\%$ standard deviation). For $z<0.1$
clusters, and if the $\rass$\ detection is reasonably good a more
precise procedure is recommended, such as an adapted growth curve
analysis, on a case-by-case basis.

\subsection{Scaling relations between SZ and X-ray quantities}\label{sec:scal}

\begin{figure*}[t]
\begin{center}
\resizebox{0.9\textwidth}{!} {
\includegraphics[]{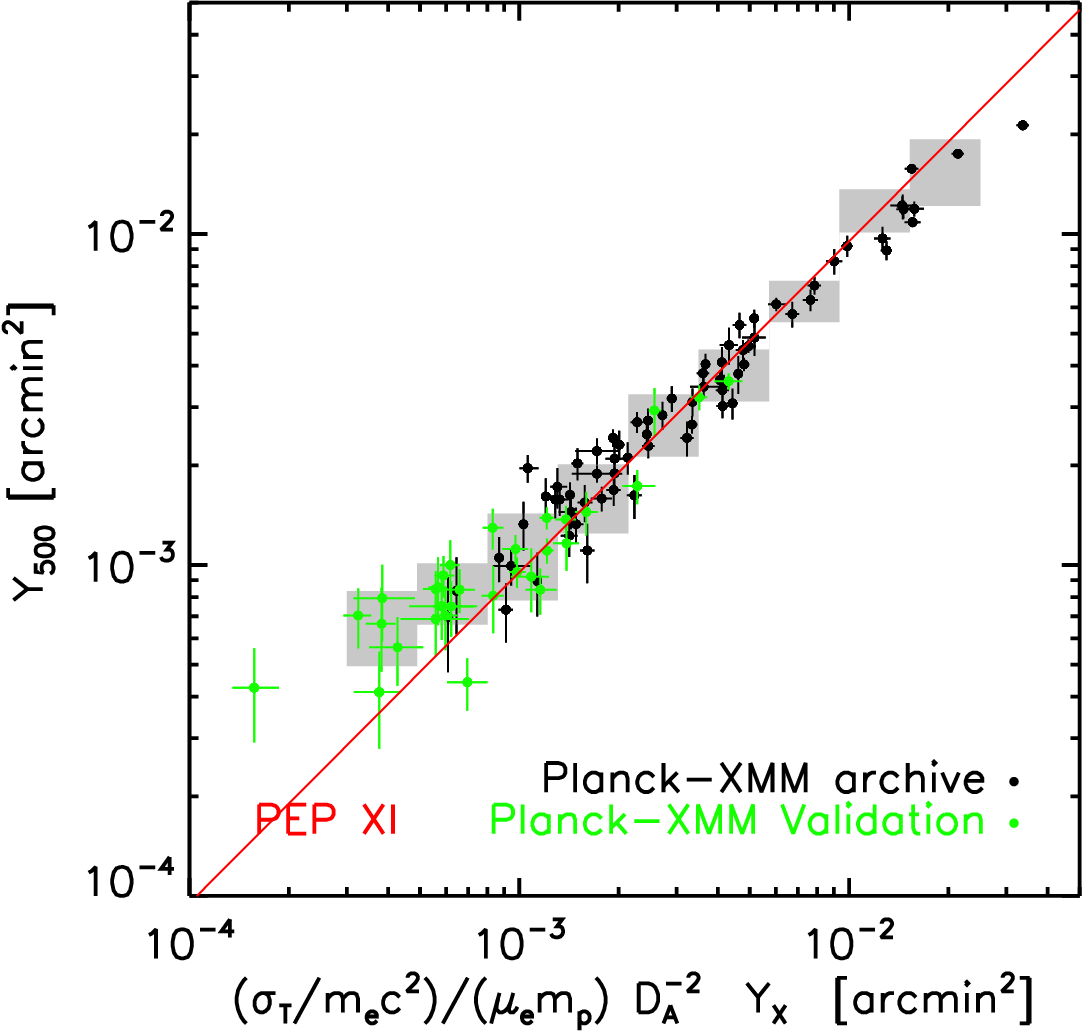}
\hspace{3cm}
\includegraphics[]{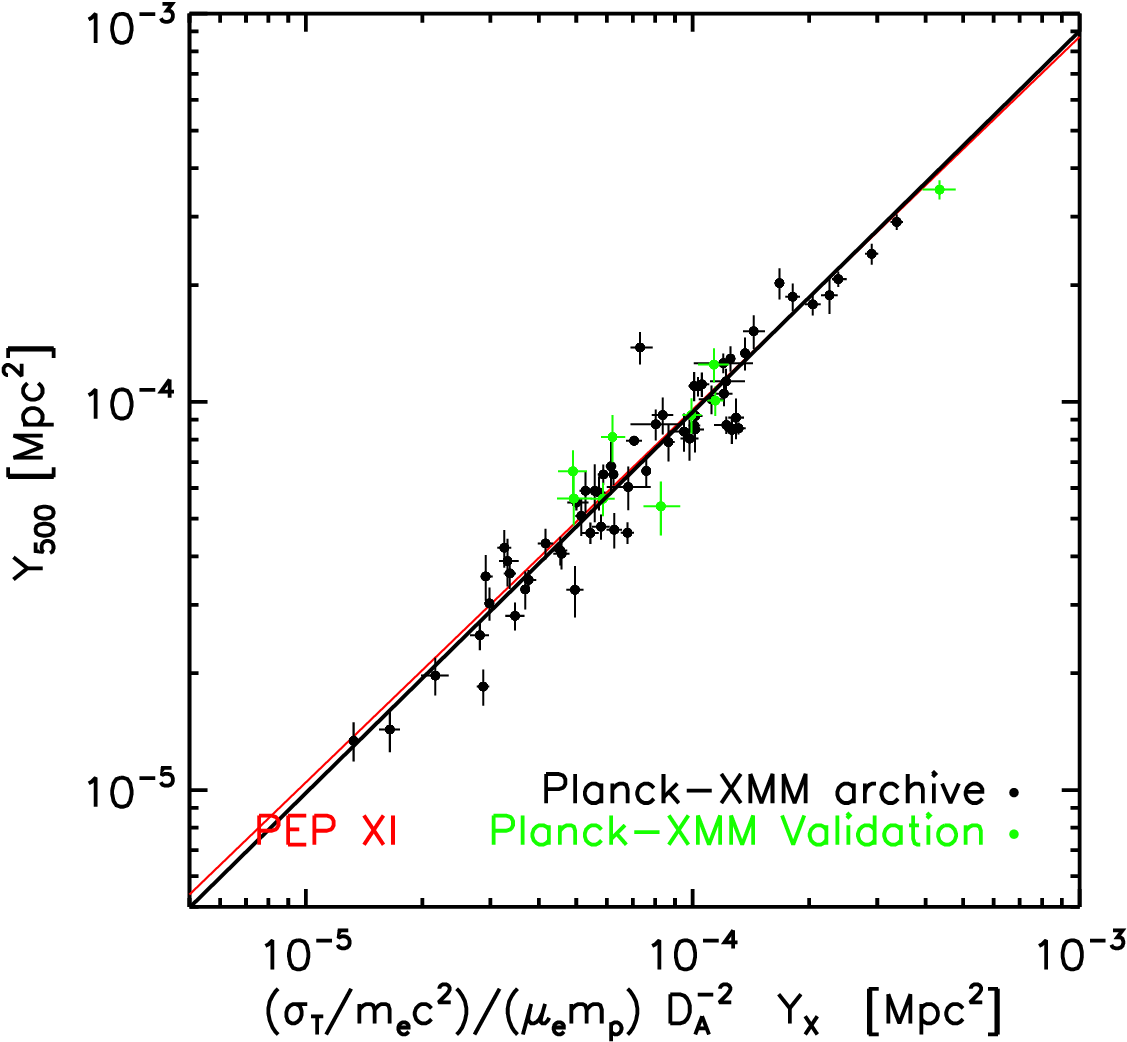}}
\caption{Relation between the Comptonization parameters $\YSZ$, and
  the normalized $\YX$ parameter for a sub-sample of the present
  catalogue.  Black points show clusters in the \Planck-ESZ sample
  with \xmm\ archival data presented by \citet{planck2011-5.2b} and
  additional LoCuSS clusters studied by \citet{planck2012-III}.  Green
  points represent new \Planck\ clusters confirmed with
  \xmm\ \citep{planck2011-5.1b, planck2012-I, planck2012-IV}. The red
  line denotes the scaling relations of \citet{planck2011-5.2b}. {\it
    Left panel:} relation in units of arcmin$^2$ where $\YSZ$ is
  extracted using the \citet{arn10} pressure profile.  The grey area
  corresponds to median $\YSZ$ values in $\YX$ bins with $\pm\sigma$
  standard deviation.  {\it Right panel:} scaling relation between the
  intrinsic Compton parameter, $D^2_{\mathrm{A}} \YSZ$, and $\YX$ for
  the sub-sample of S/N $>7$ clusters used in the cosmological
  analysis. The data are corrected for Malmquist bias, and $\YSZ$ is
  extracted using the \planck\ pressure profile (see text). The black
  line is the best-fit power-law relation. }
\label{fig:yszyx}
\end{center}
\end{figure*}

\begin{table*}[t]
\caption{{The \YSZYX\ relation.  ``MB'' is the Malmquist bias
    correction and ``Profile'' is the shape used in $\YSZ$ extraction.
    Parameters are given for the fit $\YSZ/Y_{\mathrm{p}}=
    A\,(\YX/Y_{\mathrm{p}})^\alpha$ using BCES orthogonal regression
    with pivot $Y_{\mathrm{p}} =10^{-4}$Mpc$^2$, along with the
    intrinsic and raw scatter around the best-fit relation.  The mean
    ratio is $\Delta \log{Q} = \log(\YSZ/\YX)$, with corresponding
    intrinsic and raw scatter.  Scatters are error-weighted values.
    The best estimate is in bold type. The \rexcess\ prediction is
    from \citet{arn10}.}}
\label{tab:yszyx}
\tiny
\setbox\tablebox=\vbox{
\newdimen\digitwidth
\setbox0=\hbox{\rm 0}
\digitwidth=\wd0
\catcode`*=\active
\def*{\kern\digitwidth}
\newdimen\signwidth
\setbox0=\hbox{+}
\signwidth=\wd0
\catcode`!=\active
\def!{\kern\signwidth}
\halign{\hbox to 1.0in{#\leaderfil}\tabskip=1em&
    \hfil#\hfil\tabskip=0.5em& 
    \hfil#\hfil& 
    \hfil#\hfil\tabskip=2em& 
    \hfil#\hfil\tabskip=0.7em& 
    \hfil#\hfil& 
    \hfil#\hfil& 
    \hfil#\hfil\tabskip=2em& 
    \hfil#\hfil\tabskip=1em& 
    \hfil#\hfil\tabskip=0.7em& 
    \hfil#\hfil\tabskip=0pt\cr
\noalign{\doubleline}
\omit&&&&\multispan4\hfil Power-law fit\hfil&\multispan3\hfil Mean ratio\hfil\cr
\noalign{\vskip -3pt}
\omit&&&&\multispan4\hrulefill&\multispan3\hrulefill\cr
\omit\hfil Sample\hfil&MB&Profile&$N$&$100A$&$\alpha$&$100\sigma^{\mathrm{log}}_{\mathrm{int}}$&$100\sigma^{\mathrm{log}}_{\mathrm{raw}}$&$\Delta\log{Q}$&$100\sigma_{\mathrm{int}}$&$100\sigma_{\mathrm{raw}}$\cr 
\noalign{\vskip 3pt\hrule\vskip 5pt}
PEPXI&   N&    A10&$62$& $-2.0\pm1.0$& $0.960\pm0.040$& $10.0\pm1.0$&\dots& $ -0.022\pm0.014$&\dots&\dots\cr 
ESZ&     N&    A10&$62$& $-2.2\pm1.1$& $0.966\pm0.034$& $*7.2\pm1.1$& $8.2\pm1.0$& $-0.023\pm0.011$& $7.3\pm1.1$& $8.5$\cr 
ESZ&     Y&    A10&$62$& $-3.0\pm1.1$& $0.975\pm0.035$& $*7.1\pm1.1$& $8.2\pm1.0$& $-0.031\pm0.011$& $7.2\pm1.1$& $8.4$\cr 
S/N $>7$&Y&    A10&$78$& $-2.4\pm1.0$& $0.972\pm0.029$& $*6.9\pm1.1$& $8.1\pm0.9$& $-0.024\pm0.010$& $6.9\pm1.0$& $8.3$\cr 
Cosmo&   Y&    A10&$71$& $-1.9\pm1.1$& $0.990\pm0.032$& $*7.2\pm1.2$& $8.3\pm1.0$& $-0.021\pm0.010$& $6.9\pm1.0$& $8.3$\cr 
Cosmo&   Y&A10+err&$71$& $-1.9\pm1.1$& $0.987\pm0.031$& $*6.3\pm1.1$& $7.9\pm0.9$& $-0.019\pm0.010$& $6.5\pm1.1$& $8.2$\cr 
\bf Cosmo&\bf Y&\bf PIP-V&$\mathbf{71}$& $\mathbf{-2.6\pm1.0}$& $\mathbf{0.981\pm0.027}$& $\mathbf{*6.6\pm1.2}$& $\mathbf{7.8\pm1.0}$& $\mathbf{-0.027\pm0.010}$& $\mathbf{6.6\pm1.0}$&$\mathbf{8.0}$\cr 
\noalign{\vskip 5pt}
\multispan5REXCESS X-ray prediction\leaderfil&&&&$-0.034\pm0.002$\cr   
\noalign{\vskip 5pt\hrule\vskip 3pt}}}
\endPlancktablewide
\end{table*}

A fundamental scaling relation is that between $\YSZ$ and its X-ray
analogue, $\YX$. Introduced by \citet{kra06}, $\YX$ is the product of
$\Mgv$, the gas mass within $\Rv$, and $\TX$, the spectroscopic
temperature outside the core.\footnote{Here we use the temperature
measured in the [0.15--0.75]\,$\Rv$ aperture.} From the fact that
the gas density profile used to compute $\Mgv$ is derived from
deprojection of the X-ray surface brightness profile, and that the
X-ray emission depends on the square of the density, the ratio of
these two quantities is
\begin{eqnarray}
\frac{D_{\mathrm{A}}^2\,\YSZ }{C_{\mathrm{XSZ}}\,\YX} &= &
\frac{1}{Q}\frac{\langle \ne T\rangle_{\Rv}} {\langle\ne\rangle_{\Rv}
\TX}\\
\nonumber Q & =&\frac{\sqrt{\langle \ne^2 \Lambda_{\mathrm{x}}(T)\rangle}}
  {\sqrt{\langle\ne\rangle^2\Lambda_{\mathrm{x}}(T)}}\,,
\label{eq:yszyx}
\end{eqnarray}
where the angle brackets denote volume-averaged quantities, and $Q$ is
the {emissivity-weighted clumpiness factor, which affects the
density profile derived from X-ray data}
radial bins used to derive the density profile \citep{zhu13}. {
  $\Lambda_{\mathrm{x}}(T)$ is the temperature-dependent emissivity in
  the considered X-ray band.} The numerical constant
$C_{\mathrm{XSZ}}=\sigma_{\mathrm{T}}/(m_{\mathrm{e}}\,c^2\,
\mu_{\mathrm{e}}\,m_{\mathrm{p}})=1.416 \times
10^{-19}\,{\mathrm{Mpc}}^2\,(\msol\, \keV)^{-1}$.  The ratio thus
depends only on the internal structure of the intra-cluster medium.
{ Note that the conversion of $Y_{500}/Y_{\mathrm{x}}$ into the
  amplitude of density/temperature variations depends on the
  correlation between variations of these thermodynamic properties,
  which differ between isobaric and adiabatic cases \citep[see][for
    more details]{khe13}. }

The properties of the $\YX$--$\YSZ$ relation, in particular its
variation with mass and redshift and the dispersion about the mean
relation, are important probes of the physics of cluster
formation. 

\subsubsection{Data set}

Here we extend the study of a sample of 62 clusters from the
\planck--ESZ sample with good quality \xmm\ archive data presented
in \citet[][hereafter PEPXI]{planck2011-5.2b}. This study found
$D_{\mathrm{A}}^2\,\YSZ/C_{\mathrm{XSZ}}\,\YX = 0.95\pm0.03$, in a good
agreement with \rexcess\ prediction, $0.924\pm0.004$, of \citet{arn10}.

All 62 objects in the PEPXI sample are included in the present
catalogue. We further add 40 clusters from the catalogue, including
nine additional objects from the \xmm\ archival study of \planck-detected
LoCuSS systems presented by \citet{planck2012-III}, and the 31
\Planck-discovered clusters with good redshift estimates
($Q_{\mathrm{z}} = 2$) confirmed with the \xmm\ 
\citep[][]{planck2011-5.1b,planck2012-I,planck2012-IV}. The total
sample thus consists of 102 clusters.

For each object, $\YX$ and the corresponding $\Rv$ value were
estimated simultaneously by iteration about the $\Mv$--$\YX$ relation
of \citet{arn10},
\begin{equation}
E^{2/5}(z)\Mv = 10^{14.567}
\left[\frac{\YX}{2\times10^{14}\,\msol \,\keV}\right]^{0.561}\,\msol \,. 
\label{eq:myx} 
\end{equation}
In the present study, we focus on the physical relation between $\YSZ$
and $\YX$.  While these quantities must be estimated within the same
radii, the exact value of $\Rv$ is irrelevant as the radial dependence
of the $\YSZ/\YX$ ratio is negligible.  We thus propagated only the
measurement uncertainties on the temperature and gas mass profiles,
fixing the aperture to $\Rv$. We ignored the statistical and
systematic uncertainties on the \MY\ relation itself.\footnote{These
must however be taken into account when using $\YSZ$ or $\YX$ as
a mass proxy, e.g., when calibrating the \YM\ relation from combining
the \MYX\ relation and the relation between $\YSZ$ and $\YX$ (or
equivalently $\Mv$). This calibration is extensively addressed in
the \citet{planck2013-p15}.} Similarly $\YSZ$ was re-extracted at the
X-ray position with size fixed to X-ray size.  Its uncertainty
corresponds to the statistical error on the SZ signal.  The results
are summarized Table~\ref{tab:yszyx}, with the best estimate
indicated in bold face.

\subsubsection{The best-fit \YSZYX\ relation}

The \YSZYX\ scaling relation for the full sample is shown in units of
arcmin$^2$ in Fig.~\ref{fig:yszyx}.  At high flux the points follow
the PEP XI relation.  The slope and normalization are determined at
slightly higher precision, due to the better quality SZ data. The
derived intrinsic scatter (Table~\ref{tab:yszyx}) is significantly
smaller, a consequence of the propagation of gas mass profile errors
in the $\YX$ error budget, which was neglected in our earlier study.

The relation levels off at around $\YX = 5 \times
10^{-4}\,{\mathrm{arcmin}}^2$, with a bin average deviation increasing
with decreasing $\YX$ (Fig.~\ref{fig:yszyx} left panel). This is an
indication of Malmquist bias, as noted by
\citet[][]{planck2012-I}. Full correction of this bias when fitting
scaling relations involves drawing mock catalogues according to the
cluster mass function, to which the sample selection criteria are then
applied. The present sample is a small subset of the full S/N $\ge4.5$
\planck\ catalogue and thus such a
procedure cannot be applied.  To minimize bias effects we will only
consider high S/N detections, S/N $>7$. To correct for the residual
bias, we adapted the approach proposed by \citet{vik09b}.  Before
fitting the \YSZYX\ relation, each individual $Y$ value was divided by
the mean bias, $b$, given by
\begin{equation}
\ln b=\frac{\exp\left(-x^2/2\sigma^2\right)}{\sqrt{\pi/2}\,\mathrm{erfc}\left(x/\sqrt{2}\sigma\right)}\,\sigma \,,
\end{equation}
where $x=-\log(Y/Y_{\mathrm{min}})$, $Y_{\mathrm{min}}$ being the flux
threshold corresponding to the S/N cut,
(S/N)$_{\mathrm{cut}}$. At the location of the cluster,
$Y/Y_{\mathrm{min}}$ = (S/N)/(S/N)$_{\mathrm{cut}}$. Here $\sigma$ is the
log-normal dispersion at fixed $\YX$. We took into account both the
intrinsic dispersion $\sigma_{\mathrm{int}}$, estimated iteratively,
and the statistical dispersion, given by
$\sigma=\sqrt{\ln{\left[(({\mathrm{S/N}})+1)/({\mathrm{S/N}})\right]}^2 +
\left[\ln{10}\,\sigma_{\mathrm{int}}\right]^2}$.
The correction decreases the effective $\YSZ$ values at a given $\YX$,
an effect that is larger for clusters closer to the S/N threshold;
i.e., low-flux objects. The net effect on the scaling relation is
small, giving a $0.7\,\sigma$ decrease of the normalization and a slight
steepening of the power-law slope (Table~\ref{tab:yszyx}).

The slope and normalization of the relation are robust to the
inclusion of newly-discovered \planck\ clusters. The results derived from the
extended sample of $78$ clusters with S/N $>7$ agree with those
obtained for the updated ESZ-XMM sample within $0.5\,\sigma$
(Table~\ref{tab:yszyx}). They are also in agreement with the
sub-sample of 71 S/N $>7$ clusters included in the cosmological sample
discussed by \citet{planck2013-p15}. We measured a significant intrinsic
scatter of $\sigma_{\mathrm{int}} = 0.07\pm0.01\,$ dex. There is one
spectacular outlier with an $\YSZ/\YX$ ratio nearly twice as big as
the mean.  This is the \planck\ ESZ cluster identified
with A2813 or RXC~J0043.4-2037 in the \reflex\ catalogue, located at
$z=0.29$. Its high ratio is very puzzling.  It cannot result from an
inaccurate redshift measurement, as this is based on spectroscopic
data for several cluster galaxies \citep{boe04}. There is no evidence
of a peculiar dynamical state from the X-ray morphology, and there is
no evidence of contamination in the SZ data.

Part of the dispersion could be due to the use of an inappropriate
fixed pressure profile in the $\YSZ$ extraction.  When including
possible errors on $\YSZ$ due to dispersion around the mean
\citet{arn10} profile, the scatter is decreased to
$\sigma_{\mathrm{int}} = 0.06$, a decrease at the $1\,\sigma$ level. To
further assess the effect of the choice of the pressure profile, we
re-extracted the SZ signal using the \planck+\xmm\ profile shape
measured for ESZ clusters by \citet[][hereafter
PIPV]{planck2012-V}. Individual profiles are used for \planck\ ESZ clusters,
and the mean profile is used for the other clusters.  This should give
the most reliable estimate of the \YSZYX\ relation, since it is based
directly on measured profile shapes. In this case, the slope and
scatter remain unchanged but the normalization is slightly decreased
(at the $0.5\,\sigma$ level). This is a result of the more inflated
nature of the PIPV profile as compared to the \citet{arn10}
\rexcess\ profile.  The relation derived using PIPV pressure profiles
is plotted in the right-hand panel of Fig.~\ref{fig:yszyx} together
with the corresponding data points.

The relation does not exhibit significant evidence of variance of the
$\YSZ/\YX$ ratio with mass, the slope is consistent with unity, as
expected from strong self-similarity of pressure profile shape.
However, we found an intrinsic scatter about three times larger than
the results of \citet{kay12}. Partly this is due to the presence of
outliers in our data set (as discussed above), or it may be due to
projection effects in observed data sets \citep{kay12}.  The mean
ratio is very well constrained with a precision of $2.5\%$,
$\log(\YSZ/\YX)=-0.027\pm0.010$. This confirms at higher precision the
strong agreement between the SZ and X-ray measurements (within $\Rv$)
of the intra-cluster gas properties found by PEP XI and other studies
\citep[]{and11,sif12,mar12,roz12}.  {The ratio is 
  consistent with the X-ray prediction. In the simplest scenario of
  pure density variations in an isothermal ICM at the scale of the radial
  bin, this suggests a low clumpiness factor. However there are still
  large systematics that are discussed in
  Appendix~\ref{sec:scal_syst}.  We can translate those into an upper
  limit of order 30\%.}

\section{Summary}

\planck's all-sky coverage and broad frequency range are designed to
detect the SZ signal of galaxy clusters across the sky. We provide,
from the first 15.5 months of observations, the largest ensemble of
SZ-selected sources detected from an all-sky survey. The
\planck\ catalogue of SZ sources contains 1227 detections. This
catalogue, statistically characterized in terms of completeness and
statistical reliability, was validated using external X-ray and
optical/NIR data, alongside a multi-frequency follow-up programme for
confirmation. A total of 861 SZ detections are confirmed associations
with bona fide clusters, of which 178 are brand-new
clusters. The remaining 366 cluster candidates are divided into three
classes according to their reliability, i.e., the quality of evidence
that they are likely to be bona fide clusters.

A total of 813 \planck\ clusters have measured redshifts ranging from
$z=0.01$ to order one, with one-third of the clusters lying above
$z=0.3$. The brand-new \planck\ clusters extend the redshift range
above $z=0.3$. For all the \planck\ clusters with measured redshift, a
mass can be estimated from the Compton $Y$ measure. We provide a
homogeneous mass estimate ranging from ($0.1$ to $1.6)\times
10^{15}\,\msol$.  Except at low redshifts, the \planck\ cluster
distribution exhibits a nearly redshift-independent mass limit and
occupies a unique region in the $M$--$z$ space of massive,
$M\ge5\times 10^{14}\,\msol$, and high-redshift ($z\ge 0.5$)
clusters. Owing to its all-sky nature, \Planck\ detects new clusters
in a region of the mass--redshift plane that is sparsely populated by
the RASS catalogues. It detects the rarest clusters, i.e., the most
massive clusters at high redshift in the exponential tail of the
cluster mass function that are the most useful clusters for
cosmological studies. With the presently confirmed \planck\ SZ
detections, \planck\ doubles the number of massive clusters above
redshift 0.5, as compared to other surveys. The \Planck\ SZ catalogue
is, and will be for years to come, the deepest all-sky SZ catalogue
spanning the broadest cluster mass range.

The \planck\ SZ catalogue should motivate multi-wavelength follow-up
efforts. The confirmation of the cluster candidates will reveal
clusters at higher redshifts than the present distribution. Such
follow-up efforts will further enhance the value of the \planck\ SZ
catalogue as the first all-sky SZ selected catalogue. It will serve as
a reference for studies of cluster physics (e.g., galaxy properties
versus intra-cluster gas physics, dynamical state, evolution,
etc.). Using an extended sub-sample of the \planck\ SZ clusters with
high-quality \xmm\ data, the scaling relations between SZ and X-ray
properties were reassessed and updated. With better-quality
data and thus higher precision, we show excellent agreement between
SZ and X-ray measurements of the intra-cluster gas properties. We
have thus derived a new up-to-date reference calibrated local relation
between $Y$ and $Y_{\mathrm{X}}$. \\
The \planck\ SZ catalogue will also serve to define samples for
cosmological studies. A first step in this direction is already taken
in \citet{planck2013-p15}, where an analysis of the SZ cluster
abundance to constrain the cosmological parameters is performed using
a sub-sample selected from the PSZ catalogue consisting of 189
clusters detected above a S/N of 7 with measured
redshifts.  The value-added information derived from the validation of
the \planck\ SZ detections, in particular the SZ-based mass estimate,
increases even further the value of the \planck\ SZ catalogue.

The combination of the \Planck\ all-sky SZ data with near future and
planned observations of the large-scale structure by surveys such as
PAN-STARRS, LOFAR, Euclid, LSST, and RSG/e-ROSITA will revolutionize
our understanding of large-scale structure formation and evolution.

\begin{acknowledgements}

The development of \Planck\ has been supported by: ESA; CNES and
CNRS/INSU-IN2P3-INP (France); ASI, CNR, and INAF (Italy); NASA and DoE
(USA); STFC and UKSA (UK); CSIC, MICINN, JA and RES (Spain); Tekes,
AoF and CSC (Finland); DLR and MPG (Germany); CSA (Canada); DTU Space
(Denmark); SER/SSO (Switzerland); RCN (Norway); SFI (Ireland);
FCT/MCTES (Portugal); and PRACE (EU). A description of the
\Planck\ Collaboration and a list of its members, including the
technical or scientific activities in which they have been involved,
can be found at
\url{http://www.sciops.esa.int/index.php?project=planck}\\ 
\url{&page=Planck_Collaboration}. The
authors thank N. Schartel, ESA {\it XMM-Newton} project scientist, for
granting the Director Discretionary Time used for confirmation of SZ
\Planck\ candidates. The authors thank TUBITAK, IKI, KFU and AST for
support in using RTT150 (Russian-Turkish 1.5-m telescope, Bakyrlytepe,
Turkey); in particular we thank KFU and IKI for providing significant
amount of their observing time at RTT150. We also thank BTA 6-m
telescope Time Allocation Committee (TAC) for support of optical
follow-up project.  The authors acknowledge the use of the INT and WHT
telescopes operated on the island of La Palma by the Isaac Newton
Group of Telescopes at the Spanish Observatorio del Roque de los
Muchachos of the Instituto de Astrof\'isica de Canarias (IAC); the
Nordic Optical Telescope, operated on La Palma jointly by Denmark,
Finland, Iceland, Norway, and Sweden, at the Spanish Observatorio del
Roque de los Muchachos of the IAC; the TNG telescope, operated on La
Palma by the Fundacion {\it Galileo Galilei} of the INAF at the Spanish
Observatorio del Roque de los Muchachos of the IAC; the GTC telescope,
operated on La Palma by the IAC at the Spanish Observatorio del Roque
de los Muchachos of the IAC; and the IAC80 telescope operated on the
island of Tenerife by the IAC at the Spanish Observatorio del Teide of
the IAC. Part of this research has been carried out with telescope
time awarded by the CCI International Time Programme. The authors
thank the TAC of the MPG/ESO-2.2m telescope for support of optical
follow-up with WFI under {\it Max Planck} time.  Observations were
also conducted with ESO NTT at the La Silla Paranal
Observatory. This research has made use of
SDSS-III data. Funding for SDSS-III \url{http://www.sdss3.org/} has
been provided by the Alfred
P. Sloan Foundation, the Participating Institutions, the National
Science Foundation, and DoE. SDSS-III is managed by the Astrophysical
Research Consortium 
for the Participating Institutions of the SDSS-III Collaboration. \\ 
This research has made use of the following databases: the NED and IRSA
databases, operated by the Jet Propulsion Laboratory, California
Institute of Technology, under contract with the NASA; 
SIMBAD, operated at CDS, Strasbourg, France; SZ cluster database
operated by  Integrated Data and Operation Center (IDOC) 
operated by IAS under contract with
CNES and CNRS. The authors acknowledge the use of software provided by
the US National Virtual Observatory.  

\end{acknowledgements}

\bibliographystyle{aa}

\bibliography{szcat,Planck_bib}

\appendix

\section{Selection of Frequency Channel Maps}\label{Sec:freq_cho}

An assessment of which combination of \planck\ frequency channels to
use was performed using the \texttt{MMF1} implementation of the
matched multi-filter described in Sect.~\ref{ss_mmf}. The HFI and
LFI channel maps were preprocessed as described in
Sect.~\ref{ss_preprocessing}, with the only difference being that
the point-source mask contained, in addition, detections from the LFI
channel maps with S/N $\ge 10$.  Five different combinations of
frequency channels were investigated, all \Planck\ channels
(30--857 GHz), all HFI channels plus the 70~GHz channel map from LFI
(70--857 GHz), all HFI channels (100--857 GHz), the five lowest
frequency HFI channels (100--545 GHz) and the four lowest frequency HFI
channels (100--353 GHz). For each combination of frequency channels a
catalogue of SZ sources was extracted, resulting in five different
catalogues; the only differences between them must be entirely due to
the choice of channels in the combination.

The first four of these catalogues are in good agreement in terms of
the clusters detected, with all the differences amongst them being due
to detections with S/N $<5$. The (100--353 GHz) catalogue, however, contains
significantly more detections, resulting in a poor agreement between it
and the other catalogues that is not limited to low S/N
detections. This is interpreted as being due to the lack of a
dust-dominated channel in this combination, without which it is more
difficult to constrain contamination due to dust emission.

In order to assess any improvement in the S/Ns of detected
clusters with the inclusion of extra data, a robust sample of reliable
sources is required. To produce this, only clusters outside the 65\%
dust mask and with S/N $\ge 8$ were kept from each combination.  The
differences in the S/N of the same sources detected using different
frequency channel combinations can then be examined.  The ratio
between the S/N values of the common detections in each combination to those
of the (100--857 GHz) combination was then found; the mean of this ratio is
shown in Table~\ref{tab:fchl_combinations}.  This approach clearly
shows the (100--353 GHz) combination to be considerably noisier than the
other combinations, which is consistent with the observations reported
above. Neither the inclusion of the LFI frequency channels or just the
70~GHz channel brings any significant improvement in the S/N of the
clusters. Using the six HFI channel combination results in marginally
better S/N than the (100--545 GHz) combination.  The frequency channel
combination chosen therefore is (100--857 GHz) since this gives the
highest S/N with the smallest data-set. Reducing the S/N threshold
from 8 to 6 and hence doubling the number of SZ sources used to
evaluate the mean ratio does not change the conclusions of this
analysis.

\begin{table*}
\begingroup
\newdimen\tblskip \tblskip=5pt
\caption{{Effect of frequencies used in the extraction on the
    S/N of the detections.  The set of frequencies
    used is specified as a range, e.g., $100\rightarrow353$ (in GHz).  For a
    given cluster detected in two sets of frequencies, the ratio of
    S/N for the two detections is written as, e.g.,
    $(100\rightarrow353)/(100\rightarrow857)$.  The improvement in the S/N
    of the detected clusters between the $100\rightarrow353$ and
    $100\rightarrow857$ combinations is clearly demonstrated, as is
    the lack of significant improvement when 30 or 70\,GHz data are
    included. The improvement between the $100\rightarrow545$ and
    $100\rightarrow857$ combinations is smaller, in the range 1 to
    2\%.}}
\label{tab:fchl_combinations}
\nointerlineskip \vskip 1mm \footnotesize
\setbox\tablebox=\vbox{
\newdimen\digitwidth
\setbox0=\hbox{\rm 0}
\digitwidth=\wd0
\catcode`*=\active
\def*{\kern\digitwidth}
\newdimen\signwidth
\setbox0=\hbox{+}
\signwidth=\wd0
\catcode`!=\active
\def!{\kern\signwidth}
\halign{\hbox to 1.2in{#\leaderfil}\tabskip=2em&
    \hfil#\hfil& 
    \hfil#\hfil& 
    \hfil#\hfil& 
    \hfil#\hfil\tabskip=0pt\cr
\noalign{\doubleline}
\omit&\multispan4\hfil Mean ratio of detection S/N\hfil\cr
\noalign{\vskip -3pt}
\omit&\multispan4\hrulefill\cr
\noalign{\vskip 2pt}
\omit\hfil Selection criterion\hfil&$\frac{100\rightarrow353}{100\rightarrow857}$&
     $\frac{100\rightarrow545}{100\rightarrow857}$&
     $\frac{70\rightarrow857}{100\rightarrow857}$&
     $\frac{30\rightarrow857}{100\rightarrow857}$\cr
\noalign{\vskip 3pt\hrule\vskip 5pt}
S/N $\ge 6$& 0.86& 0.99& 1.00& 1.00\cr
S/N $\ge 8$& 0.84& 0.98& 1.00& 1.00\cr
\noalign{\vskip 5pt\hrule\vskip 3pt}}}
\endPlancktable 
\endgroup
\end{table*}

\section{Extract from the \planck\ catalogue of SZ sources}\label{ap:cat_cont}

We describe here the \planck\ catalogue of SZ sources delivered by the
collaboration and available together with the individual lists from
all three detections methods, the union mask used by these methods and
the ensemble of notes on individual clusters\footnote{
\url{http://www.sciops.esa.int/index.php?page=}\\
\url{Planck\_Legacy\_Archive\&project=planck}}. 

The union \planck\ SZ catalogue contains the coordinates and the
S/N of the SZ detections and a summary of the
validation information, including external identification of the
cluster and redshifts if they are available. The external
identification quoted in the delivered product corresponds to the
first identifier as defined in the external validation hierarchy,
namely identification with MCXC clusters followed by Abell and Zwicky,
followed by SDSS-based catalogues, followed by SZ catalogues, followed
finally by searches in NED and SIMBAD. Due to the size--flux degeneracy
discussed in Sect.~\ref{sec:dege}, no reference flux quantity is
outputted for the union catalogue.

The individual catalogues from the three detection methods,
\texttt{MMF1}, \texttt{MMF3}, and \texttt{PwS}, contain the
coordinates and the S/N of the detections, and information on
the size and flux of the clusters. The size is given in terms of
$\theta_\mathrm{s}$ and the flux is given in terms of the total
integrated Comptonization parameter, $Y = \yfrfh$. The full
information on the degeneracy between $\ts$ and $Y$ is provided in the
form of the two-dimensional marginal probability distribution for each
cluster.  \\
The degeneracy information is provided in this form so that
it can be combined with a model or external data to produce tighter
constraints on the parameters.  For example, combining it with an
X-ray determination of the size can be done by taking a slice through
the distribution at the appropriate $\theta_\mathrm{s}$. This is what
is done in Sect.~\ref{sec:sizeF} and the refined measurement using
X-ray information can be found in Table~\ref{tab:valid_info}.

\begin{table*}
\begingroup
\newdimen\tblskip \tblskip=5pt
\caption{Extract from the \Planck\ catalogue of SZ sources.  The first
  rows of the online table are shown. The online table contains
  additional columns as described in the Explanatory Supplement and in
  the text.}
\label{tab:PSZ}
\nointerlineskip \vskip 1mm \footnotesize
\setbox\tablebox=\vbox{
\newdimen\digitwidth
\setbox0=\hbox{\rm 0}
\digitwidth=\wd0
\catcode`*=\active
\def*{\kern\digitwidth}
\newdimen\signwidth
\setbox0=\hbox{+}
\signwidth=\wd0
\catcode`!=\active
\def!{\kern\signwidth}
\halign{\hbox to 1.7in{#\leaderfil}\tabskip=1em& 
   \hfil#\hfil& 
   \hfil$#$\hfil& 
   \hfil#\hfil& 
   \hfil#\hfil& 
        #\hfil& 
   \hfil#\hfil&  
   \hfil#\hfil\tabskip=0pt\cr
\noalign{\doubleline}
\omit\hfil Name\hfil&RA&\omit\hfil Dec\hfil&S/N& Validation&\omit\hfil ID$_{\mathrm{EXT}}$\hfil& $z$& Comments\cr
\noalign{\vskip 3pt\hrule\vskip 5pt}
PSZ1 G000.08+45.15&229\pdeg19790&*-0\pdeg979280&4.60&20&RXC J1516.5$-$0056&0.1198&F\cr
PSZ1 G000.42$-$41.84&316\pdeg06990&-41\pdeg339730&5.99&20&RXC J2104.3$-$4120&0.1651&F\cr
PSZ1 G000.77$-$35.67&307\pdeg93571&-40\pdeg595198&5.30&20&RXC J2031.8$-$4037&0.3416&F\cr
PSZ1 G001.00+25.71&244\pdeg58411&-13\pdeg070074&6.04&*3&\omit\hfil $\dots$\hfil& $\dots$& F\cr
PSZ1 G002.24$-$68.27&  349\pdeg60728&-36\pdeg278003&4.50&20&ACO S 1109 &0.1400&F\cr
PSZ1 G002.77$-$56.16&334\pdeg65975&-38\pdeg880540&7.84&20&RXC J2218.6$-$3853&0.1411&F\cr
PSZ1 G002.80+39.24&234\pdeg99997&*-3\pdeg292940&7.03&20&RXC J1540.1$-$0318&0.1533& F\cr
PSZ1 G003.09$-$22.51&  292\pdeg16440&-35\pdeg711064&4.92&*3&\omit\hfil $\dots$\hfil& $\dots$& F\cr
\noalign{\vskip 5pt\hrule\vskip 3pt}}}
\endPlancktable
\endgroup
\end{table*}

Table~\ref{tab:PSZ} presents an extract of the PSZ catalogue, in terms of
the first rows of the online table and the following selected columns:\\
\noindent
Name: name of cluster.\\ 
RA, Dec: right ascension (J2000) and declination (J2000). \\  
S/N: signal-to-noise ratio of the SZ detection. \\ 
Validation: status of the SZ detection from external validation:
20 = previously-known cluster; 10 = new confirmed \Planck\ cluster;
1 = {\sc 
  class1}  candidate ;  2 = {\sc class2} candidate; 3 = {\sc class3} 
 candidate.\\
ID$_{\mathrm{EXT}}$: first external identifier of the known clusters. \\
$z$: redshift of the cluster as reported from the external
validation.\\  
Comments: F = no comment; T = comment. Comments are readable in
an external file.\\ 
The complete version of the PSZ catalogue also contains the additional
columns:\\
\noindent
Index: index of the detection,  determined by
the order of the clusters in the union catalogue and sorted into order
of ascending Galactic longitude.\\
GLON, GLAT: Galactic coordinates.\\
POS\_ERR: errors on the position.\\
Pipeline: pipeline from which information is taken; namely 1 =
\texttt{MMF1}; 2 = \texttt{MMF3}; 3 = \texttt{PwS}.\\ 
PIPE\_DET: pipeline making the detection, with the following
order in  bits: 1st = \texttt{MMF1}; 2nd = \texttt{MMF3}; 3rd =
\texttt{PwS}. \\ 
PCCS: flag for a match with sources from the PCCS catalogue. \\
COSMO: flag for those clusters that are included in the
sample used for the cosmological analysis of \citet{planck2013-p15}.\\

\section{Outstanding outputs from the external validation}\label{ap:valid}

Based on the ancillary data used for the validation of the \planck\ SZ
catalogue, we provide value-added information to the \planck\ SZ
detections. \\ 
Namely, we provide, in addition to the first external
identifier, possible other common identifiers, {\tt IDs}. \\ 
We report
the redshift information associated with the \planck\ clusters ($z$)
and specify its source, ({\tt scr}).\\ 
For clusters with measured
redshifts, we compute the SZ-proxy $Y_z$ and the mass estimate
($M_{500}^{Y_z}$) and associated errors. For the clusters identified
with MCXC clusters we provide the SZ signal, $Y_{500,{\mathrm{PSX}}}$,
re-extracted fixing the size to the X-ray size provided in the MCXC
catalogue at the X-ray position. {We also provide the associated
  S/N in the {\it Planck} data}. Note that the 
X-ray positions used in the
present study are those quoted in the MCXC meta-catalogue. The
positions reported in the ESZ sample were taken from a sampled grid of
coordinates with a pixel size of 1.71 arcmin. Due to this sampling,
the reported MCXC positions in the ESZ sample exhibit an average
offset of 70 arcsec (less than a pixel, which varies depending on the
position of the object on the sphere). \\ 
For \planck\ SZ detections
not associated with a previously-known X-ray cluster and with
(S/N)$_{\mathrm{RASS}} \ge 1\,\sigma$, we provide the
unabsorbed X-ray flux, $S_{\mathrm{X}}$ (and error), measured in an
aperture of 5 arcmin in the band [0.1-2.4] keV. We only provide an
upper limit in the case of (S/N)$_{\mathrm{RASS}}<1\,\sigma$, except
for three SZ detections for which RASS exposure is very low and
(S/N)$_{\mathrm{RASS}}<-5\,\sigma$. The aperture is centred on the
\planck\ position, except for candidates associated with a BSC source
for which we adopt the X-ray position. These clusters are flagged.

\begin{table*}
\begingroup
\newdimen\tblskip \tblskip=5pt
\caption{{Information from external validation.  The ``Src'' for
    the cluster redshift is a code expanded in the readme file. $Y_z$
    is the SZ signal with asymmetric errors, computed within $R_{500}$.
    $M_{500}^{Y_z}$ is the derived mass with asymmetric
    errors. $S_{\mathrm{X}}$ is the unabsorbed X-ray flux measured in
    an aperture of 5\arcm\ in the band [0.1--2.4]\,keV. The aperture is
    centred on the \Planck\ position, except for candidates associated
    with a BSC source, for which we adopt the X-ray position.  For
    sources with (S/N)$_{\mathrm{RASS}}<1\,\sigma$, we quote an upper
    limit.  ``{\tt ID}'' gives other names for previously-known
    clusters.  $Y_{500,{\mathrm{PSX}}}$ is the SZ signal re-extracted after
    fixing size and position to the values given in the MCXC X-ray
    catalogue, if available. S/N$_{\mathrm{PSX}}$ is the associated
    S/N in the {\it Planck} data. The full table and
    the readme file are available at
    \url{http://www.sciops.esa.int/index.php?page=Planck\_Legacy\_Archive\&project=planck.}}
}
\label{tab:valid_info}
\nointerlineskip \vskip 1mm
\footnotesize
\setbox\tablebox=\vbox{
\newdimen\digitwidth
\setbox0=\hbox{\rm 0}
\digitwidth=\wd0
\catcode`*=\active
\def*{\kern\digitwidth}
\newdimen\signwidth
\setbox0=\hbox{+}
\signwidth=\wd0
\catcode`!=\active
\def!{\kern\signwidth}
\halign{\hbox to 1.5in{#\leaderfil}\tabskip=0.4em&
   \hfil$#$\hfil\tabskip=0.1em&
   \hfil$#$\hfil\tabskip=0.5em&
   \hfil$#$\hfil&
   \hfil$#$\hfil&
   \hfil$#$\hfil&
          #\hfil\tabskip=0.1em&
   \hfil$#$\hfil\tabskip=0.5em&
    \hfil#\hfil\tabskip=0pt\cr
\noalign{\doubleline}
\omit&&&Y_z&M_{500}^{Y_z}&S_{\mathrm{X}}&&Y_{500,\mathrm{PSX}}\cr
\noalign{\vskip 3pt}
\omit\hfil Name\hfil&z&\omit\hfil Src\hfil&[10^{-4}{\mathrm{arcm}}^2]&[10^{14}\,\msol]&[{\mathrm{erg}}\,{\mathrm{s}}\mo\,{\mathrm{cm}}^2]&\omit\hfil ID\hfil&[10^{-4}{\mathrm{arcm}}^2]&S/N$_{\mathrm{PSX}}$\cr
\noalign{\vskip 3pt\hrule\vskip 5pt}
PSZ1\,G000.08+45.15&   0.1198&!(1)& 12.35^{+3.43}_{-3.33}& 3.10^{+0.45}_{-0.50}&                      \dots& RXC\,J1516.5$-$0056, A2051&12.43& 4.34\cr
\noalign{\vskip 4pt}
PSZ1\,G000.42$-$41.84& 0.1651&!(1)& 14.05^{+2.78}_{-2.70}& 4.46^{+0.47}_{-0.50}&                      \dots& RXC\,J2104.3$-$4120, A3739&14.35& 6.18\cr
\noalign{\vskip 4pt}
PSZ1\,G000.77$-$35.67& 0.3416&!(1)& *9.14^{+1.98}_{-1.93}& 6.20^{+0.72}_{-0.77}&                      \dots& RXC\,J2031.8$-$4037&*7.89& 4.37\cr
\noalign{\vskip 4pt}
PSZ1\,G001.00+25.71&    \dots&(-1)& \dots&                \dots&    \le !1.35\times10^{-12}&       \omit\hfil\dots\hfil&\dots&\dots\cr
\noalign{\vskip 4pt}
PSZ1\,G002.24$-$68.27& 0.1400&!(2)& *7.43^{+2.71}_{-2.61}& 2.69^{+0.51}_{-0.58}&(1.74\pm0.65)\times10^{-12}& ACO\,S\,1109&\dots&\dots\cr
\noalign{\vskip 4pt}
PSZ1\,G002.77$-$56.16& 0.1411&!(1)& 18.29^{+2.92}_{-2.85}& 4.49^{+0.39}_{-0.41}&                      \dots& RXC\,J2218.6$-$3853, A3856&15.09& 6.56\cr
\noalign{\vskip 4pt}
PSZ1\,G002.80+39.24&   0.1533&!(1)& 26.14^{+4.68}_{-4.53}& 5.91^{+0.57}_{-0.60}&                      \dots& RXC\,J1540.1$-$0318, A2104&22.13& 6.41\cr
\noalign{\vskip 4pt}
PSZ1\,G003.09$-$22.51&  \dots&(-1)& \dots&                \dots&    \le -0.07\times10^{-12}&       \omit\hfil\dots\hfil&\dots&\dots\cr
\noalign{\vskip 5pt\hrule\vskip 3pt}}}
\endPlancktablewide
\endgroup
\end{table*}


\section{Systematic effects on the X-ray versus SZ scaling
  relation}\label{sec:scal_syst} 
Both X-ray and SZ measurements are likely affected by systematic
effects linked to e.g., background estimation and subtraction methods,
calibration issues, etc. One sign of the impact of these effects is
the fact that the slope of the relation between $\YSZ$ flux and
$\YX/D_{\mathrm A}^{2}$ in units of arcmin$^2$ is
$\alpha=0.91\pm0.02$, which is significantly smaller than unity even
after Malmquist bias correction. As this is not the case for the
relation in physical units (Mpc$^2$), the observed slope cannot be due
to a true physical variation in the ratio (e.g., with mass).

SZ fluxes are subject to uncertainties due to systematic differences
between measurement methods. From the comparison between \texttt{PwS}
and \texttt{MMF} 
photometry (Sect.~\ref{sec:dege}), we estimate that the net effect is
typically $0.03$ dex. The effect is independent of SZ flux, thus
cannot explain the shallower than expected slope.

Uncertainties in the X-ray measurements are dominated by temperature
uncertainties due to calibration systematics. We can investigate the
magnitude of these effects by examining the relation between the $\YX$
values obtained with \xmm\ by \citet[][hereafter the PEP XI ESZ-XMM
sample]{planck2011-5.2b} to those obtained with \chandra\ in a study
of 28 clusters from the same sample by \citet{roz12} (hereafter the
ESZ--\chandra\ sample). The \chandra\ values are larger, with a mean
offset of $0.02$ dex. However, there is no significant evidence of
variation with $\YX$, thus X-ray calibration issues again cannot
explain the observed slope.

A further source of uncertainty in X-ray measurements concerns the
X-ray analysis method (e.g., due to background estimation and
subtraction of point sources and substructure). \citet{roz12} noted
the difference between the ratio obtained with ESZ--\chandra\ and
ESZ-XMM samples and suggested that it might be due to \xmm\ data
analysis issues. The PEP XI ESZ-XMM sample was analysed by two
independent methods depending on the cluster extension in the
field-of-view. Sub-sample A consisted of 19 nearby clusters that
extend beyond the \xmm\ field--of--view, and for which direct
background estimates are not possible, while the background for the
remaining 43 objects was estimated using a region external to the
cluster. The ESZ--\chandra\ sample studied by \citet{roz12} consists
mostly sub-sample A objects. While systematic effects due to
background estimation are certainly more important for sub-sample A
than for sub-sample B, these effects cannot fully explain the observed
behaviour of the $\YSZ/\YX$ ratio. Indeed, excluding sub-sample A
clusters, the slope of the $\YSZ$--$\YX/D_{\mathrm A}^{2}$ relation is
$\alpha=0.89\pm0.04$, still significantly smaller than unity. The
origin of the systematic differences between sub-sample A and B
objects is unclear.

The variation of the $\YSZ/\YX$ ratio with flux remains largely
unexplained. It may be due to residual Malmquist bias, in addition to
a complex combination of systematic effects in SZ and X--ray
measurements. For instance, we note that higher flux clusters
correspond to nearby objects that have larger angular sizes. The
background estimate in both X-ray and SZ signals is subject to larger
uncertainty in this case.

The lack of a complete explanation for the observed slope of the
$\YSZ$--$\YX$ relation, and its ultimate correction, has several
implications. Firstly, the shallower slope in units of arcmin$^2$
translates into an over-estimate of the dispersion about the relation
when measured in Mpc$^2$. From the difference in intrinsic scatter
about the relation in both physical and arcmin units, we estimate that
this effect contributes at the level of about $0.01$ dex to the
scatter seen in the physical \YSZYX\ relation.

Secondly, the $\YSZ/\YX$ ratio will depend on the exact sample
definition, via the range of fluxes probed. The observed slope of
$\alpha=0.91\pm0.02$ translates into a variation of about $\pm 0.06$
dex of the $\YSZ/\YX$ ratio over the range of SZ fluxes studied
here. The ESZ--\chandra\ objects studied by \citet{roz12} lie
preferentially at high fluxes, with a median flux two times higher
than the PEP XI-XMM sample. For $\alpha=0.91$, this will translate
into a roughly $0.03$ dex difference in the $\YSZ/\YX$ ratio. The
$\YSZ/\YX$ ratio found by \citet{roz12},
$\log(\YSZ/\YX)=-0.088\pm0.012$, is significantly lower than our value
of $-0.027\pm0.010$. However, it can be explained by a combination of
their sample definition, a neglect of Malmquist bias, and the
aforementioned calibration issues between \xmm\ and \chandra.

In summary, uncertainties on the $\YSZ/\YX$ ratio are dominated by
systematic effects in both X--ray and SZ measurements. This
unfortunately precludes any definitive statement on the magnitude of
the gas clumpiness within $\Rv$. Follow-up of well-defined sub-samples
(e.g., above a given S/N) should help to disentangle biases due to
sample selection and measurement of the different quantities.

\raggedright
\end{document}